\documentclass[10pt]{thesis}
\usepackage{epsfig, float, amsmath,amssymb,xr}
\usepackage{mydefs, deluxetable}
\usepackage{aastex_hack, chapterbib}
\usepackage{aas_macros}
\usepackage{natbib}
\begin{document}

\title{THE ROLE OF MAGNETIC HELICITY IN THE STRUCTURE \\AND HEATING OF THE SUN'S CORONA}
\author{Kalman J. Knizhnik}
\doctorphilosophy
\dissertation 
\copyrightnotice
\degreemonth{June}\degreeyear{2016}
\maketitle

\begin{frontmatter}

\begin{dedication} 
\vspace*{100px}
\begin{center}
\end{center}
\begin{center}
\Large To my parents, Boris and Irina Knizhnik. \newline
\quote{Sometimes when you sacrifice something precious, you're not really losing it. \newline You're just passing it on to someone else. \par --Mitch Albom, The Five People You Meet In Heaven}
\end{center}
\end{dedication}


\begin{acknowledgement}     
Although my name is the only one that appears on the front of this thesis, at its heart, science is a collaborative discipline. For me, this collaboration extends beyond those who have worked with me on a day-to-day basis to conduct research, but includes also those who have influenced my education and career, as well as those who have been there for me as I coped with setbacks. As a result, this thesis would not have been possible without the scientific advice and moral support of a lot of people. I owe tremendous thanks to:  \par

Spiro Antiochos, for taking me on as a second year graduate student who knew virtually nothing about solar physics, MHD, computation, science, or life in general. Spiro is an absolutely brilliant physicist, and beyond just teaching me about physics, he taught me how to think about it and how to approach solving new problems. He taught me how to give a proper research presentation and write a fellowship proposal. He always found time to meet with me to discuss science, and never criticized me for making stupid mistakes. I learned a tremendous amount from him, and I am extremely grateful for his time, patience, and support over the last several years. Conspicuously absent from my acknowledgements is Spiro's cat, who was supposedly capable of solving almost any `basic' problem in solar MHD, yet never wanted to share his results. \par

Rick DeVore, for being a fantastic mentor. Although Rick was not my `official' advisor, in my mind I have been lucky effectively to have had two amazing advisors over the last four years.  This thesis would not have been possible without him, so I consider him my advisor just as much as Spiro. Beyond just being the author of the code I used to perform all of my simulations, Rick took time out of his schedule every single week to meet with me, frequently multiple times, to go over any problems that I had. In addition to a whole lot of science and computation, the main thing I learned from Rick was how to do sanity checks. If something wasn't working for me, Rick would always suggest a test case in which the answer was known, in order for me to test my technique. This proved invaluable in debugging multiple codes, and also taught me always to think about what I expect the answer to look like, and to investigate any discrepancies between my expectations and the outcome. Rick was also absolutely fantastic at carefully reading my writing, providing excellent and prompt feedback that consistenly improved the final result. \par

Julian Krolik, for helping me navigate graduate school while conducting research remotely. I often heard horror stories of students conducting thesis research at a location outside their graduate school. Frequently, these students were left out to dry by their graduate school advisor, who was unresponsive and unhelpful at making sure deadlines were met. I was very fortunate that Julian was none of these things, instead frequently meeting with Spiro and me to discuss my progress. He wrote me a letter of recommendation for my winning NASA Earth and Space Science Fellowship application, and a question he asked at one of our meetings led to two of the chapters in this thesis and two soon-to-be-published papers. \par

Marc Swisdak, who first introduced me to scientific computation and plasma physics, and was unbelievably helpful and remarkably patient as I navigated my research. I can't recall a single time he was unable to talk when I walked into his office, and it is hard to overstate how important this was for my success. He was a tremendous mentor, and he is the main reason I pursued this area of research.\par 

Aleida Higginson, who consistently provided encouragement, support, and feedback, and was always willing to discuss physics with me. She frequently provided useful advice and gave me ideas about my work that allowed me to keep moving forward. I am extremely grateful to have had an officemate and fellow graduate student who was also my close friend. This is especially true since I did not do my research at the Johns Hopkins campus, where I would have been surrounded by other graduate students, but at Goddard surrounded by scientists, which is quite intimidating for graduate students. Having someone to talk to every day and whose company I enjoyed made up for missing out on the `typical' graduate student experience.  \par

The other members of the group here at Goddard: Carrie Black, Silvina Guidoni, Judy Karpen, Sophie Masson, and Peter Wyper, for frequent scientific discussions, and for occasionally dragging me out of my hole and forcing me to interact with other people (thanks to Aleida for this as well). \par 

Holly Gilbert, for bringing me to Goddard after I emailed her out of the blue. I am unbelievably grateful she took a chance on me. \par 

Manuel Luna, for expertly guiding me through my first data driven solar physics project. This was a remarkable accomplishment especially considering he did it remotely. His dedication and passion for science was evident from the start, and it made my transition to solar physics a whole lot easier. It was fascinating work, and I really enjoyed it. \par

Everyone else I interacted with at Goddard, including, but not limited to: Anand Bhatia, Lars Daldorff, Richard Drachman, Jim Klimchuk, Terry Kucera, Yi-Hsin Liu, Karin Muglach, Barbara Thompson, Aaron Tempkin, Pete Schuck and Vadim Uritsky. \par

All of the other students, postdocs, and scientists that I interacted with who were not at Goddard, but nevertheless provided me with insightful comments and helpful discussions over the course of my time as a graduate student: Joel Dahlin, Kevin Dalmasse, Ben Lynch, Alex Russell, Sam Schonfeld, Mark Stakhiv, and Micah Weberg.\par

Panayiotis Tzanavaris, for being an amazing research mentor, inspiring me to continue in science, and helping me apply to graduate school. \par

All of the research advisors I have had over the years, from high school onwards:  Jim Drake, Spiros Boutlokos, Ann Hornschemeier, Mona Kessel, Ron Lepping, Cole Miller, Misha Sitnov and Adam Szabo. They guided me through my many mistakes, and nurtured the interest I had in pursuing science. \par

Carter Hall, who was far and away the best professor I have ever had, who first taught me Electricity and Magnetism and later Quantum Mechanics. Perhaps not coincidentally, I now consider those my best subjects.\par

Eran ``\emph{Did you look in the book??}'' Barnoy, who taught me a lot of physics. The main thing I learned from him was to look in the book. \par

The teachers who first sparked my interest in science: Joshua Rapps, Helen Raucher, and Barney Trams. \par

Those who helped me make it through undergrad: Becca, Ben, Josh, Jen, Wes, Sarah, Matt, and Ed. \par

The Silver Spring crowd: I am so fortunate that I am able to get away after work and over the weekend and hang out with an amazing group of friends: Julia, Joey, Lila, Bin, Yitzchak, Scott, Shira, Shimon, Malka, Amy, Max, Paulina, Ken, Erica, Nomi, Dani, Shoshi, Sivan, Dee, Joe, Ariel and Steve. Thank you all for frequently having me over and feeding me and giving me something to do with my time when I am not doing research. \par

Ami, Nathan, Chaim, JoJo, and Raphael for sticking with me through my awkward years. Shockingly, I was not always as cool as I am now. \par

\emph{My family: you have been with me every step of the way on this journey. Thank you so much for your support and encouragement. I am so lucky to have you all. A special thank you to my parents. You uprooted your lives to give me an opportunity that you never had. This is all possible because of you. Finally, a special thank you to my wife, Jessica, for your love and support, for putting up with my quirks, and for being the real shining star in my life.} \par

\end{acknowledgement}

\listoffigures
\end{frontmatter}

\begin{abstract}        
Two of the most important and interesting features of the solar atmosphere are its hot, smooth coronal loops and the significant concentrations of magnetic shear, known as filament channels, that reside above photospheric polarity inversion lines (PILs; locations where the line-of-sight component of the photospheric magnetic field changes sign). The shear that is inherent in filament channels represents magnetic helicity, a topological quantity describing the amount of linkage in the magnetic field. The smoothness of the coronal loops, on the other hand, indicates an apparent lack of magnetic helicity in the rest of the corona. At the same time, models that attempt to explain the high temperatures observed in these coronal loops require magnetic energy, in the form of twist, to be injected at the photospheric level, after which this energy is converted to heat through the process of magnetic reconnection. In addition to magnetic energy, the twist that is injected at the photospheric level also represents magnetic helicity. Unlike magnetic energy, magnetic helicity is conserved under reconnection, and is consequently expected to accumulate and be observed in the corona. However, filament channels, rather than the coronal loops, are the locations in the corona where magnetic helicity is observed, and it manifests itself in the form of shear, rather than twist. This naturally raises the question: if magnetic helicity needs to be injected to heat coronal loops, why is it only observed in filament channels, with the rest of the corona, and coronal loops in particular, observed to be laminar and smooth? This thesis addresses this question using a series of numerical simulations that explore the effect of helicity injection into the coronal magnetic field. These simulations demonstrate that magnetic helicity is transported throughout the solar corona by magnetic reconnection in such a way that it accumulates above PILs, forming filament channels, and leaving the rest of the corona generally smooth. In the process, it converts magnetic energy into heat throughout the `loop' portion of the corona, accounting for the large observed temperatures. As a result of these simulations, this thesis presents a model for 1) the formation of filament channels in the solar corona and 2) the presence of smooth, hot coronal loops, and shows how the transport of magnetic helicity throughout the solar corona by magnetic reconnection is responsible for both of these phenomena.
\newline\newline
\textbf{Advisors:} Spiro K. Antiochos \& Julian H. Krolik
\newline\newline
\textbf{Thesis Committee:} Spiro Antiochos, James Drake, Thomas Haine, Marc Kamionkowski, Julian Krolik

\end{abstract}

\chapter{Introduction}\label{Intro}


\section{The Solar Atmosphere}
\begin{figure}[!h]
\centering\includegraphics[scale=0.3]{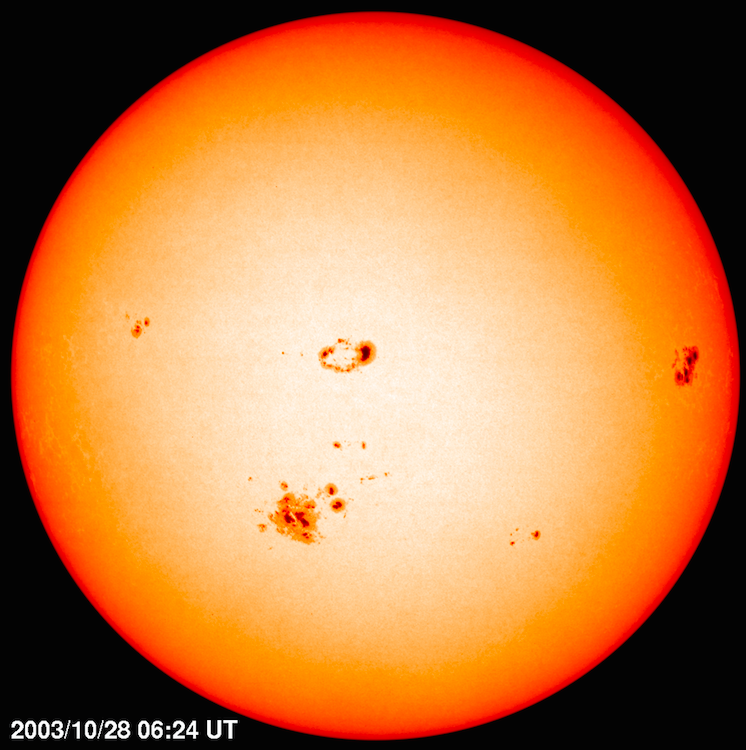}
\caption[The Solar Photosphere]{The solar photosphere as observed by the \emph{Solar and Heliospheric Observatory} (SOHO). The dark spots are regions of intense magnetic fields, known as sunspots.}
\label{fig:photosphere}
\end{figure}

The solar atmosphere is loosely defined as that part of the Sun which can be observed directly through remote observations. It consists of four distinct regions, each possessing different plasma and magnetic field properties. Listed by increasing distance from the center of the Sun, these are the photosphere, chromosphere, transition region, and corona. The photosphere, seen in Fig. \ref{fig:photosphere}, is an extremely thin ($\sim 500 \; \mathrm{km}$), dense ($\sim 10^{-7} \; \mathrm{g\; cm^{-3}}$) layer of plasma with a temperature of approximately $10^4 \; \mathrm{K}$. This is the visible `surface' of the Sun, i.e., it is this layer that we would be seeing if we were (foolishly) to look directly at the Sun with our naked eye. 
\par
\begin{figure*}[!h]
\centering\includegraphics[scale=0.2]{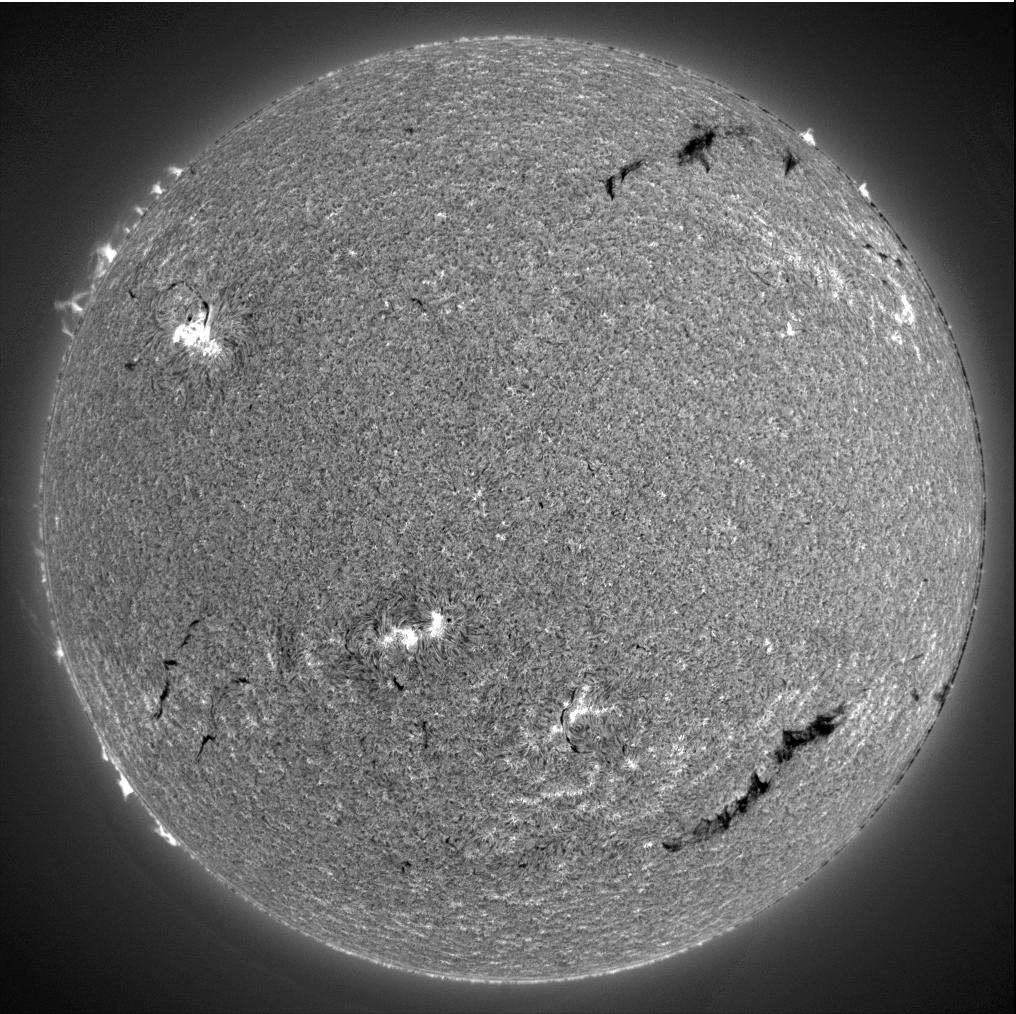}
\centering\includegraphics[scale=0.269]{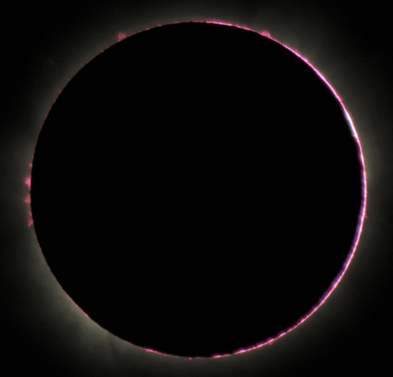}
\caption[The Solar Chromosphere]{Left: The solar chromosphere as observed in H$\alpha$ by the Big Bear Solar Observatory. The dark structures are known as filaments, which will be discussed in \S \ref{sec:prominences}. Right: The pink glow of the chromosphere, seen during a solar eclipse (Alson Wong).}
\label{fig:chromosphere}
\end{figure*}

 Above the photosphere is the chromosphere, seen in Fig. \ref{fig:chromosphere}. It is thicker ($\sim 1500 \; \mathrm{km}$) and rarer ($\sim 10^{-11} \; \mathrm{g\; cm^{-3}}$) than the underlying photosphere, though its temperature is approximately the same. During a solar eclipse, the chromosphere can be seen as a pink glow around the Sun. This glow is caused by emission from electron transitions from the $\mathrm{n=3}$ to the $\mathrm{n=2}$ state of hydrogen (known as Balmer H$\alpha$ emission). 
\par
\begin{figure}[!h]
\centering\includegraphics[scale=0.25]{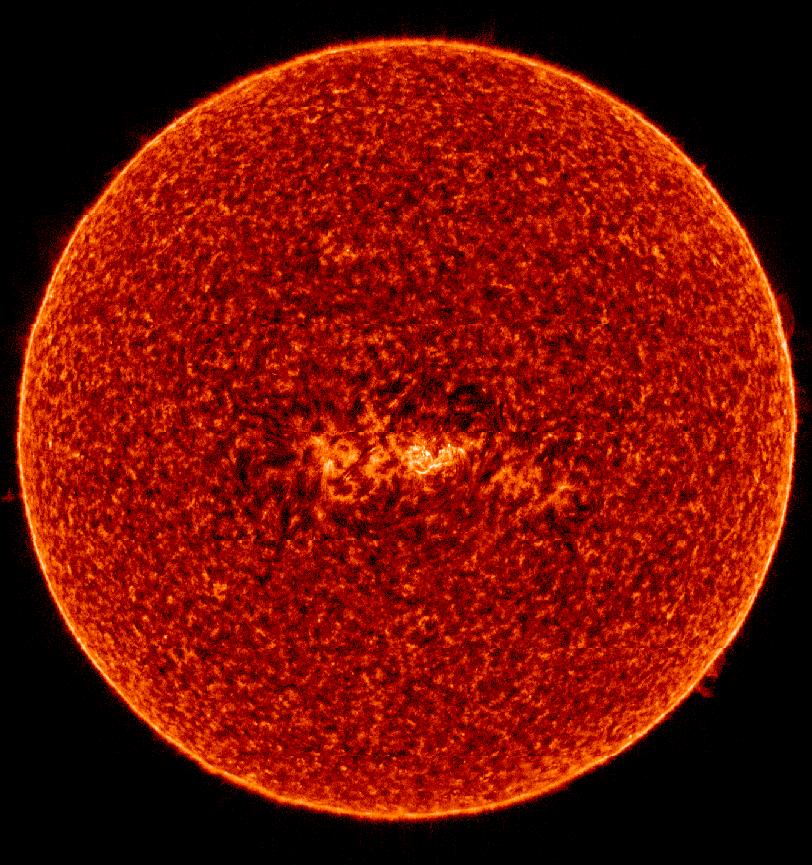}
\caption[The Transition Region]{The transition region as observed by the SUMER instrument on SOHO.}
\label{fig:transition_region}
\end{figure}

 Above the chromosphere is an extremely thin ($\sim 100 \; \mathrm{km}$) transition region, seen in Fig. \ref{fig:transition_region}, over which the temperature increases by four orders of magnitude and the density decreases by an order of magnitude. 
\par
\begin{figure*}[!h]
\centering\includegraphics[scale=0.217]{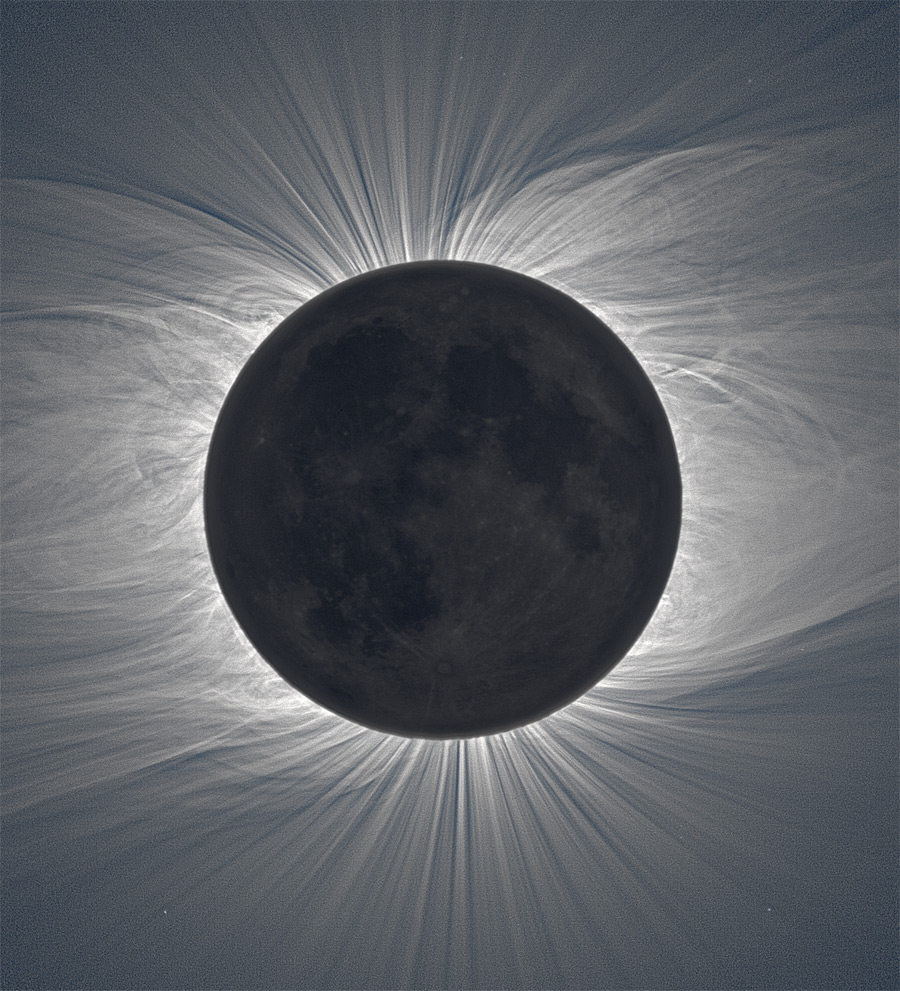}
\centering\includegraphics[scale=0.21]{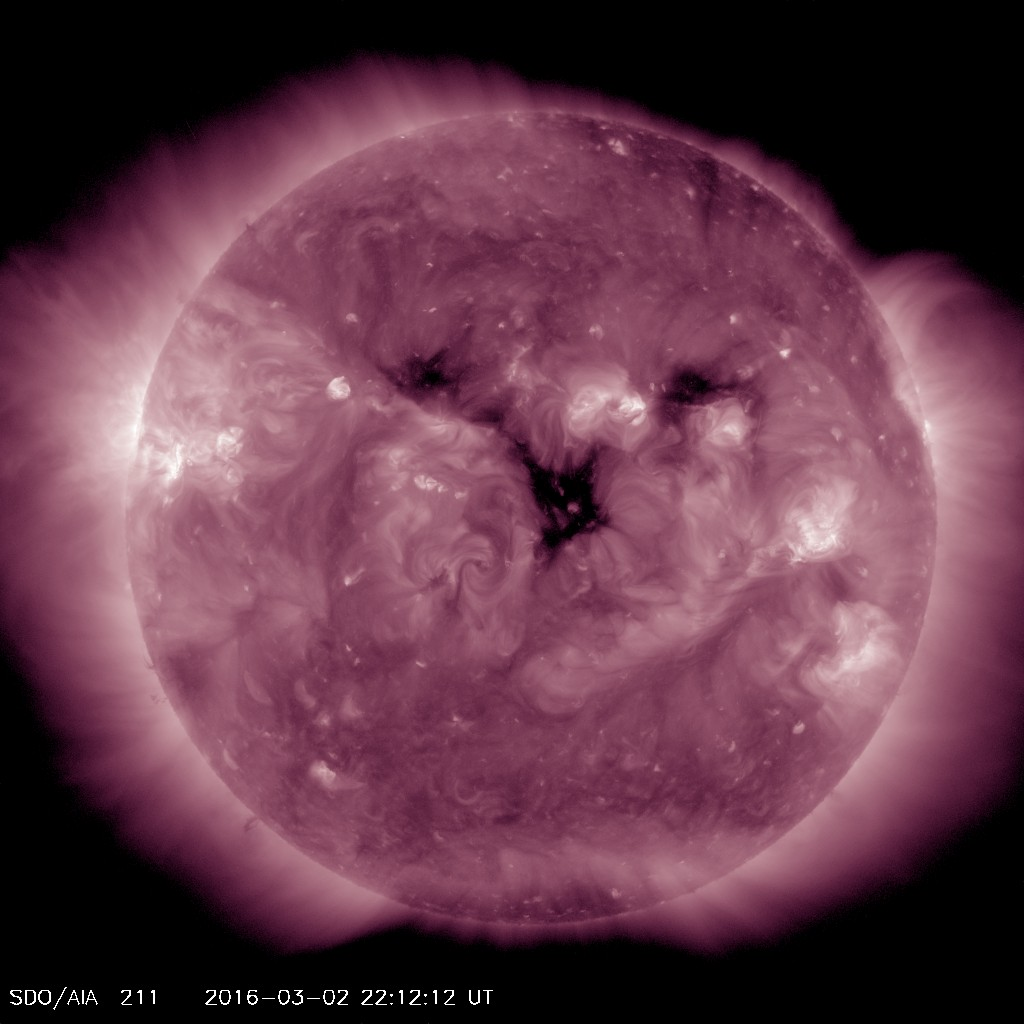}
\caption[Solar Corona]{Left: The solar corona observed during a total eclipse. Straight, open magnetic field lines located at the poles extend out into the heliosphere, while short closed loops are located at lower heights near the equator. Long, extended streamers, which are the Y-shaped structures situated high in the corona represent regions of more complex magnetic topology. All of these features combine to give the corona its dipole-like appearance (Courtesy M. Druckmuller). Right: The solar corona seen in Fe XIV emission by the \emph{Solar Dynamics Observatory} (SDO). Bright spots are regions of strong emission, indicating a higher density, while the dark spots, known as coronal holes, are indicative of low density, open magnetic field regions.}
\label{fig:corona}
\end{figure*}

 The outermost layer of the Sun's atmosphere, the corona (Fig. \ref{fig:corona}), extends from the top of the transition region out into the space and is filled with hot ($>10^6 \; \mathrm{K}$), low density($< 10^{-16} \; \mathrm{g\;cm^{-3}}$) plasma. 
\par
\begin{figure}[!h]
\centering\includegraphics[scale=0.6]{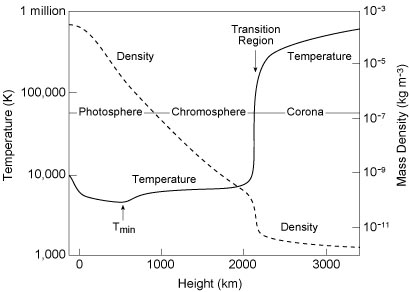}
\caption[Solar Properties vs. Height]{Solar temperature (solid line) and density (dashed line) vs. height from the solar surface. (Figure courtesy E. Avrett)}
\label{fig:SolarProfile}
\end{figure}

Fig. \ref{fig:SolarProfile} shows the variation of the temperature and density throughout the solar atmosphere. The temperature is relatively constant throughout the photosphere and chromosphere, before increasing very rapidly in the transition region to the hot coronal temperatures described above. The density, meanwhile, decreases gradually from the photosphere to the chromosphere, before dropping rapidly from the transition region to the extremely low density corona.\par
\begin{figure*}[!h]
\centering\includegraphics[scale=0.251]{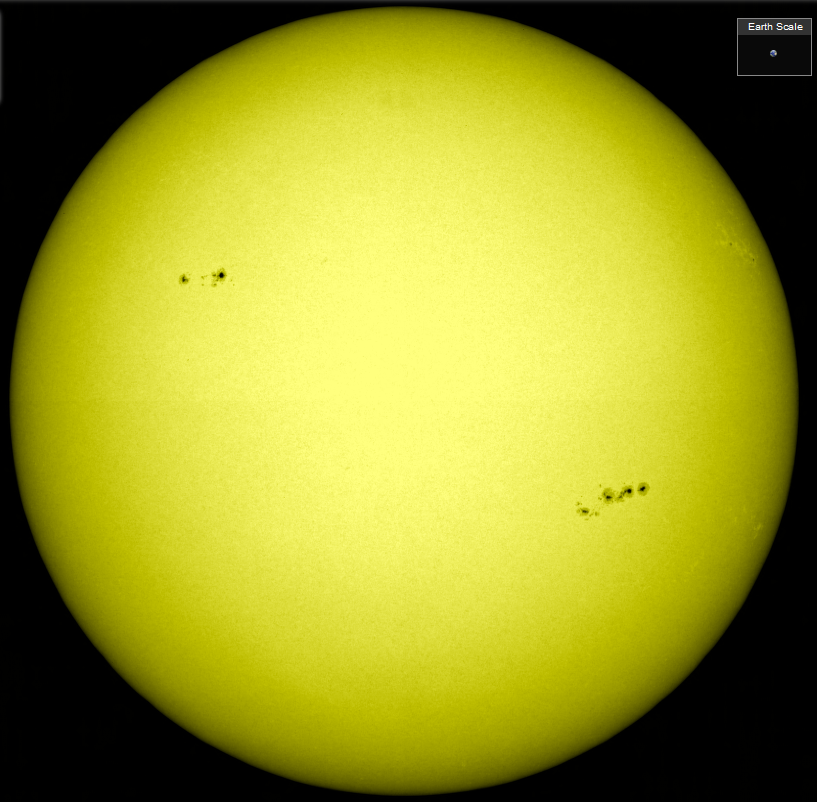}
\centering\includegraphics[scale=0.25]{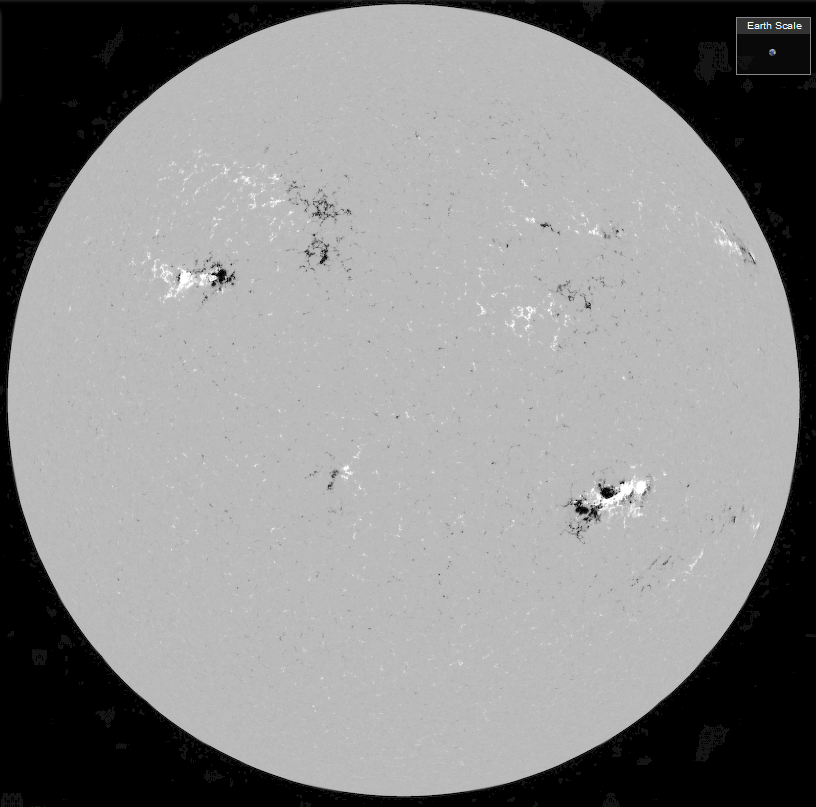}
\caption[Photospheric Image and Magnetogram]{Left: The photosphere, as seen by the AIA instrument on board SDO, showing several sunspots. Right: a photospheric magnetogram, taken by the HMI instrument on board SDO, showing a map of the line-of-sight component of the photospheric magnetic field. Black/white colors show positive/negative polarity. The sunspots on the left are seen as strong magnetic field regions on the right.}
\label{fig:photospheric_magnetogram}
\end{figure*}

The entire solar atmosphere is threaded by magnetic fields that are generated deep inside the Sun by a dynamo process. Once they are created, bundles of magnetic field, known as magnetic flux tubes, rise buoyantly through the inner layers of the Sun before breaking through the photosphere, where especially strong flux tubes are visible as sunspots. At the photospheric level, the magnetic fields are so strong that atomic spectral lines experience Zeeman splitting, allowing measurements of the photospheric magnetic field to be made. Fig. \ref{fig:photospheric_magnetogram} shows an image of the photosphere and the associated magnetogram, a map of the line-of-sight component of the magnetic field, taken by the \emph{Solar Dynamics Observatory} (SDO). Sunspots, regions where the strongest magnetic flux tubes penetrate the photosphere, are visible as dark spots in continuum images (left), and as black and white opposite polarity regions in the photospheric magnetogram (right).  
Once they penetrate the photosphere, magnetic flux tubes expand to fill the entire atmosphere, creating many beautiful structures, some of which will be discussed below. \par
Unlike photospheric magnetic fields, the magnetic field in the rest of the solar atmosphere cannot be measured directly. The exceedingly high temperature of the corona causes thermal broadening of atomic spectral lines to dwarf any line splitting due to the magnetic field. As a result, the coronal magnetic field can only be inferred, indirectly, through observations of enhanced plasma density. As will be described below, in the upper atmosphere plasma is aligned along the magnetic field, allowing regions of enhanced density to be used as tracers of the magnetic field. In Fig. \ref{fig:corona}, the magnetic field of the corona is visible during a total solar eclipse since the enhanced plasma density scatters light from below.\par
While all four regions of the solar atmosphere are interesting and important in their own right, the focus of this thesis will mostly be on the interplay between the photosphere, where energy and magnetic helicity are injected, and the corona, where they are stored. Qualitatively, magnetic helicity describes the amount of linkage, twist, or shear in the magnetic field. It will be given a more formal definition in Chapter \ref{SimsHelicity}. The following sections will briefly describe the relevant observations of the photosphere, describing how energy and magnetic helicity are injected at the photospheric level, and of the corona, describing where energy and magnetic helicity are (and are not) observed.\par 

\section{Photosphere}\label{sec:photosphere}
\begin{figure}[!h]
\centering\includegraphics[scale=0.25]{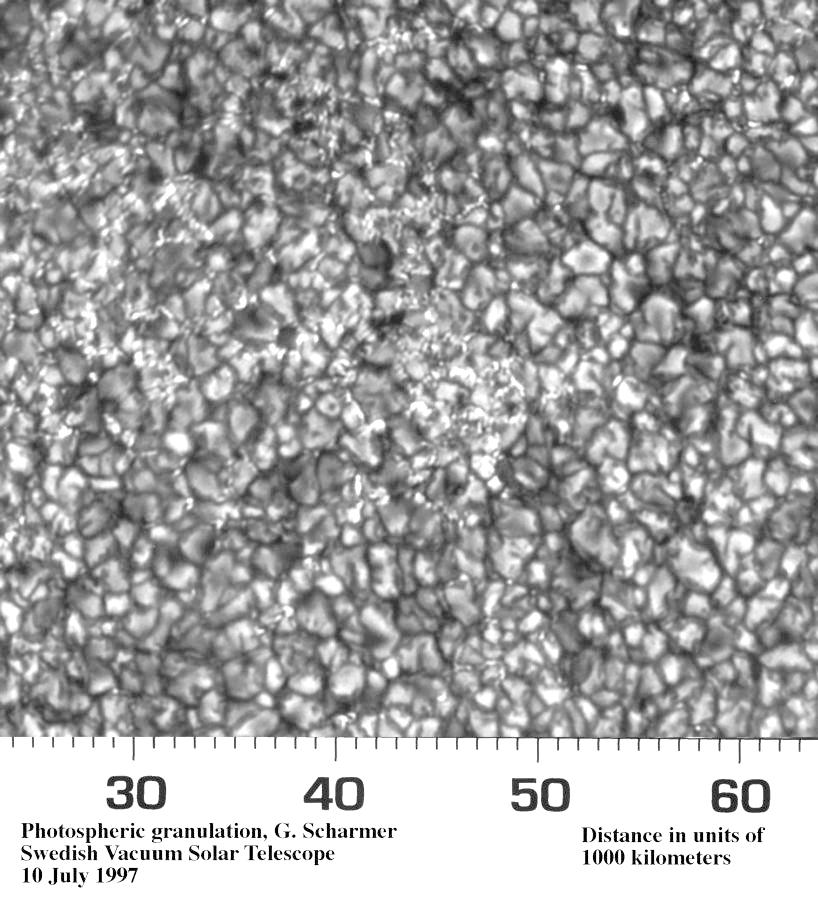}
\caption[Photospheric Convection]{Convection cells on the solar photosphere. Convection cells are split into two types: granules, shown here, and supergranules. Supergranules are typically bigger and have longer lifetimes.}
\label{fig:convection}
\end{figure}

The photosphere is the innermost layer of the Sun's atmosphere. It is characterized by having a plasma $\beta\gg1$ (here $\beta$ is the ratio of gas to magnetic pressure), so that the plasma motion dominates the dynamics, and magnetic fields penetrating the photosphere are jostled around by the plasma motion on the photosphere. The magnetic field itself, in contrast, has very little influence on the motion of the photospheric plasma. This photospheric motion is dominated by convection, so the solar surface, shown in a close-up view in Fig. \ref{fig:convection}, somewhat resembles a boiling pot. Hot plasma is constantly rising, cooling, and falling back down to the surface. The dominant types of convection are called granulation and supergranulation, which differ both in the size of the convective cells and in their lifetimes.
\subsection{Granules}
Granules have a typical diameter of about $1\;\mathrm{Mm}$ and a mean lifetime of $\sim10\;\mathrm{min}$. The velocity of this plasma motion is $\sim 10^5\;\mathrm{cm\;s^{-1}}$ and they cover almost the entire $6\times10^6\;\mathrm{Mm}^2$ photosphere. Granules consist of a central region of upflowing plasma, which then falls radially outward toward the granule boundaries. 
\subsection{Supergranules}
Supergranules are similar to granules, though they are much larger and evolve much more slowly. They have sizes of $\sim 30\;\mathrm{Mm}$ and mean lifetimes of about a day. As in granules, plasma rises in the center of the cell and then falls radially outward, but with a smaller velocity of $\sim 3\times 10^4 \;\mathrm{cm\;s^{-1}}$. \newline\par

The flows that radially diverge from both granules and supergranules meet along boundaries of adjacent convection cells. Since the photosphere is high $\beta$, photospheric convection carries magnetic field lines away from the center of the cells, where the field strength is several hundred $\mathrm{G}$, to the cell boundaries, where the field frequently reaches values of several $\mathrm{kG}$. The net effect of all of this motion is that bundles of magnetic field lines are continually emerging via the convective upflows, and are randomly jostled around by the convective cells. Measurements of the photosphere have decomposed the velocity field into a dominant radial component \citep[e.g.,][]{Schmieder14}, corresponding to the radial outflow from the center of the convection cells, and a smaller vortical component \citep{Brandt88, Duvall00, Gizon03, Komm07, Bonet08, Attie09, Bonet10, VD11, Seligman14, VD15}, corresponding to the result of the interaction of the radial outflows from many convection cells. As will be described in more detail in Chapter \ref{SimsHelicity}, these vortical motions inject both magnetic energy and magnetic helicity into the magnetic field by twisting and braiding field lines. This thesis will explore the consequences of this energy and helicity injection. 
\section{The Solar Corona}\label{sec:corona}
Perhaps the most beautiful portion of the Sun's atmosphere is its hot ($\mathrm{>10^6\; K}$)  corona, the outermost layer of the Sun's atmosphere, which starts at about 3,000 $\mathrm{km}$ above the photosphere. Though it is the hottest part of the solar atmosphere, with a temperature two orders of magnitude higher than the underlying chromosphere, the corona is also the rarest, its particle number density being two orders of magnitude smaller than the underlying chromosphere (which itself is four orders of magnitude rarer than the photosphere). Since the density is so much smaller than that of the photosphere, the corona is extremely faint in visible wavelengths relative to the solar disk. As a result, it is most easily observed during a total solar eclipse, such as the one seen in Fig. \ref{fig:corona}.  \par

The solar corona is a magnetic field dominated environment, with $\beta\ll 1$. While in the photosphere the plasma drags the magnetic field around with it wherever it goes, in the corona it is the magnetic field that tells the plasma what to do. However, since the magnetic field in the photosphere is ultimately connected to the magnetic field in the corona, this creates an interesting dynamical interaction between these two regions of the solar atmosphere. This struggle between photospheric plasma and coronal magnetic fields creates some of the Sun's most beautiful structures, but also generates some of its most violent behavior. This will be described in sections \S \ref{sec:prominences}, \S \ref{sec:loops}, and \S \ref{sec:motivation}. \par
Detailed observations of the Sun's corona by both ground- and space-based observatories have exhibited two main types of structures. High-resolution EUV and X-ray images of the closed-field corona, such as those taken by the \emph{Transition Region and Coronal Explorer} (TRACE), reveal a collection of smooth inverted-U shaped loops, known as coronal loops \citep{Schrijver99}. At the same time, observations also show dark, dense structures, known as prominences or filaments \citep{Vourlidas10}. Both of these structures will be discussed below.

\subsection{Filaments and Prominences}\label{sec:prominences}
\begin{figure}[!h]
\centering\includegraphics[scale=1.0]{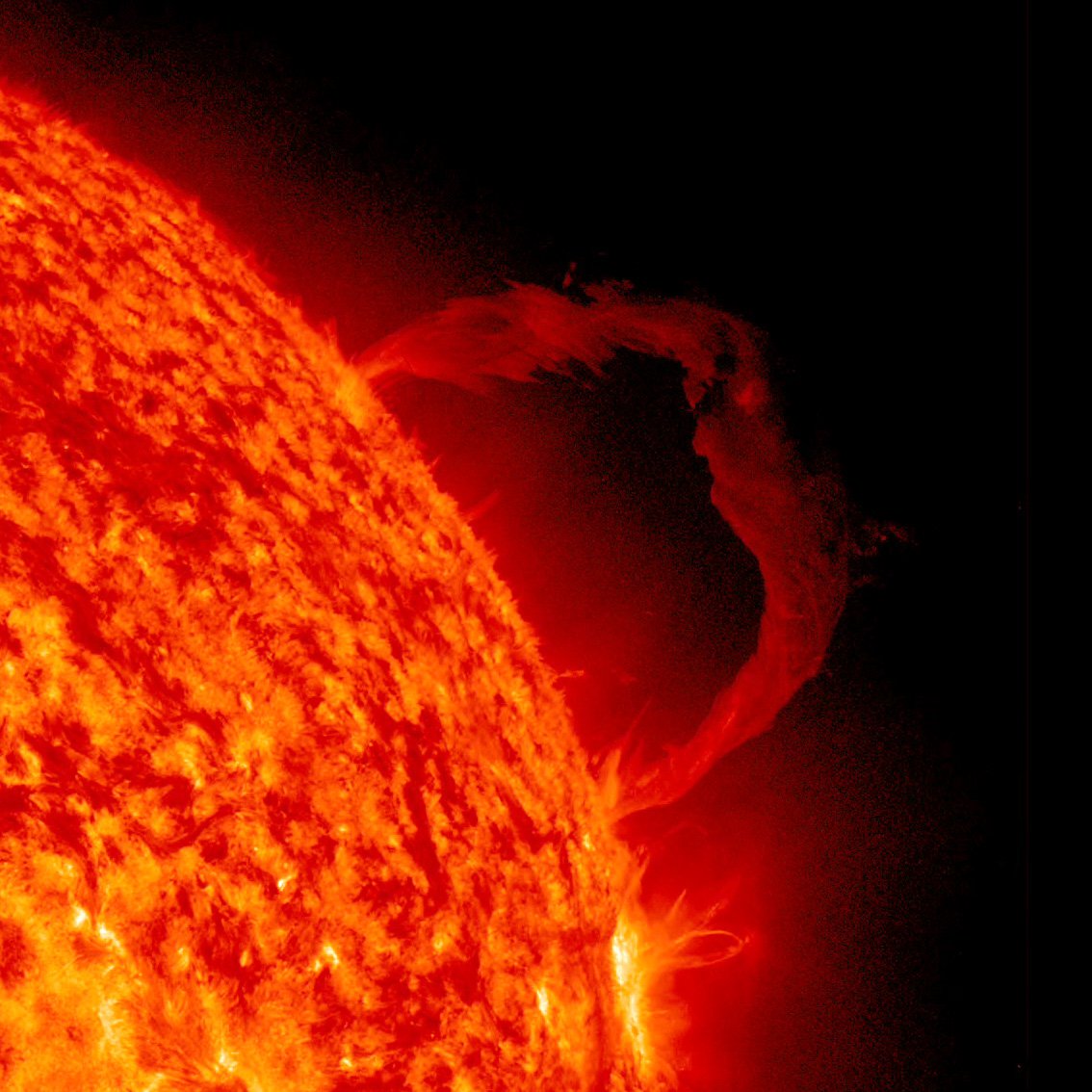}
\caption[Solar Prominence]{A solar prominence as seen by SDO in the 304 \AA\;wavelength. }
\label{fig:prominence}
\end{figure}

\emph{Solar and Heliospheric Observatory} (SOHO) and SDO observations of the solar corona reveal giant structures, known as prominences, suspended in the corona. As seen in Fig. \ref{fig:prominence}, these are long, thin structures, with length of order several hundred $\mathrm{Mm}$, heights of order tens of $\mathrm{Mm}$ and width of several $\mathrm{Mm}$. They have temperatures a factor of 100 lower than the surrounding plasma, and densities a factor of 100 higher. Prominences are sometimes called filaments, with the former name being given to structures seen against the background of the sky, and the latter name given to structures seen against the background of the solar disk. These names will be used interchangeably throughout this work. \par
\begin{figure}[!h]
\centering\includegraphics[scale=0.3]{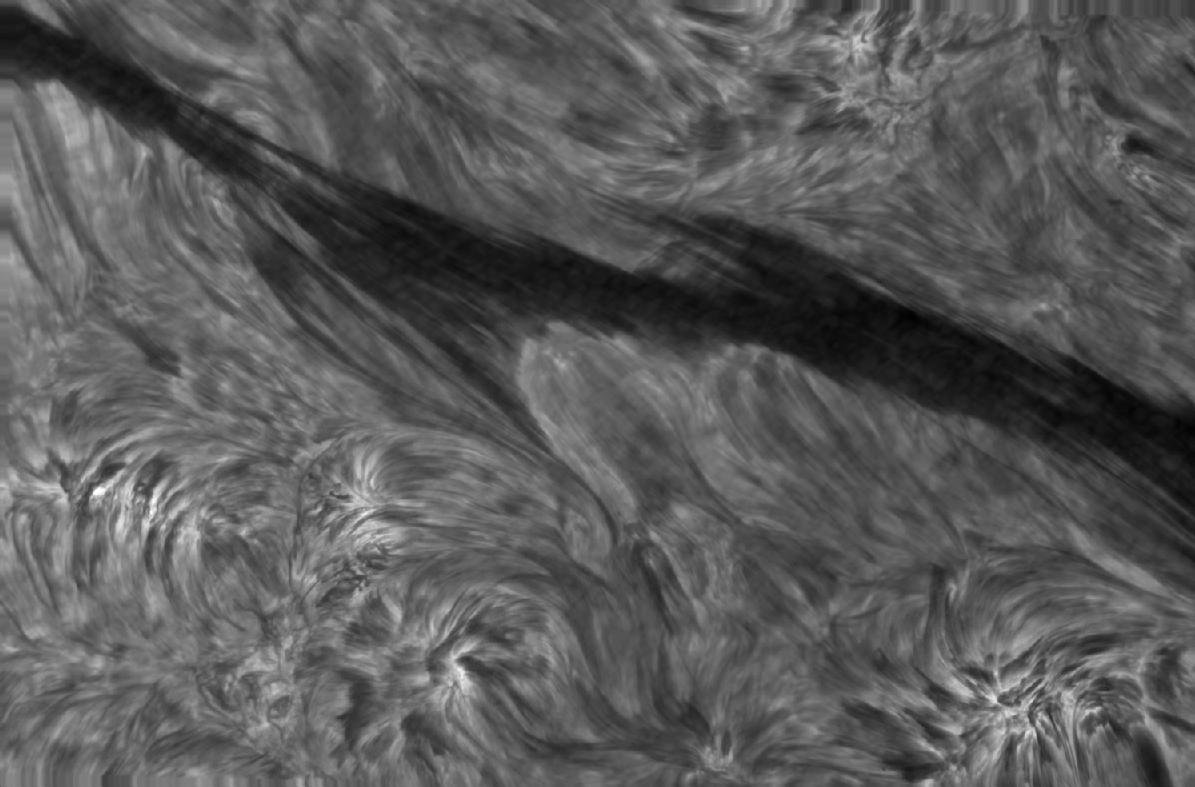}
\caption[Filament seen in H$\alpha$]{A solar filament observed in H$\alpha$ by the Swedish Vacuum Solar Telescope. This filament is dextral, since it has right-leaning barbs.}
\label{fig:Ha_filament}
\end{figure}

Filaments are comprised of two components: the spine, which runs along its axis, and barbs, or legs, which extend down to the chromosphere and look like huge tree trunks. Occasionally, filaments will have more than the two barbs seen in Fig. \ref{fig:prominence}. When observed on the disk in H$\alpha$, as in Fig. \ref{fig:Ha_filament}, filaments are either sinistral (left-handed) or dextral (right-handed). Dextral filaments have barbs that protrude out to the sides in the direction of American traffic, whereas sinistral filaments have barbs that protrude out to the sides in the direction of British traffic. Dextral filaments have been found to dominate in the northern hemisphere, while sinistral filaments dominate the southern hemisphere \citep{Martin94}.\par


Prominences and filaments can be further subdivided into quiescent- and active-region prominences. Quiescent prominences have mainly horizontal magnetic fields, with a magnitude of about $10 \mathrm{G}$ \citep{Leroy89}. Active-region prominences have footpoints rooted in regions of strong magnetic activity, and are generally smaller in size than quiescent ones, though they contain much stronger magnetic fields, sometimes of order $100 \mathrm{G}$ \citep{Kuckein09}. \par
\begin{figure*}[!h]
\centering\includegraphics[scale=1.30]{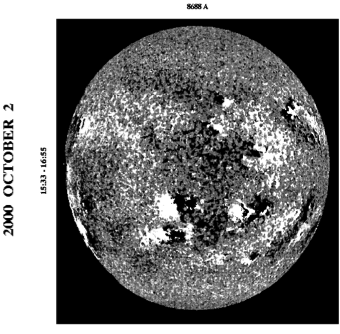}
\centering\includegraphics[scale=1.30]{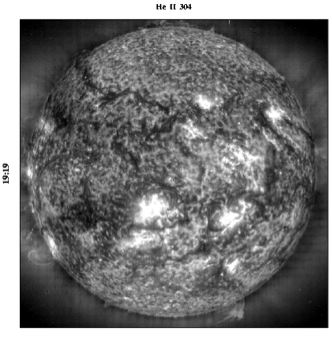}
\caption[Magnetogram-EUV comparison]{Shown above is a Kitt Peak line-of-sight magnetogram, in which magnetic fields of opposite polarity are shown in black and white. A SOHO UV image, showing the locations of numerous filaments (dark, against the disk) and prominences (bright, above the limb). These structures can be seen to lie along lines separating regions of opposite magnetic polarity, known as polarity inversion lines (PILs)}
\label{fig:comparison_intro}
\end{figure*}

 
Comparison of photospheric magnetograms and coronal images indicates that filaments are invariably located above photospheric polarity inversion lines (PILs). Fig. \ref{fig:comparison_intro} shows a photospheric magnetogram side by side with a SOHO EUV image of the solar atmosphere. 
PILs are the lines separating the opposite (black and white) line-of-sight magnetic polarities in the left panel. Filaments are the dark structures visible in the right panel. Careful comparison of the two images reveals that filaments are oriented almost precisely along the PILs. \citet{Leroy83} and \citet{Tandberg-Hanssen95} found that filament spines form an angle of $\sim 25^\circ$ , on average, with respect to the PIL. The angle between the footpoints of the magnetic field and the PIL is known as the shear angle. Shear is defined as a misalignment between the line connecting the footpoints of a magnetic field line on the photosphere and the line perpendicular to the PIL. The importance of shear will be described more fully in Chapters \ref{SimsHelicity} and \ref{Sunspot}. \par

\subsection{Filament Channels}
The plasma that comprises prominences and filaments cannot be present in the corona without a source of support against gravity. Since the corona is a low $\beta$ environment, the magnetic field is able to provide the upward force to prevent plasma from falling down, and the magnetic structure that supports the plasma is known as a filament channel. For a filament to form, the filament channel must first be present. If plasma is present in the filament channel, it may be visible as a filament. Thus, filament channels can exist without filaments, but not vice versa. The structure of filament channels can be inferred from high resolution images of filaments themselves, such as the one shown in Fig. \ref{fig:Ha_filament}. 

\begin{figure}[!h]
\centering\includegraphics[scale=0.5]{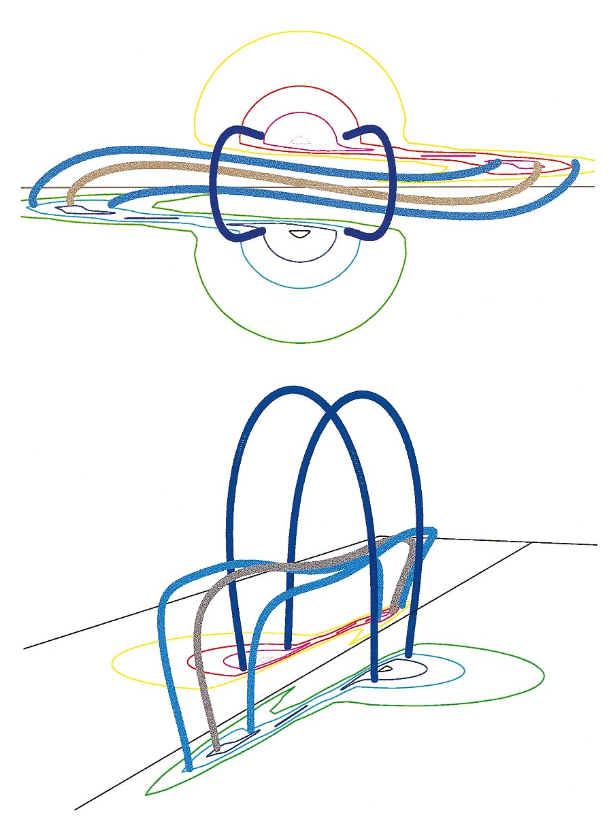}
\caption[Filament channel structure]{Overhead (top) and perspective (bottom) views of a sample of representive field lines comprising the filament channel. Color contours on the bottom surface represent lines of constant normal magnetic field. The light blue and grey/gold field lines are highly sheared with respect to the PIL (straight black line separating the two polarities). They have a slight dip in the middle, which is where filament plasma is expected to accumulate. Dark blue field lines are overlying, quasi-potential loops that constrain the underlying filament channel. Figure taken from \citet{DeVore00}.}
\label{fig:filchan_model}
\end{figure}

The filament, and therefore the filament channel, is characterized by an incredibly smooth, laminar structure, with little, if any, twist. However, the magnetic field must have an upward curvature, which is not seen from a top-down view, in order to provide the necessary upward force, in the form of a magnetic tension, to support the filament plasma. Thus, the filament channel is expected to look something like the model shown in Fig. \ref{fig:filchan_model}. In this model, a set of relatively low lying field lines that are highly sheared with respect to the PIL, known as a sheared arcade, is constrained from above by overlying quasi-potential loops (i.e., loops that are very close to their lowest energy - potential - state), which are concave down and provide a tension force on the underlying field. The field lines comprising the sheared arcade develop a dip due to this tension force, and this dip cradles the prominence plasma that accumulates in the filament channel.\par
\begin{figure}[!h]
\centering\includegraphics[scale=0.7, trim=0.0cm 1.75cm 0.0cm 0.0cm,clip=true]{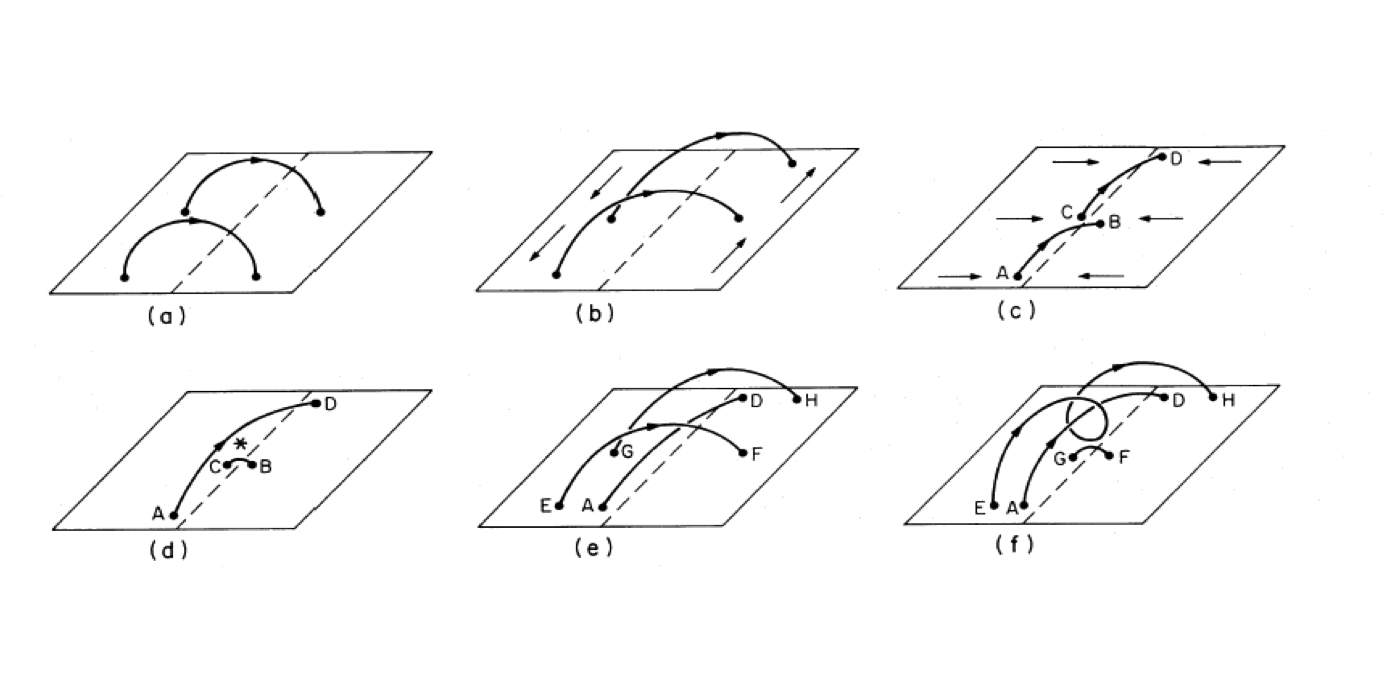}
\caption[Flux cancellation model of filament channel formation]{Flux cancellation model of filament channel formation. Field lines are represented as solid lines, and the PIL is the dashed line separating their footpoints. Shearing and converging flows on the photosphere drive oppositely directed magnetic field lines together, forming a twisted, sheared structure. Figure taken from \citet{VanB89}.}
\label{fig:flux_cancellation}
\end{figure}

While the general structure of filament channels is taken to be a set of field lines arranged in a sheared arcade, the question of how the field lines become sheared in the first place is an open one. Until recently, there were two main models for the formation of filament channels. The first model, known as flux cancellation \citep[e.g.,][]{VanB89}, is shown in Fig. \ref{fig:flux_cancellation}.
Initially potential field lines (a) are sheared in opposite directions on either side of the PIL by motions on the photosphere (b). Converging flows (c) push the footpoints of the field lines towards the PIL. Field lines AB and CD come into contact, and the oppositely directed magnetic fields permit reconnection to change the topology of the magnetic field lines, creating two new field lines AD and BC (d). Field line BC soon submerges due to its short length and downward curvature, but two trailing field lines, EF and GH, go through the same procedure (e), which results in a short field line, GF, that soon submerges, and a twisted field line EH (f). \par
While the flux cancellation model creates a dip in the magnetic field which can provide an upward tension force, the resulting filament channel is highly twisted, and looks nothing like the structure observed in high resolution images such as Fig. \ref{fig:Ha_filament}. It is unlikely, therefore, that the flux cancellation model can explain the formation of filament channels.\par
\begin{figure*}[!h]
\centering\includegraphics[scale=0.8]{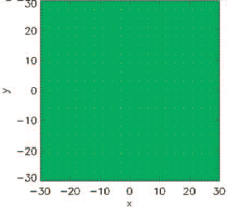}
\centering\includegraphics[scale=0.74]{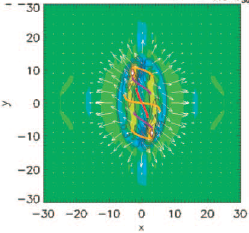}
\centering\includegraphics[scale=0.82]{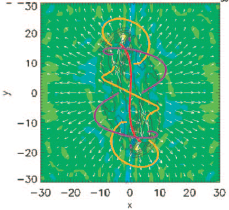}
\caption[Flux emergence model of filament channel formation]{Top-down view of a sequence of frames from the flux emergence simulation of \citet{Magara03}. An initially submerged flux rope (left) slowly rises through the photosphere (middle) into the corona (right). Color contours on the bottom surface represent horizontal velocity magnitude. Yellow and maroon field lines represent overlying potential field lines, while the red field line represents the highly sheared filament channel.}
\label{fig:flux_emergence}
\end{figure*}

A second model, flux emergence \citep[e.g.,][]{Fan01,Manchester01,Magara03}, is shown in Fig. \ref{fig:flux_emergence}.
In the flux emergence model, a subphotospheric flux rope, i.e., a highly twisted set of field lines, emerges through the photosphere into the corona. Rather than manifesting itself as a twisted structure, the flux rope appears in the corona as a highly sheared arcade, with the twist component appearing as the concave-down overlying potential loops, and the axial component as the highly sheared filament channel.\par
Although the flux emergence model produces a structure that is consistent with observations of filaments, filament channels are frequently observed to form in quiet regions where there is no flux emergence occurring \citep{Mackay10}. So, while flux emergence may be responsible for the formation of some filament channels, it cannot be responsible for all of them.\par
It appears, therefore, that both models of filament channel formation are incompatible with observations, since neither model is able to adequately explain how filament channels acquire their shear. As mentioned above, shear is a form of magnetic helicity, meaning that filament channels are structures with large amounts of magnetic helicity. This thesis summarizes and rigorously tests a new model of filament channel formation, magnetic helicity condensation \citep{Antiochos13}, that describes how they obtain this magnetic helicity.

\subsection{Coronal Loops}\label{sec:loops}
\begin{figure*}[!h]
\centering\includegraphics[scale=0.46]{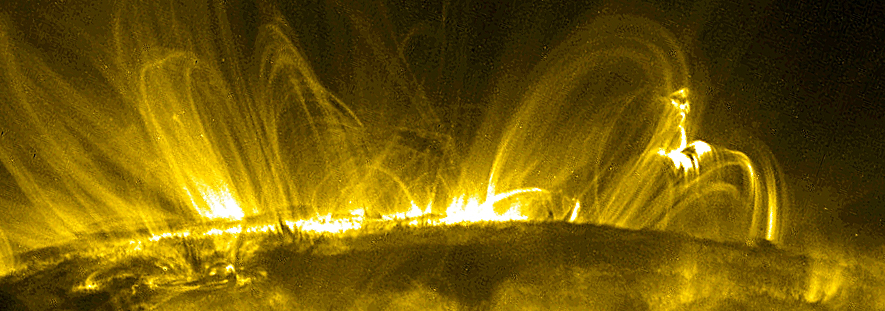}
\centering\includegraphics[scale=0.772]{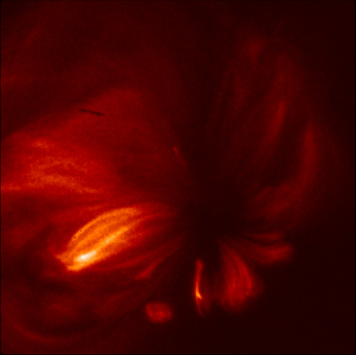}
\centering\includegraphics[scale=0.555]{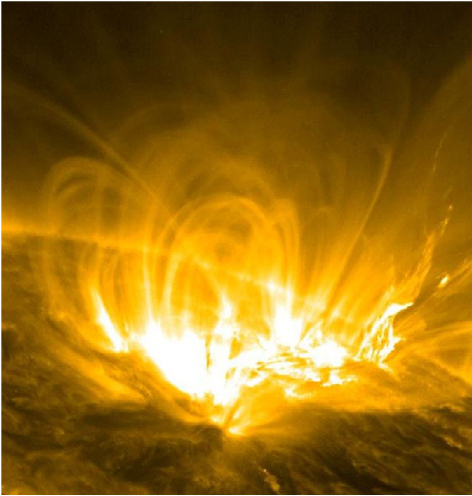}
\caption[Coronal Loops]{\emph{TRACE} EUV (top), \emph{Hinode} X-ray (left), and SDO EUV (right) images of various smooth coronal loops.}
\label{fig:loops}
\end{figure*}

Coronal loops, shown in Fig. \ref{fig:loops}, are hot, quasi-potential structures that can be detected in X-ray, ultraviolet, and even white light wavelengths when they occasionally rapidly brighten and gradually dim due to an impulsive heating event. They have heights of up to $\mathrm{10^5 \; km}$, temperatures of anywhere from $\mathrm{10^6 \; K}$ (X-ray loops) up to $\mathrm{10^7 \; K}$ (flaring loops) and densities of about $\mathrm{10^{9.5} \; cm^{-3}}$ in X-ray loops to $\mathrm{10^{10.5} \; cm^{-3}}$ in flaring loops. The strong anisotropy in the thermal conductivity of the magnetized solar corona means that thermal energy is conducted mostly along magnetic field lines, making coronal loops excellent tracers of the coronal magnetic field. Coronal loops are striking for their apparent smoothness. There is no discernable twist observed over the entire length of the loop, so coronal loops are frequently considered to have very little, if any, helicity. \par
In addition to their importance as tracers of the coronal magnetic field, hot coronal loops are manifestations of the so-called coronal heating problem. Spontaneous brightenings of the coronal loops are indicative of an impulsive event that heats the plasma at the footpoints of the coronal loop, causing the plasma on the loops to glow. The famous coronal heating problem attempts to explain what these impulsive events are that heat these loops to their $\mathrm{10^6\;K}$ temperatures, two orders of magnitude hotter than the underlying photosphere. Almost all models of coronal heating argue that the energy for heating comes from the motion of convective cells, such as those described in section \S \ref{sec:photosphere}, at the footpoints of coronal loops. \par
\begin{figure}[!h]
\centering\includegraphics[scale=0.7]{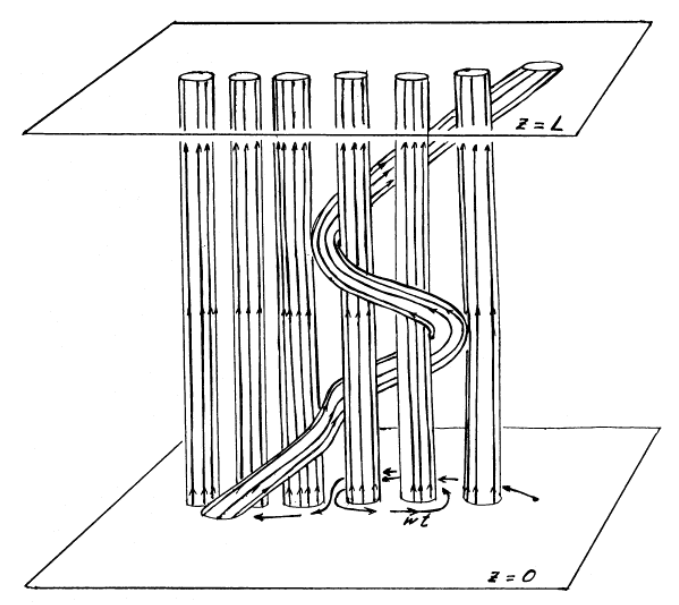}
\caption[Parker's model of coronal heating]{A plane-parallel Parker corona, in which magnetic fields are tangled in a complicated manner, is highly susceptible to the formation of numerous current sheets, the dissipation of which leads to coronal heating. Figure taken from \citet{Parker83}.}
\label{fig:braiding}
\end{figure}

 
Heating models are classified into two types: DC and AC heating models. In DC heating models, the slow shuffling of the footpoints of coronal loops gradually injects energy  into the coronal magnetic field, leading to a complex magnetic topology. As the stress builds up, current sheets form between adjacent magnetic field strands. Recent high resolution observations have placed an upper limit on the diameter of these strands of less than $\mathrm{15\;km}$ \citep{Peter13}. As stress continues to build up, the current sheets thin, allowing magnetic reconnection to occur, a process that is known to convert magnetic energy into thermal \citep[e.g.][]{Parker83} and kinetic energy of both ions \citep{Knizhnik11} and electrons \citep{Dahlin15}. \citet{Parker83}, in particular, modelled the coronal magnetic field as a uniform field between two parallel plates, each of which represented the photosphere (Fig. \ref{fig:braiding}). In Parker's model, the apex of coronal loops was in the middle of the domain, between the two plates. Photospheric motions then twisted and tangled the coronal magnetic field, forming and dissipating numerous current sheets, leading to significant heating via Ohmic dissipation. In these models, therefore, it is the slow buildup of magnetic energy from footpoint motions, followed by the conversion of magnetic energy into heat through magnetic reconnection that is responsible for impulsively heating the corona.\par

\begin{figure}[!h]
\centering\includegraphics[scale=1.0]{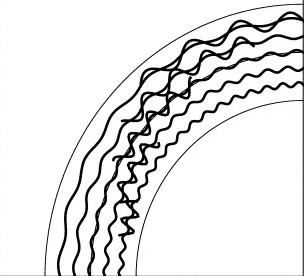}
\caption[Wave model of coronal heating]{In wave models of coronal heating, waves generated by fast photospheric motions deposit their energy along coronal loops due to the temperature and density gradient along the loops. Figure taken from \citet{Aschwanden01}.}
\label{fig:waveheating}
\end{figure}

In AC (also called wave) heating models, fast motions on the photosphere generate many types of waves, such as the Alfv\'en \citep{Osterbrock61} or magnetosonic \citep{Pekunlu01} modes. Wave energy is then transfered into coronal heating via damping or absorption of these waves as they travel up from the photosphere through strong temperature and density gradients along the coronal part of the loop (Fig. \ref{fig:waveheating}).\par
Both DC and AC models are frequently able to reproduce various observed heating diagnostics \citep[e.g.][]{Kumar06, Viall11, Bradshaw12, Hahn14}, but, to date, there has been little consideration of the topological implications of either type of model. Regardless of whether the photospheric motions are fast or slow, they rely on magnetic energy, in the form of twist, to be injected into the coronal magnetic field. However, twist, a form of magnetic helicity, is not observed in coronal loops (c.f. Fig. \ref{fig:loops}). Since magnetic helicity is a conserved quantity in the solar corona (the arguments of \citet{Taylor74} and \citet{Berger84b} will be summarized in Chapter \ref{SimsHelicity}), it is surprising that it is not observed in the coronal loops. Addressing this problem will be the goal of Chapters \ref{Structure} and \ref{Heating}.
\section{Research Questions}\label{sec:researchquestions}
The observations described above paint a picture of the Sun's corona in which some structures, hot coronal loops, have little if any magnetic helicity, while other structures, filament channels, contain a tremendous amount of magnetic helicity. At the same time, the observations of the photosphere indicate that helicity is constantly being injected by convective cells, along with a large amount of energy that goes into heating the coronal loops. In other words, both energy and helicity are injected throughout the photosphere, but only energy is observed in the entire corona, in the form of hot coronal loops; the helicity, meanwhile, is concentrated only at specific locations, in the form of sheared filament channels. These observations motivate the question that we seek to answer in this thesis: If energy and helicity are both injected by convective cells at the photospheric level, why is magnetic helicity only observed in filament channels and not in coronal loops, while the energy, in the form of heating, is observed everywhere in the corona? To answer this question, we address each aspect separately. In particular, we address the questions: 

\begin{enumerate}
  \item How do filament channels obtain their helicity? 
  \item Why do coronal loops lack helicity, appearing unsheared and quasipotential?
  \item What process converts energy into heat throughout the corona? 
\end{enumerate}
Answering these questions is the focus of this thesis.

\section{Motivation}\label{sec:motivation}
Filament channels are extremely important objects to understand. Since filament channels are highly sheared structures, they contain a tremendous amount of both magnetic energy and magnetic helicity. Under certain conditions that are still not fully understood, they can release this energy and helicity by erupting in an explosive event known as a coronal mass ejection (CME). The release of magnetic energy and helicity by the filament channel is accompanied by the release of the plasma that was supported by the filament channel. This plasma, comprised of charged particles, is expelled into space and has the potential to damage spacecraft, including both GPS and defense satellites, harm astronauts, and destroy power grids. Studying filament channels, as well as their manifestations in the solar atmosphere, filaments and prominences, and specifically their formation and eruption mechanisms, is vital to mankind's safety, security, and livelihood. \par
\begin{figure}[!h]
\centering\includegraphics[scale=0.4]{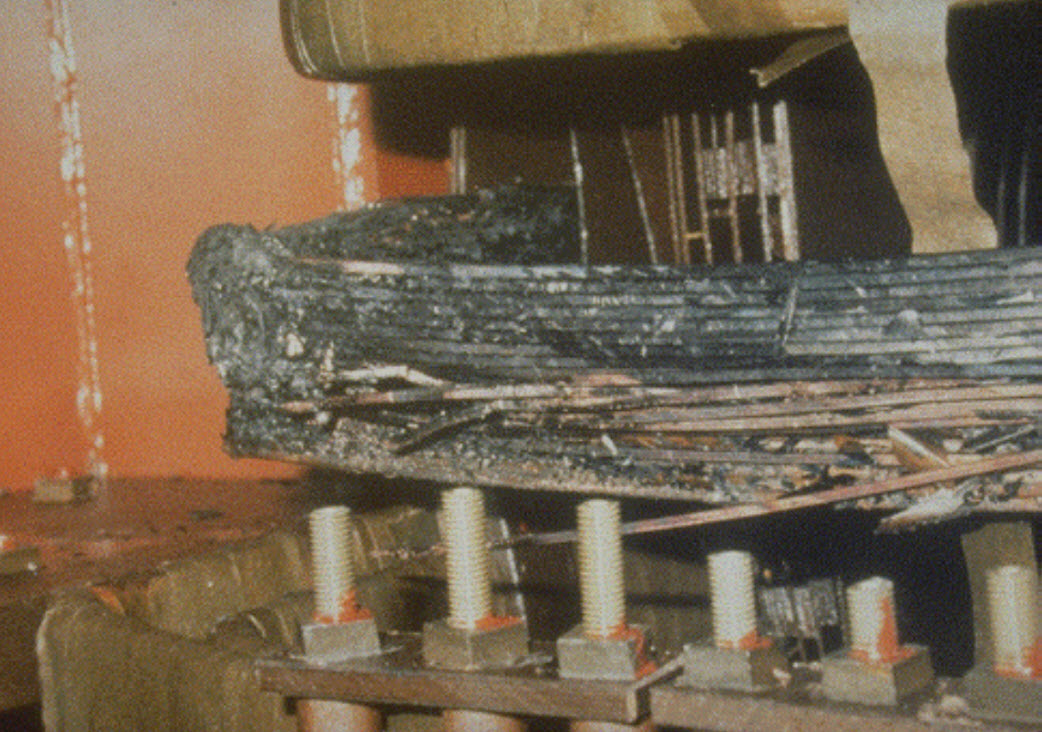}
\caption[Transformer damaged by a solar storm]{The March 1989 geomagnetic storm caused damage to power grids from Quebec, Canda all the way to New Jersey, USA. Shown here is a power transformer at the Salem Nuclear Generating Station in New Jersey run by Public Service Electric and Gas. Image courtesy of Peter Balma and PSE\&G.}
\label{fig:transformer}
\end{figure}

The explosive behavior exhibited by prominences was first realized after the solar storm known as the Carrington Event. On September 1, 1859, a CME caused a geomagnetic disturbance that caused auroras that in the Northern hemisphere reached as far south as the Caribbean, causing telegraph systems all over the world to fail. It was also experienced first-hand by the citizens in and around Montreal, Quebec, in March of 1989, when a massive solar storm destroyed the entire Quebec power grid, and even affected parts of the United States' power supply (Fig. \ref{fig:transformer}) leaving over $6$ million people without power for over $9$ hours. In the 21\textsuperscript{st} century, as society has become more and more technologically dependent, the stakes are even higher, and GPS, communications, and defense satellites are still constantly at risk of damage from solar storms. Understanding and predicting such events is the ultimate goal of most of solar physics. Incredible effort is spent, therefore, on observing and studying the coronal magnetic field, where these spectacular events originate.\par
Being able to predict when prominences and filaments erupt starts with understanding filament channels, the magnetic structure that supports the dense plasma suspended in the corona. In order to make accurate predictions of filament eruptions, therefore, it is important to understand how filament channels obtain their shear.

\section{Thesis Outline}\label{sec:thesisoutline}
This thesis is organized as follows. In Chapter \ref{SimsHelicity}, the question of filament channel formation will be addressed. We will present a model, magnetic helicity condensation \citep{Antiochos13}, that explains the formation of filament channels via the transport of magnetic helicity throughout the solar corona by magnetic reconnection.  Using helicity conserving numerical simulations of helicity injection into a model corona, we demonstrate that as a result of magnetic reconnection occurring throughout the corona, magnetic helicity accumulates above photospheric PILs, forming filament channels, leaving almost no helicity in coronal loops. We derive analytical formulae for the formation rate of filament channels and show that simulation, theory and observations are in excellent agreement. In Chapter \ref{Structure} the lack of helicity in coronal loops will be explored, as will the role of helicity preference in determining the formation rate of filament channels and the amount of structure in coronal loops. Using a set of helicity conserving numerical simulations that have varying helicity injection preferences, we study how magnetic helicity is transported throughout the corona when helicity of opposite signs is injected. We will show that the filament channel formation rate is proportional to the helicity injection rate, and argue that for helicity preferences in the range observed on the Sun, filament channels will form in about a day or so, in excellent agreement with observations. Meanwhile, the amount of structure in the `loop' portion of the corona, is weakly dependent on helicity preference. In Chapter \ref{Sunspot}, we extend the scope of the results presented in Chapters \ref{SimsHelicity} and \ref{Structure} by making our simulation configuration much more representative of the real corona. We will introduce a real PIL into the simulation, and demonstrate the formation of sheared filament channels at the PIL, in agreement with the helicity condensation model, our previous results, and most importantly, observations. In Chapter \ref{Heating} the question of how coronal loops are heated in the magnetic helicity condensation process is studied using energy and helicity conserving numerical simulations. We will demonstrate that the magnetic reconnection that is responsible for transporting magnetic helicity throughout the corona converts magnetic energy into heat in the process. Since this magnetic reconnection occurs everywhere throughout the corona, the result is that coronal loops are heated relatively uniformly throughout the majority of the corona. We will quantitatively compare the results of our simulations with observations and find that the heat flux resulting from this process is more than sufficient to account for the observed heating. Furthermore, we will explain why our simulations indicate that the amount of heating is completely independent of the helicity injection preference.  The results will be summarized in Chapter \ref{Conclusions} along with a description of the future direction of this research.
\chapter{Filament Channel Formation Via Magnetic Helicity Condensation}\label{SimsHelicity}

\section{Introduction}\label{sec:intro}

Prominences and filaments are among the most stunning features of the solar corona. They are long, thin structures, with lengths of order several hundred Mm, heights of order tens of Mm, and widths of several Mm. Prominences and filaments are known to be the same phenomenon, called by the former name when seen in emission at the solar limb, the latter when seen in absorption on the disk. Filaments form in filament channels \citep{Martin98,Gaizauskas00}, which are located above and concentrated adjacent to polarity inversion lines (PILs), where the radial component of the magnetic field changes sign. Filament channels are strongly sheared magnetic structures that can support substantial mass against solar gravity. A filament comes into existence if sufficient cool plasma accumulates in the coronal magnetic field of the channel. \par

Because filament channels are highly sheared, their nonpotential magnetic fields contain substantial amounts of free energy. This free energy is converted into kinetic and thermal energy of the gas and nonthermal energy of accelerated particles when filaments erupt in coronal mass ejections (CMEs). The shear that is inherent in the filament channels represents not only free energy, but magnetic helicity, which is carried away by the ejected flux ropes formed during CMEs. Filament channels are the only locations in the corona where significant magnetic free energy and helicity are observed. High resolution EUV and X-ray images of the closed-field corona taken by the \emph{Transition Region And Coronal Explorer} (TRACE) mission show a collection of smooth loops everywhere except in these filament channels \citep{Schrijver99}. A fundamentally important question then is: How do filament channels form? \par

Two general classes of mechanisms have been proposed to explain filament channels. One mechanism is flux cancellation \citep{VanB89}. In this model, the coronal magnetic field is sheared by the Sun's large-scale differential rotation. The shear collects and strengthens due to converging flows at the PIL, which cancel and reconnect opposite-polarity fluxes at the photosphere. This reconnection forms low-lying concave-down loops that disappear below the surface and concave-up loops that rise into the corona, generally occurring at multiple locations along the PIL. Thus, flux cancellation invariably produces a twisted coronal flux rope. Many theoretical investigations of filaments assume that the underlying structure is a twisted flux rope \citep[e.g.,][]{Malherbe83, Aulanier98, VanB00}. However, high-resolution observations of filaments \citep{Lin05, Vourlidas10} indicate that filaments are laminar and smooth, showing little evidence of substantial twist. This suggests that flux cancellation, although routinely observed in filament channels \citep[e.g.,][]{Martin98, Wang07}, is unlikely to be responsible for their development. \par

A second mechanism for filament channel formation is flux emergence, in which a rising sub-photospheric twisted flux rope breaches the surface to produce the filament channel directly. First-principles numerical simulations \citep[e.g.,][]{Fan01, Manchester01, Magara03} show that flux emergence results in the formation of a strongly sheared arcade of coronal loops. The axial field of the flux rope comprises the sheared field of the filament channel, the concave-down portion above forms the overlying quasi-potential arcade, and the concave-up portion below remains submerged beneath the photosphere. The resultant structure has the defining properties of a filament channel, and this mechanism is plausibly responsible for the formation of filament channels in solar active regions. On the other hand, filament channels are observed routinely to form in quiet regions where no significant flux emergence is occurring \citep{Mackay10}.  Moreover, Solar Dynamics Observatory measurements of helicity injection show that shearing and twisting motions frequently dominate flux emergence in mature active regions \citep{Liu12}. Therefore, it seems probable that flux emergence produces some, but almost certainly not all, solar filament channels.\par

\citet{Antiochos13} proposed a third mechanism for filament channel formation, helicity condensation. A key difference between this model and the previous two is that filament channel formation is due primarily to evolutionary processes occurring in the corona rather than in the photosphere. There are compelling arguments favoring coronal over photospheric processes for filament channel formation. Filaments channels always occur along PILs, and over all types of PILs, from the strongest active region to the quietest high-latitude region. PILs, however, are irrelevant to the high-beta photosphere, especially in weak-field quiet regions, so it seems unlikely that there is anything special about the photospheric motions there. On the other hand, PILs are absolutely critical to the low-beta corona; they define the various magnetic flux systems and, to a large extent, the topology of the field. Consequently, it seems much more likely that, whatever process is giving rise to filament channels, it operates in the corona rather than in the photosphere. This is exactly the idea behind our model: reconnection in the corona is responsible for forming filament channels at PILs.  

The helicity condensation model is motivated by two long-established results, one theoretical and one observational. The theoretical result underpinning helicity condensation is that unlike magnetic free energy, helicity is conserved even in the presence of copious reconnection \citep{Woltjer58,Taylor74,Taylor86,Berger84b}. Consequently, determining the helicity evolution can provide the most useful insights into the evolution of a system that is undergoing turbulent reconnection \citep{Taylor86}. The corona is expected to be such a system. Beginning with the seminal work of \citet{Parker72}, many authors have argued that the corona is continuously heated by numerous small-scale reconnection events termed nanoflares \citep{Parker83,Parker88,vanB86,Mikic89,Berger91,Rappazzo08}. These reconnection events are believed to be due to the formation of ubiquitous small-scale current sheets in the coronal magnetic field. The photospheric convection constantly twists and tangles the field, thereby injecting both free energy and helicity into the corona. Whereas the energy heats the coronal plasma, the helicity persistently accumulates there because it is conserved.\par

For helicity injection, the most important flows are those that twist the field. Direct observations have been made of photospheric vortical flows associated with the granular and supergranular convection. Helioseismic measurements of near-surface motions have revealed such flows at the scales of supergranulation cells \citep[e.g.,][]{Duvall00, Gizon03, Komm07, Seligman14}. Using the Swedish Solar Telescope, \citet{Brandt88} discovered photospheric vortical flows at the scales of the supergranular boundary lanes, $\sim$ 5,000 km, with typical photospheric velocities, $\sim$ 1 km s$^{-1}$. Similar results were obtained with Hinode observations by \citet{Attie09}.  Furthermore, many authors have observed vortical motions on smaller scales, down to those of granules \citep[e.g.][]{Bonet08, Bonet10, VD11, VD15}. Such motions at the intersections of convective cells are theoretically expected and invariably observed in numerical simulations of solar convection \citep[e.g.,][]{Nordlund85, Danilovic10}.\par

The evidence indicates that vortical photospheric motions are often present on the Sun. However, it is important to recognize that the overturning convective flows, due to their random nature, also will twist and tangle the coronal magnetic field. As is well known, the horizontal flows of any single convective cell, whether granule or supergranule, are primarily nonsolenoidal, consisting of a divergence from the cell center and a convergence onto the cell boundary \citep[e.g.,][]{Bray09, Schmieder14}. While the full 3D flow field may be incompressible, the surface horizontal flows certainly are not. Flows that are purely diverging and/or converging are not expected to impart any helicity or free energy into the coronal field, however. Consequently, they would not produce the hot corona that is observed. The key aspect of the photospheric convection is that the horizontal flows are not fixed in time, but vary quasi-chaotically, as granules and supergranules appear, mutate, and disappear randomly over the solar surface. These random motions will twist and tangle the coronal field, thereby adding helicity and free energy. As argued in \citet{Antiochos13} and demonstrated in numerous numerical simulations \citep[][]{Zhao15, WilmotSmith10, Pontin11, DelSordo10}, reconnection destroys the higher-order topological features produced by the random motions, such as field-line braiding, but conserve the injected helicity. From the viewpoint of determining helicity injection and evolution, therefore, it is adequate to model the random convective motions as a set of twisting motions \citep[][]{WilmotSmith10}. We adopt this approximation in this paper.\par
  
The major observational result underpinning the helicity condensation model is that the coronal magnetic field exhibits a strong hemispheric preference in magnetic helicity. \citet{Martin92} discovered that filament channels have a clear hemispheric preference for so-called dextral chirality (negative helicity) in the north solar hemisphere and sinistral chirality (positive helicity) in the south. This finding has been confirmed by many other observers \citep[e.g.,][]{Rust94b, Zirker97, Pevtsov03}. Clearly, such a global organization must be due to the photospheric emergence and driving of the coronal magnetic flux. Indeed, this pattern is seen in many other signatures of the field, ranging from interplanetary coronal mass ejections (so-called magnetic clouds), to sunspot whorls, to quiet-Sun X-ray bright points \citep[see review by][]{Pevtsov14}. It is evident, therefore, that the large-scale emergence and subsequent evolution of coronal magnetic flux injects a preferred sense of helicity into each hemisphere.\par

One source of helicity is the differential rotation, but it does not produce the correct sense of observed helicity for many prominences. The problem is that the differential rotation, with or without flux cancellation, injects helicity along an east-west polarity inversion line that is opposite to the observed helicity \citep{Mackay12}. This disagreement with data has led investigators to propose other sources of helicity injection, in particular, the emergence of large-scale active regions with an internal helicity given by the hemispheric preference. The most detailed model studies of the large-scale emergence and evolution were performed by Yeates and collaborators \citep{Yeates07, Yeates08, Yeates09}. These authors found that although active-region emergence helped alleviate some of the disagreement between model and observations, it did not eliminate all. In particular, if the initial helicity of a filament channel is ejected by a prominence eruption, then any subsequent re-formation of the filament channel, which is commonly observed, tends to produce the incorrect sign of helicity \citep{Yeates09}.\par

These difficulties in understanding filament-channel formation, and also in understanding the observed laminar structure of coronal loops \citep[e.g.,][]{Schrijver99}, led \citet{Antiochos13} to propose that small-scale flux emergence and photospheric motions inject helicity into the corona consistent with the hemispheric preference. The key point is that these small-scale processes operate continuously, thereby providing an ongoing source of helicity for the corona to replenish that lost to eruptions. \citet{Mackay14} demonstrated that the inclusion of this small-scale helicity injection in global models corrects the disagreement with observations. Importantly, however, global models do not include such small-scale processes as nanoflare reconnection, which is believed to play a critical role in this evolution. The question remains, therefore, as to exactly how the helicity injected at small scales evolves in the corona. The answer has been suggested by numerous laboratory experiments and studies of driven turbulent systems: the helicity undergoes an inverse cascade and condenses at the largest system scale \citep{Biskamp93}. \citet{Antiochos13} argued that such a helicity cascade is exactly what happens in the corona, and that it explains both the formation of filament channels and the smoothness of coronal loops.\par

\begin{figure}[!h]
\includegraphics[scale=0.4]{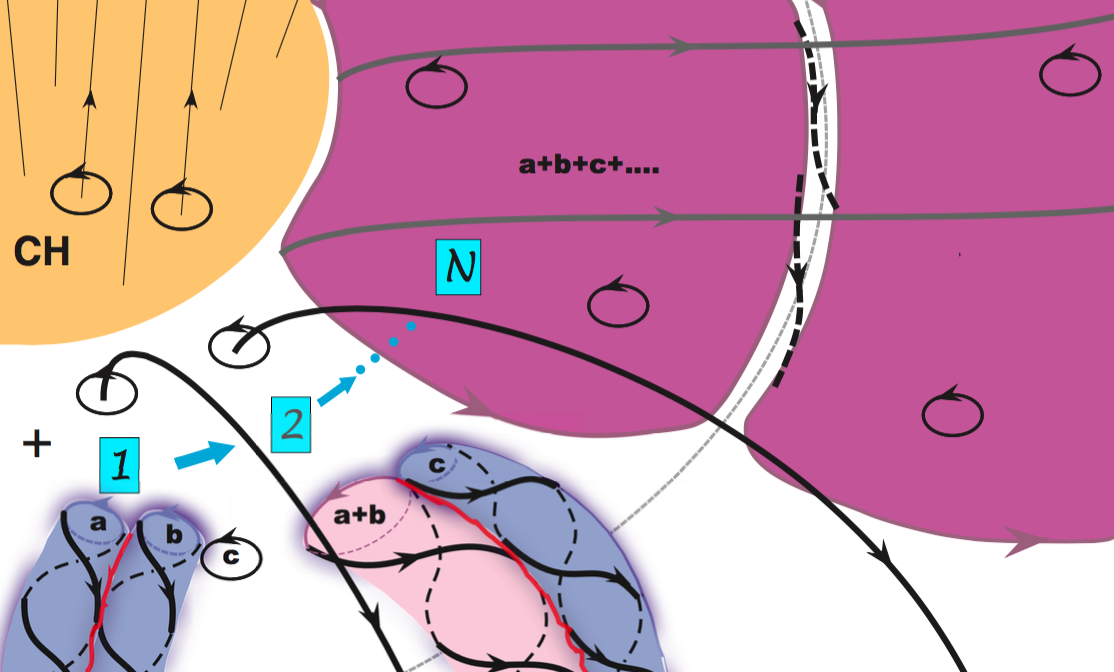}
\caption[Helicity Condensation Model]{Helicity condensation model, after \citet{Antiochos13}. The yellow region represents the photosphere within a coronal hole (`CH'). The thin gray dashed line is a polarity inversion line (`PIL'). The counter-clockwise circular arrows represent a few individual rotation cells among the whole, densely distributed population of such cells on the surface of the Sun. Thick black curves represent magnetic field lines, drawn solid on the top side and dashed on the underside of the magnetic flux tubes to which they belong. Yellow curves indicate contact points where the azimuthal fields of adjacent tubes are antiparallel and so can reconnect. Individual flux tubes are colored light blue (`a', `b', and `c') or light purple (`a+b'). Step numbers (`1', `2', `$N$') indicate the progressive transport of azimuthal magnetic flux from the injection scale of the rotation cells to the global scales of the CH and PIL. Ultimately, the azimuthal flux either condenses at the PIL to form a filament channel (thick black dashed lines linking the dark purple regions) or is released into the solar wind after propagating into the open field at the CH boundary.
\label{fig:condensation}}
\end{figure}

This coronal helicity cascade is illustrated schematically in Figure \ref{fig:condensation}. A bipolar magnetic region with a polarity inversion line (`PIL'; gray dashed line) contains an embedded coronal hole (`CH'; yellow shading). Thick black curves are magnetic field lines, which are drawn as dashed curves when on the underside of the magnetic structure to which they belong. The ubiquitous twisting flows are represented by the black counter-clockwise circular arrows distributed across the surface. At step 1, two neighboring flux tubes (`a' and `b'; light blue) with like-polarity axial fields and similarly twisted azimuthal fields have come into contact. Because the azimuthal fields are anti-parallel at contact points (yellow lines) between tubes, they can reconnect and cancel. The result at step 2 is a newly merged, single flux tube (`a+b'; light purple) containing the combined axial fluxes of the original pair, enclosed by the azimuthal flux that previously wrapped around each individual tube. Helicity is conserved under reconnection in the highly conducting corona, so this merging process has transferred the helicity injected at some elemental scale to the next larger scale. This process continues with further mergers, such as that at step 2 between the newly formed, larger flux tube and a neighboring elemental tube (`c'; light blue) at the injection scale.\par

The transfer of helicity to ever-larger scales (`a+b+c+...') continues until the azimuthal flux attains the scales of the encircling PIL and the embedded CH at step $N$. At the PIL, this flux (thick dashed curves with arrowheads, linking the dark purple regions) is trapped on low-lying field lines, where it can only accumulate until it is liberated by a CME. This concentration of azimuthal flux at the PIL, where the helicity is said to `condense,' is precisely the signature of a filament channel. At the CH, in contrast, the azimuthal flux is imparted to high-lying field lines that can open easily into the solar wind, so the field there remains untwisted. Note from the figure that the sense of twist of the azimuthal flux propagating toward the CH (in flux tubes `a', `b', and `c') is clockwise; this is opposite to the sense of twist imparted to the open flux tubes residing within the coronal hole by the counter-clockwise rotations. Thus, the helicity condensation model makes an observational prediction that helicity fluxes measured within the interior of a coronal hole and at its perimeter will have opposite signs \citep{Antiochos13}. \par

An initial numerical investigation of basic predictions of the helicity condensation model has been reported in \citet{Zhao15}. The ansatz of \citet{Parker72}, in which the corona is modeled by a uniform magnetic field between two horizontal plates, was adopted. A simulation of two neighboring like-polarity, like-twisted flux tubes verified the reconnection and merger of two small tubes into one larger tube, as shown schematically in steps 1 to 2 of our Figure \ref{fig:condensation}. A companion simulation showed that reconnection does not occur if the two flux tubes are twisted in opposite senses: because their azimuthal fields are parallel rather than anti-parallel, the tubes remain separate and distinct instead of merging, as predicted by \citet{Antiochos13}. A second pair of simulations assumed a close-packed, regular hexagonal array of seven rotation cells. In one case, the array was kept spatially fixed through a dozen temporal cycles of turning the flows on and off; in the other, the array of cells was randomly translated and rotated between cycles, to emulate the ever-shifting pattern of the Sun's surface convection. Both simulations demonstrated that the injected azimuthal flux was transferred via reconnection to the perimeter of the region of flows, again as predicted by the helicity condensation model. Only minor, quantitative differences between the fixed and randomized cases were found, indicating that the evolution of the system is insensitive to the details of the photospheric driving motions so long as the motions are sufficiently complex. For computational simplicity, therefore, we use only a fixed pattern of driving flows in the calculations discussed in this paper. \par

We examine in detail the processes of magnetic helicity injection, transport, and condensation within a plane-parallel corona for a far larger ensemble of rotation cells than that assumed by 
\citet{Zhao15}. In addition, and more fundamentally, we include an interior region that is free of rotation cells, as an elementary model for a coronal hole embedded within a unipolar region of closed field. This revised configuration is a simplified representation of the scenario shown here in Figure \ref{fig:condensation}. Our simulations enable us to test the predictions of the helicity condensation hypothesis with greater generality and complexity than in previous work. We also develop and verify several quantitative, analytic estimates of the early- and late-time behaviors in our simulation. Those results aid our understanding of the plane-parallel system that we have adopted here. They also prepare the way for future studies of more geometrically realistic scenarios having true polarity inversion lines and coronal holes with open fields. \par

The remainder of our paper is organized as follows. In \S \ref{sec:simulations}, we describe the numerical simulation model that we used to test the helicity condensation hypothesis. Our analytical deductions and numerical diagnostics for magnetic twist and helicity are presented in \S \ref{sec:helicity}.  The core of the paper consists of the results and analyses of our numerical simulation results, which are given in \S \ref{sec:Results}. Finally, in \S \ref{sec:discussion} we discuss the implications of our findings for understanding filament channel formation on the Sun. \par

\section{Uniform Magnetic Field Between Two Plates}\label{sec:simulations}
\subsection{The Numerical Model}
We use the Adaptively Refined Magnetohydrodynamics Solver \citep[ARMS; e.g.,][]{DeVore08} to solve the equations of magnetohydrodynamics (MHD) in three Cartesian dimensions. 
The equations have the form
\beg{cont}
\pd{\rho}{t}+\divv{\left( \rho\vecv \right)}=0,
\done
\beg{momentum}
\pd{\rho\vecv}{t} + \divv{\left( \rho\vecv\vecv \right)} = - \nabla P + \frac{1}{4\pi} \left( \curl{\vecB} \right) \times \vecB,
\done
\beg{energy}
\pd{T}{t} + \divv{\left( T\vecv \right)} = \left( 2 - \gamma \right) T \divv{\vecv},
\done
\beg{induction}
\pd{\vecB}{t} = \curl{ \left( \vecv \times \vecB \right)}.
\done
Here $\rho$ is mass density, $T$ is temperature, $P$ is thermal pressure, $\gamma$ is the ratio of specific heats, $\vecv$ is velocity, $\vecB$ is magnetic field, and $t$ is time. We close the equations via the ideal gas equation
\beg{idealgas}
P=\rho RT,
\done
where $R$ is the gas constant. ARMS employs Flux-Corrected Transport algorithms \citep{DeVore91} and finite-volume representation of the variables to obtain its solutions. Its minimal, but finite, numerical dissipation allows reconnection to occur at electric current sheets associated with discontinuities in the direction of the magnetic field. \par

\begin{figure*}[!h]
\centering\includegraphics[scale=0.3,trim=0cm 0.0cm 0cm 0.0cm, clip=true]{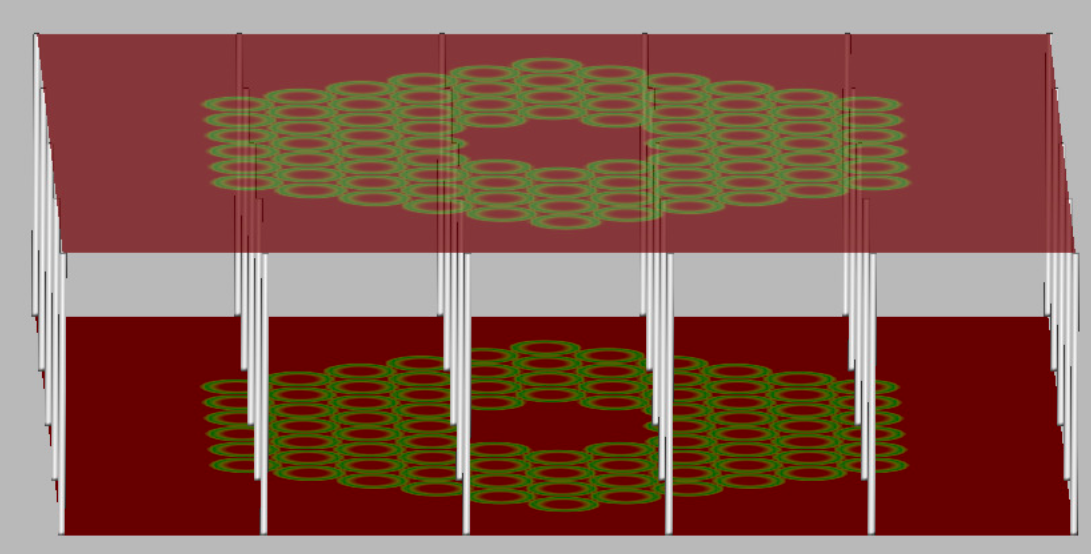}
\centering\includegraphics[scale=0.27]{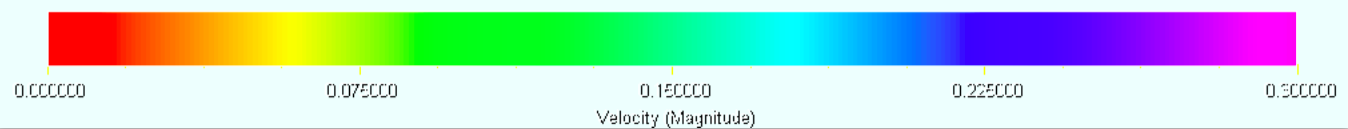}
\caption[Uniform Field Simulation Setup]{Initial simulation setup having a uniform magnetic field between two plates and 84 rotation cells -- locations where the field is twisted by line-tied flows -- in a hexagonal array on each plate. White lines are magnetic field lines. Color shading on the top and bottom planes indicates the magnitude of the flow velocity: red shading indicates locations where the flow speed is zero, green where it is nonzero. The outer boundary of the flow region represents a PIL; the inner flow-free region represents a CH (compare with Figure \ref{fig:condensation}).
\label{fig:initial}}
\end{figure*}

Our simulation configuration is shown in Figure \ref{fig:initial}. We model the coronal magnetic field as initially straight and uniform ($\vecB = B_0 \hat{\vecx}$) between two plates \citep{Parker72}. Straight flux tubes therefore represent coronal loops, with the apex of each `loop' positioned in the center of the domain. Each of the two boundary plates represents the photosphere. The domain extent 
in $(x,y,z)$ is $[0,L_x]\times[-L_y,+L_y]\times[-L_z,+L_z]$, with $x$ the vertical direction (normal to the photosphere), $L_x=1$, and $L_y=L_z=1.75$. \par
We employ zero-gradient conditions at all times at all six boundaries,
\beg{boundaries}
\begin{split}
\pd{\rho}{n}=0,\\
\pd{T}{n}=0,\\
\pd{\vecv}{n}=0,\\
\pd{\vecB}{n}=0,
\end{split} 
\done
where $n=x,y,z$ is the normal coordinate. The four side boundaries are all open to accommodate lateral expansion of the stressed magnetic field. The top and bottom boundaries are closed, where the magnetic field is line-tied. The footpoints of the field lines move only in response to imposed boundary flows, emulating the slow driving at the dense solar photosphere, rather than in response to coronal magnetic forces. \par

The initial, uniform values used in our dimensionless simulation are $\rho_0=1$, $T_0=1$, $P_0=0.05$, and $B_0=\sqrt{4\pi}$. These choices set the gas constant, $R=0.05$, the Alfv\'en speed, $c_{A0}=B_0 / \sqrt{4\pi\rho_0} = 1$, and the plasma beta, $\beta_0=8\pi P_0/B_0^2=0.1$. The regime $\beta\ll1$ corresponds to a magnetically dominated plasma, which is generally true of the corona. We note that the simulation time therefore is normalized to the time required for an Alfv\'en wave at unit speed ($c_{A0}=1$) to propagate between the top and bottom plates separated by unit distance ($L_x=1$). \par

To model the ever-present twisting of the photospheric footpoints of a finite flux system, we confine the helicity injection to the hexagonal annulus defined by the pattern of surface flows shown in Figure \ref{fig:initial}. The PIL is defined by the outer boundary of this pattern, while the CH is defined by the untwisted region within the inner boundary of the pattern. $N = 84$ identical, circular rotation cells of radius $a_0$ are positioned in the hexagonal array shown, on both the top and bottom plates. Our CH thus has radius $a_c=3a_0$ at the photosphere, while the PIL has radius $a_p=10a_0$. \par

We emphasize that the velocity pattern of Figure \ref{fig:initial} is not in any way meant to resemble the observed convective flows at the photosphere. As discussed above, rotational motions similar to those in the figure are sometimes observed at the photosphere, but only in isolated locations, primarily at the intersections of convective cells. The key point, however, is that even though a single granule or supergranule exhibits very little rotation, the random appearance and disappearance of the convective cells does inject twist into the coronal magnetic field. Because the granular/supergranular flows are approximately uniform over the photosphere, the twist injection also should be roughly uniform and with a preferred scale given by the size of the cells. Furthermore, we assume that this small-scale twist injection has the same hemispheric helicity preference observed in large-scale structures. Figure \ref{fig:initial} is merely the simplest possible (and, consequently, easiest to simulate) flow pattern for injecting such a twist into the coronal magnetic field. Future studies will consider more realistic and complex driving flows. 

The adaptive mesh refinement capability of ARMS was used to resolve very finely that part of the domain volume where the photospheric flows are imposed and the coronal flux tubes become twisted. Each elemental block of the grid contained $8\times8\times8$ uniform, cubic grid cells. Four such blocks, or 32 grid cells, were used to span the radius $a_0$ of each of our photospheric rotations. The full pattern of rotations, the lanes between them, the coronal hole in the interior, and a buffer region around the outer perimeter of the pattern were covered uniformly by these high-resolution grid blocks. This finely gridded region extended vertically throughout the corona. Outside of this region toward the side walls, the grid was allowed to coarsen by two levels, the grid spacing increasing by a factor of two at each change of refinement level. The resulting grids were about $25\%$ of the sizes of equivalent uniform grids throughout the domain. \par

We set the vertical velocity $v_x\vert_S=0$ at the top and bottom boundaries. The $x$-component of the induction Equation (\ref{induction}) can be written
\beg{inductionx}
\pd{B_x}{t}=-(\vecv_\perp\cdot\nabla_\perp)B_x-B_x(\nabla_\perp\cdot\vecv_\perp),
\done
where $\perp$ represents the $y$ and $z$ directions. At $t=0$, $B_x$ is uniform on the boundary, so the first term on the right-hand side vanishes. If the second term also vanishes, because the flows are incompressible, then $\partial B_x /\partial t = 0$ at all times $t$. The latter condition is satisfied identically if, for each rotation cell,
\beg{flowv}
\vecv_\perp = \hat{\vecx} \times \nabla \chi(r,t).
\done
Here $\chi$ is any scalar function, taken to depend only upon time $t$ and the radial coordinate $r$ centered on the rotation cell.
We set 
\beg{flowchi}
\chi(r,t)=\Omega_0 a_0^2 f(t) g(r).
\done
The temporal profile $f(t)$ is given by
\beg{foft}
f(t) = \frac{1}{2} \left[ 1 - \cos \left( 2\pi \frac{t}{\tau} \right) \right].
\done
The flows ramp up from zero and then back down to zero over the period $\tau$, with the corresponding displacement being proportional to the integral of $f(t)$,
\beg{foftI}
F(t) = \frac{1}{2} \left[ t - \frac{\tau}{2\pi} \sin \left( 2\pi \frac{t}{\tau} \right) \right].
\done
The spatial profile $g(r)$ is given by
\beg{gofr}
g(r)=\frac{1}{6}\left[1-\left(\frac{r}{a_0}\right)^{6}\right] - \frac{1}{10}\left[1-\left(\frac{r}{a_0}\right)^{10}\right].
\done
Outside of $r=a_0$, we fix $g(r)=0$. With this form of $\chi(r,t)$, the angular rotation rate for $r \le a_0$ is given by 
\beg{omega}
\begin{split}
\Omega(r,t)&=\Omega_0 a_0^2 f(t) \frac{1}{r}\frac{dg}{dr}\\
&=- \Omega_0 f(t) \left(\frac{r}{a_0}\right)^4 \left[1-\left(\frac{r}{a_0}\right)^4\right].
\end{split}
\done
For the flow parameters, we choose $a_0=0.125$, $\Omega_0=7.5$, and $\tau=3.35$. These set $\vert \vecv_\perp \vert_{\rm max}=0.200$ and a maximum angle $\vert \Delta \phi \vert_{\rm max} = \pi$ of the clockwise rotation within each cell over the period $\tau$ of each cycle. \par

Our idealized model includes only the small-scale motions that are critical for changing the coronal helicity: the twisting flows. For computational simplicity, we omitted the dominant radial outflow seen in granules and supergranules because, as noted above, this motion does not inject helicity into the corona. Consequently, one of our elemental rotations is not meant to represent an actual convective cell, whether granule or supergranule, but only the twisting motions produced by the random nature of the convective flows. We examine below the injection of helicity due to these vortical motions and its subsequent transport through the corona via reconnection. \par

\subsection{Magnetic Helicity and Twist}\label{sec:helicity}
\subsubsection{Magnetic Helicity Injection}\label{sec:helinj}
In a volume $V$ bounded by magnetic flux surfaces $S$, so that $\vecB\cdot\hat{\vecn}|_S=0$ with $\hat{\vecn}$ the unit vector normal to $S$, the magnetic helicity is simply defined as 
\beg{helicity}
H=\int_V{\vecA\cdot\vecB \; dV,}
\done
where $\vecB=\curl{\vecA}$ and $\vecA$ is the vector potential \citep{Berger99}. This integral measures the general topological property of field line linkages. In the corona and in our simulation domain, where magnetic field lines enter from or exit to the photosphere, the bounding surface is not a flux surface, and the relative magnetic helicity \citep{1984JFM...147..133B} must be used in place of Equation 
(\ref{helicity}).  We adopt the gauge-invariant form of \citet{Finn85}, 
\beg{relativehelicity}
H=\int_V{(\vecA+\vecA_P)\cdot(\vecB-\vecB_P) \; dV},
\done
where $\vecB_P = \curl{\vecA_P}$ is a current-free field ($\curl{\vecB_P} = 0$) satisfying $\vecB_P\cdot\hat{\vecn}|_S=\vecB\cdot\hat{\vecn}|_S$. It can be shown through integrations by parts that, under ideal evolution, the time derivative of Equation (\ref{relativehelicity}) leads to (see Appendix for derivation):
\beg{dHdt}
\frac{dH}{dt}=2\oint_S{\left[ \left( \vecA_P \cdot \vecv \right) \vecB - \left( \vecA_P \cdot \vecB \right) \vecv \right] \cdot \vecdS.}
\done
The first term represents the effects of twisting or shearing motions on the boundary, while the second term represents helicity injected (removed) by the emergence (submergence) of helical field across the boundary. Thus, the only contributions to $dH/dt$ are due to motions on or through the bounding surface $S$. Absent such motions, the helicity $H$ is conserved perfectly in the volume $V$. In highly conducting plasmas that are almost ideal, the helicity is conserved even in the presence of a small localized resistivity that gives rise to magnetic reconnection \citep{Woltjer58,Taylor74,Taylor86,Berger84b}. \par

To monitor the magnetic helicity in our simulations, we calculate the volume integral and surface injection rate using Equations (\ref{relativehelicity}) and (\ref{dHdt}), respectively. The vector potential for the uniform, current-free initial field is
\beg{potA}
\vecA_p = \frac{B_0}{2} (y\hat{\vecz}-z\hat{\vecy}),
\done 
so that $\vecB_p = \nabla \times \vecA_p = B_0 \hat{\vecx}$. Using $\vecA_p$ as the boundary value at the photosphere at $x=0$, and adopting the gauge $A_x=0$, the vector potential at all times $t$ and heights $x$ becomes 
\beg{vecpotA}
\begin{split}
\vecA(x,y,z,t) = \vecA_p(y,z) &+ \hat{\vecy}\int_0^x {dx'B_z(x',y,z,t)}\\
&- \hat{\vecz}\int_0^x {dx'B_y(x',y,z,t).}
\end{split}
\done
This expression was used to evaluate the volume integral in Equation (\ref{relativehelicity}) for $H$. Because the helicity injected by each rotational cell is independent of its position on either plate, we can evaluate Equation (\ref{dHdt}) for the helicity injection rate due to a single flux tube centered at $(x,y,z)=(0,0,0)$. We rewrite Equation (\ref{potA}) as 
\beg{vecAcyl}
\vecA_p=\frac{B_0}{2}r\hat{\phi}.
\done
Here $r$ is the radial coordinate centered on the cell. Because there is no motion through the boundary, Equation (\ref{dHdt}) simplifies to 
\beg{calcdhdt}
\begin{split}
\frac{dH'}{dt} &= - \int_{S'} {r \hat{\phi} \cdot \vecv B_0^2 dS} \\
&= - \int_{S'} {r^2 \Omega(r,t) B_0^2 dS}.
\end{split}
\done
$H'$ is the helicity contributed by the footprint $S'$ of the flux tube, and we have used $\vecv\cdot\hat{\phi} = r\Omega(r,t)$. Substituting Equation (\ref{omega}) for $\Omega$, performing the area integral, and doubling the result to include the rotation cells at both the top and bottom plates yields for the helicity injection rate per flux tube 
\beg{dhdtanalytic}
\frac{dH_{f}'}{dt} = + \frac{\pi}{6} \Omega_0 a_0^4 B_0^2 f(t).
\done
Integrating Equation (\ref{dhdtanalytic}) using Equation (\ref{foftI}), we obtain for the time history of the helicity contributed per flux tube 
\beg{hanalytic}
\begin{split}
H_{f}'(t) &= \frac{\pi}{12} \Omega_0 a_0^4 B_0^2 \left[t - \frac{\tau}{2\pi} \sin \left(2\pi\frac{t}{\tau}\right) \right] \\
&= \langle \frac{dH_{f}'}{dt} \rangle \left[t - \frac{\tau}{2\pi} \sin \left(2\pi\frac{t}{\tau}\right) \right].
\end{split}
\done
The average rate of helicity injection and the resultant injected helicity per flux tube over one cycle of duration $\tau$ are 
\beg{avgdHdt}
\langle \frac{dH_{f}'}{dt} \rangle = \frac{\pi}{12} \Omega_0 a_0^4 B_0^2 = 6.0 \times 10^{-3}
\done
and 
\beg{changeH}
\Delta {H_{f}'} = \frac{\pi}{12} \Omega_0 a_0^4 B_0^2 \tau = 2.0 \times 10^{-2},
\done
respectively, after substituting the numerical values of the parameters given previously. \par 

\begin{figure}[!h]
\centering\includegraphics[scale=0.3]{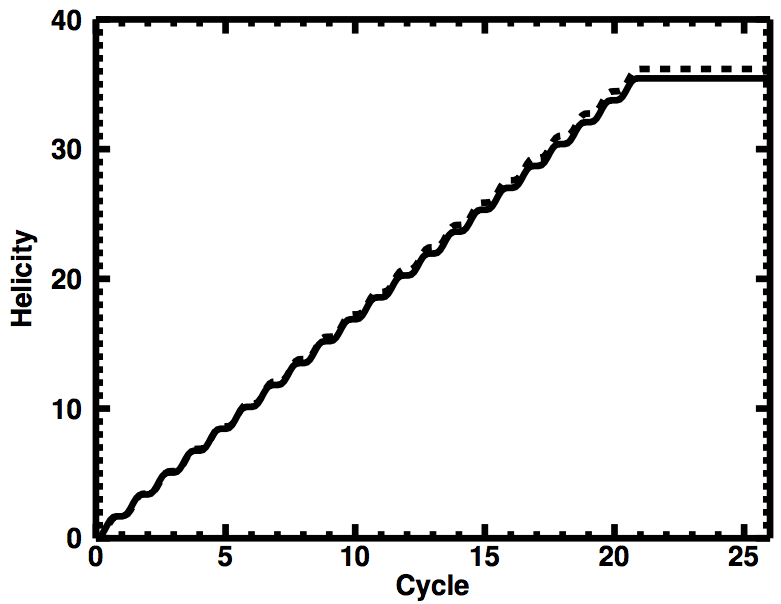}
\caption[Magnetic Helicity Conservation]{Accumulated magnetic helicity in the domain as calculated numerically (solid line) from Equation (\ref{relativehelicity}) and analytically (dashed line) from Equation (\ref{heltime}), demonstrating magnetic helicity conservation to a high degree of accuracy. 
\label{fig:heldiag}}
\end{figure}

We performed 21 cycles of duration $\tau$ of the flows, each of which twists the field by a maximum angle of $\pi$ on both the top and bottom boundaries, thus yielding up to one full rotation within each flux tube. At the end of the 21 twisting cycles, we performed 5 more cycles during which the field was allowed to relax, i.e., we fixed ${\vecv_\perp}\vert_S=0$, so no new helicity was injected into the system. During the 21 twisting cycles, we can multiply Equations (\ref{avgdHdt}) and (\ref{changeH}) by the $N=84$ flux tubes in the hexagonal pattern to obtain the total average helicity injection rate and helicity injected per cycle, 
\beg{gvals}
\begin{split}
\langle \frac{dH}{dt} \rangle &= 5.0 \times 10^{-1}, \\
\Delta {H} &= 1.7 \times 10^{0}.
\end{split}
\done
Figure \ref{fig:heldiag} compares the numerical results from the volume integral in Equation (\ref{relativehelicity}) with the theoretical prediction obtained by multiplying Equation (\ref{hanalytic}) by 
$N=84$,
\beg{heltime}
H(t) = 5.0 \times 10^{-1} \left[t - \frac{\tau}{2\pi} \sin \left(2\pi\frac{t}{\tau}\right) \right],
\done
for 21 cycles. Clearly, the numerically calculated helicity matches the analytically calculated value to high accuracy. This is a very important diagnostic, because conservation of magnetic helicity under reconnection is crucial to the helicity condensation model. The agreement between the numerically and analytically calculated helicities demonstrates that we are calculating accurately the transport of helicity and twist flux throughout the domain. The algorithms employed by ARMS conserve magnetic flux to machine accuracy, so that the minimal numerical diffusion allows the twist flux to spread out without decreasing. As a result, evidently, a slightly larger axial flux is enclosed by the twist flux, so that the numerically calculated helicity is slightly larger than the analytic value. In any case, the figure demonstrates conclusively that, in spite of the large amount of reconnection that occurs inside the volume, the only significant change in magnetic helicity is due to the twisting motions imposed on the boundaries. \par
\subsubsection{Twist Flux Generation}\label{sec:twist}

For our physical system we expect that all the helicity is due to the linkages between the uniform background field and the twist flux generated by the photospheric motions. Consequently, in order to understand the helicity evolution, it is highly instructive to follow the evolution of the twist flux.  In fact, the surface helicity integral, Equation (\ref{calcdhdt}), can be recast as a relationship between the rates of helicity injection and twist-flux generation. This result is useful within the individual flux tubes formed early in our simulation and, in addition, when applied across the entire hexagonal array of rotation cells during the late stages of evolution to measure the helicity condensation. 

Over the time interval $dt$, rotation at the rate $\Omega(r,t)$ conveys the vertical field $B_0$ along a strip of length $r \Omega(r,t) dt$ at radius $r$ through the base of the plane $\phi={\rm constant}$. This motion increments the twist flux per unit length in $r$ that passes through that plane, $\Psi_{tw}(r,t)$, by the amount of axial flux that is conveyed across its base,
\beg{dPsitw}
\frac{\partial \Psi_{tw}(r,t)}{\partial t} = - r \Omega(r,t) B_0. 
\done
The total change of the (net) twist flux across the footprint $S'$ of the rotational motion is the radial integral 
\beg{dPhitw}
\frac{d\Phi_{tw}}{dt} = \int_{S'} {dr \; \frac{\partial \Psi_{tw}(r,t)}{\partial t}}.
\done
We note that the minus sign in Equation (\ref{dPsitw}) ensures that a clockwise rotation ($\Omega < 0$) of a positive vertical field ($B_0 > 0$) induces a positive change in the twist flux ($d\Psi_{tw} / dt > 0$). This increment in the twist flux wraps around the axial (vertical) flux that is enclosed within radius $r$,
\beg{Psiax}
\Psi_{ax}(r) = 2\pi \int_0^r {dr' \; r' B_0} = \pi r^2 B_0.
\done
Using Equations (\ref{dPsitw}) and (\ref{Psiax}), the surface helicity integral in Equation (\ref{calcdhdt}) therefore can be rewritten 
\beg{calcdhdt3}
\frac{dH'}{dt} = 2 \int_{S'} {dr \; \Psi_{ax}(r) \frac{\partial \Psi_{tw}(r,t)}{\partial t}}.
\done
In this form, magnetic helicity as a measure of the linkages between the axial and twist fluxes is explicit. We rewrite the twist flux increment, Equation (\ref{dPsitw}), in terms of the corresponding axial flux increment using Equation (\ref{Psiax}),
\beg{dPsitw2}
\frac{\partial \Psi_{tw}(r,t)}{\partial t} = - \frac{\Omega(r,t)}{2\pi} \frac{d\Psi_{ax}(r)}{dr}.
\done
The helicity integral, Equation (\ref{calcdhdt3}), then can be recast further into the alternate expression 
\beg{calcdhdt4}
\frac{dH'}{dt} = - \frac{1}{\pi} \int_{S'} {dr \; \Omega(r,t) \Psi_{ax}(r) \frac{d\Psi_{ax}(r)}{dr}}. 
\done
This version can be integrated immediately for any axisymmetric flux distribution, for the special case of rigid-body rotation ($\Omega$ independent of $r$). \par

The rate of twist-flux generation in a single flux tube is calculated by evaluating the integral in Equation (\ref{dPhitw}). Substituting from Equation (\ref{omega}) for $\Omega$ into Equation (\ref{dPsitw}), and doubling the result to include the contributions from both plates, we obtain 
\beg{dtwdtanalytic}
\frac{d\Phi_{tw,f}}{dt} = +\frac{2}{15} \Omega_0 a_0^2 B_0 f(t).
\done
The corresponding average rate of twist flux injection and the resultant twist flux injected per flux tube over one cycle of duration $\tau$ are 
\beg{avgdtwdt}
\langle \frac{d\Phi_{tw,f}}{dt} \rangle = \frac{1}{15} \Omega_0 a_0^2 B_0 = 2.8 \times 10^{-2}
\done
and
\beg{changetw}
\Delta \Phi_{tw,f} = \frac{1}{15} \Omega_0 a_0^2 B_0 \tau = 9.3 \times 10^{-2},
\done
respectively, after substituting the numerical values of the parameters. \par

The surface helicity integral as written in Equation (\ref{calcdhdt3}) can be used to relate the rate of helicity injection to the rate of generation of large-scale twist flux by reconnection. We make the approximation that the twist flux is concentrated into one or more bands of radial extent $\left[ r_{i_-}, r_{i_+} \right]$ within which $\Omega$ is quasi-uniform. Note that for the large-scale twist flux, 
$\Omega$ is only an effective rotation due to reconnection, not a true rotation due to mass flow.  Equations (\ref{dPhitw}) and (\ref{dPsitw2}) then yield 
\beg{dPhitw2}
\begin{split}
\frac{d\Phi_{tw,r_i}}{dt} &= - \frac{1}{2\pi} \int_{r_{i_-}}^{r_{i_+}} {dr \; \Omega(r,t) \frac{d \Psi_{ax}(r)}{dr} dr} \\
&= - \frac{1}{2\pi} \langle \Omega \rangle_{i} \left[ \Psi_{ax} \left( r_{i_+} \right) - \Psi_{ax} \left( r_{i_-} \right) \right].
\end{split}
\done
This is the rate of twist-flux generation in the $i$th band, with $\langle \Omega \rangle_{i}$ the average angular rotation rate in the band. The associated helicity generation rate is, from Equation (\ref{calcdhdt4}),
\beg{calcdhdt5}
\begin{split}
\frac{dH'_{r_i}}{dt} &= - \frac{1}{\pi} \int_{r_{i_-}}^{r_{i_+}} {dr \; \Omega(r,t) \Psi_{ax}(r) \frac{d\Psi_{ax}(r)}{dr}} \\
&= - \frac{1}{2\pi} \langle \Omega \rangle_{i} \left[ \Psi_{ax}^2 \left( r_{i_+} \right) - \Psi_{ax}^2 \left( r_{i_-} \right) \right] \\
&= + 2 \langle \Psi_{ax} \rangle_{i} \frac{d\Phi_{tw,r_i}}{dt}.
\end{split}
\done
The average axial flux encircled by the band is 
\beg{avgpsiax}
\langle \Psi_{ax} \rangle_{i} = \frac{1}{2} \left[ \Psi_{ax} \left( r_{i_+} \right) + \Psi_{ax} \left( r_{i_-} \right) \right].
\done
We can substitute the numerical rates of helicity and twist-flux generation for a single flux tube, from Equations (\ref{avgdHdt}) and (\ref{avgdtwdt}) respectively, into Equation (\ref{calcdhdt5}) to determine $\langle 
\Psi_{ax} \rangle_{i}$. The procedure yields an effective radius $r_i \approx 0.8a_0 = 0.10$ where the axial flux $\langle \Psi_{ax} \rangle_{i}$ is evaluated; this location is very near the peak angular rotation within the individual flux tubes. For application to the late-time flux distribution in our simulations, we now sum the contributions from  Equation (\ref{calcdhdt5}) for multiple bands, obtaining 
\beg{dhdttotal}
\frac{dH}{dt} = \sum_i \frac{dH'_{r_i}}{dt} = 2 \sum_i \langle \Psi_{ax} \rangle_{i} \frac{d\Phi_{tw,r_i}}{dt}.
\done
This relationship links the total rate of helicity injection into the domain, on the left-hand side, to the twist-flux generation rates in the bands, mediated by the enclosed axial fluxes, on the right-hand side. \par

\subsection{Results}\label{sec:Results}

\subsubsection{Twist Flux Condensation}\label{sec:condensation}

\begin{figure*}[!h]
\centering\includegraphics[scale=0.65,trim=0.3cm 8.5cm 0.3cm 8.5cm, clip=true]{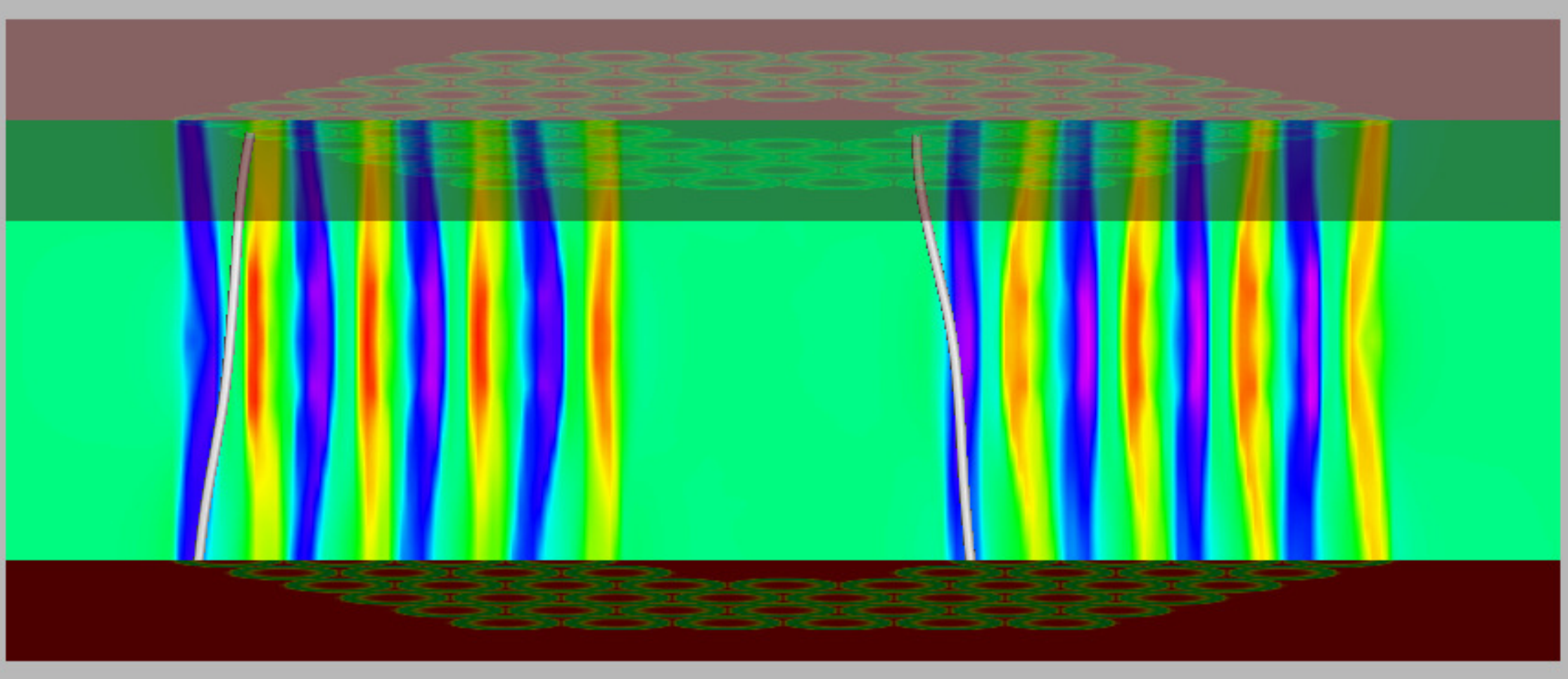}
\centering\includegraphics[scale=0.57]{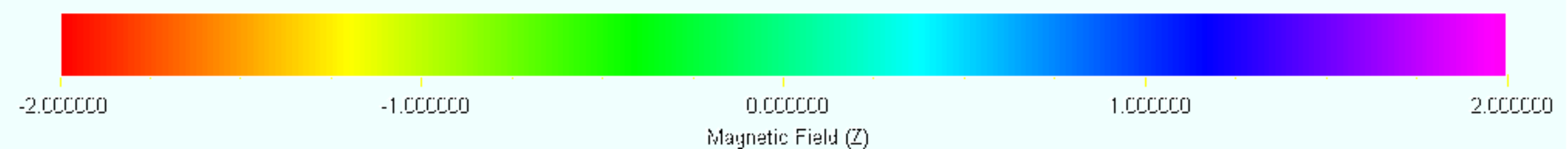}
\centering\includegraphics[scale=0.65,trim=0.3cm 8.5cm 0.3cm 8.5cm, clip=true]{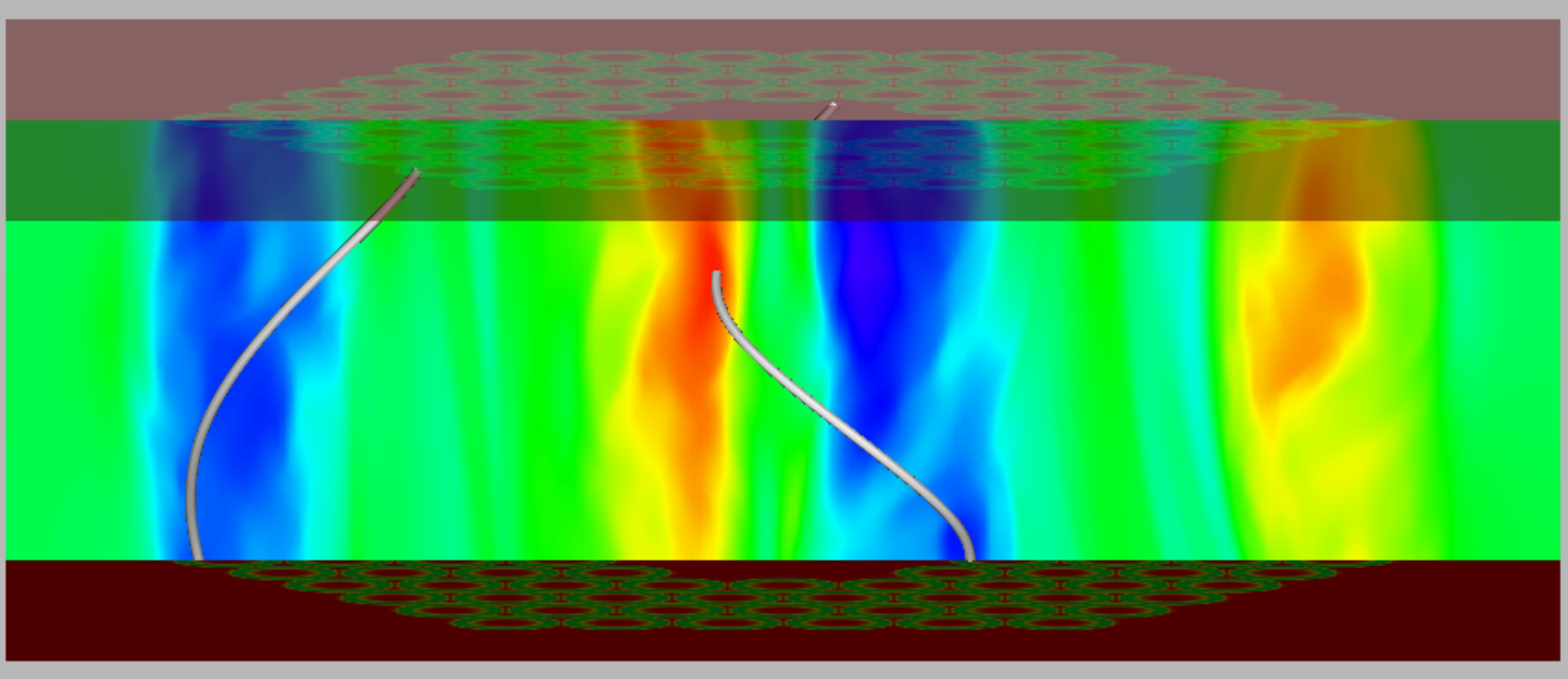}
\centering\includegraphics[scale=0.57]{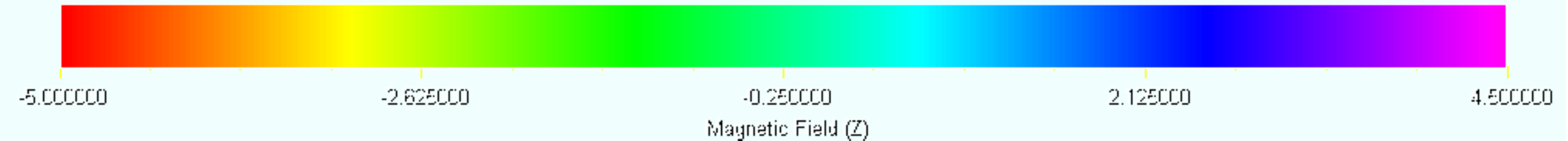}
\caption[$B_z$ through the $z = 0$ plane]{Horizontal field $B_z$ (color shading) through the $z = 0$ vertical plane between the top and bottom boundaries. Red/yellow (blue/teal) represents field pointing into (out of) the plane. Two field lines (white) indicate how much twist is present. Top: After 1 twist cycle. Bottom: After 21 twist cycles and 5 relaxation cycles.
\label{fig:fluxandfieldlines}}
\end{figure*}

We begin the presentation of our results by describing qualitatively the global evolution of the system. As will be seen, the magnetic field transitions from numerous individual, twisted flux tubes at early times to a configuration dominated at late times by large-scale flux accumulations at the perimeter of the hexagonal annulus where the rotational flows are imposed. The outer boundary of the flux system, the PIL, is demarcated by the transition to the flow-free region outside the annulus; the inner boundary, the CH, consists of the flow-free region in the center of the annulus. In Figure \ref{fig:fluxandfieldlines}, we show the $y$ component of the magnetic field, $B_y$, on a vertical plane cut through the center of the domain, along with two field lines positioned at the PIL and CH boundaries, at two times in the simulation. After one twist cycle (top), every pair of rotation cells on the top and bottom planes is the footprint of an associated flux tube whose $y$ component of magnetic field goes into and comes out of the vertical plane. The field lines contain slightly less than one full turn of twist. The field line at the PIL, on the left side of the image, is confined within one flux tube; the field line at the CH, on the right side, in contrast, has reconnected already, linking one rotation cell at the bottom boundary to the cell next to its partner at the top.  After 21 twist cycles and 5 relaxation cycles (bottom), the twist flux has condensed almost completely to the PIL and CH boundaries. The field line at the right wraps about three-fourths of the way around the CH, crossing about 10 rotation cells; the field line at the left wraps just under one-sixth of the way around the PIL, crossing only four cells. \par

\begin{figure*}[!h]
\centering\includegraphics[scale=0.35, trim=1.0cm 6cm 0.75cm 6cm, clip=true]{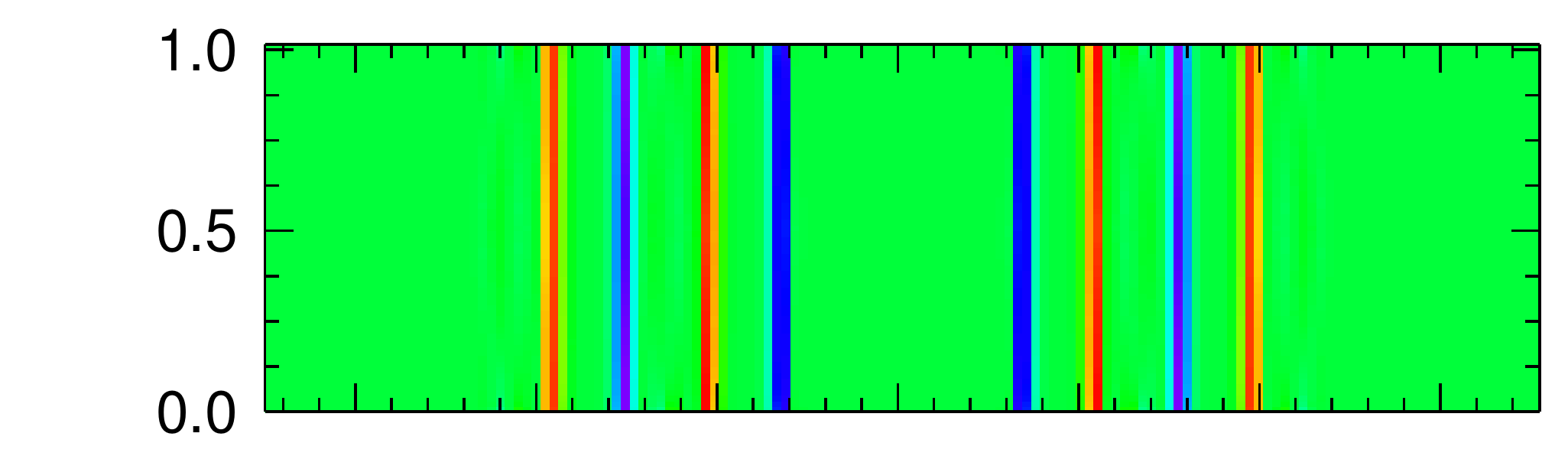}
\centering\includegraphics[scale=0.35, trim=1.0cm 6cm 0.75cm 6cm, clip=true]{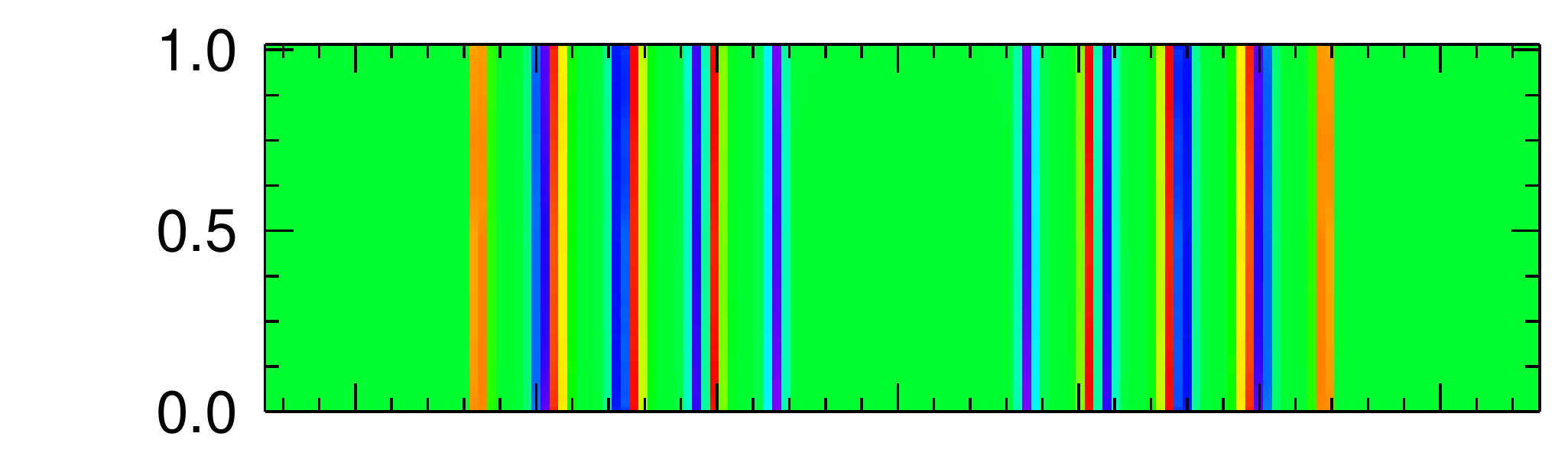}
\centering\includegraphics[scale=0.35, trim=1.0cm 6cm 0.75cm 6cm, clip=true]{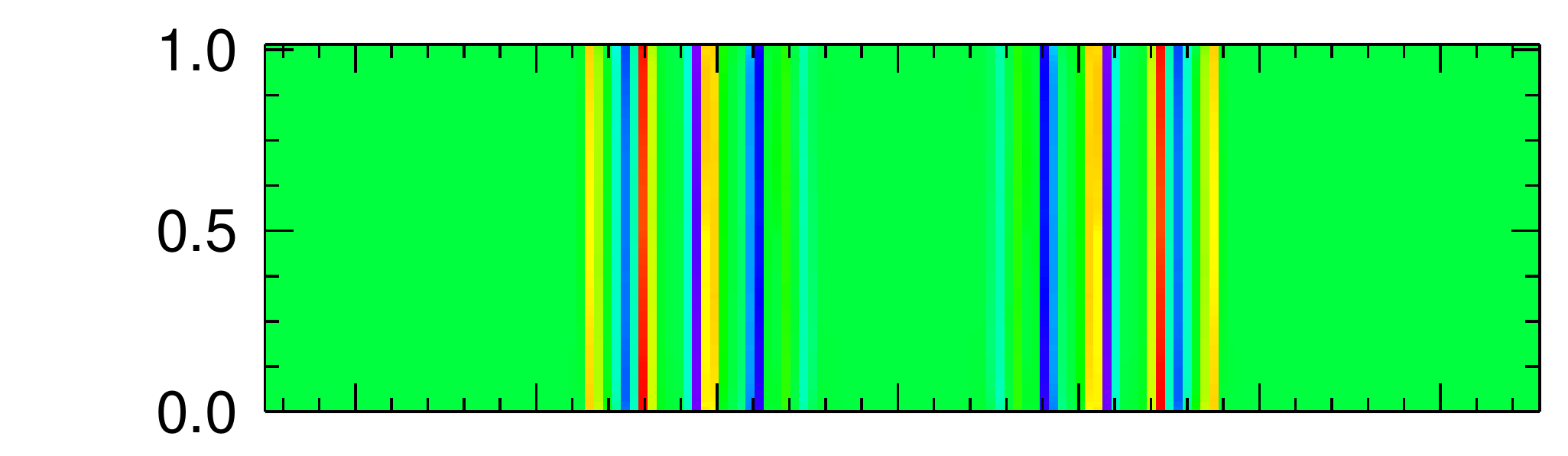}
\centering\includegraphics[scale=0.35, trim=1.0cm 6cm 0.75cm 6cm, clip=true]{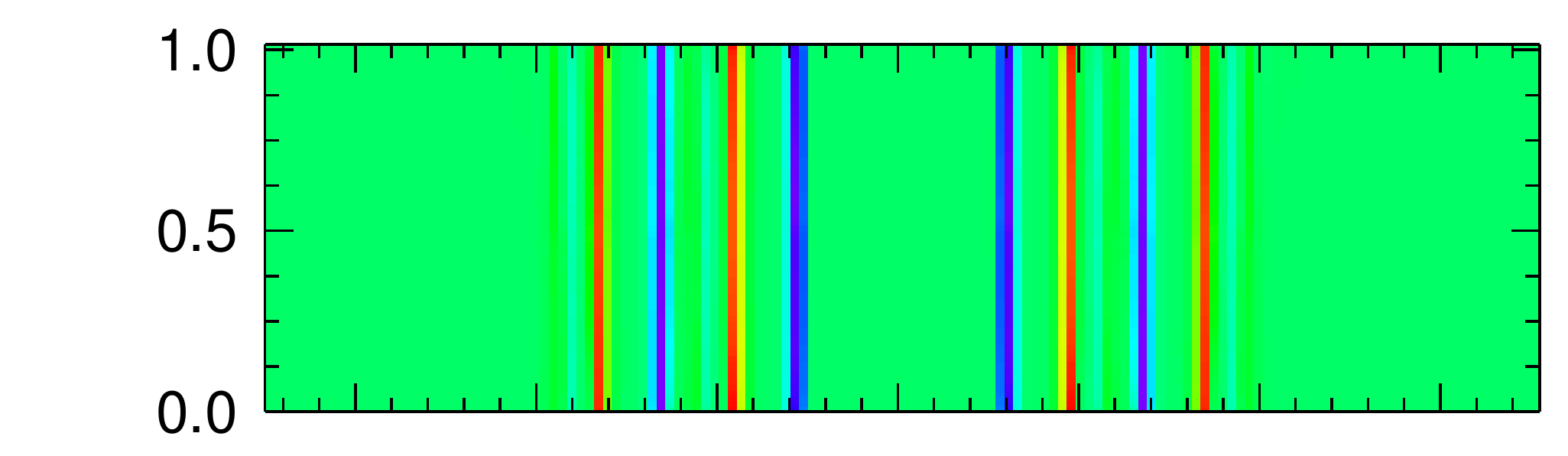}
\centering\includegraphics[scale=0.35, trim=1.0cm 3cm 0.75cm 6cm, clip=true]{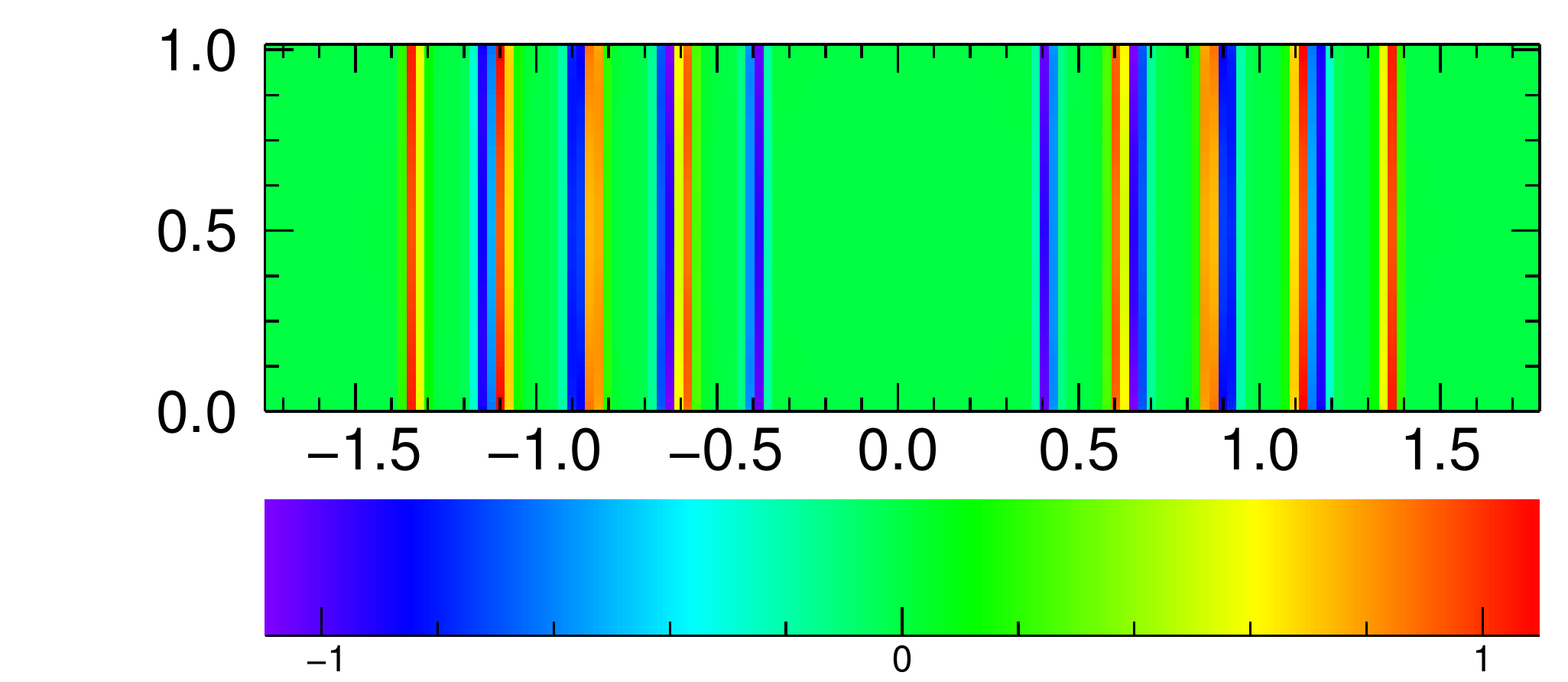}
\caption[Twist field at early times]{Vertical plane cuts of the twist component of the magnetic field, $B_\phi$ (color shading), after one Alfv\'en time ($t = 1$) at angles $\phi=$ $0^\circ$, $30^\circ$, $45^\circ, 60^\circ$, and $90^\circ$ (with respect to the $y$ axis). Red/yellow (blue/teal) represents field pointing in the $+\hat{\phi}$ ($-\hat{\phi}$) direction. At this early time, each pair of rotation cells hosts a clearly delineated individual flux tube.
\label{fig:early}}
\end{figure*}

In Figure \ref{fig:early} we plot the twist component of the magnetic field, i.e., the azimuthal ($\hat{\phi}$) component, $B_\phi$, at azimuthal angles $\phi=0^\circ$, $30^\circ$, $45^\circ$, $60^\circ$, and $90^\circ$ after one Alfv\'en time, $t=1$. Red/yellow (blue/teal) contours show where the field points in the $+\hat{\phi}$ ($-\hat{\phi}$) direction. At this early time, the twist field is concentrated along the boundaries of the individual flux tubes. At the contact point between any two adjacent tubes, all of which are twisted in the same sense, the twist field switches sign. Due to the nonaxisymmetric layout of the velocity pattern, the number of contact points varies with the angle at which the cut is taken. The cut at $\phi=0^\circ$ passes through the centers of two flux tubes between the CH and the PIL on both sides, the cuts at $\phi=30^\circ$ and $90^\circ$ pass through the centers of four tubes, and the cuts at $\phi=45^\circ$ and $60^\circ$ pass through the centers of two tubes and parts of two others. \par

\begin{figure*}[!h]
\centering\includegraphics[scale=0.35, trim=1.0cm 6cm 0.75cm 6cm, clip=true]{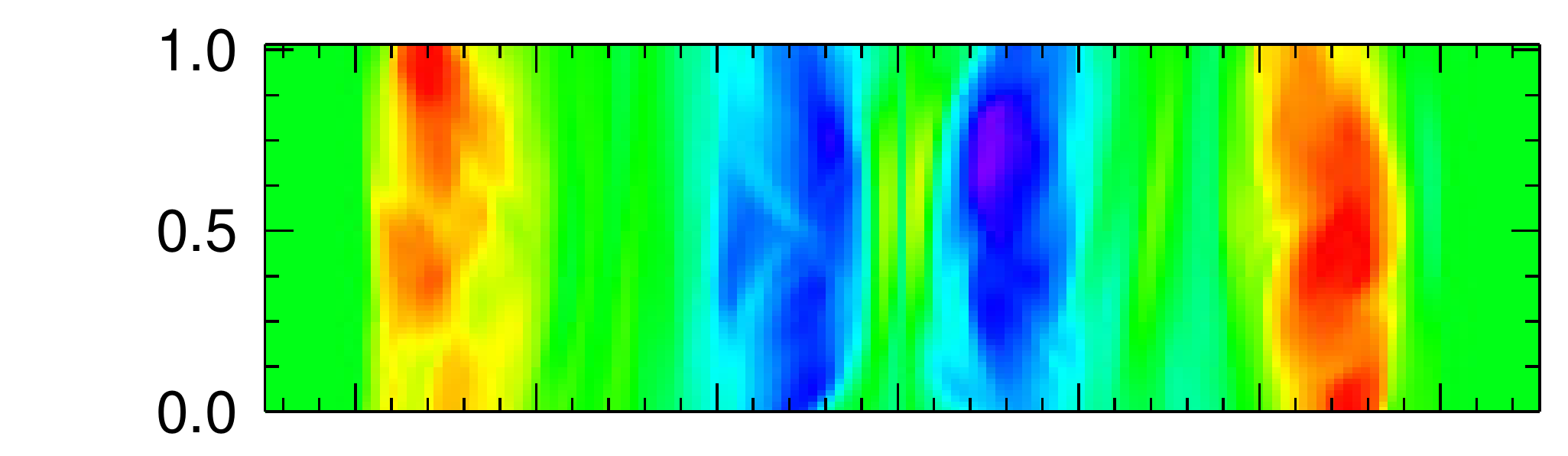}
\centering\includegraphics[scale=0.35, trim=1.0cm 6cm 0.75cm 6cm, clip=true]{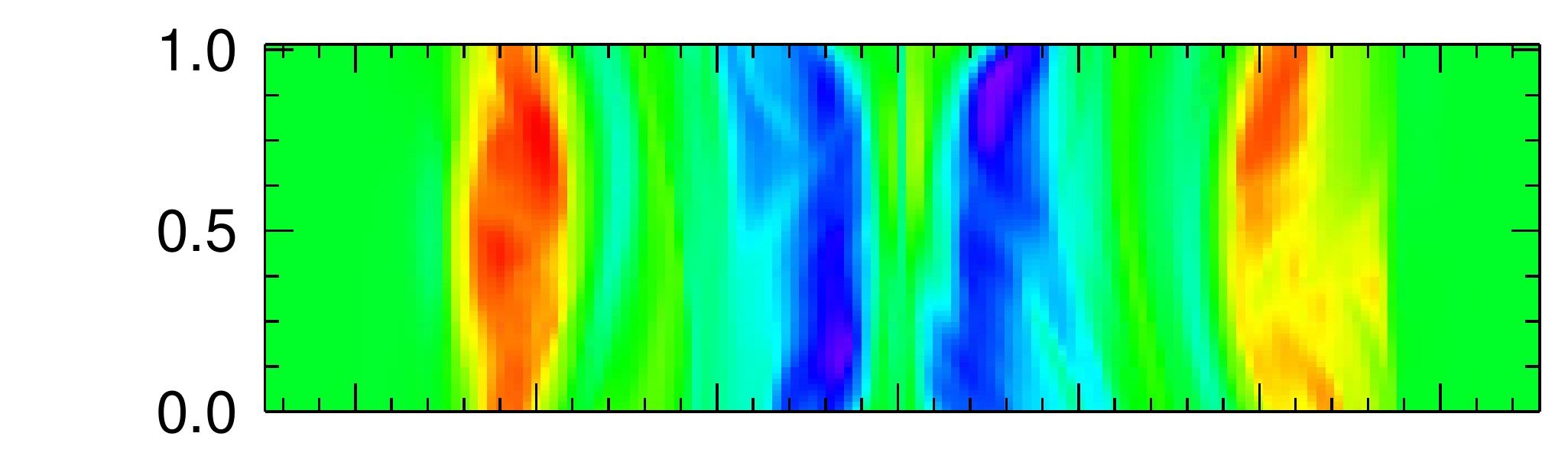}
\centering\includegraphics[scale=0.35, trim=1.0cm 6cm 0.75cm 6cm, clip=true]{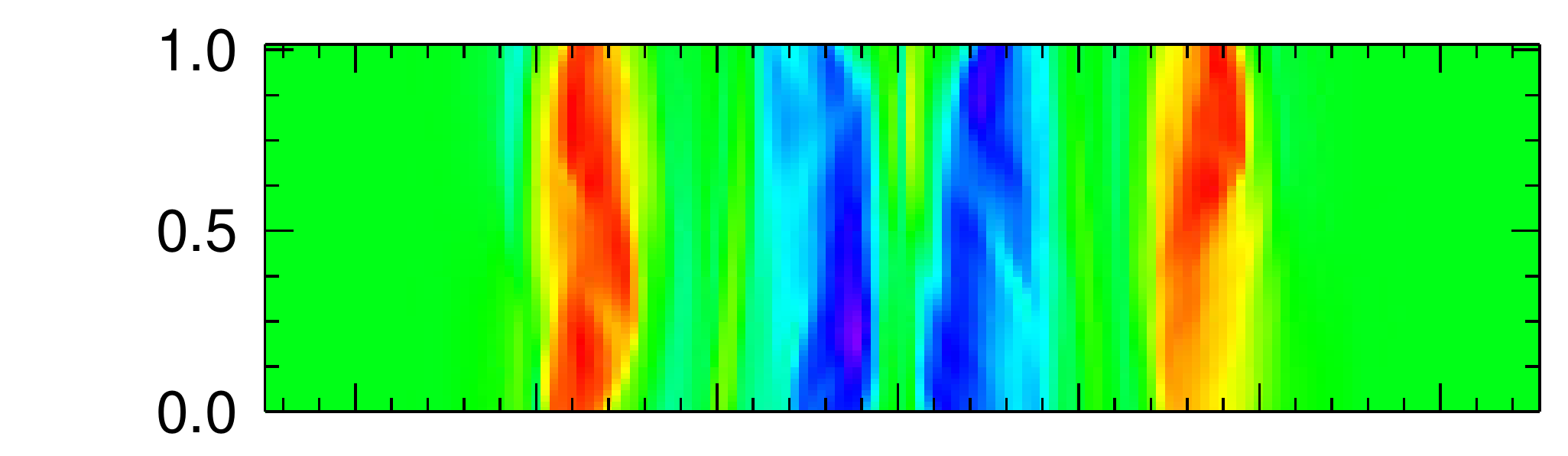}
\centering\includegraphics[scale=0.35, trim=1.0cm 6cm 0.75cm 6cm, clip=true]{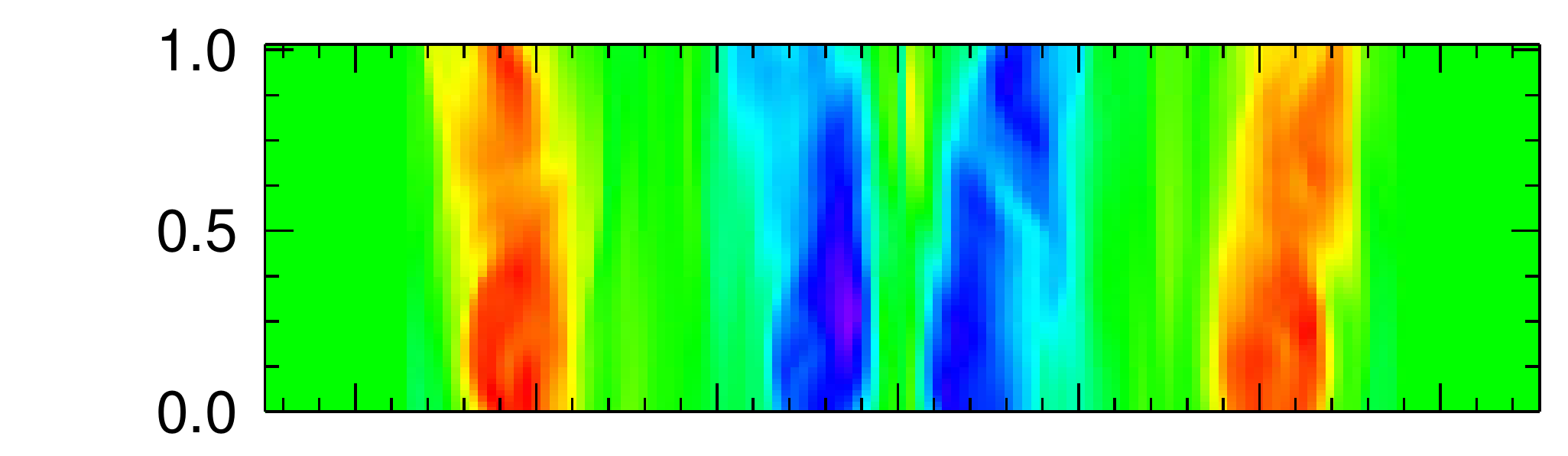}
\centering\includegraphics[scale=0.35, trim=1.0cm 3cm 0.75cm 6cm, clip=true]{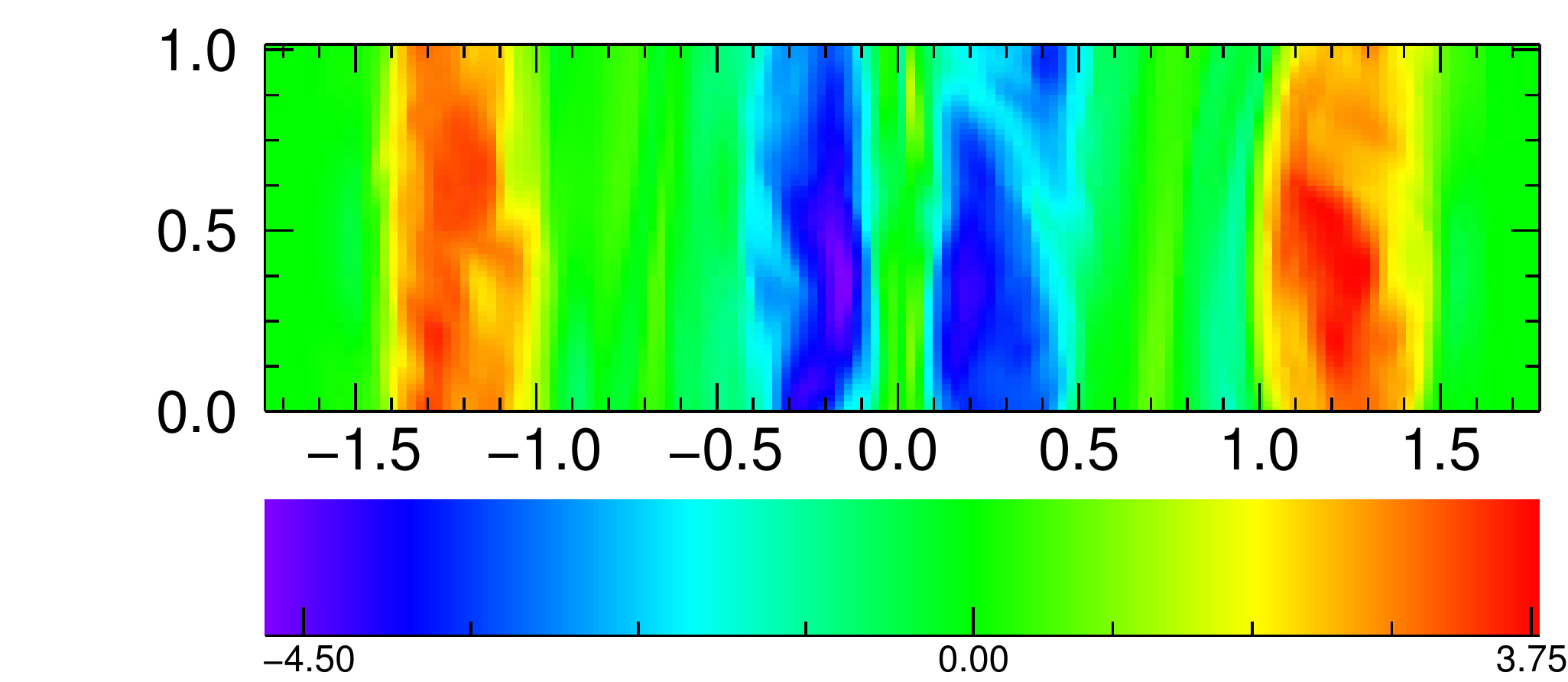}
\caption[Twist field at late times]{Same vertical cuts as in Figure \ref{fig:early} at the end of the simulation. At this late time, the twist flux has condensed to the boundaries of the flux system, with opposite signs at the PIL (outer stripes, red/yellow) and CH (inner stripes, blue/teal).
\label{fig:late}}
\end{figure*}

Figure \ref{fig:late} shows the same angle cuts after 21 twist cycles and 5 relaxation cycles. By this time, magnetic reconnection between flux tubes has allowed the magnetic helicity to condense at the boundaries of the system: the PIL and the CH, as shown schematically in Figure \ref{fig:condensation}. Note that, regardless of how many rotation cells are bisected by these cuts, the final distribution of twist field contains just two bands of opposite polarity between, and concentrated near, the PIL and the CH. All of these features are consistent with the helicity condensation model \citep{Antiochos13}. \par
\subsubsection{Spatial Distribution}\label{sec:distribution}

\begin{figure}[!h]
\centering\includegraphics[scale=0.55]{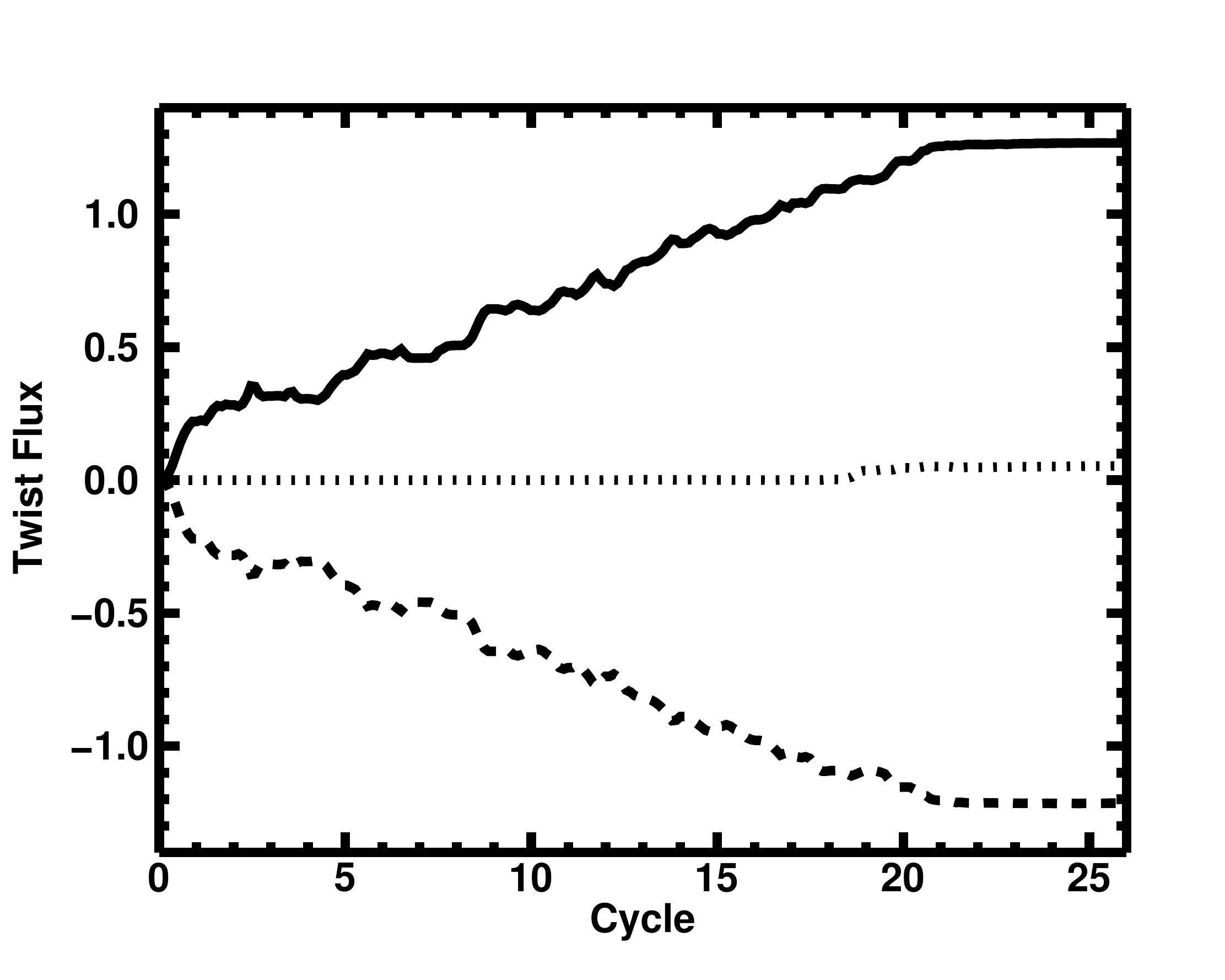}
\caption[Twist fluxes]{Positive and negative twist fluxes, $\Phi_{tw}^+$ and $\Phi_{tw}^-$, from Equation (\ref{twistfluxes}), along the half plane $y=0$, $z\ge0$ ($\phi=0^\circ$). The solid (dashed) curve represents the positive (negative) twist flux, which accumulates at the outer (inner) boundary of the hexagonal flow pattern adjacent to the PIL (CH). The dotted curve is the sum of the two, i.e., the net twist flux through the half plane.  Similar results were obtained on half planes at other angles $\phi$ shown in Figures \ref{fig:early} and \ref{fig:late}.
\label{fig:signedflx}}
\end{figure}

To understand better the quantitative features of our results, we now develop some measures of the spatial distribution of the twist flux through the half plane $y=0$, $z\ge0$. In Figure \ref{fig:signedflx} we plot the signed twist fluxes, $\Phi_{tw}^+$ (solid curve) and $\Phi_{tw}^-$ (dashed curve), 
\beg{twistfluxes}
\begin{split}
{\Phi_{tw}^+} &= \int_0^{L_x} dx \int_0^{L_z} dz \; {B_{tw}^+}, \\
{\Phi_{tw}^-} &= \int_0^{L_x} dx \int_0^{L_z} dz \; {B_{tw}^-}.
\end{split}
\done
We also plot their sum, the net signed twist flux, $\Phi_{tw} = \Phi_{tw}^+ + \Phi_{tw}^-$ (dotted curve). The corresponding signed twist fields are defined by 
\beg{twistfields}
\begin{split}
B_{tw}^+ &= \frac{1}{2} \left( B_\phi + \vert B_\phi \vert \right) \ge 0, \\
B_{tw}^- &= \frac{1}{2} \left( B_\phi - \vert B_\phi \vert \right) \le 0.
\end{split}
\done

In this particular half plane, $B_\phi = B_y$. At all times, there are almost exactly equal amounts of positive and negative twist flux, owing to the symmetry of the imposed rotational flows and the uniformity of the vertical magnetic field, as we now show. By integrating the $y$ component of the induction equation, we obtain for the net signed twist flux, $\Phi_{tw}$, the result 
\beg{nettwistflux}
\begin{split}
\frac{d\Phi_{tw}}{dt} &= \int_0^{L_x} dx \int_0^{L_z} dz \; {\frac{\partial B_y}{\partial t}} \\
&= \int_0^{L_z} dz \; \left[ v_y B_x - v_x B_y  \right]_{x=0}^{x=L_x} \\
&\phantom{=} \; + \int_0^{L_x} dx \; \left[ v_y B_z - v_z B_y \right]_{z=0}^{z=L_z} \\
&\approx \int_0^{L_z} dz \; \left[  v_y B_x \right]_{x=0}^{x=L_x} \\
&= -2 \int_0^{L_z} dz \; \left[ v_y B_x \right]_{x=0} \\
&= 0.
\end{split}
\done
The $v_x$ terms vanish identically at the top and bottom $x$ boundaries, while the $v_y$ and $v_z$ terms vanish approximately at the inner and outer $z$ boundaries. Only the $v_y B_x$ integrals remain. 
In those integrals, $B_x$ is uniform, and although $v_y$ is antisymmetric at the top and bottom $x$ boundaries, it averages to zero along $z$. Hence, the net twist flux approximately vanishes. \par

To a very good approximation, therefore, throughout the simulation the signed twist fluxes obey 
\beg{twistfluxequality}
\begin{split}
\frac{d\Phi_{tw}^-}{dt} &= - \frac{d\Phi_{tw}^+}{dt}, \\
\Phi_{tw}^- &= - \Phi_{tw}^+.
\end{split}
\done
That is, the fluxes are opposite in sign and equal in magnitude. At early times in Figure \ref{fig:signedflx}, the twist fluxes increase together rapidly; at later times, the oppositely signed twist fields at the interface between the two flux tubes in the $y=0$, $z\ge0$ half plane largely cancel. This cancellation moderates the rise of the accumulated twist fluxes, leaving primarily the residual fluxes that aggregate at the inner and outer boundaries of the hexagonal pattern to contribute to $\Phi_{tw}^-$ and $\Phi_{tw}^+$, respectively. We observed qualitatively similar behaviors of the signed twist fluxes through all the other plane cuts shown in Figures \ref{fig:early} and \ref{fig:late}. Those results will be examined in more detail and analyzed quantitatively in the next subsection. \par

\begin{figure}[!h]
\centering\includegraphics[scale=0.45]{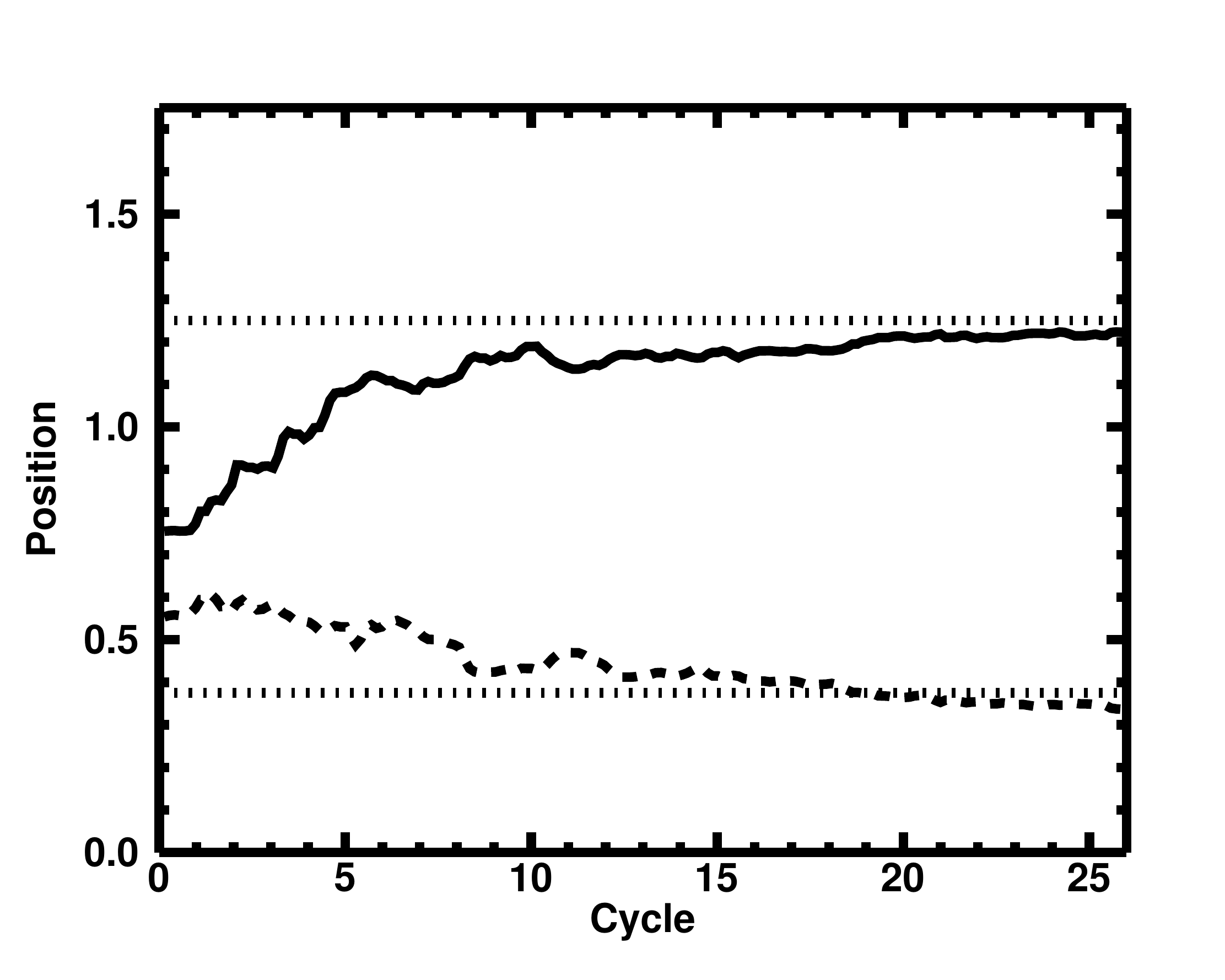}
\caption[Flux-weighted average positions]{Flux-weighted average positions, $\langle s_+ \rangle$ and $\langle s_- \rangle$, from Equation (\ref{weightpos}), of the signed twist fluxes shown in Figure \ref{fig:signedflx}. The solid (dashed) curve represents the position of the positive (negative) twist flux, which accumulates at the outer (inner) boundary of the hexagonal flow pattern adjacent to the PIL (CH). The dotted lines at $s = 3a_0 = 0.375$ and $s = 10a_0 = 1.25$ mark the boundaries of the hexagonal annulus of rotational flows. Similar results were obtained on half planes at other angles $\phi$ shown in Figures \ref{fig:early} and \ref{fig:late}.
\label{fig:weighted}}
\end{figure}

In Figure \ref{fig:weighted} we plot the flux-weighted positions, $\langle s_+ \rangle$ and $\langle s_- \rangle$, where $s = \sqrt{y^2+z^2}$, of the signed twist fluxes in Equation (\ref{twistfluxes}) 
for the same half plane $y=0$, $z\ge0$,
\beg{weightpos}
\begin{split}
\langle s_+ \rangle &= \frac{1}{\Phi_{tw}^+} \int_0^{L_x} dx \int_0^{L_z} dz \; {s B_{tw}^+}, \\
\langle s_- \rangle &= \frac{1}{\Phi_{tw}^-} \int_0^{L_x} dx \int_0^{L_z} dz \; {s B_{tw}^-}.
\end{split}
\done
At early times, the separation of the weighted positions is about $2r_i = 1.6a_0 = 0.20$, very close to the diametric separation of the locations of maximum angular velocity within each rotation cell. The absolute positions agree well with the observed average locations of the thin ribbons of twist flux in the first panel of Figure \ref{fig:early}. As time passes and reconnection enables the inverse cascade of magnetic helicity, on the other hand, the weighted positions of the signed fluxes migrate toward the boundaries of the hexagonal annulus, with the positive (negative) flux approaching the PIL (CH). The inner and outer limits of the hexagonal flow pattern, at radii $3a_0 = 0.375$ and $10a_0 = 1.25$, respectively, are indicated by the dotted lines in Figure \ref{fig:weighted}. Both are close to the calculated flux-weighted positions at late times in the simulation, which converge to approximately 0.30 and 1.20, respectively. Similar behaviors of the weighted positions were observed for the other plane cuts shown in Figures \ref{fig:early} and \ref{fig:late}. \par

\begin{figure}[!h]
\centering\includegraphics[scale=0.45]{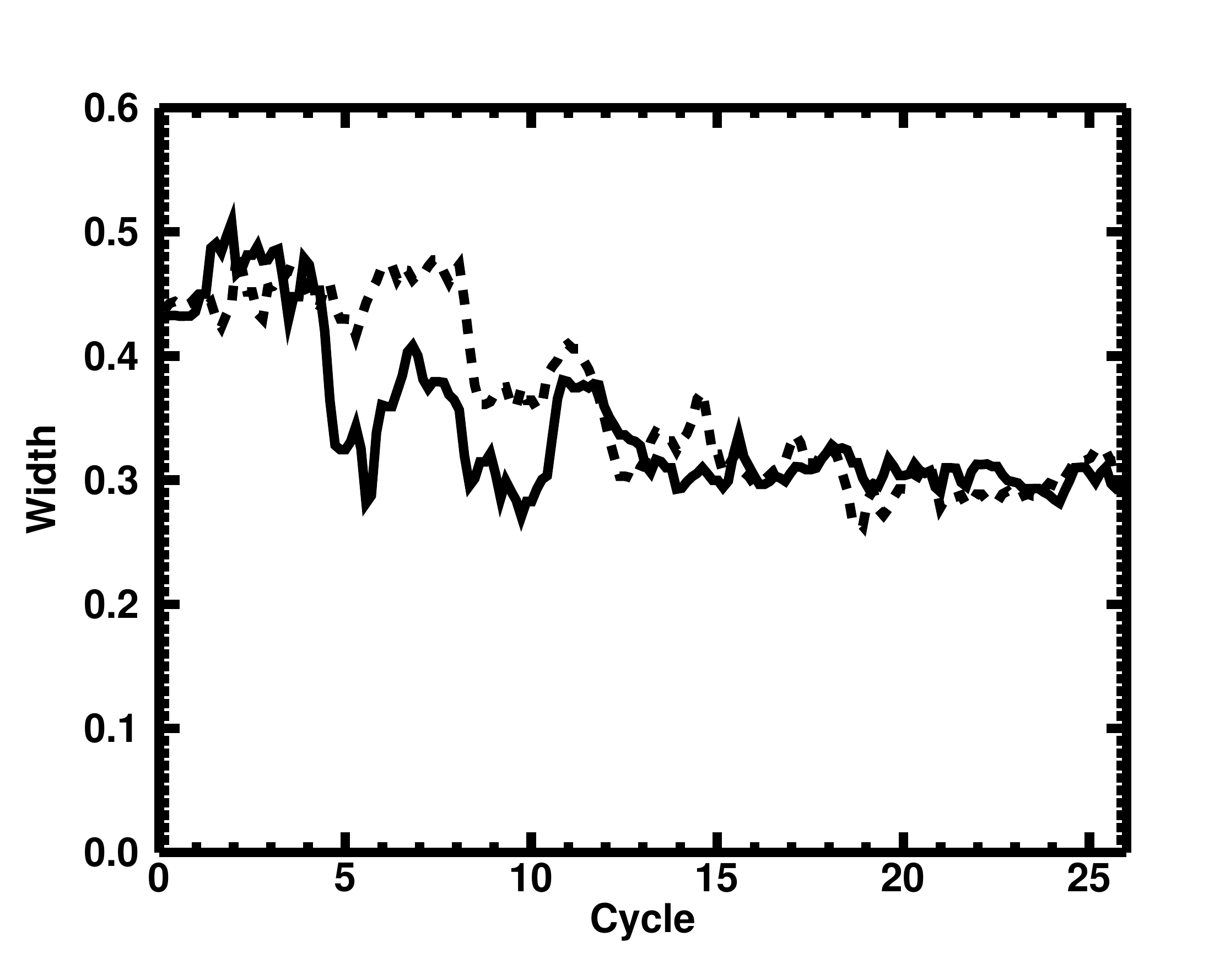}
\caption[Flux-weighted average widths]{Flux-weighted average widths, $\langle w_+ \rangle$ and $\langle w_- \rangle$, from Equation (\ref{weightwidth}), of the signed twist fluxes shown in Figure \ref{fig:signedflx}. The solid (dashed) curve represents the width of the positive (negative) twist flux, which accumulates at the outer (inner) boundary of the hexagonal flow pattern adjacent to the PIL (CH). Similar results were obtained on half planes at other angles $\phi$ shown in Figures \ref{fig:early} and \ref{fig:late}.
\label{fig:widths}}
\end{figure}

We also calculated the flux-weighted full widths, $\langle w_+ \rangle$ and $\langle w_- \rangle$, of the signed twist fluxes in the half plane, 
\beg{weightwidth}
\begin{split}
\langle w_+ \rangle &= 2 \sqrt{\langle s^2_+ \rangle - \langle s_+ \rangle^2}, \\
\langle w_- \rangle &= 2 \sqrt{\langle s^2_- \rangle - \langle s_- \rangle^2}. \\
\end{split}
\done
Here 
\beg{weightpossq}
\begin{split}
\langle s^2_+ \rangle &= \frac{1}{\Phi_{tw}^+} \int_0^{L_x} dx \int_0^{L_z} dz \; {s^2 B_{tw}^+}, \\
\langle s^2_- \rangle &= \frac{1}{\Phi_{tw}^-} \int_0^{L_x} dx \int_0^{L_z} dz \; {s^2 B_{tw}^-}.
\end{split}
\done
The results are shown in Figure \ref{fig:widths}. At early times, the widths reflect the separation between the concentrated bands within the two flux tubes, which is $2\sqrt{3}a_0 = 0.43$ for both the positive and negative fluxes in this $y = $ plane cut.  At late times, the widths measure primarily the radial extent of the condensed bands of twist flux near the PIL and CH, respectively. The late-time widths are essentially equal to each other, and have decreased to about 0.30. Note that, on average, the width decreases steadily with time, but appears to be close to saturation by the end of the 21 rotations. Furthermore, we find a similar behavior for a simulation with only $N = 30$ flux tubes, indicating that the final width of the condensed region is independent of its radial position. Note also that by the end of the simulation, the width approximately equals the diameter of our rotation cells, $2a_0 = 0.25$, as conjectured by \citet{Antiochos13}. We observed similar behaviors of the weighted widths for the other plane cuts shown in Figures \ref{fig:early} and \ref{fig:late}. \par

\subsubsection{Accumulation Rate}\label{sec:rate}

\begin{figure}[!h]
\centering\includegraphics[scale=0.55]{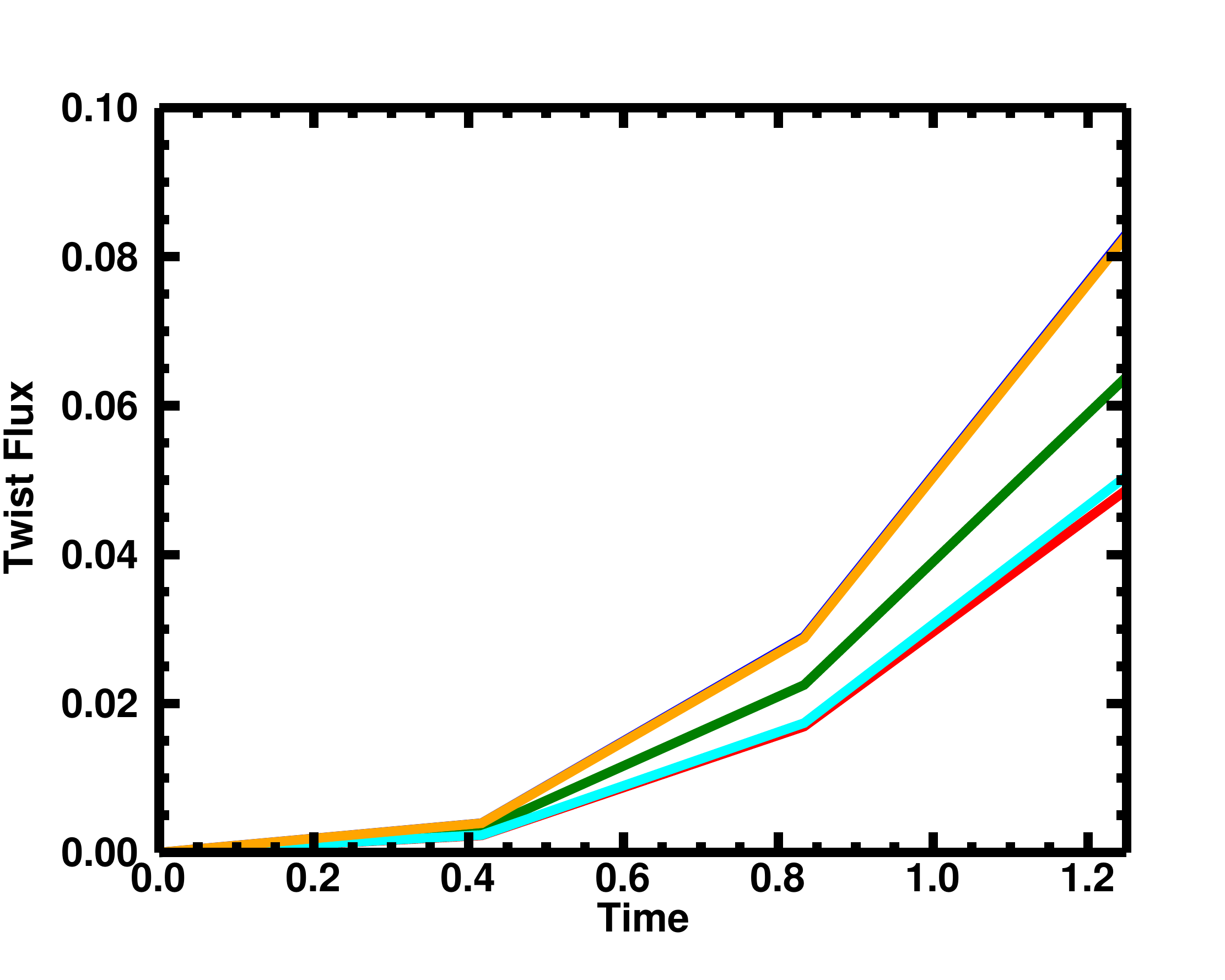}
\caption[Positive twist flux during first cycle]{Positive twist flux $\Phi_{tw}^+$ early in the first twist cycle through vertical half planes at azimuthal angles $\phi = 0^\circ$ (red), $30^\circ$ (blue), $45^\circ$ (green), $60^\circ$ (cyan), and $90^\circ$ (orange). The individual curves reflect the number of flux tubes cut by the planes early in the simulation.
\label{fig:twistflx1}}
\end{figure}

In analogy to Equation (\ref{twistfluxes}) for the $\phi=0$ half plane, we calculated the signed twist fluxes versus time for each of the cuts shown in Figures \ref{fig:early} and \ref{fig:late}. Results for the positive twist flux, $\Phi_{tw}^+$, through the half planes early in the first twist cycle are displayed as colored curves in Figure \ref{fig:twistflx1}. The rates of flux increase vary significantly among the different planes, owing to changes in the number of flux tubes that are cut through versus the azimuthal angle $\phi$ (cf.\ Figure \ref{fig:early}). The fluxes along the planes at 30$^\circ$ (blue) and 90$^\circ$ (orange) represent four flux tubes each and rise fastest; the flux along the 0$^\circ$ plane (red) represents two flux tubes and rises slowest, while that along the 60$^\circ$ plane (cyan) is similar; and the flux along the 45$^\circ$ plane (green) rises at an intermediate rate. 
Prior to reconnection setting in, the evolution of the twist flux is perfectly ideal. Inspection of the plane cuts like those in Figure \ref{fig:early} reveals that reconnection of opposite-polarity fluxes from adjacent flux tubes begins to occur at around $t=1.12$ at 30$^\circ$, but does not yet commence at this time at 90$^\circ$. There also is evidence for buckling of some of the flux tubes, leading to partial penetrations of certain planes. Such features can be observed already at time $t=1$ in Figure \ref{fig:early} at 45$^\circ$ and 60$^\circ$. \par

\begin{figure}[!h]
\centering\includegraphics[scale=0.55]{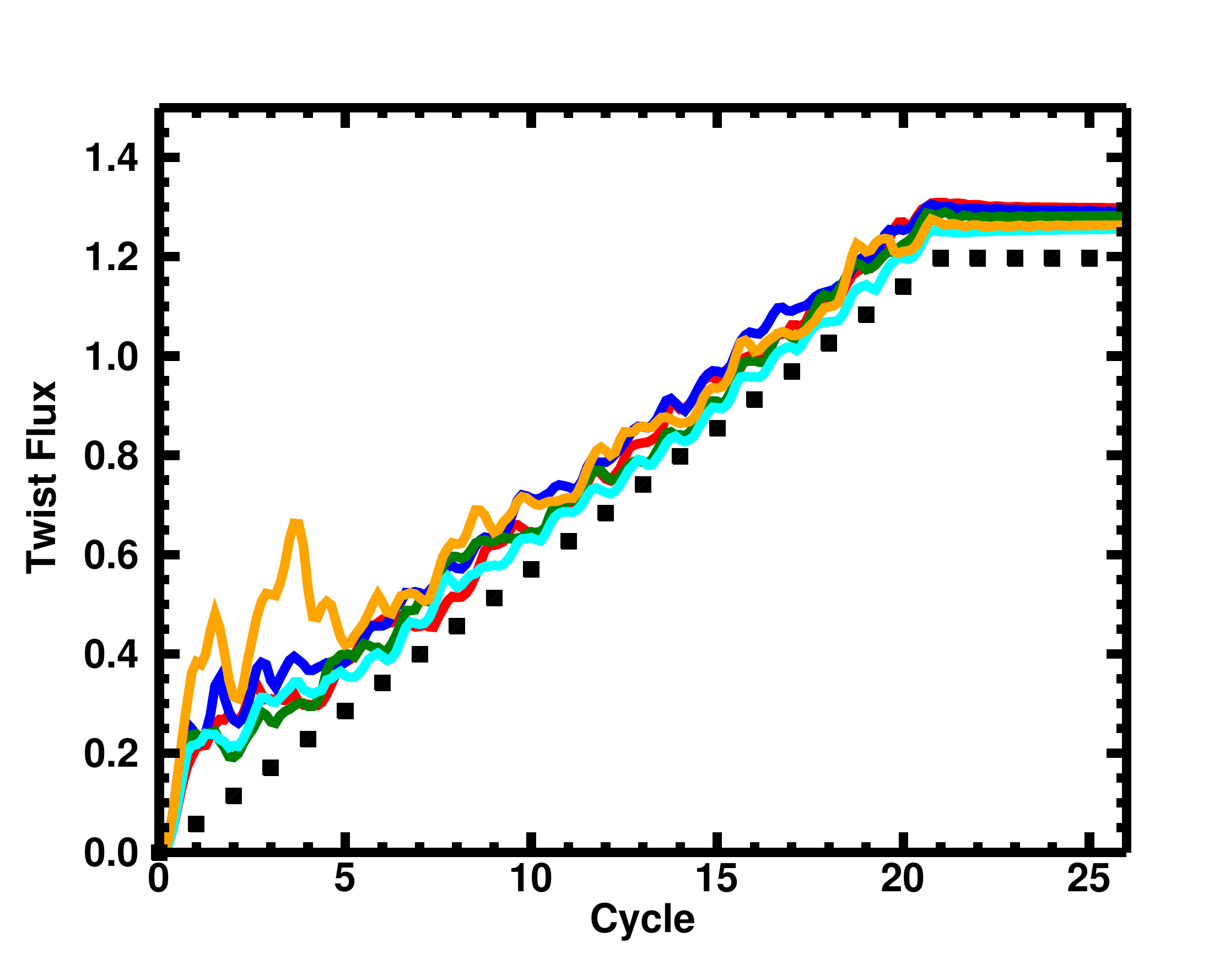}
\caption[Positive twist flux during simulation]{Positive twist flux $\Phi_{tw}^+$ versus twist cycle through vertical half planes at azimuthal angles $\phi = 0^\circ$ (red), $30^\circ$ (blue), $45^\circ$ (green), $60^\circ$ (cyan), and $90^\circ$ (orange). Also shown is the accumulated twist flux (filled squares) predicted by Equation (\ref{phitwist}), which is based on the helicity condensation model.
\label{fig:twistflx2}}
\end{figure}

In contrast to the early evolution of the twist flux shown in Figure \ref{fig:twistflx1}, which depends strongly upon azimuthal angle $\phi$, at late times in the simulation the twist flux is approximately independent of angle. This result, suggested qualitatively by Figure \ref{fig:late}, is confirmed quantitatively in Figure \ref{fig:twistflx2}. 
Clearly, the flux distribution is becoming more nearly azimuthally symmetric as reconnection transfers the twist fluxes toward the PIL and the CH, where they condense. We showed in \S \ref{sec:helicity} that, if the twist flux is concentrated in discrete bands that enclose a known amount of axial flux, then the helicity injection rate is given by Eq. (\ref{dhdttotal}). Here, that relationship takes the explicit form 
\beg{dhdtglobal1}
\begin{split}
\frac{dH}{dt} &= 2 \sum_i \langle \Psi_{ax} \rangle_{r_i} \frac{d\Phi_{tw}(r_i)}{dt} \\
&= 2 \left( \Phi_{CH} \frac{d\Phi_{tw,CH}}{dt} + \Phi_{PIL} \frac{d\Phi_{tw,PIL}}{dt} \right).
\end{split}
\done
Because the twist fluxes are equal in magnitude and opposite in sign, as shown in \S \ref{sec:distribution}, we have 
\beg{dtwdtglobal}
\frac{d\Phi_{tw,CH}}{dt} = - \frac{d\Phi_{tw,PIL}}{dt}.
\done
The preceding relation (\ref{dhdtglobal1}) then simplifies to 
\beg{dhdtglobal2}
\frac{dH}{dt} = 2 \left( \Phi_{PIL} - \Phi_{CH} \right) \frac{d\Phi_{tw,PIL}}{dt}.
\done
\par

As described in \S \ref{sec:helicity}, the total helicity injection rate is simply the sum of all of the individual rates within the $N = 84$ flux tubes, so
\beg{HinjTotal}
\frac{dH}{dt} = N {\frac{dH_{f}'}{dt}}.
\done
In addition, the total axial flux in the annular region between the CH and the PIL is approximately the sum of the individual axial fluxes within the $N$ flux tubes that essentially fill the region, so we also have 
\beg{FluxTotal}
\Phi_{PIL} - \Phi_{CH} \approx N \Phi_f.
\done
Here $\Phi_f = \pi a_0^2 B_0$ is the total axial magnetic flux per tube. Substituting from the above two expressions into Equation (\ref{dhdtglobal2}), and solving for the twist-flux generation rate, we obtain 
\beg{dphidt}
\frac{d\Phi_{tw,PIL}}{dt} = \frac{1}{2 \Phi_f} \frac{dH_{f}'}{dt}.
\done
Thus, we find that the helicity condensation model predicts that the rate of twist flux accumulation at the boundaries of the flux system equals the effective rate of twist flux generation at the outer edge of an individual flux tube. \par

The helicity injection rate per tube is given by Equation (\ref{dhdtanalytic}), hence, the rate of generation of twist flux at the PIL is 
\beg{dphidt2}
\frac{d\Phi_{tw,PIL}}{dt} \approx \frac{1}{12} \Omega_0 a_0^2 B_0 f(t).
\done
The average twist-flux generation rate and the amount of twist flux generated in one twist cycle are, respectively, 
\beg{avgdphidt}
\begin{split}
\langle \frac{d\Phi_{tw,PIL}}{dt} \rangle &\approx \frac{1}{24} \Omega_0 a_0^2 B_0 \approx 1.7 \times 10^{-2}, \\
\Delta \Phi_{tw,PIL} &\approx \frac{1}{24} \Omega_0 a_0^2 B_0 \tau \approx 5.8 \times 10^{-2}.
\end{split}
\done
These values are somewhat smaller than the realized rate of twist flux generation in the individual flux tubes, given in Equations (\ref{avgdtwdt}) and (\ref{changetw}), by the numerical coefficient 
$15/24=0.625$. This is due to the concentration of the twist flux in the interior of the flux tubes by the assumed angular rotation profile. \par

Now integrating Equation (\ref{dphidt2}) over time, we obtain for the accumulated twist flux
\beg{phitwist}
\Phi_{tw,PIL}(t) = 5.8 \times 10^{-2} \left[t - \frac{\tau}{2\pi} \sin \left(2\pi\frac{t}{\tau}\right) \right].
\done
Therefore, the total twist flux accumulated at the PIL at the end of 21 twist cycles is predicted to be $\Phi_{tw,PIL} = 1.22$. This is in excellent agreement with our numerical results, as shown by the data points (filled squares) from the analytic expression, Equation (\ref{phitwist}), included in Figure \ref{fig:twistflx2}. \par

We have further confirmed the prediction of Equation (\ref{dphidt}) that the rate of twist flux accumulation is equivalent to that at the edge of an individual flux tube. A second simulation was performed that included only $N=30$ flux tubes, corresponding to a much smaller polarity region, in which we removed the outermost two layers of rotation cells shown in Figure \ref{fig:initial}. The late-time twist-flux accumulation rate agreed very well with that shown in Figure \ref{fig:twistflx2}. \par

\subsubsection{Magnetic Shear}\label{sec:shear}

A key indicator of the amounts of free energy and helicity contained in the magnetic field of a filament channel is its shear. Observationally, a standard way to quantify the shear is to measure the angle between a field line of the channel and the PIL. Because our model configuration lacks a true PIL, however, this measurement cannot be made. Instead, we determine the displacement of the footpoints of the magnetic field lines along our equivalent PIL. We adopt this quantity, which is zero for the initial potential field, as our proxy for the magnetic shear. \par

The equation describing the trajectory of a magnetic field line in space is 
\beg{fieldlineeq1}
\frac{d\vecr}{d\ell} = \frac{\vecB}{B},
\done
where $\ell$ denotes the length along the field line. Assuming that the magnetic field is approximately cylindrically symmetric, the field-line equation can be rewritten 
\beg{fieldlineeq2}
\frac{\rho d\phi}{dx} = \frac{B_\phi(\rho)}{B_x(\rho)},
\done
where $\rho$ is the cylindrical radial coordinate. This equation can be integrated immediately to obtain the angular displacement $\Delta \phi$ from one footpoint to the other,
\beg{shearangle1}
\Delta \phi (\rho) = \frac{B_\phi(\rho)}{B_x(\rho)} \frac{L_x}{\rho}.
\done
We evaluate this expression at the center of the band of twist flux at radius $\rho=a$, 
\beg{shearangle2}
\Delta \phi (a) = \frac{B_\phi(a)}{B_x(a)} \frac{L_x}{a}.
\done
This can be recast into a relationship between the twist and axial fluxes in the band, 
\beg{shearflux1}
\begin{split}
\Phi_{tw} &\approx w L_x B_\phi(a), \\
\Phi_{ax} &\approx 2\pi a w B_x(a),
\end{split}
\done
respectively, where $w$ is the width of the band. The result is 
\beg{shearangle3}
\Delta \phi (a) \approx 2\pi \frac{\Phi_{tw}}{\Phi_{ax}}.
\done
This expression is just the time integral of Equation (\ref{dPhitw2}): the axial flux in the band, $\Phi_{ax}$, is simply the difference between the total enclosed axial fluxes, $\Psi_{ax}$, at the outside and inside of the band, and $\Omega$ is the effective angular rotation rate associated with the angular displacement $\Delta \phi$. \par

\begin{figure}[!h]
\centering\includegraphics[scale=0.55]{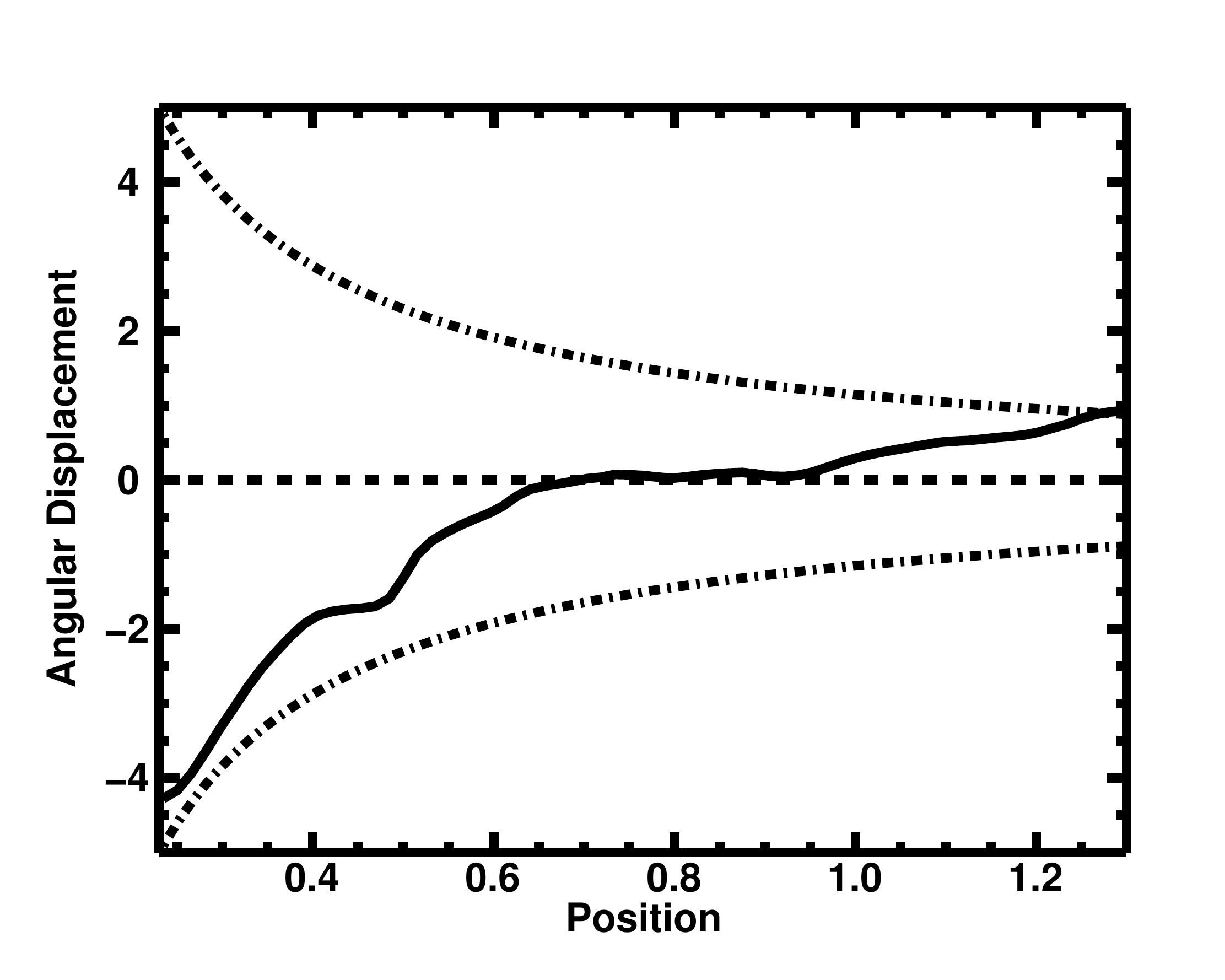}
\caption[Shear angular displacement]{The measured shear angular displacement $\Delta \phi$ from Equation (\ref{shearangle1}) in the half plane $y=0$, $z\ge0$ (solid curve), in the region between the two twist bands. Also shown (dot-dash curves) is the predicted displacement from Equation (\ref{shearangle4}), which is based on the helicity condensation model. Untwisted field corresponds to no angular displacement (dashed line).
\label{fig:shearangle}}
\end{figure}

We can estimate the displacement in Equation (\ref{shearangle3}) by substituting the initial uniform field value, $B_0$, for the axial field strength, $B_x(a)$. After using $B_0=\sqrt{4\pi}$, $\Phi_{tw}=1.22$, 
$w=0.3$, and $L_x=1$ in the above equations, we obtain
\beg{shearangle4}
\Delta \phi (a) \approx \frac{1.15}{a}.
\done
The predicted angular displacements are 0.95 (55$^\circ$) and 3.80 (220$^\circ$) at the PIL and CH ($a$ = 1.2 and 0.3), respectively. Both values agree quite well with the field-line rotations observed in Figure \ref{fig:late}. More generally, we plot as the solid curve in Figure \ref{fig:shearangle} the displacement $\Delta \phi (\rho)$, from Equation (\ref{shearangle1}), in the $y=0, z\ge0$ half plane. For comparison, the prediction from Equation (\ref{shearangle4}) is displayed as the dashed curves for both positive and negative twist fields. The agreement between the predicted and observed shear displacements clearly is very good. \par
\subsection{Application to Filament Channels}\label{sec:channels}

We now use the above results to determine the time scales for filament-channel formation and magnetic-helicity condensation in the solar atmosphere. To do this, we must assume some physical scale for the helicity injection. Let us take our model angular rotation constant $\Omega_0$ to be given by the vorticity $\omega_0$ associated with supergranular cells. Equating the vorticity to the average rotation rate of our cells, after integrating Equation (\ref{omega}) for the rotation profile we find 
\beg{vorticity}
\omega_0 = \frac{2}{a_0^2} \int_0^{a_0} dr \; r \Omega(r) = \frac{2}{15} \Omega_0.
\done
Substituting into Equation (\ref{dtwdtanalytic}) gives for the peak rate of twist-flux generation 
\beg{dtwdtchannel1}
\frac{d\Phi_{tw}}{dt} = \omega_0 a_0^2 B_0.
\done
Taking the scale to be that of a typical supergranule radius, $a_0 = 1.4 \times 10^9$ cm, and the angular velocity to be a typical supergranular flow speed, $\omega_0 a_0  = 5 \times 10^{4}$ cm s$^{-1}$, we find that 
\beg{dtwdtchannel2}
\frac{d\Phi_{tw}}{dt} = 7 \times 10^{13} B_0 \; {\rm Mx\; s}^{-1}.
\done
If we assume that the twist flux is generated in a semicircular volume of width $w = 2a_0$, then the rate of accumulation of twist field is 
\beg{dbdtchannel}
\frac{dB_{tw}}{dt} = \frac{2}{\pi a_0^2} \frac{d\Phi_{tw}}{dt} = \frac{2}{\pi} \omega_0 B_0.
\done
Defining the channel formation time, $\tau_{fc}$, to be the time required for the twist field $B_{tw}$ to acquire the same strength as the vertical field $B_0$, we obtain 
\beg{tchannel}
\tau_{fc} = \frac{\pi}{2} \omega_0^{-1} = 4.5 \times 10^4 \; {\rm s},
\done
or approximately one-half day. This is more than fast enough to explain the formation and maintenance of filament channels \citep[e.g.,][]{Tandberg-Hanssen95,Anderson05,Mackay10}. We emphasize that this half-day estimate is only a lower limit on the formation time, because all of the rotational flows above were assumed to inject the same sense of twist into the corona. That is, we imposed a perfect helicity bias due to the small-scale motions and flux emergence.  In reality, there is only a hemispheric preference, so some non-negligible fraction is injecting the opposite sense of twist, slowing down the rate of shear buildup. Consider instead a 70\% bias (i.e., in each hemisphere 70\% of the injected helicity is of one sign and 30\% of the other), which is typical of the observed filament bias. This would imply a net helicity injection rate in each hemisphere that is 40\% of the maximum. In this case the shear buildup time would increase by factor of 2.5, to roughly one day, which is in somewhat better agreement with observations. \par

The peak rate of magnetic helicity injection per rotational cell, given by one-half of Equation (\ref{dhdtanalytic}), becomes 
\beg{dhdtsg1}
\frac{dH_{sg}}{dt} = \frac{5\pi}{8} \omega_0 a_0^4 B_0^2.
\done
Here, we used Equation (\ref{vorticity}) for the vorticity. Substituting for the fixed parameters, we obtain the numerical value 
\beg{dhdtsg2}
\frac{dH_{sg}}{dt} = 2.7 \times 10^{32} B_0^2 \; {\rm Mx}^2 \; {\rm s}^{-1}.
\done
Assuming that 50$\%$ of the solar surface is covered by closed magnetic flux where the injected helicity can be stored, we obtain for the full-Sun rate of helicity injection 
\beg{dhdtsol}
\frac{dH_{\odot}}{dt} = 2 \frac{R_{\odot}^2}{a_0^2} \frac{dH_{sg}}{dt} = 1.4 \times 10^{36} B_0^2 \; {\rm Mx}^2 \; {\rm s}^{-1}.
\done
Over the duration $3.5 \times 10^8$ s of the sunspot cycle, the total injected helicity is 
\beg{dhsol1}
\Delta H_{\odot} \approx 3.5 \times 10^{44} B_0^2 \; {\rm Mx}^2.
\done
Assuming an average flux density $B_0 = 10$ G \citep{Harvey93}, we find for the total helicity injected over the cycle 
\beg{dhsol2}
\Delta H_{\odot} \approx 3.5 \times 10^{46} \; {\rm Mx}^2.
\done
This result agrees well with the estimate by \citet{Antiochos13}. The value is similar to estimates of the total magnetic helicity expelled by the corona in the solar wind and as coronal mass ejections, and stored in the corona by the Sun's differential rotation \citep{DeVore00b}. \par

\section{Discussion}\label{sec:discussion}

In this Chapter, we modeled the injection and transport of magnetic helicity in a plane-parallel model of the solar corona \citep{Parker72} using helicity-conserving numerical simulations. Helicity was injected into the corona at the photospheric boundary by numerous rotation cells, emulating the small-scale injection due to flux emergence and convective flows. This helicity was then transported by magnetic reconnection throughout the coronal volume. We found that the reconnection cancelled the opposite-polarity twist fluxes from adjoining flux tubes twisted at the scale of the individual rotation cells, leaving a large-scale residual twist flux that condensed at the inner and outer boundaries of the region of imposed photospheric flows. These boundaries constituted the coronal hole (CH) and polarity inversion line (PIL) of our model system, and the twist fluxes condensing at the two locations were opposite in sign. All of these qualitative features of the simulation results agree well with the helicity condensation model developed by \citet{Antiochos13}. \par

The simple initial and boundary conditions imposed in our simulations also enabled us to quantify certain aspects of our results. Owing to the uniformity of the magnetic field and the rotation speeds in the model, the twist fluxes condensing at the CH and PIL were equal in magnitude, as well as opposite in sign (\S \ref{sec:distribution}). This leads to the prediction that the rate of accumulation of twist flux, into two bands localized at the CH and PIL boundaries, is equivalent to the rate of accumulation at the outer edge of a single rotation cell (\S \ref{sec:rate}). We calculated the latter exactly and used it to verify the simulation data. The resulting expression for the twist flux accumulation rate was converted into a prediction for the shear angular displacement of field lines rooted near the CH and PIL boundaries (\S \ref{sec:shear}), which we confirmed by comparing it with measured values from the simulation. We emphasize that all of these demonstrations rely heavily upon the conservation of magnetic helicity by our high-fidelity numerical-simulation model, ARMS \citep[e.g.,][]{DeVore08}. \par

Our results have important implications for filament-channel formation on the Sun. The simulations confirm that the inverse cascade of magnetic helicity from small to large scales indeed occurs, as has been found in other MHD simulations generally \citep{Biskamp93} and postulated for the helicity condensation model specifically \citep{Antiochos13}. This cascade culminates in the twist flux accumulating at the PIL, yielding a global magnetic shear concentrated at the location where filament channels are known to form \citep{Martin98,Gaizauskas00}. The quantitative application of our analytical and numerical results to filament channel formation (\S \ref{sec:channels}) yields numbers for the accumulation of twist flux, helicity, and magnetic shear that are generally in accord with solar observations. We note that, as in the flux emergence and flux cancellation models and in agreement with observations \citep{Parenti14}, helicity condensation predicts that shear should continue to build up at the PIL until it is eventually ejected by eruption. Our model does not explicitly constrain the eruption mechanism. However, as discussed at length in \citet{Antiochos13}, helicity condensation does have important implications for eruption: it predicts the formation of a sheared arcade rather than a twisted flux rope. Our results above are in full agreement with this prediction. Obviously, eruption is not possible in our simulations, because the field is confined between the two plates. We plan to perform calculations in a spherical open geometry and determine whether eruption eventually occurs, but such calculations are well beyond the scope of this Chapter.  \par

The simplified geometry of our system also caused twist flux to condense at the CH boundary, where it accumulated much like the twist flux at our PIL. Unlike coronal holes on the Sun, the field lines in our model CH are line-tied at both ends. Therefore, they are effectively closed, rather than open; they are not free to expand outward, but compress together at the center of our domain. These artificial constraints imposed by our simulation geometry can be alleviated by adopting a truly bipolar magnetic field above a single planar photosphere with an open top boundary. This extension of our current modeling is underway. The results presented here nevertheless demonstrate conclusively that, as predicted by \citet{Antiochos13}, the twist flux condensing at the CH boundary is opposite in sign to that at the PIL. Furthermore, it is opposite in sign to that generated by the rotational motions that are present in the center of solar coronal holes (but are omitted from the CH in our current study). Our follow-up simulations, therefore, will test rigorously the prediction of the helicity condensation model that the flux of magnetic helicity at the boundary of a coronal hole, carried away by the slow solar wind, should be of opposite sign to the helicity flux from the interior of the CH, carried away by the fast solar wind \citep{Antiochos13}. \par

Adopting a more realistic magnetic geometry with a true PIL also will introduce photospheric gradients of the vertical magnetic field into our system. The uniformity of the vertical field and of the rotation speeds in our current study implies that all of the twisted flux tubes are formed with equal twist fluxes. In a nonuniform vertical field, or in a region where the rotation speeds are nonuniform, this equality of the twist fluxes on adjacent tubes will break down. The resulting imperfect cancellation of adjacent twist fluxes is anticipated to lead to a broader distribution of the twist flux condensing at the PIL, compared to that formed in the simulations of this paper. Thus, an investigation of the effect of gradients in the background vertical magnetic field will be a part of our future studies. This also is true of other factors influencing the distribution of the filament channel flux, such as the size of the rotation cells at the photosphere. \par

The primary conclusion to be drawn from this paper is that the helicity condensation model developed by \citet{Antiochos13} is in excellent agreement with the results of our numerical experiments, both qualitatively and quantitatively. This is a very encouraging, albeit far from final, step forward toward demonstrating that the helicity condensation model explains the heretofore mysterious origin of solar filament channels. \par

\chapter{The Role of Helicity Injection in Determing Coronal Structure}\label{Structure}

\section{Introduction}\label{sec:intro}
A well-known feature of the solar magnetic field is the presence of filament channels at photospheric polarity inversion lines (PILs). These structures, situated in the upper chromosphere and lower corona, underlie and support the cool plasma that comprises prominences and filaments \citep{Martin98,Gaizauskas00}. Filament channels are highly sheared structures, containing tremendous amounts of free energy, that ultimately gets converted into kinetic and thermal energy of the plasma, as well as nonthermal particle energies when filament channels erupt in coronal mass ejections (CMEs). The shear inherent in the filament channels is a form of magnetic helicity, and filament channels are known as dextral if they have negative helicity and sinistral if they have positive helicity. Observations indicate that dextral (sinistral) filament channels dominate in the northern (southern) hemisphere \citep[e.g.][]{Martin92, Rust94b, Zirker97, Pevtsov03}. This hemispheric helicity rule has also been observed in quiet Sun magnetic fields \citep{Pevtsov01b}, sigmoids \citep{Rust96}, active region magnetic fields \citep{Seehafer90}, coronal mass ejections, and sunspot whorls \citep{Pevtsov14}. The strength of the preference ranges from about 55\% in active region filaments \citep{Martin94}, to $>80\%$ in quiescent filaments \citep{Pevtsov03} and does not seem to change with solar cycle \citep{Hale27, Martin94,Hagino02}. \par
\begin{figure*}[!h]
\centering\includegraphics[scale=0.65,trim=0.0cm 0.0cm 0.0cm 0.0cm,clip=true]{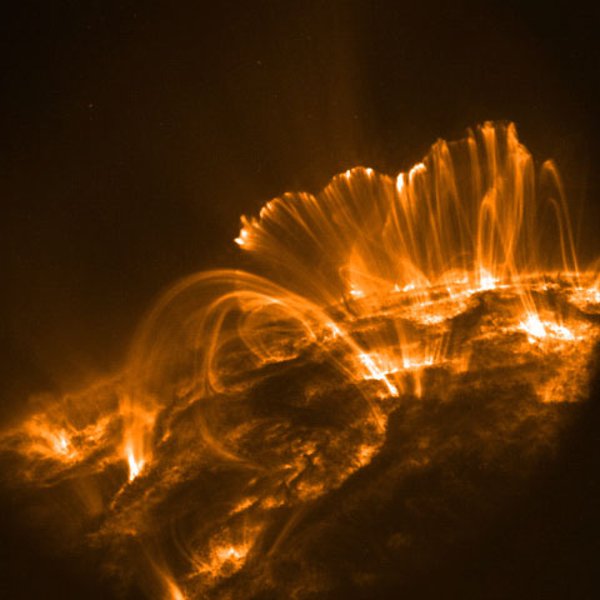}
\caption[Coronal loops as observed by TRACE]{Coronal loops as observed by TRACE, showing the quasi-potential structure, with little to no tangling.}
\label{fig:obs}
\end{figure*}

A second, seemingly unrelated, feature of the solar magnetic field is the preponderance of coronal loops in the closed field corona. These loops have been observed in high resolution in \emph{Transition Region and Coronal Explorer (TRACE)} XUV and X-ray images, such as the one in Fig. \ref{fig:obs}, where they are seen to be incredibly smooth and laminar, with little to no tangling \citep{Schrijver99}. In other words, there is very little magnetic helicity observed in these loops. The picture of the corona that emerges, therefore, is one in which magnetic helicity manifests itself at specific locations, namely above PILs, while leaving the rest of the corona generally smooth and quasi-potential.\par
This global organization of the magnetic field cannot be due to processes occurring in the low-$\beta$ corona; rather, all of the helicity must be injected at, or below, the photospheric level and the high conductivity of the corona ensures that no new helicity is injected above the photospheric footpoints. \emph{Solar Dynamics Observatory (SDO)} measurements of helicity injection into the coronal field indicate that shearing and twisting motions on the photospheric level dominate the flux emergence \citep{Liu12}, meaning that the helicity budget of the corona is primarily due to the jostling of the field at the surface, rather than due to new flux coming into the corona. Nevertheless, observations of photospheric convection show that these footpoint motions are extremely complex \citep{Schmieder14}. Convective cells appear randomly all over the photosphere, and have different lifetimes, of order minutes for granules, and days for supergranules \citep{Hirzberger08}. From the standpoint of helicity injection, however, the only flows that matter are the ones that twist the field. Compression of the field caused by converging flows is not expected to impart any helicity into the field, and so these flows cannot be responsible for the shear observed in filament channels. The flows that twist up the field, in contrast, do inject a net helicity into the corona. We therefore conclude that it is sufficient to model the helicity injection into the corona with simple twisting motions \citep{WilmotSmith10, Rappazzo13}. Such flows have routinely been observed in helioseismic measurements \citep{Duvall00,Gizon03,Komm07,Seligman14}. Vortical flows on the scale of granules \citep[e.g.][]{Bonet08, Bonet10, VD11, VD15} and supergranules \citep{Brandt88,Attie09} have also been observed. \par
Unfortunately, these considerations make it even harder to understand the simultaneous presence in the corona of both filament channels and coronal loops, since magnetic helicity is, apparently, injected throughout the solar photosphere, yet is almost nowhere to be found in the corona, except in filament channels. Understanding why this occurs has proven to be a challenge.\par
Recently, \citet{Antiochos13} presented a new model for the formation of filament channels, magnetic helicity condensation, based on the well-known inverse cascade of mangetic helicity, that answers this question. In this helicity condensation model, photospheric convection imparts helicity into the coronal field, and this helicity is then transported throughout the corona by magnetic reconnection, before accumulating in filament channels. Surface convection imparts the same sense of twist to adjacent flux tubes, which are then able to undergo component reconnection at their contact point. This component reconnection produces a single flux tube with an axial flux equal to the sum of the two original axial fluxes, but encircled by the same twist field present on each of the two original flux tubes. In this way the helicity, in the form of twist, inverse-cascades to larger and larger scales. The PIL forms a natural boundary of the flux system, so that when the twist reaches this boundary, it cannot proceed further, since all of the flux has already reconnected. The end result of this process is a mostly axial (untwisted) internal field, and a highly sheared (twisted) field at the PIL, precisely what is observed as a filament channel. At the same time, the untwisted internal field corresponds to the laminar coronal loops. In this way, the helicity condensation model provides a natural mechanism for the simultaneous formation of both highly sheared filament channels and relatively untwisted coronal loops. In this model, these two seemingly unrelated features of the solar atmosphere are actually created by the same process.\par 
This model was initially investigated by \citet{Zhao15}, who injected magnetic helicity into a plane-parallel Parker corona \citep{Parker72}, and found that for photospheric motions that inject the same helicity everywhere, filament channels will form at the PIL. Further, they found that randomizing the photospheric motions while keeping the same helicity injection rate did not qualitatively affect the accumulation of twist flux at the PILs. They also tested the effect of injecting helicity of opposite signs, and found that adjacent flux tubes could not reconnect due to their twist components being co-aligned.\par 
In our previous work, \citet{Knizhnik15}, we rigorously tested the helicity condensation model and found that the model not only qualitatively reproduced results consistent with properties of filament channels, but demonstrated that the inverse cascade of magnetic helicity due to magnetic reconnection produces a twist flux at the PIL that is in quantitative agreement with that predicted by the helicity condensation model. With this result, we estimated that with the helicity preference observed on the Sun, filament channels will form in about a day or so, in line with observations of filament channel formation \citep{Martin98, Gaizauskas00}. We showed that helicity condensation agreed both qualitatively and quantitatively with observed properties of filament channels, and that the process produced relatively untwisted coronal loops everywhere except at the PIL. These results, however, were obtained with an $100\%$ helicity rule, meaning that all of the helicity injected into the corona was of the same sign. An obvious question that can be raised is: what happens if a fraction of the injected helicity has the opposite sign? Indeed, this is a more realistic scenario since, as described above, the corona has a hemispheric helicity preference, rather than a rule, so that some helicity of the non-preferred sign is injected into the corona. \par 
While it stands to reason that injecting helicity of the opposite sign into the corona will simply slow down the helicity condensation process, this may not be as straightforward as it seems. The results of \citet{Zhao15} indicate that adjacent flux tubes will have difficulty reconnecting if they are twisted in opposite senses. As a result, the twist will be unable to inverse-cascade to larger scales as it needs to do in order to form filament channels and smooth coronal loops. Even if reconnection between the adjacent flux tubes is eventually achieved as a result of some ideal instability - such as kinking - driving the interaction, the twist cancellation will not be perfect, and some residual twist may remain in what would otherwise have been smooth, untwisted coronal loops. It is therefore important to test whether filament channels and coronal loops will form with both signs of magnetic helicity being injected into the corona.\par
In this paper, we study the effect of varying helicity preferences on the structure of the corona using helicity-conserving numerical simulations that inject helicity into a plane-parallel Parker corona, as was done in \citet{Knizhnik15} (hereafter KAD15). In this work, however, we vary the fraction of helicity of each sign that is injected into the corona. We study three cases: 1) $100\%$ of the injected helicity is positive, 2) $75\%$ ($25$\%) of the injected helicity is positive (negative), 3) $50$\% ($50$\%) of the injected helicity is positive (negative). To make the simulations as realistic as possible, we also randomize the pattern of helicity injection, and compare the simulations with a fixed pattern of helicity injection to those with a randomized pattern of helicity injection. \par
The paper is organized as follows. In \S \ref{sec:model} we discuss the setup of our numerical simulations, how magnetic helicity is injected into the domain and how the helicity preference is determined. In \S \ref{sec:Results} we discuss the results of our simulations, exploring both the formation of filament channels and the resulting smooth corona, and comparing the simulations with fixed and random patterns of helicity injection. Finally, in \S \ref{sec:discussion} we discuss our conclusions and the implications for future research. 

\section{Numerical Model}\label{sec:model}
The numerical model that we use is described thoroughly in KAD15, so here we merely summarize its essential features and refer readers to our previous paper for details. The new features of the simulations presented in this paper are described below.\par

\begin{figure}[!h]
\centering\includegraphics[scale=0.5,trim=0.0cm 0.0cm 0.0cm 0.0cm,clip=true]{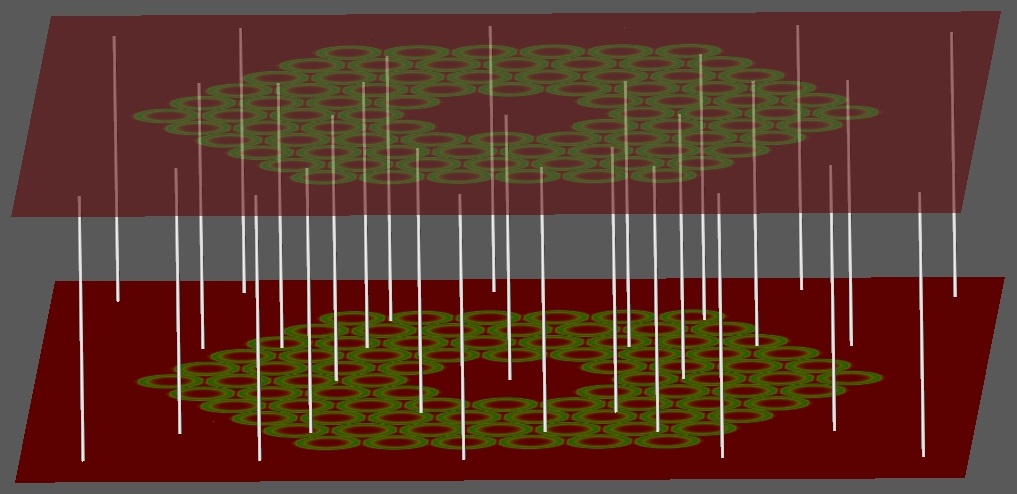}
\caption[Initial Simulation Setup]{Initial setup of our simulations. White lines represents the initial vertical magnetic field, and color shading represents velocity magnitude. The hexagonal pattern of the convective cells shown in this Figure will be shifted around randomly to emulate the random photospheric convection, and the sense of the individual convective cells will be flipped randomly.}
\label{fig:init}
\end{figure}

The numerical setup is shown in Figure \ref{fig:init}. Initially, the plasma density and temperature are uniform, and the magnetic field is uniform, straight, and vertical. The top and bottom plates present the photosphere bounding a plane-parallel corona, following \citet{Parker72}. Color shading on those plates in the figure represents velocity magnitude, highlighting the $N=84$ identical, cylindrically symmetric rotation cells that we construct on each plate in the hexagonal pattern shown. Each cell on the bottom plate is partnered by a cell on the top plate centered at the same horizontal position $(y,z)$. The rotational motions are in opposite directions at the top and bottom plates for each pair, imparting the same sense of twist to the field lines that link the cells. At the outset, the ensemble of field lines within each pair of cells represents a straight coronal loop whose apex is located in the center of the domain along the vertical ($x$) direction. The footpoint motions are ramped up and then down smoothly in successive cycles of fixed duration. Each cycle represents the lifetime of a single convective cell on the surface of the Sun, while the imposed rotational motion emulates the vortical twisting component of convection within such a cell.\par
We advance our system in time by solving the ideal, adiabatic magnetohydrodynamics equations in three Cartesian dimensions. The top and bottom boundaries $(x)$ are closed and line-tied, while the four side boundaries $(y,z)$ are open and free-slip. Our domain extent is $[0,L_x] \times [-L_y,L_y] \times [-L_z,L_z]$ with $L_x=1$ and $L_y=L_z=1.75$.\par

The magnetic helicity $H_0$ injected into one flux tube (i.e. one top/bottom pair of rotation cells) over one cycle is calculated in KAD15. If the sense of rotation is clockwise, the resulting positive helicity injected is 
\beg{helicity1cell}
H_0 = 2\times10^{-2}.
\done
Trivially, the net helicity $H_\Sigma$ injected into $N$ such flux tubes all twisted in the clockwise sense, as in KAD15, is 
\beg{NH1KAD15}
H_\Sigma = N H_0.
\done
In this paper, we generalize to cases in which $N_+$/$N_-$ cells rotate clockwise/counter-clockwise and inject positive/negative helicity, with $N_+ + N_- = N$. The net helicity injected into the corona then becomes 
\beg{helicityallN}
H_\Sigma = \left( N_+ - N_- \right) H_0.
\done
The case studied in KAD15 has $N_+ = N$ and $N_- = 0$, so Equation (\ref{helicityallN}) reduces to Equation (\ref{NH1KAD15}). If, on the other hand, $N_+ = N_-$, equal numbers of cells rotate in each sense and the net injected helicity vanishes, $H_\Sigma = 0$. In our simulations described below, we allowed $N_+$ and $N_-$ to vary from cycle to cycle. The net helicity injected into the corona after $M$ cycles therefore is 
\beg{Htotal}
H_\Sigma = \sum_{i=1}^{M} \left( N_{+,i} - N_{-,i} \right) H_0,
\done
where $N_{+,i}/N_{-,i}$ is the number of cells that rotate clockwise/counter-clockwise during cycle $i$.\par

In each simulation, we assign a probability $k$ that any individual top/bottom pair of cells injects positive helicity and $1-k$ that the pair injects negative helicity. A random number $\kappa_j \in [0,1]$ is generated for each pair of cells $j \in [1,N]$ during each cycle, and $\kappa_j$ is compared with $k$ to determine whether the sense of rotation is clockwise or counter-clockwise over that cycle. The helicity $H_j$ injected by the $j$th pair of cells is 
\beg{Hj}
H_j = H_0 \times 
\begin{cases}
+1 & \text{if } \kappa_j \le k; \\
-1 & \text{if } \kappa_j > k. \\
\end{cases}
\done
On average, the expectation is that during each cycle, a fraction $2k-1$ of the maximum positive helicity $N H_0$ will be injected into the corona, 
\beg{Hk}
\langle H_\Sigma \rangle = (2k-1) N H_0 = f N H_0.
\done
Throughout the paper, we will refer to $k$ as the {\rm helicity preference} of each simulation, and to $f = 2k-1$ as the {\rm net fractional helicity} associated with $k$.\par

For this paper, the cases $k=0.75$ and $k=0.5$ were simulated to complement the $k=1$ case previously presented in KAD15. For reference, the expectation values of the net helicity injected per cycle are 
\beg{helicityk}
\langle H_\Sigma \rangle = f N H_0 = N H_0 \times
\begin{cases}
1.0 & \text{if } k=1.0; \\ 
0.5 & \text{if } k=0.75; \\
0.0 & \text{if } k=0.5. \\
\end{cases}
\done
Below, we use the precise number of cells injecting each sign of helicity during each cycle to evaluate $H_\Sigma$ in Equation (\ref{Htotal}). That prediction is compared to the instantaneous value $H(t)$ calculated directly from the \citet{Finn85} volume integral for the relative magnetic helicity in the simulation, as described in KAD15.\par

The helicity preference $k$ introduces one aspect of randomness into our simulations through the assignment of a clockwise/counter-clockwise sense of rotation (and positive/negative helicity) to each pair of rotation cells during each cycle of rotation. In order to emulate the stochastically shifting spatial pattern of convection on the solar surface and investigate its effect on coronal structure, we also introduce a second aspect of randomness into a separate set of simulations. After each cycle of rotations, we displace the entire hexagonal cellular pattern shown in Figure \ref{fig:init} by a randomly chosen angle $\theta \in [0^\circ,60^\circ]$ about its central vertical axis $(y,z) = (0,0)$. For simplicity, the same angular displacement is applied to both the top and bottom plates, so that the top/bottom pairs of rotation cells remain aligned as before. Now, however, the random displacement means that the rotation cells will, in general, encompass parts of multiple neighboring flux tubes that were twisted during the previous cycle of rotations. The ensuing cycle therefore introduces braiding, as well as twisting, into the coronal magnetic field between the plates. As we will show, however, this displacement-induced braiding has no effect on the rate of helicity accumulation in the corona, and has only a minor influence on the smoothness of the induced magnetic structure. This result concurs with the qualitative conclusions of \citet{Zhao15} from a much simpler simulation setup and is analyzed quantitatively here for the first time.\par

\section{Results}\label{sec:Results}

In this section, we first describe the results of our simulations with both fixed and randomized patterns and for the various helicity preferences. Then, we analyze those results in the context of filament-channel formation and the smoothness of coronal loops.\par

\subsection{Fixed Cellular Pattern}

The first set of simulations holds the cellular pattern fixed in the orientation shown in Figure \ref{fig:init}, randomizing only the sense of rotation of the individual cells as described in \S \ref{sec:model}. To compare the $k=0.75$ simulation most consistently with the $k=1.0$ case presented in KAD15, we ran it for twice as many cycles (42 vs.\ 21). As shown by Equation (\ref{Htotal}), the expectation value for the net injected helicity is the same (50\% of $42 \times N$ vs.\ 100\% of $21 \times N$). The $k=0.5$ case, in contrast, accumulates a net helicity only due to statistical fluctuations away from its average value of zero. For that case, therefore, we simply ran the simulation for $21$ cycles. All three simulations then were extended for $5$ additional cycles without imposing any rotational motions, to allow transients to die down and the system to relax toward a quasi-equilibrium final state.\par



\begin{figure}[!h]
\centering\includegraphics[scale=0.35,trim=0.0cm 0.0cm 0.0cm 0.0cm]{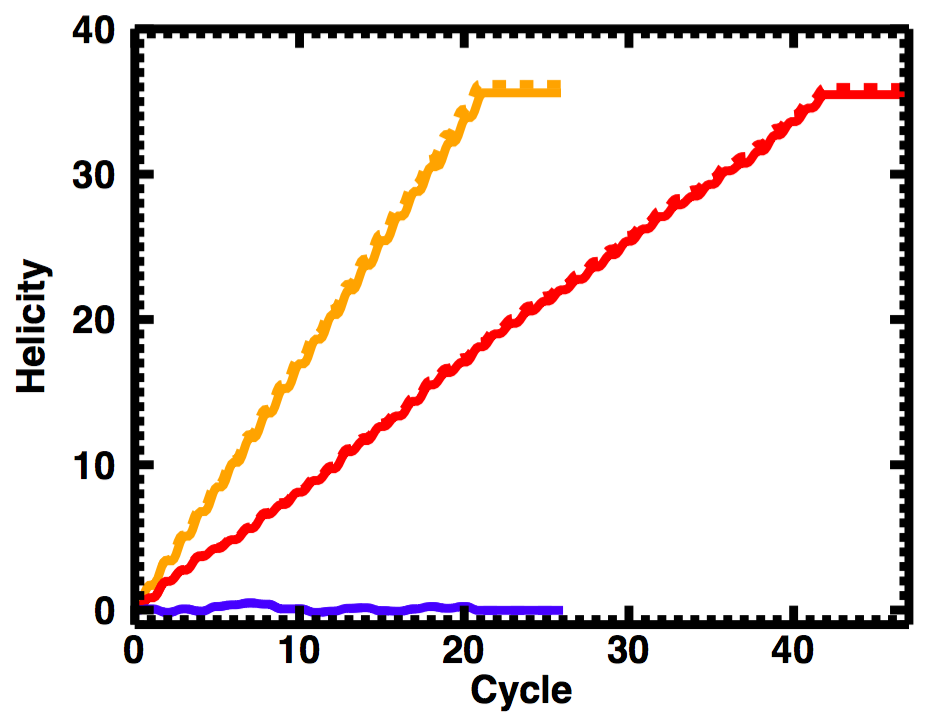}
\caption[Helicity for fixed pattern cases]{Helicity for fixed pattern cases. Solid curves represent the analytically expected helicity based on the number of convective cells injecting positive and negative helicity during each cycle, while the dashed curves represent the numerically calculated values. The orange, red and blue curves represent the $k=1$, $k=0.75$, and $k=0.5$ cases, respectively.}
\label{fig:helicityf}
\end{figure}

Figure \ref{fig:helicityf} shows the analytically expected (solid line) and numerically calculated (dashed line) helicities for each simulation. The orange ($k=1.0$) curve is the same as that presented in KAD15. After $21$ cycles, the rotation cells have injected $H=36$ units of helicity. The red ($k=0.75$) curve shows that approximately the same $H=36$ units are injected over twice the time (cf.\ Equation \ref{helicityk}) for the 75\% preference. The blue ($k=0.5$) curve shows, as expected, that almost no net helicity is accumulated in the simulation with 50\% preference over its first $21$ cycles. All of the numerically calculated curves match very well the analytical values at each cycle, demonstrating that our simulations conserve helicity to a very high degree of accuracy. Therefore, the evolution of the magnetic field in our simulations is due predominantly to convection and reconnection, rather than to numerical diffusion that would dissipate helicity.\par

\subsection{Randomly Displaced Patterns}

The second set of simulations is identical to the first, except that we also displace the entire hexagonal cellular pattern through a random angle after each cycle as described in \S \ref{sec:model}. In these setups, different flux tubes wrap around each other, creating a braided field, in addition to being twisted by the rotation cells. We ran each simulation for the same number of cycles as in the fixed-pattern cases and for the same values of $k$.\par

\begin{figure*}[!h]
\centering\includegraphics[scale=0.3,trim=0cm 0cm 4.5cm 0cm, clip=true]{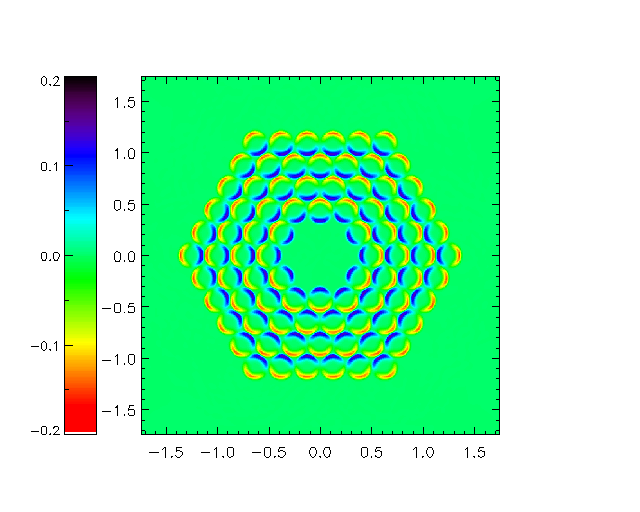}
\centering\includegraphics[scale=0.3,trim=0cm 0cm 4.5cm 0cm, clip=true]{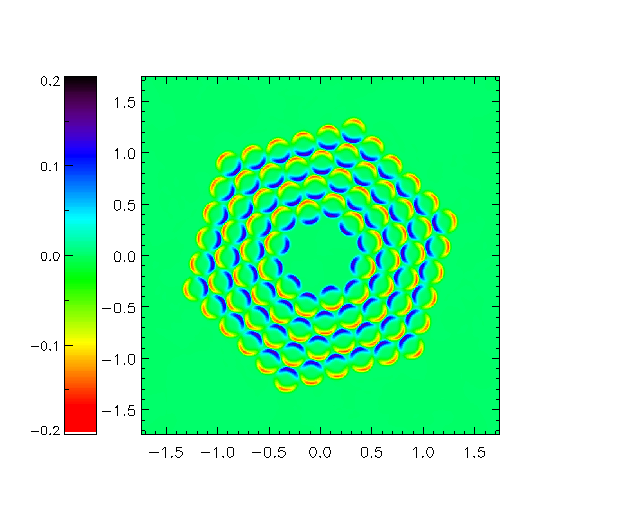}
\centering\includegraphics[scale=0.3,trim=0cm 0cm 4.5cm 0cm, clip=true]{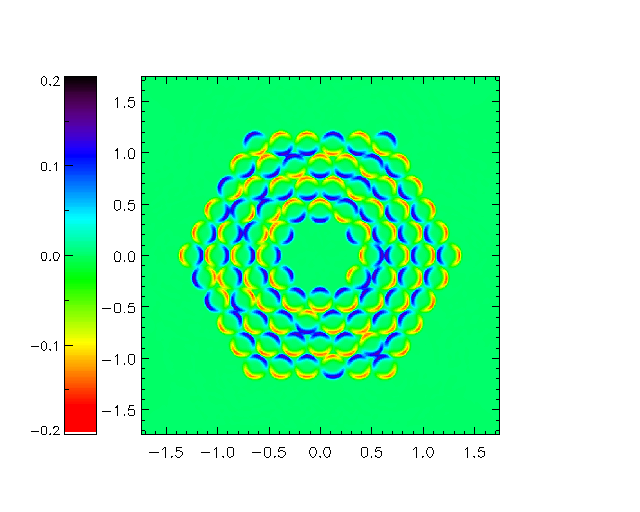}
\centering\includegraphics[scale=0.3,trim=0cm 0cm 4.5cm 0cm, clip=true]{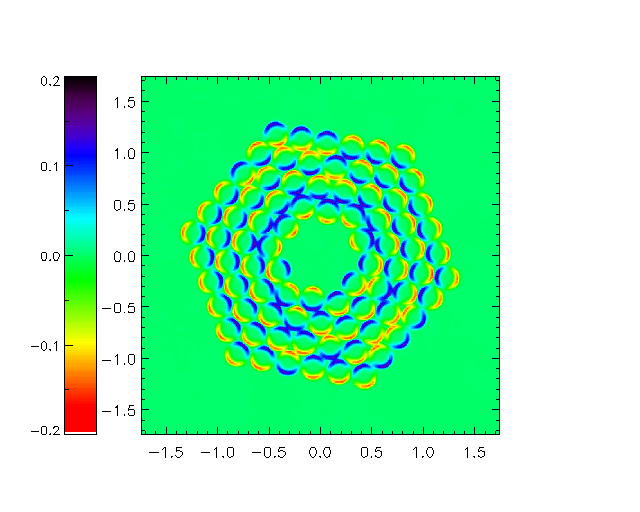}
\centering\includegraphics[scale=0.3,trim=0cm 0cm 4.5cm 0cm, clip=true]{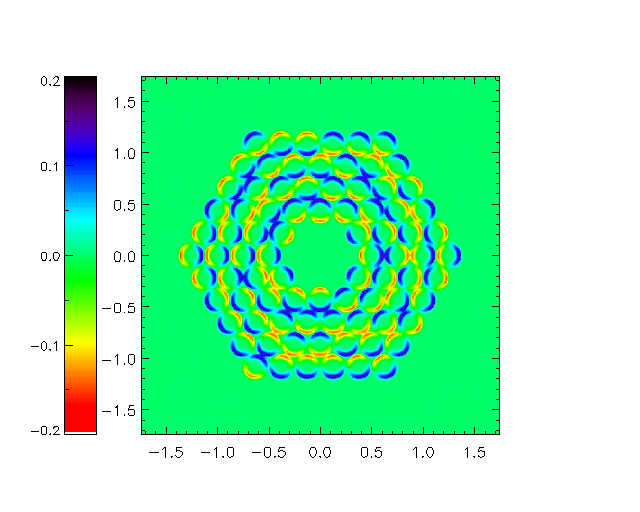}
\centering\includegraphics[scale=0.3,trim=0cm 0cm 4.5cm 0cm, clip=true]{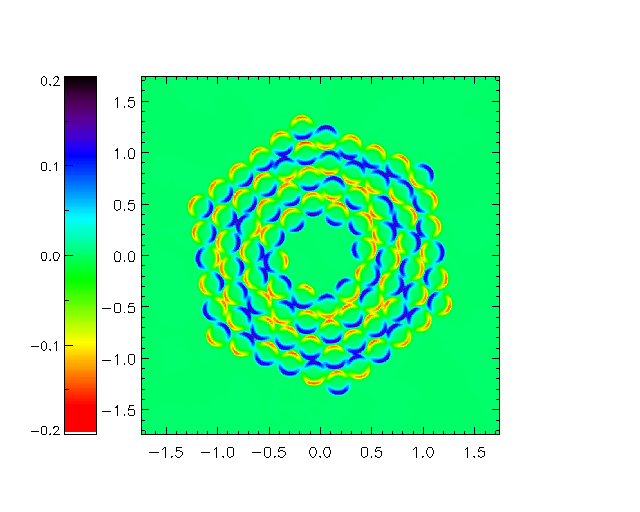}
\caption[$V_{\phi}$ for random pattern at early times]{$V_\phi$ on the bottom plate for the random pattern $k=1$ (top row), $k=0.75$ (second row) and $k=0.5$ (bottom row) cases during the first (left) and second (right) twist cycle.}
\label{fig:vphi_r}
\end{figure*}

Figure \ref{fig:vphi_r} shows $V_\phi(x=0,y,z)$ during the first two cycles. Except for the $k=1.0$ case, the randomized patterns (Fig.\ \ref{fig:vphi_r}) exhibit different distributions of color than the fixed patterns (Fig.\ \ref{fig:vphi_f}). We used different sequences of random numbers $\kappa_j$ to set the clockwise/counter-clockwise sense of rotation of the individual rotation cells in the two sets of simulations. The random angular displacements of the cellular pattern between the first and second cycles are evident by comparing the left and right columns for each helicity preference.\par

\begin{figure}[!h]
\centering\includegraphics[scale=0.35,trim=0.0cm 0.0cm 0.0cm 0.0cm]{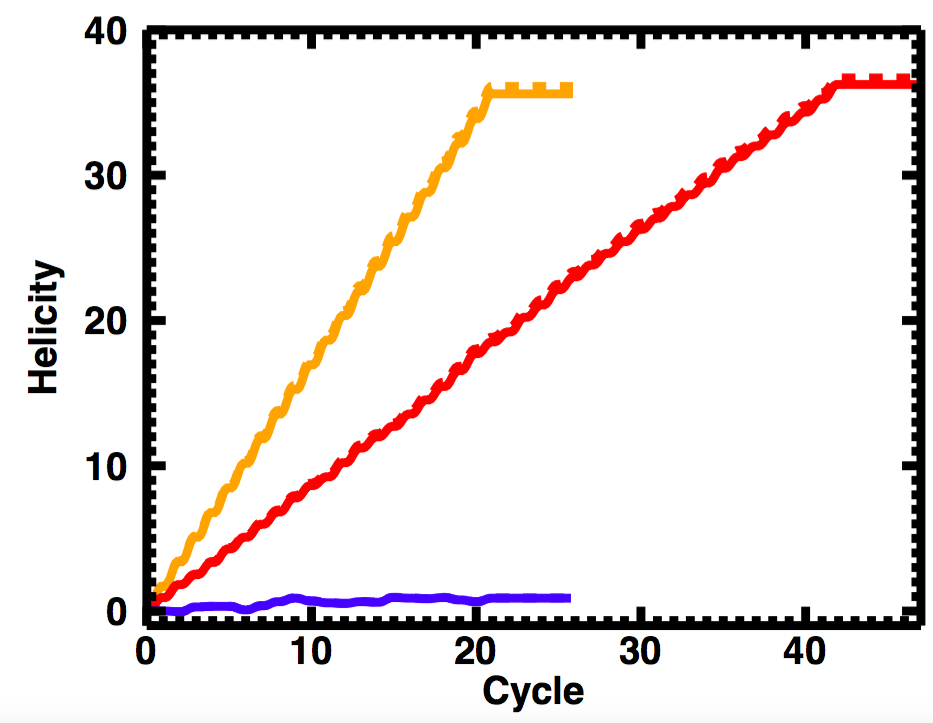}
\caption[Helicity for random pattern cases]{Helicity for random pattern simulations. Solid curves represent the analytically expected helicity based on the number of convective cells injecting positive and negative helicity during each cycle, while the dashed curves represent the numerically calculated values. The orange, red and blue curves represent the $k=1$, $k=0.75$, and $k=0.5$ cases, respectively.}
\label{fig:helicityr}
\end{figure}

Figure \ref{fig:helicityr} shows the analytically expected (solid line) and numerically calculated (dashed line) helicities for the various cases. Although the average helicities (Eq.\ \ref{helicityk}) injected into the corona are identical for each value of $k$, the precise helicities (Eq.\ \ref{Htotal}) actually injected differ between the fixed and randomized cases due to statistical fluctuations. Thus, the curves in Figures \ref{fig:helicityf} and \ref{fig:helicityr} are slightly different for $k \ne 1.0$. The orange ($k=1.0$) curve shows that the helicity injected for the 100\% preference is identical for the fixed and randomized patterns, as expected. The red ($k=0.75$) curve shows that the 75\% preference injects slightly more helicity in twice the time. The blue ($k=0.50$) curve shows that, as before, almost no net helicity is injected in this case. In this randomized-pattern simulation, the residual net helicity for the 50\% preference is small and positive, whereas in the fixed-pattern simulation, it is negative. In all cases, we again find excellent agreement between the numerically calculated and analytically expected helicities.\par

\subsection{Formation of Filament Channels}

\begin{figure*}[!h]
\centering\includegraphics[scale=0.3,trim=0cm 0cm 4.5cm 0cm, clip=true]{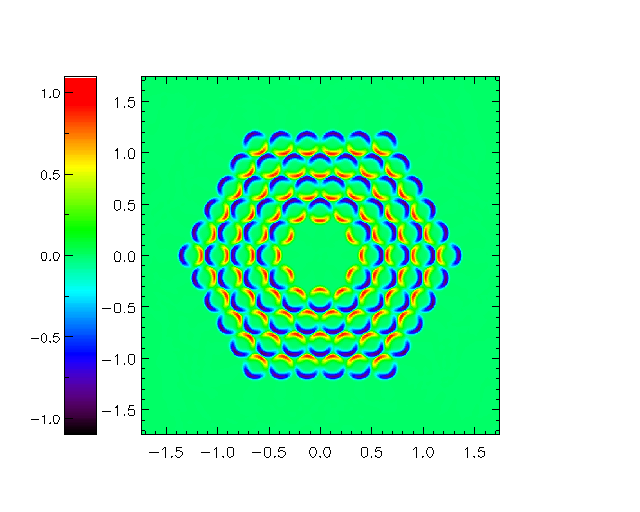}
\centering\includegraphics[scale=0.3,trim=0cm 0cm 4.5cm 0cm, clip=true]{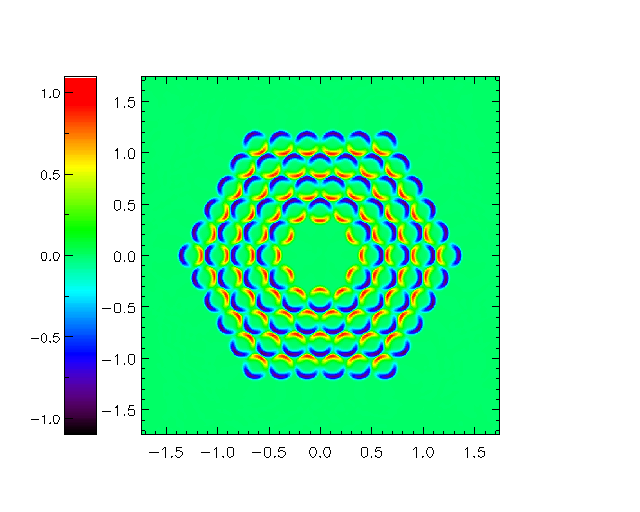}
\centering\includegraphics[scale=0.3,trim=0cm 0cm 4.5cm 0cm, clip=true]{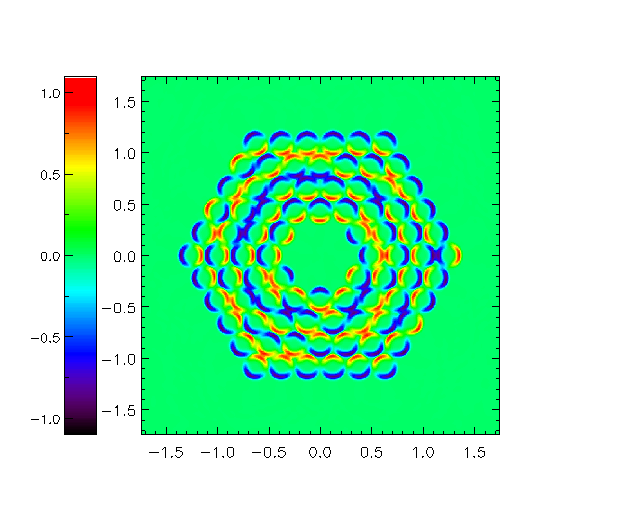}
\centering\includegraphics[scale=0.3,trim=0cm 0cm 4.5cm 0cm, clip=true]{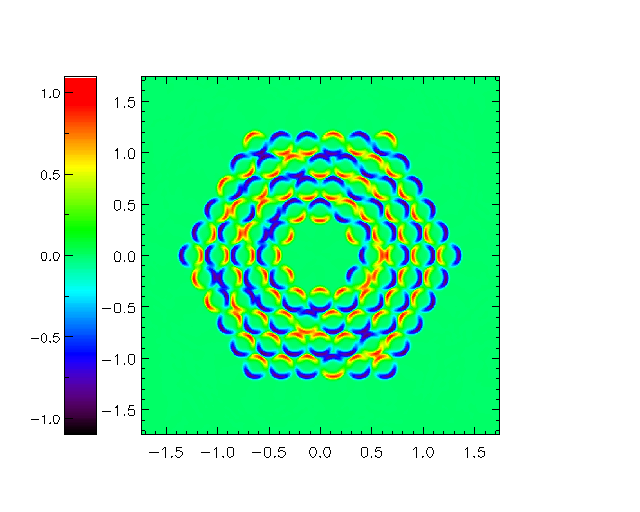}
\centering\includegraphics[scale=0.3,trim=0cm 0cm 4.5cm 0cm, clip=true]{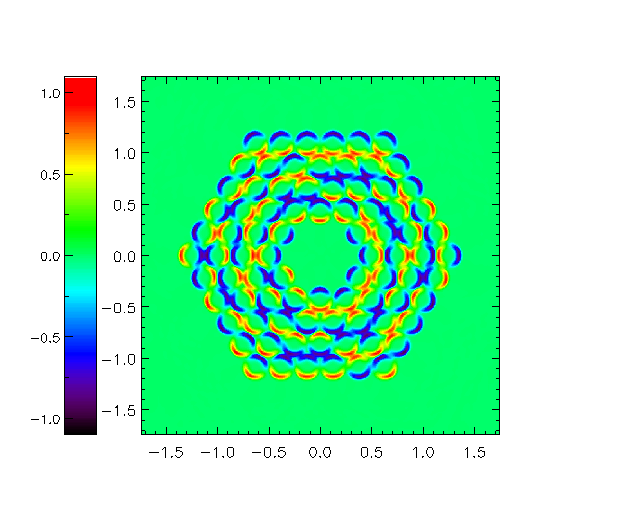}
\centering\includegraphics[scale=0.3,trim=0cm 0cm 4.5cm 0cm, clip=true]{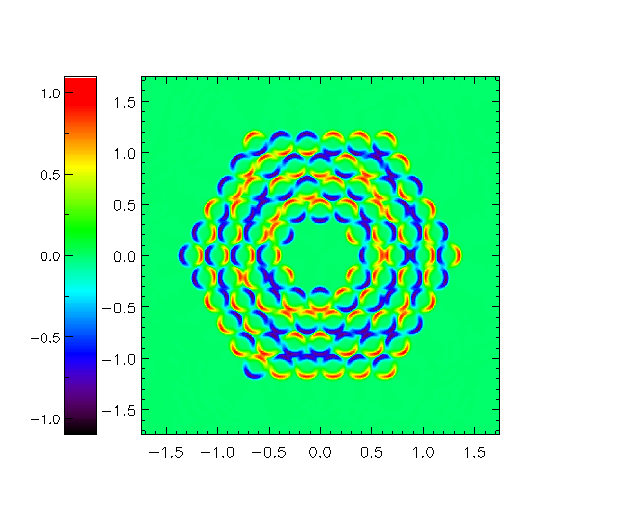}
\caption[Twist field at early times]{Twist field $B_\phi$ (color shading) in the midplane ($x=0.5$) halfway through the first cycle of the fixed pattern (left) and random pattern (right) simulations. From top to bottom are the $k=1$, $k=0.75$, and $k=0.5$ cases, respectively. Red/yellow (blue/teal) represents clockwise (counterclockwise) twist.}
\label{fig:Bphi_beg}
\end{figure*}

Figure \ref{fig:Bphi_beg} shows the azimuthal component of the magnetic field in the horizontal mid-plane, $B_\phi(x=0.5,y,z)$, halfway through the first cycle of twist for each simulation. At this early stage, each case exhibits the characteristic hexagonal pattern of rotation cells. For $k=1.0$, every cell injects the same sign of $B_\phi$, so adjacent flux tubes always have oppositely directed twist fields and are able to reconnect readily. For the $k=0.75$ and $k=0.5$ cases, in contrast, adjacent twist fields sometimes are in the same direction. On average, this is true half the time in the $k=0.5$ case, suppressing reconnection between adjacent flux tubes whose twist fields are parallel rather than anti-parallel.\par

\begin{figure*}[!h]
\centering\includegraphics[scale=0.3,trim=0cm 0cm 4.5cm 0cm, clip=true]{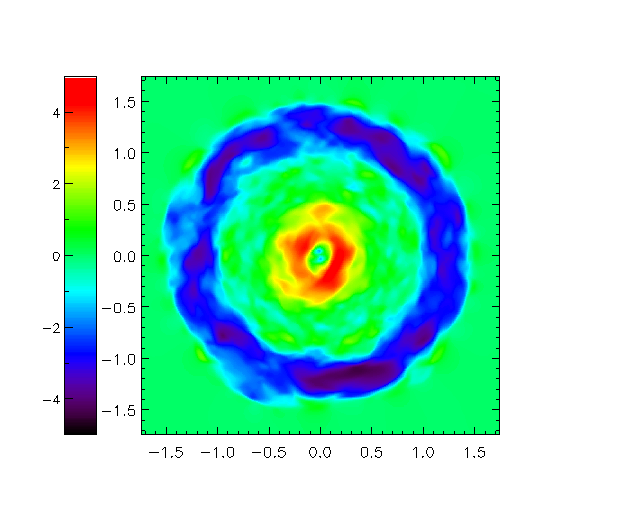}
\centering\includegraphics[scale=0.3,trim=0cm 0cm 4.5cm 0cm, clip=true]{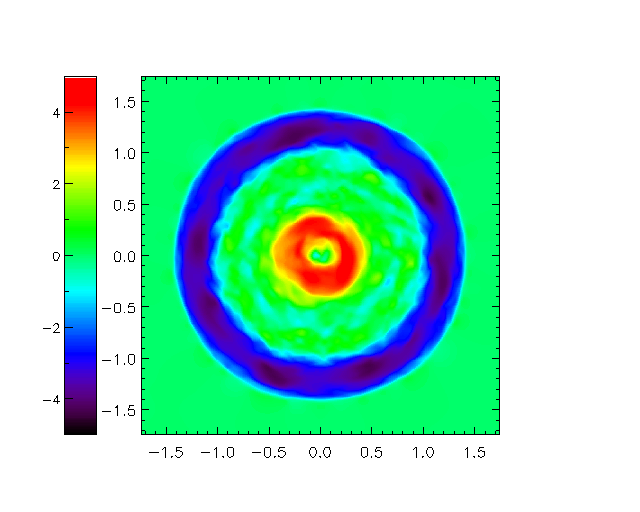}
\centering\includegraphics[scale=0.3,trim=0cm 0cm 4.5cm 0cm, clip=true]{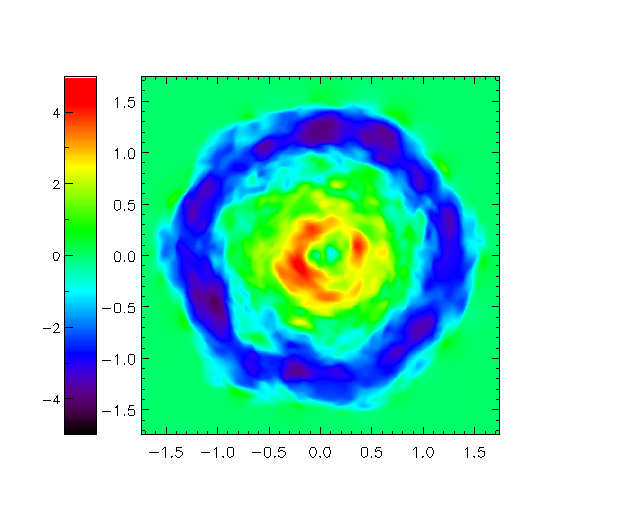}
\centering\includegraphics[scale=0.3,trim=0cm 0cm 4.5cm 0cm, clip=true]{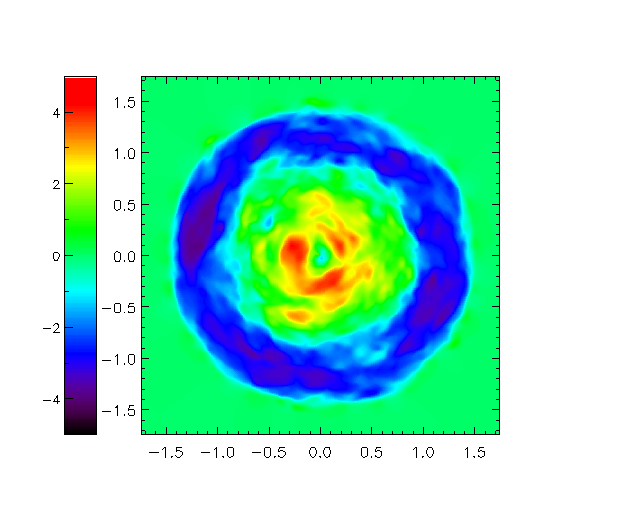}
\centering\includegraphics[scale=0.3,trim=0cm 0cm 4.5cm 0cm, clip=true]{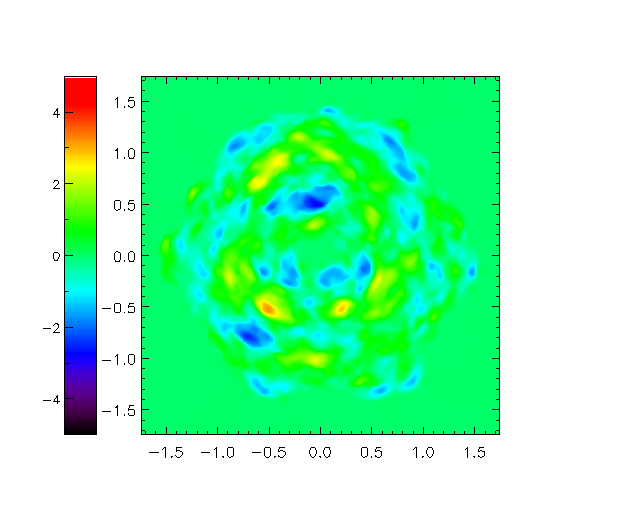}
\centering\includegraphics[scale=0.3,trim=0cm 0cm 4.5cm 0cm, clip=true]{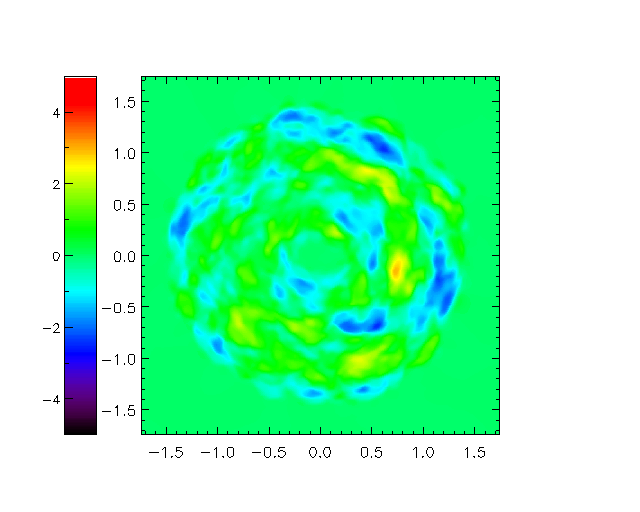}
\caption[Twist field at late times]{Twist field $B_\phi$ (color shading) in the midplane ($x=0.5$) at the end of the fixed pattern (left) and random pattern (right) simulations. From top to bottom are the $k=1$, $k=0.75$, and $k=0.5$ cases, respectively. Red/yellow (blue/teal) represents clockwise (counterclockwise) twist.}
\label{fig:Bphi_end}
\end{figure*}

The effect of the helicity preference on the formation of filament channels can be seen clearly in the final-time $B_\phi$ maps in Figure \ref{fig:Bphi_end}. The $k=1.0$ case has accumulated oppositely signed bands of twist flux at the outer and inner boundaries of the hexagonal pattern, as described previously in KAD15. These bands result from the inverse-cascade of twist flux from small to large scales due to reconnection, collecting at the boundaries to form filament channels according to the helicity-condensation model \citep{Antiochos13}. The $k=0.75$ case has been run out twice as long in order to accumulate roughly the same helicity, and it has acquired similar bands of twist at the outer and inner boundaries of the hexagonal pattern. Thus, despite the one-third (25\%/75\%) of twist fields on neighboring flux tubes that are parallel rather than anti-parallel in this simulation, sufficient reconnection has occurred to enable the helicity to condense at the flux-system boundaries here, as well. The shapes of the filament channels differ slightly in the $k=1.0$ and $k=0.75$ cases. The contrast is most evident in the randomized-pattern simulations, where the twist flux has a very uniform, circular appearance in the $k=1.0$ case, while the structure is more ragged, especially at the inner boundary, in the $k=0.75$ case. At the largest scales, however, these two cases yield qualitatively identical outcomes: the twist flux forms two bands of opposite polarity at the boundaries of the hexagonal pattern of rotations. In the corona, such bands would manifest themselves as extended, sheared filament channels.\par

The sharply contrasting $k=0.5$ case, on the other hand, displays a very different final-time appearance. No long, coherent bands of twist flux have accumulated at either the outer or inner boundaries of the hexagonal pattern. Instead, there are localized concentrations of twist flux dispersed across the interior of the pattern, as well as at its boundaries. Because zero net helicity is injected into this system, on average, zero net twist flux is available to be transported by reconnection to the hexagonal boundaries where it can accumulate. Turning this argument around, if the net condensed twist flux were finite, then the net helicity would be finite as well. We demonstrated this result analytically in KAD15. Consequently, the helicity-condensation process does not form filament channels in the case of a 50\% helicity preference.\par

These examples demonstrate that the helicity preference plays a major role in the organization of the twist flux and the formation of filament channels. The $k=0.75$ case forms similarly strong, although rather more structured, bands of twist flux over twice the time as the $k=1.0$ case. As is argued below, the time scale for filament-channel formation is inversely proportional to the average net fractional helicity injected, i.e.\ to $f$. This dependence is supported further by the absence of filament-channel structure in the $k=0.5$ case, whose predicted time scale for channel formation is infinite.\par

\subsection{Accumulation of Twist Flux}\label{sec:twistflux}

\begin{figure}[!h]
\centering\includegraphics[scale=0.35, trim=0.0cm 0.0cm 0.0cm 0.0cm]{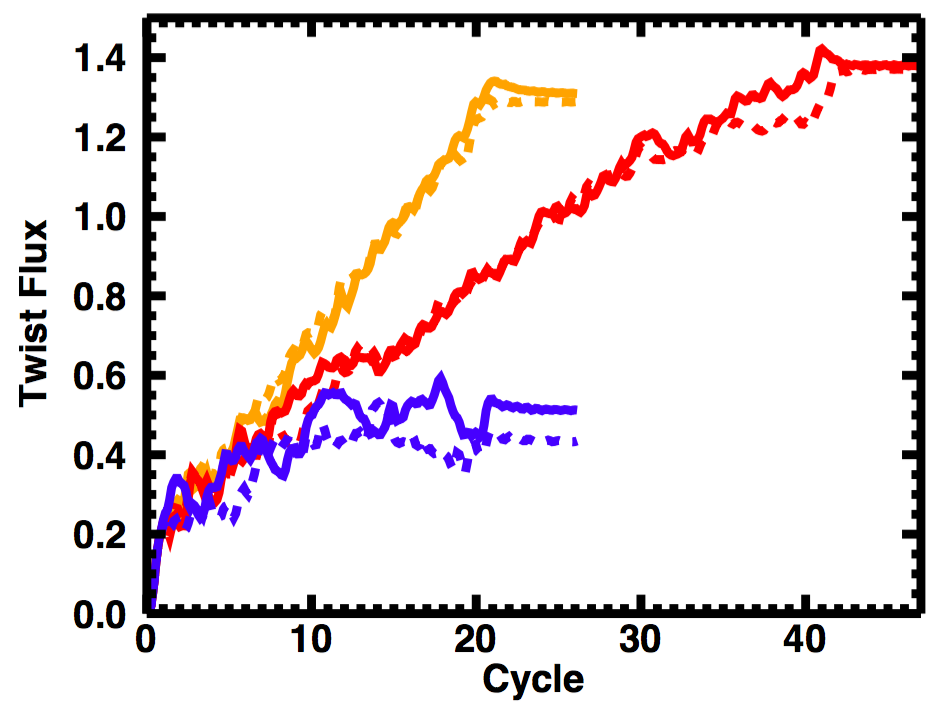}
\caption[Twist flux as a function of time]{Positive twist flux $\Phi_{tw}^+$ versus twist cycle through the vertical half plane $z=0$ for the fixed (solid curves) and random (dashed) $k=1$ (orange), $k=0.75$ (red), and $k=0.5$ (blue) cases.}
\label{fig:tw_v_time}
\end{figure}

The results above demonstrate that there are major qualitative and quantitative contrasts between the results for different helicity preferences, but more minor differences between the fixed and random patterns for a given helicity preference. We begin the quantitative analysis of our simulations by calculating the positive twist flux $\Phi_{tw}^+$ through the $z=0$ plane, 
\beg{Phitw}
\Phi_{tw}^+ = \int_0^{L_x}{dx} \int_0^{L_y}{dy \; B_{tw}^+(x,y,z=0)},
\done
where the corresponding positive twist field $B_{tw}^+$ is 
\beg{Btw}
B_{tw}^+ = \frac{1}{2} \left( B_\phi + \left\vert B_\phi \right\vert \right) \ge 0.
\done
The twist flux $\Phi_{tw}^+$ is plotted in Figure \ref{fig:tw_v_time} for both the fixed (solid curves) and random (dashed curves) patterns. All six simulations exhibit a brief initial phase of ideal evolution, of about one rotation cycle in duration, in which twist flux is injected into and stored in individual, non-interacting flux tubes. At this stage, the sense of rotation of adjacent cells is irrelevant to the accumulation of twist flux. As the twisting continues, however, the flux tubes expand laterally to compress the volume between them. This forms and strengthens electric current sheets between neighboring tubes that have anti-parallel twist fields. Reconnection between such tubes commences during subsequent twist cycles. This process, together with the randomization of the sense of rotation of individual cells (for $k \ne 1.0$) and of the orientation of the cellular pattern (for the random cases), causes the curves to deviate increasingly from one another at later times.

The two cases with nonzero net fractional helicities, $k=1.0$ (orange) and $k=0.75$ (red), show relatively small differences between the fixed and random patterns for fixed $k$. Over the full duration of the simulations, each preference accumulates essentially the same twist flux. The slightly larger values for the $k = 0.75$ case reflect the slightly larger magnetic helicities accumulated in those simulations (Figs.\ \ref{fig:helicityf} and \ref{fig:helicityr}) compared to the $k = 1.0$ case. All four of these simulations eventually accumulate twist flux at a rate per cycle that is in good agreement with the calculation by KAD15 (their Equation 4.13 and Figure 11),
\beg{DPhitw}
\Delta \Phi_{tw} = \frac{1}{2} \frac{\Delta \langle H_\Sigma \rangle}{\Phi_N} = \frac{f}{2} \frac{H_0}{\Phi_0},
\done
where each of the $N$ twisted flux tubes contains $\Phi_0$ units of magnetic flux. The expression in Equation (\ref{DPhitw}) assumes that the twist flux $\Phi_{tw}$ condenses at the outer boundary of the flux system, which occurs in our simulations with $k = 1.0$ and $k = 0.75$. In a spirit similar to KAD15 (their Equation 4.28), we calculate the filament-channel formation time $\tau_{fc}$ over which a critical amount of twist magnetic flux $\Delta \Phi_{fc}$ accumulates, 
\beg{Taufc}
\tau_{fc} = \frac{\Delta \Phi_{fc}}{\Delta \Phi_{tw}} \tau_0 = \frac{2}{f} \left( \frac{\Phi_0}{H_0} \tau_0 \right) \Delta \Phi_{fc},
\done
where $\tau_0$ is the duration of one twist cycle. Equations (\ref{DPhitw}) and (\ref{Taufc}) quantitatively express the observed factor-of-two differences in twist-flux accumulation rates and filament-channel formation times between our $k = 1.0$ ($f = 1.0$) and $k = 0.75$ ($f = 0.5$) cases. They also predict how these quantities should change for other helicity preferences $k$.\par

Our last two simulations, with $k=0.5$, have zero net fractional helicity, $f = 0.0$. For this case, the predicted accumulated twist flux $\Delta \Phi_{tw}$ vanishes and the filament-channel formation time $\tau_{fc}$ is infinite. We observe in these simulations (Figure \ref{fig:tw_v_time}) that the fluctuations in the twist flux are relatively large, and the average amount of flux saturates after about 10 cycles have elapsed. Thereafter, the average seems to be statistically quasi-steady, increasing or decreasing randomly according to the cycle-to-cycle variations of the sign of twist in individual rotation cells (in both simulations) and of the orientation of the cellular pattern (in the random-pattern simulation only). Evidently, these simulations have reached a roughly steady-state balance between the rates of twist-flux injection by the twisting motions and extraction via a combination of untwisting motions and reconnection between anti-parallel twist fields.\par

\subsection{Smoothness of Coronal Loops}

We have seen that, when it is effective, the helicity-condensation process transports twist via reconnection to the boundaries of the flux system, where it condenses. This leaves the interior of the system relatively smooth and untwisted. The final configuration then corresponds to a corona with strong shear concentrated at its PILs and laminar coronal loops in interior regions away from its PILs. This can be seen best in Figure \ref{fig:Bphi_end}. In the $k=1.0$ and $k=0.75$ cases with nonzero net fractional helicities, at a glance the interior of the flux system seems very smooth, with little twist evident. A careful comparison of the two cases reveals that the annular region between the filament channels is somewhat more structured for $k=0.75$, with localized, small-amplitude twists of both signs accumulating in the interior. As might be anticipated, this structure is somewhat less noticeable for the simulations with randomly displaced patterns compared to their fixed-pattern counterparts.\par

Although the $k=0.5$ case also follows this mild trend, its overall appearance is strikingly different from the $k=1.0$ and $k=0.75$ cases. For the fixed pattern (bottom row, left column of Figure \ref{fig:Bphi_end}) especially, small-scale, coherent concentrations of twisted field are present throughout the interior of the hexagonal flow region. In addition, the magnitude of the accumulated twist is significantly smaller than in the $k = 1.0$ and $k = 0.75$ cases. This is due to both the random untwisting of previously twisted field lines in successive cycles and the zero net twist flux that can accumulate globally and be transported to the flux-system boundaries. The local twist concentrations that are formed appear and disappear transiently as the system evolves. Each such concentration has a lifetime on the order of one rotation period of the convection cells. Taking the rotation period to be of order a day, or the lifetime of a typical supergranule, these concentrations of twist should easily survive for timescales long enough to be detected remotely. The lack of such observations indicate that the photosphere likely injects helicity with a significant preference.\par

\begin{figure*}[!h]
\centering\includegraphics[scale=0.2, trim=0.0cm 0.0cm 0.0cm 1.2cm,clip=true]{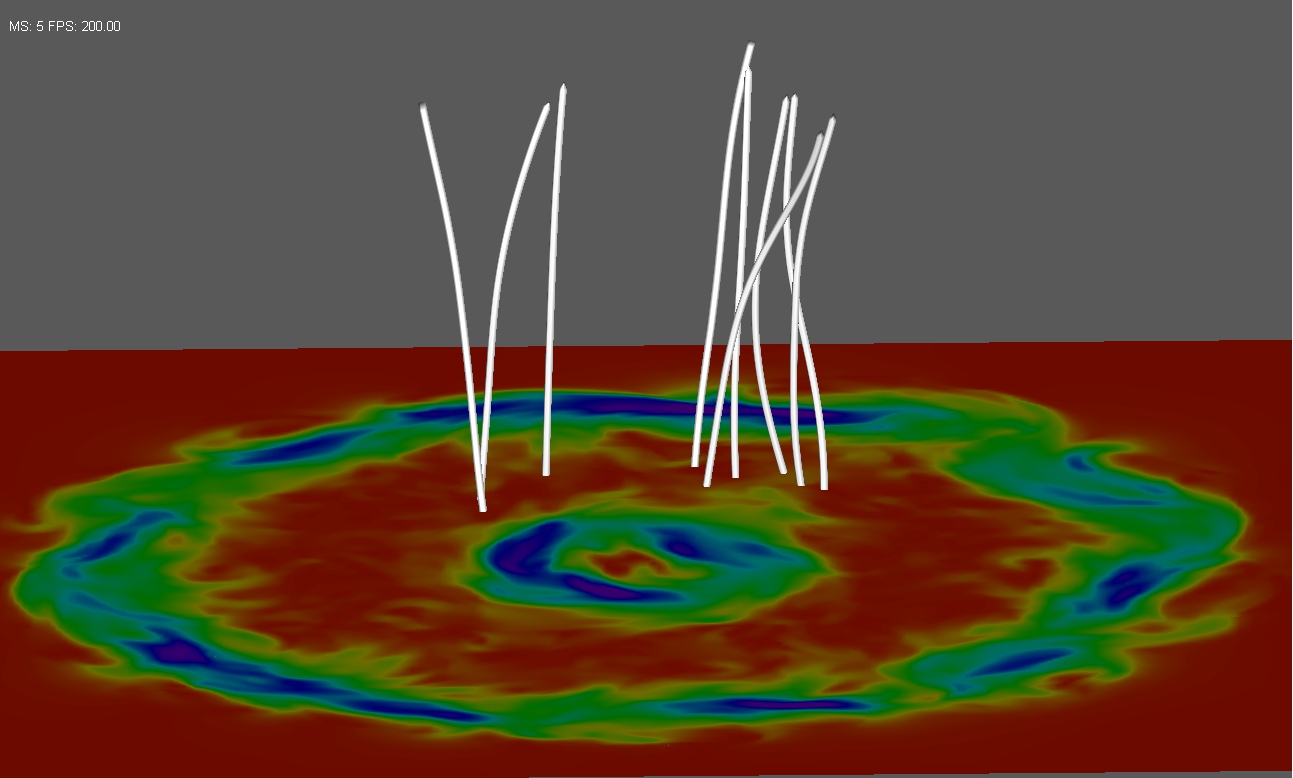}
\centering\includegraphics[scale=0.2, trim=0.0cm 0.0cm 0.0cm 1.2cm,clip=true]{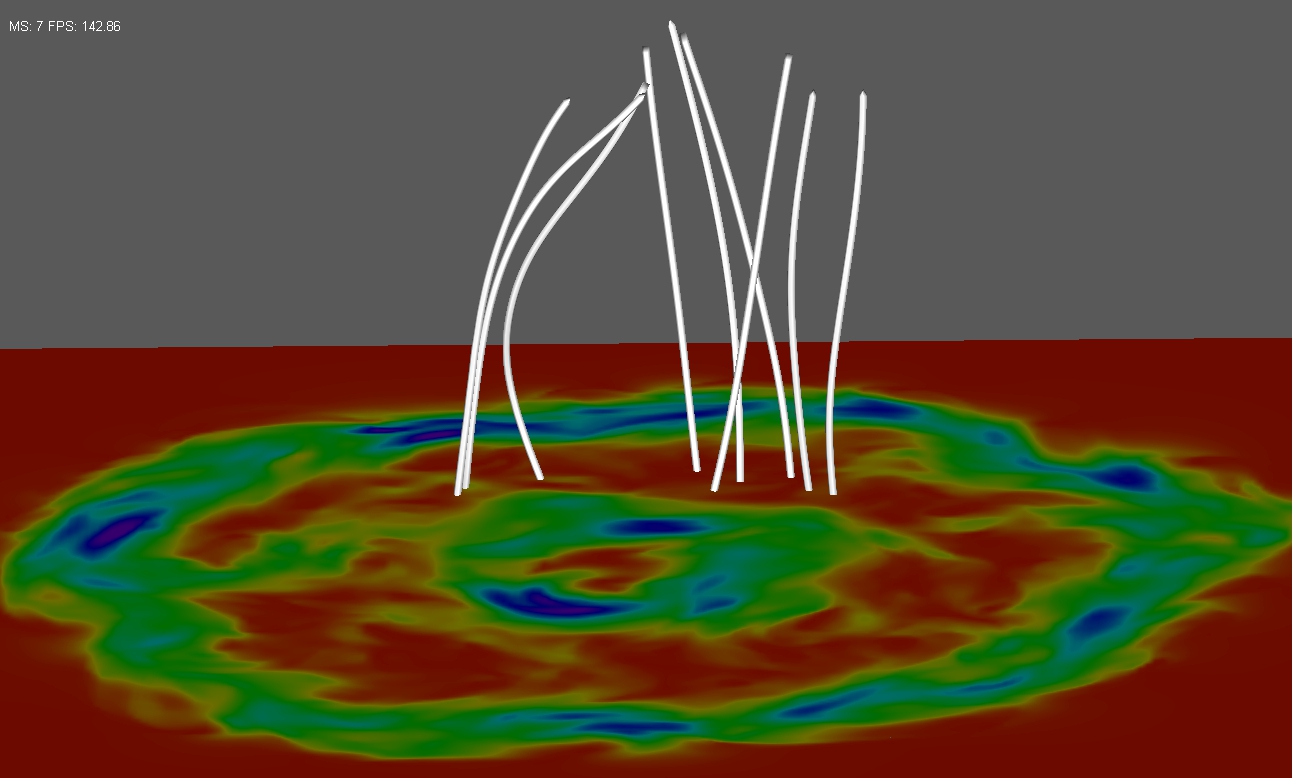}
\centering\includegraphics[scale=0.2, trim=0.0cm 0.0cm 0.0cm 1.2cm,clip=true]{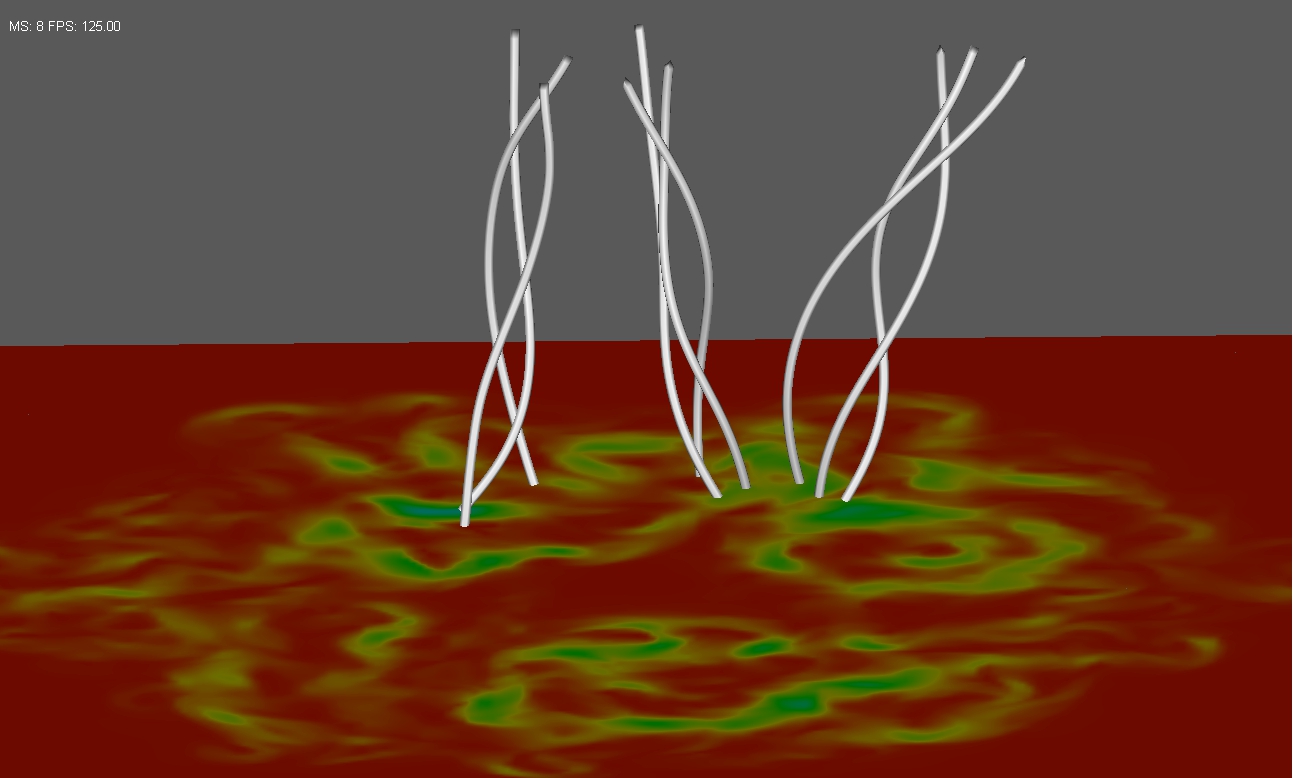}
\centering\includegraphics[scale=0.4, trim=0.0cm 0.0cm 0.0cm 0.0cm,clip=true]{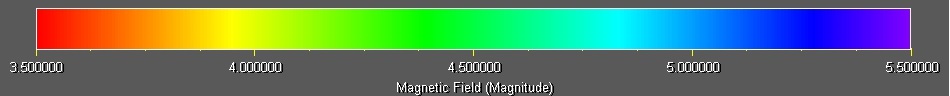}
\caption[Coronal loop field lines - fixed pattern]{A sample of `coronal loop' field lines (white) plotted at the end of the fixed-pattern $k=1$ (top) $k=0.75$ (middle) and $k=0.5$ (bottom) simulations. The field lines are plotted inside the interior of the hexagonal region on the bottom plate, which shows magnetic field magnitude (color shading). The initial (potential) field is a uniform straight field.}
\label{fig:structure_f}
\end{figure*}

\begin{figure*}[!h]
\centering\includegraphics[scale=0.2, trim=0.0cm 0.0cm 0.0cm 1.2cm,clip=true]{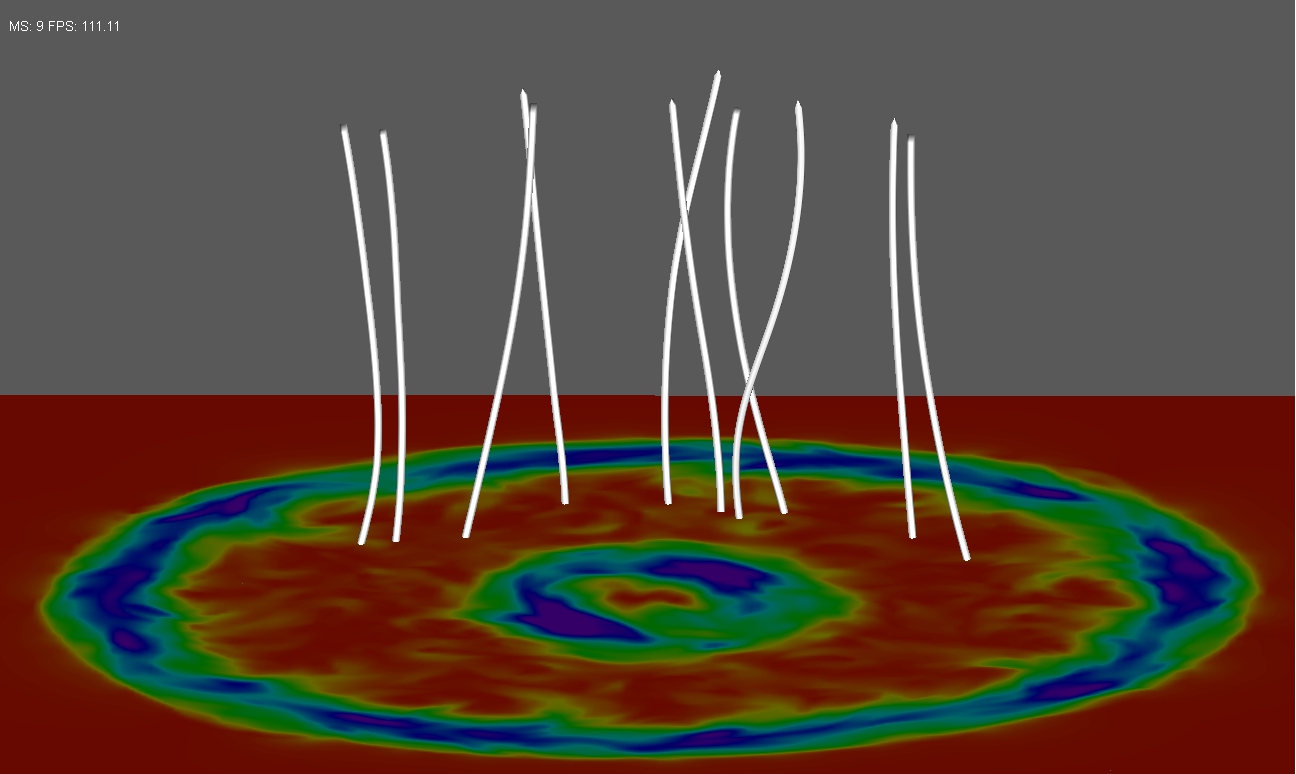}
\centering\includegraphics[scale=0.2, trim=0.0cm 0.0cm 0.0cm 1.2cm,clip=true]{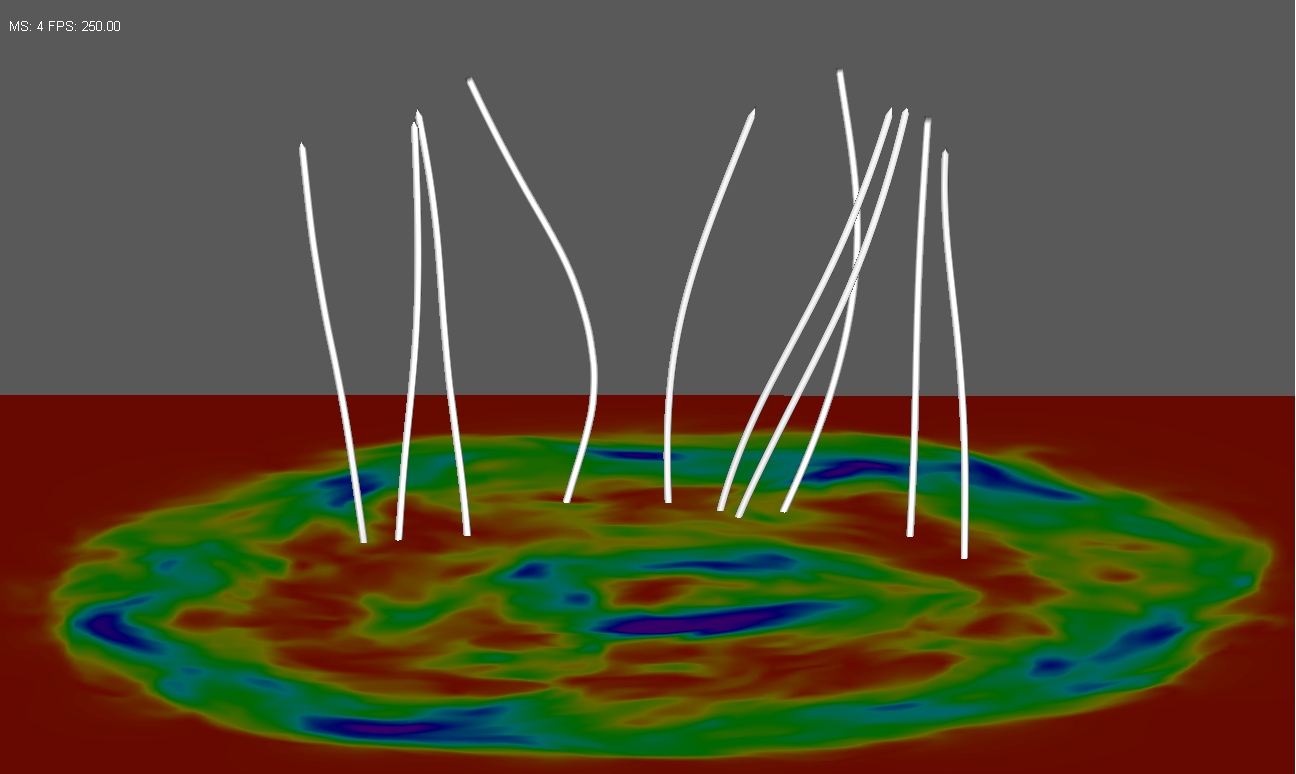}
\centering\includegraphics[scale=0.2, trim=0.0cm 0.0cm 0.0cm 1.2cm,clip=true]{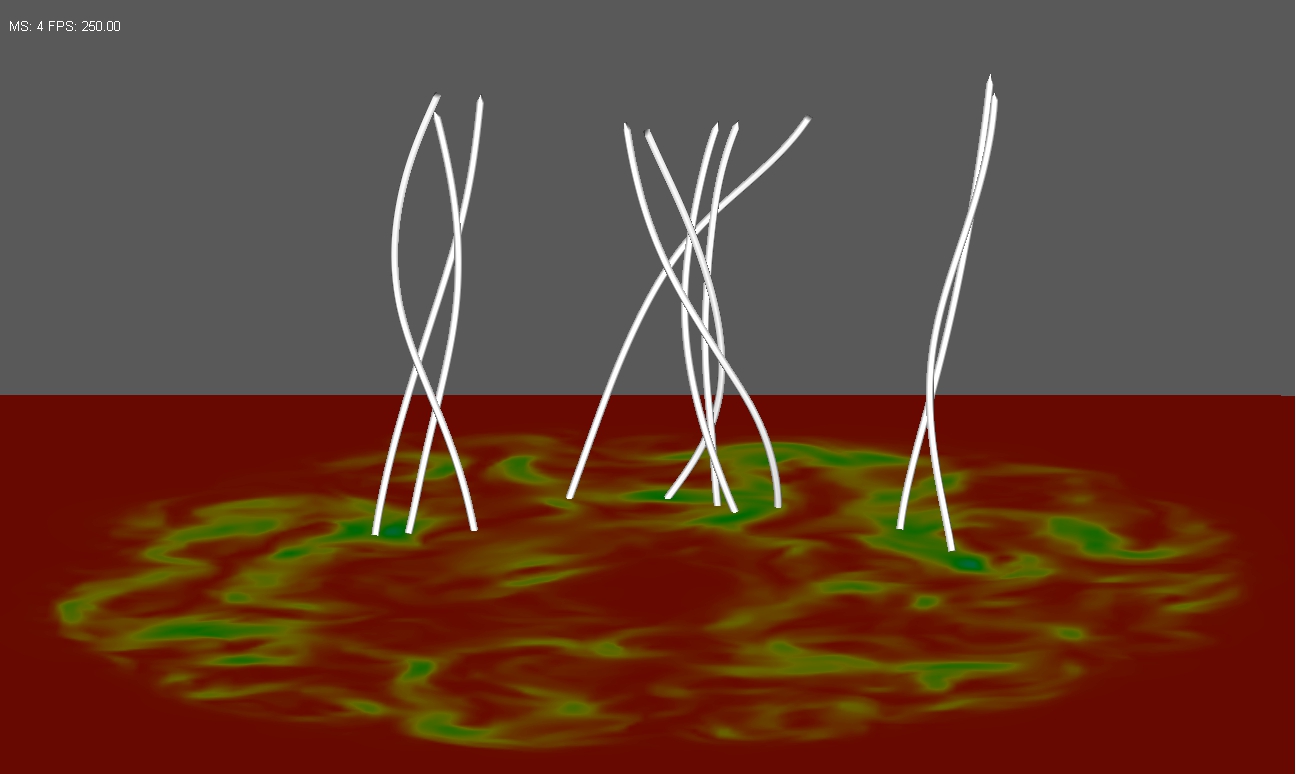}
\centering\includegraphics[scale=0.4, trim=0.0cm 0.0cm 0.0cm 0.0cm,clip=true]{colorbar.jpg}
\caption[Coronal loop field lines - random pattern]{A sample of `coronal loop' field lines (white) plotted at the end of the random-pattern $k=1$ (top) $k=0.75$ (middle) and $k=0.5$ (bottom) simulations. The field lines are plotted inside the interior of the hexagonal region on the bottom plate, which shows magnetic field magnitude (color shading). The initial (potential) field is a uniform straight field.}
\label{fig:structure_r}
\end{figure*}

To demonstrate the stark difference in the amount of structure in the different helicity preference cases we plot in Fig. \ref{fig:structure_f} a set of magnetic field lines from the same set of fixed points for the fixed-pattern $k=1$, $k=0.75$, and $k=0.5$ cases. All of the field lines are chosen from the interior of the hexagonal region, which represents the `loop' portion of the corona. The bottom plate represents magnetic field magnitude, which shows a structure similar to that shown in Fig. \ref{fig:Bphi_end}. Although only a sample of field lines is chosen, they are representative of the field lines in the rest of the `loop' portion of the corona. The striking difference in the amount of structure in the corona is immediately evident by comparing the $k=1$ and $k=0.75$ cases with the $k=0.5$ case. In the latter simulation, field lines are twisted and braided around each other in a complicated fashion. In the first two cases, although the field lines are traced from the same points, the field lines themselves are quite smooth and laminar. There may appear to be some structure due to field lines passing behind, or in front of, each other, but there is almost no significant twisting or braiding around each other. In fact, the top two figures very closely resemble the initial, uniform field configuration, and, importantly, the smooth, laminar structure observed in Fig. \ref{fig:obs}, albeit in a plane-parallel geometry. In this sense, the $k=1$ and $k=0.75$ coronal loops are quasi-potential, while the $k=0.5$ coronal loops clearly deviate quite strongly from quasi-potentiality. In Fig. \ref{fig:structure_r}, a different set of field lines is plotted for the random-pattern $k=1$, $k=0.75$ and $k=0.5$ cases. These sample field lines are traced from the same footpoints in each case, and again demonstrate the quasi-potentiality of the corona in the former two cases, and the large amount of structure in the latter case. \par

\subsection{Fluctuations of Twist Field}

\begin{figure*}[!h]
\centering\includegraphics[scale=0.3, trim=0.0cm 0.0cm 0.0cm 0.0cm]{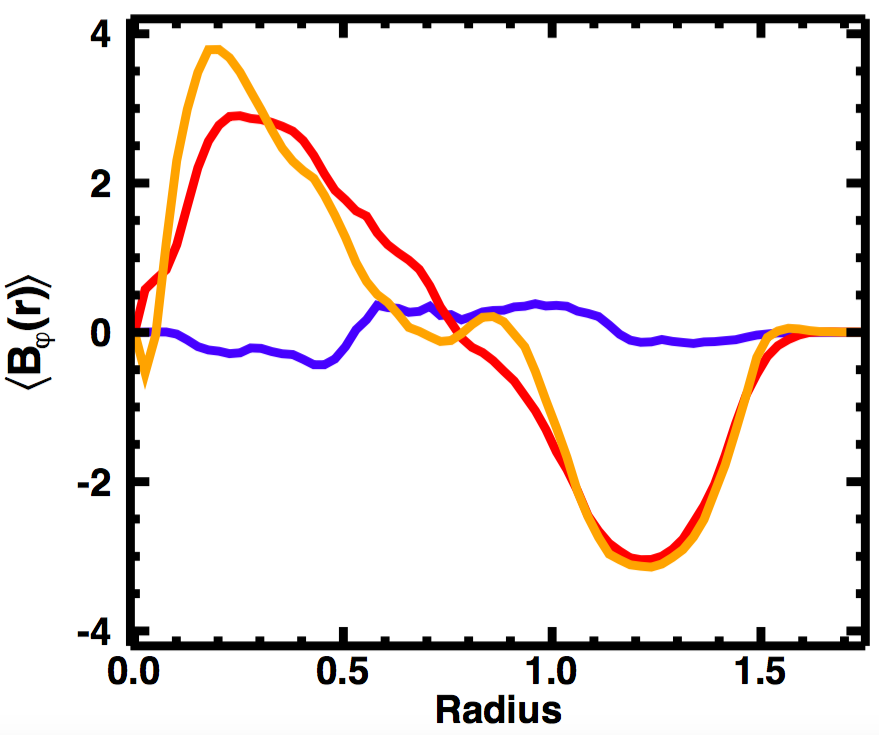}
\centering\includegraphics[scale=0.3, trim=0.0cm 0.0cm 0.0cm 0.0cm]{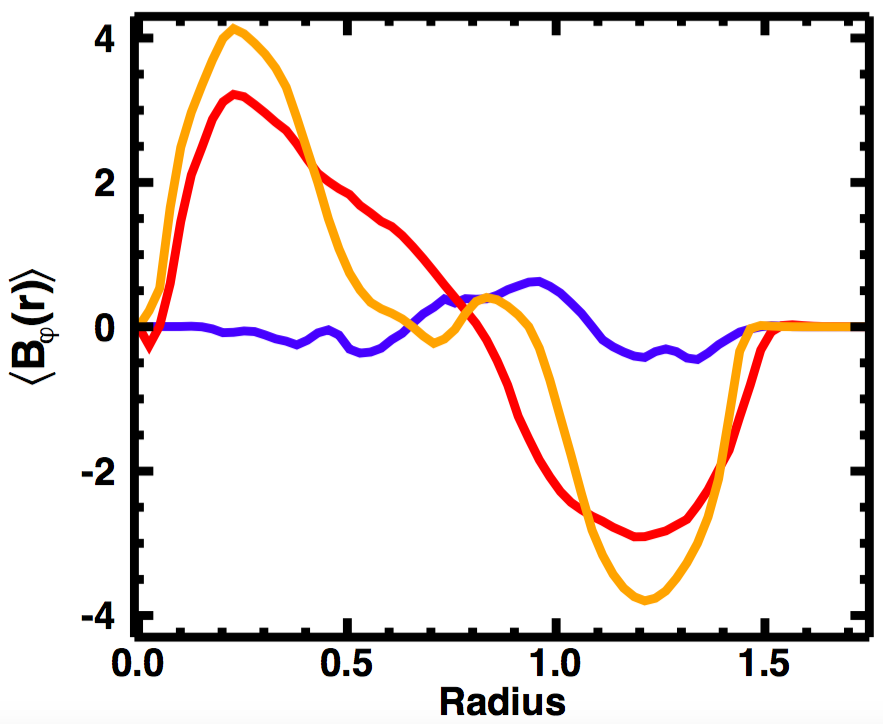}
\caption[Average $B_\phi$ at the end of the simulation for various helicity preferences]{Average azimuthal magnetic field $\langle B_\phi(r) \rangle$ versus radius in the mid-plane ($x=0.5$) at the end of the fixed-pattern (top) and random-pattern (bottom) cases with $k=1$ (orange), $k=0.75$ (red), and $k=0.5$ (blue).}
\label{fig:average}
\end{figure*}

\begin{figure*}[!h]
\centering\includegraphics[scale=0.3, trim=0.0cm 0.0cm 0.0cm 0.0cm]{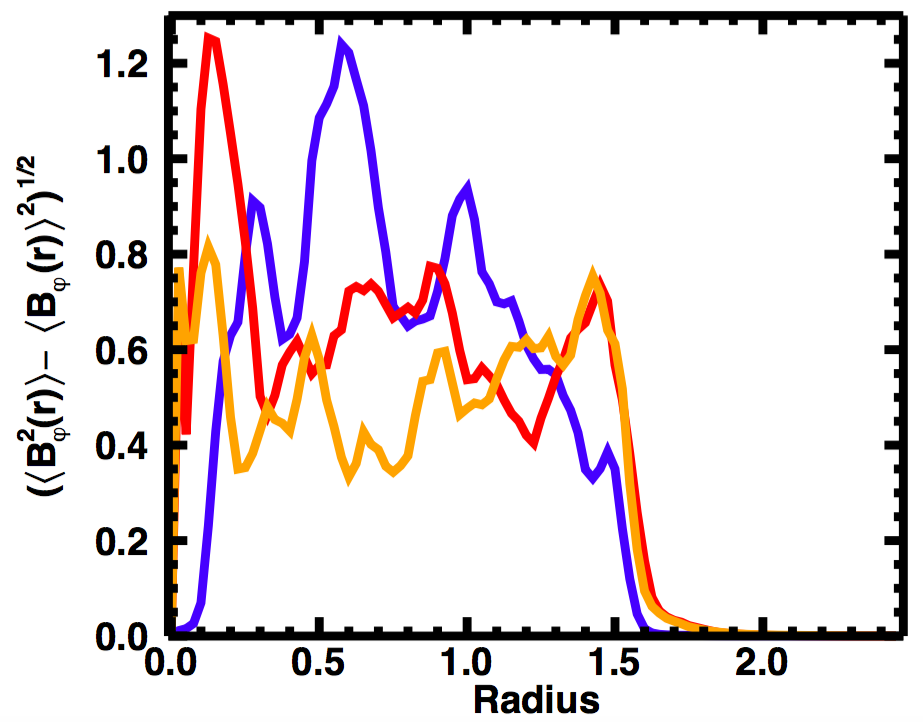}
\centering\includegraphics[scale=0.3, trim=0.0cm 0.0cm 0.0cm 0.0cm]{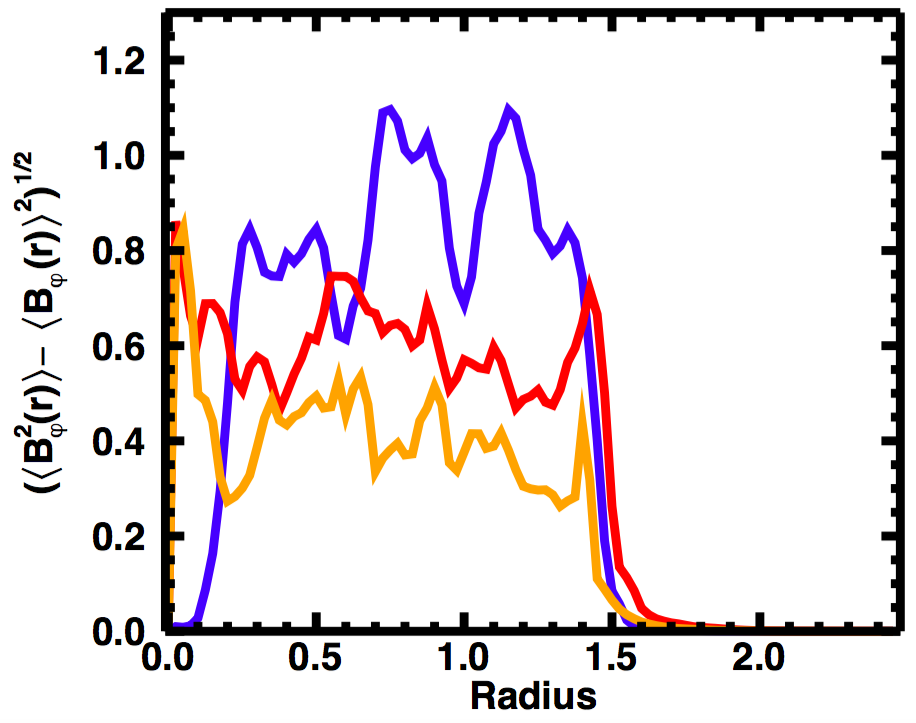}
\caption[Spread of $B_\phi(r)$]{Standard deviation of $B_\phi(r)$ as a function of radius in the midplane at the end of the fixed (top) and random (bottom) $k=1$ (orange), $k=0.75$ (red) and $k=0.5$ cases.}
\label{fig:width}
\end{figure*}

To quantify the amount of small-scale structure in the various cases, we calculated the angle-averaged azimuthal magnetic field $\langle B_\phi(r) \rangle$ and its root mean square deviation $\delta B_\phi(r)$ in the mid plane $(x = 0.5)$. Specifically, we evaluated (for $m = 1,2$) 
\begin{align}
\label{Bm}
\langle B_\phi^m(\rho) \rangle &= \frac{1}{2\pi} \int_0^{2\pi} B_\phi^m(x=0.5,y,z) d\phi,\\
\label{dB}
\delta B_\phi(\rho) &= \sqrt{\langle B_\phi^2(\rho) \rangle - \langle B_\phi(\rho) \rangle^2},
\end{align}
where $\rho = \sqrt{y^2+z^2}$. A discrete $\rho$ grid, with the same spacing as the $y$ and $z$ grids, was adopted, and all cell-center positions $(y,z)$ were grouped into corresponding $\rho$ intervals to calculate the integrals in Equation (\ref{Bm}). Figures \ref{fig:average} and \ref{fig:width} show $\langle B_\phi(r) \rangle$ and $\delta B_\phi(r)$, respectively, at the end of the simulations for each value of $k$ (color-coded) and both the fixed (top) and random (bottom) patterns. Qualitatively, the behavior is the same for the fixed and random patterns in all cases. The average field $\langle B_\phi(r) \rangle$ shown in Figure \ref{fig:average} is very similar for the cases $k = 1.0$ (orange) and $k = 0.75$ (red), with positive-polarity inner maxima and negative-polarity outer minima of comparable strengths at essentially identical positions for the two helicity preferences. In sharp contrast, for the case $k = 0.5$ (blue), the average field is much smaller in amplitude and fluctuates about zero.\par
The rms deviations $\delta B_\phi(r)$ shown in Figure \ref{fig:width} all exhibit small-scale statistical fluctuations. The average amplitude of the fluctuations is smallest for the 100\% helicity preference ($k = 1.0$, orange) and largest for the 50\% preference ($k = 0.5$, blue). In the latter case, the fluctuation amplitude substantially exceeds the average field amplitude shown in Figure \ref{fig:average}. The opposite is true for the cases with nonzero net fractional helicity ($k = 1.0$ and $0.75$), whose average fields reach much larger magnitudes than the fluctuating fields.\par

\subsection{Length of Field Lines}

\begin{figure*}[!h]
\centering\includegraphics[scale=0.3,trim=0cm 0cm 4.5cm 0cm, clip=true]{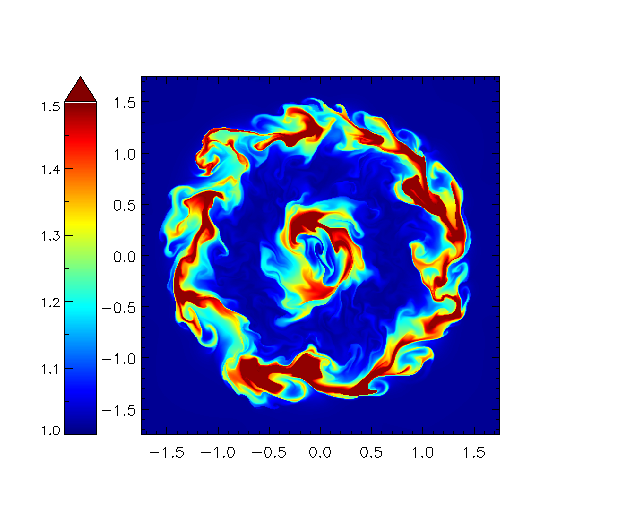}
\centering\includegraphics[scale=0.3,trim=0cm 0cm 4.5cm 0cm, clip=true]{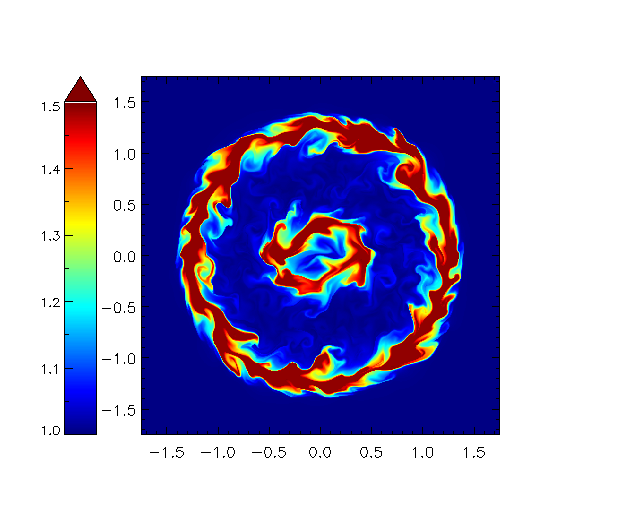}
\centering\includegraphics[scale=0.3,trim=0cm 0cm 4.5cm 0cm, clip=true]{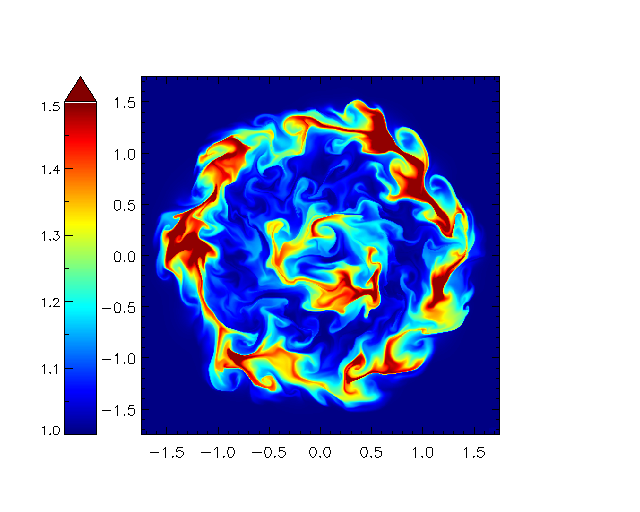}
\centering\includegraphics[scale=0.3,trim=0cm 0cm 4.5cm 0cm, clip=true]{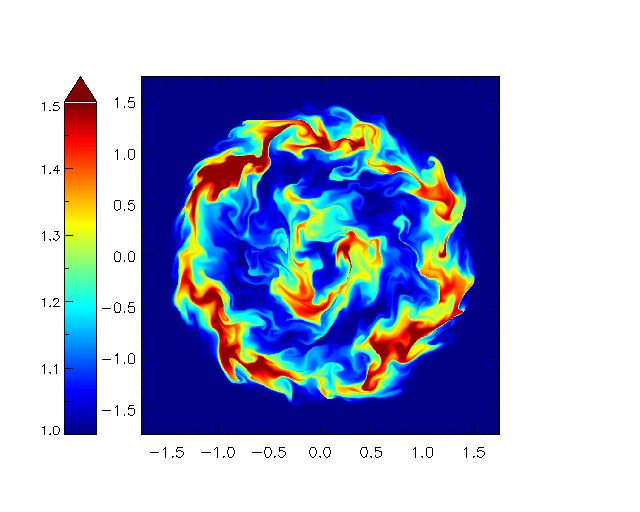}
\centering\includegraphics[scale=0.3,trim=0cm 0cm 4.5cm 0cm, clip=true]{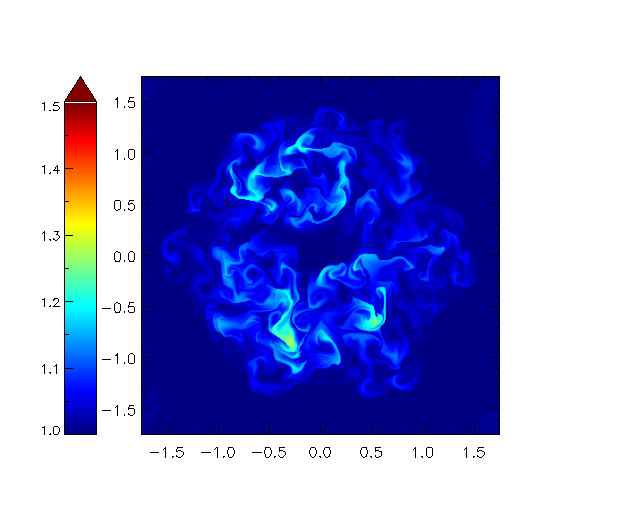}
\centering\includegraphics[scale=0.3,trim=0cm 0cm 4.5cm 0cm, clip=true]{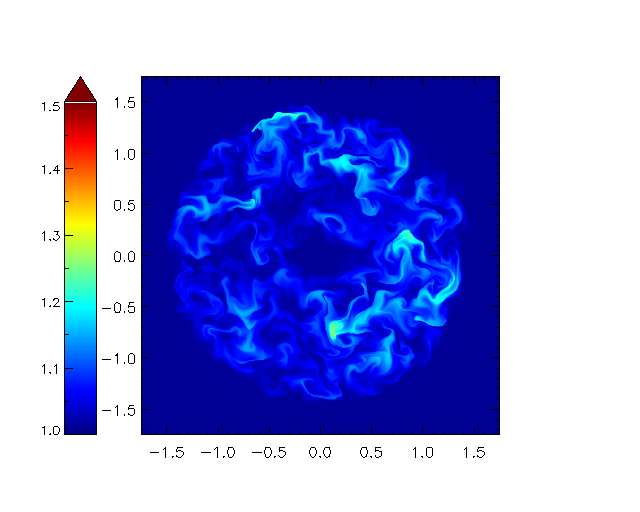}
\caption[Maps of field line length]{Maps of field line length at the end of the fixed pattern (left) and random pattern (right) $k=1$ (top row), $k=0.75$ (middle row) and $k=0.5$ (bottom row) simulations.}
\label{fig:Blength}
\end{figure*}

Further insight into the overall magnetic structure of the corona is gleaned by examining the lengths of field lines throughout the domain. In Figure \ref{fig:Blength}, we plot the length of magnetic field lines for the various helicity preferences. Here the differences between the filament channels and coronal loops is greatest in the $k = 1.0$ simulations, where substantially longer field lines reside in the filament channels. The coronal loops in the interior are quite short. Indeed, their lengths are very close to those of the untwisted field lines exterior to the hexagonal region. The two $k = 0.75$ simulations, meanwhile, display a slightly more mixed character in the interior. The coronal-loop field lines in these cases are slightly longer than the exterior untwisted field, although not nearly as long as the filament-channel field lines, which themselves are somewhat shorter than for $k = 1.0$.\par

As we found for several diagnostics described previously, the $k = 0.5$ case displays a strikingly different appearance from those with nonzero net fractional helicities $f$. As can be seen in Figure \ref{fig:Blength} (bottom row), the hexagonal region of rotation cells hosts a rather homogeneous mixture of relatively short field lines, structured at small scales. This is not dissimilar to the interior of the hexagonal region in the $k = 0.75$ case (middle row). However, no large-scale organization of the field-line length is evident, beyond the exclusion of twist from the center of the domain and the region beyond the perimeter of the hexagonal region. This is in sharp contrast to the cases $k = 1.0$ and $0.75$, where much longer field lines accrue at both the inner and outer boundaries of the flux system.\par
\par

\section{Implications for Coronal Structure}\label{sec:implications}

The results described in the preceding section have important implications for the global structure of the solar corona. Our findings demonstrate that the magnetic helicity preference $k$ plays key roles in determining how the corona is structured and the time scale over which that structure develops. The contrast is particularly strong between the cases with 100\% ($k = 1.0$) and 50\% ($k = 0.5$) preferences and randomly displaced cellular patterns shown at the top and bottom right, respectively, in Figures \ref{fig:Bphi_end} and \ref{fig:Blength}. For the 100\% preference, the twist flux condenses into two primary bands with opposite senses of twist at the inner and outer boundaries of the hexagonal region of rotation cells, with very little twist in the interior. These concentrations and dilution are reflected in the lengths of the associated magnetic field lines, which are very long near the two boundaries but minimally short in the interior. For the 50\% preference, on the other hand, the twist flux does not condense into any recognizable global-scale structure, and the field lines have an essentially homogeneous distribution of intermediate lengths. These two cases have the largest and smallest (in magnitude) net fractional helicities, $f = 1.0$ and $f = 0.0$, respectively.\par

Our intermediate case with 75\% preference ($k = 0.75$, $f = 0.5$) exhibits some features of both of the previous limiting cases but, importantly, qualitatively resembles more closely the results for the 100\% preference. The bands of condensed twist flux still form, albeit twice as slowly and with significant intrusions of twisted structures between them, and the field-line lengths are correspondingly longer at the hexagonal-region boundaries than in its interior, although with less contrast. Extrapolating to other cases with even smaller preferences but nonzero fractional helicities -- say, $k = 0.625$ and $f = 0.25$ -- we would expect these trends to continue, with a further increase in the filament-channel formation time (another doubling for $f = 0.25$) and in the amount and homogeneity of small-scale structure in the interior of the hexagonal region.\par

Perhaps the clearest example of the effect of helicity preference on the structure of the closed loop corona is evident in Figs. \ref{fig:structure_f} and \ref{fig:structure_r}. The smoothness of the $k=1$ and $k=0.75$ coronal loops is manifestly different than the complexity of the $k=0.5$ coronal loops. Observations of the coronal magnetic field, meanwhile, invariably reveal smooth, laminar loops that closely resemble those observed in the $k=1$ and $k=0.75$ cases, rather than those observed in the $k=0.5$ case. Our simulations indicate, therefore, that the photosphere must inject a significant net helicity so that structures such as those seen in the $k=0.5$ are not observed.

The simulation setups assumed in this paper are very simplified compared to the complex structure exhibited by the Sun, illustrated by Figure \ref{fig:obs}. Nevertheless, the homogeneous structure that we obtained for a 50\% helicity preference, shown in the bottom panels of Figures \ref{fig:Bphi_end} and \ref{fig:Blength}, obviously does not remotely resemble the very inhomogeneous structure observed in the corona. In contrast, our results for both 100\% and 75\% preferences, shown in the other panels of those figures, do exhibit the basic characteristics of the corona: concentrations of twist at PILs in the form of highly nonpotential magnetic shear in filament channels, and generally smooth, quasi-potential fields free of twist away from PILs in arcades of coronal loops. Therefore, a principal conclusion of our work is that the Sun must inject helicity into the corona with a significant hemispheric preference, favoring negative helicity forming left-handed structures in the north, and positive helicity forming right-handed structures in the south. These preferences are reflected in the observed statistics of solar filaments, sigmoids, and sunspot whorls. They also have been detected directly in the photospheric convection, although that measurement is very challenging, near the limits of observational resolution.\par

Our simulation is very simplified in another important way compared to the Sun: there is no source of new, weakly sheared or unsheared magnetic flux in our domain, nor is there a sink of the strongly sheared flux condensing in the filament channels. Flux emergence from below the photosphere constantly injects fresh magnetic field into the corona, and coronal mass ejections regularly eject sheared magnetic field and its entrained magnetic helicity away from the Sun into the heliosphere. The characteristic time scales for these phenomena compete directly with the filament-channel formation time $\tau_{fc}$ to establish a quasi-steady balance among these processes and the coronal magnetic structure that is observed. Such a calculation is well beyond the scope of this paper, but a first attack on the problem could be taken using global force-free modeling of the corona \citep[e.g.\ ][]{Mackay14}. We point out that $\tau_{fc}$ in Equation (\ref{Taufc}) is inversely proportional to the product of the net fractional helicity $f$ and the angular rotation rate $\omega_0$ of the twisting motions ($H_0 \propto \omega_0 \tau_0$; KAD15). If this product $f \omega_0$ is too small or too large, then the filament-channel formation time will be too long or too short compared to the emergence and ejection time scales, and the model is unlikely to replicate the Sun's observed appearance. We anticipate that global modeling of the combined processes could provide rigorous bounds on the rotation rate $\omega_0$, to complement the narrowly constrained range of values available to the net fractional helicity, $0.5 < f < 1.0$.\par

Our simulations show that random displacements of the pattern of photospheric convection have only a secondary effect on the resulting coronal structure. This also is evident in Figures \ref{fig:Bphi_end} and \ref{fig:Blength} by comparing the left (fixed-pattern) and right (random-pattern) columns for each helicity preference. The latter structures are somewhat smoother than the former, especially at small scales, but the large-scale organization is no different between them. This conclusion agrees with that reached by \citet{Zhao15}, who used a much simpler setup with far fewer rotation cells.\par

\section{Discussion}\label{sec:discussion}

The helicity condensation model of \citet{Antiochos13} accounts for both the formation of sheared filament channels adjacent to PILs and the quasi-potential, smooth character of coronal loops away from PILs. By transferring the magnetic twist injected by photospheric motions to ever larger scales, reconnection concentrates the twist at the boundaries of flux systems (i.e., at the PILs) while diluting it throughout their interiors. The remarkable implication of the model is that the global organization of the magnetic shear in the solar atmosphere is a direct consequence of local twisting of the footpoints of coronal flux tubes by surface convection.\par

In this work, we examined the consequences of injecting magnetic helicity, with various preferences and in various locations, into a plane-parallel Parker corona. Our results confirm the recently-proposed helicity condensation model \citep{Antiochos13} of filament channel formation, in which magnetic reconnection transports twist to the boundaries of the flux system, the PIL, where it accumulates to form filament channels. This produces both long, sheared field lines at the PIL and a smooth, untwisted field in the interior of the flux system, in agreement with observations. This mechanism is able to explain the apparent paradox that, though convection injects helicity throughout the photosphere, in the corona it is only observed in filament channels. We find that these results are true as long as the helicity injection preference is non-vanishing, and that the time scale of filament channel formation varies inversely with the helicity preference. In other words, doubling the helicity injection will result in filament channels forming in half the time. If there is no helicity injection preference, on the other hand, our results indicate that twist should be present throughout the corona, not simply above PILs. Thus, the helicity preference plays an important role in determining the amount of structure present in the corona.\par
Further, we tested the role that more complex, random motions play in determining the structure of the solar corona. By changing the location of the hexagonal pattern, we more accurately modelled the randomness of the Sun's convection. We find that there are no significant differences, qualitative or quantitative, with simple twisting models of helicity injection in the range of helicity preferences observed on the Sun. Even for helicity preferences larger than those observed on the Sun, the differences, which include the shape of the twist profile and field line length, are relatively minor. We conclude, therefore, that it is sufficient to model helicity injection into the corona as simple twisting motions, rather than more complex spatially varying patterns.\par
An important feature of the helicity condensation model is that it does not require helicity to be injected only at specific locations, which is a requirement of both flux cancellation and flux emergence that is frequently not observed \citep{Mackay10}. Instead, coronal structure is produced by motions that cover the photosphere, as convection does. The model does not require, as could be inferred from KAD15, that all of the photospheric motions need to be spatially fixed and injecting the same helicity. We have shown here that varying the location of helicity injection does not affect the resulting coronal structure and the injection of helicity of opposite sign merely slows down the helicity condensation process.\par
The results of our simulations justified the prediction made in KAD15, that for a helicity preference of about $75\%$, as observed on the Sun, filament channels would form in about a day, since we were able to show that the amount of twist flux accumulating at the PIL boundary scales linearly with helicity preference. \par
We demonstrated the formation of relatively untwisted coronal loops for non-negligible helicity preferences, and showed that the amount of twist in these loops is inversely related to the strength of the helicity preference. Accurate observations of one quantity could allow us to place significant constraints on the other.\par
The relatively simple configuration of these simulations has enabled us to understand the transport of helicity throughout the solar atmosphere by reconnection. A related question, which will be explored in a subsequent paper, is the role of helicity injection in coronal heating. Since the photospheric motions inject helicity, in the form of twist, into the corona, it might be expected that the energy flux injected by these motions will be sufficient to heat the closed field corona. We will use energy conserving simulations to analyze the role reconnection and helicity injection play in heating the solar atmosphere.\par
All of our simulations thus far have dealt with helicity injection into a plane-parallel Parker atmosphere. A more realistic configuration, with a dipole-like field that has a true PIL and CH, will introduce gradients in the photospheric field that will cause each convective cell to inject a different amount of helicity. As a result, the twist field in each individual flux tube will not be precisely equal to that of its neighbor, and reconnection between the two will not completely cancel the twist field between them. Follow-up simulations, performed in this geometry, will explore the role of this broader distribution of twist field, and how it affects the formation of filament channels via magnetic helicity condensation.

\chapter{Filament Channel Formation In A Coronal Magnetic Field}\label{Sunspot}

\section{Introduction}\label{sec:intro}
One of the most important features of the solar atmosphere, from the standpoint of understanding and predicting space weather, is the presence of highly non-potential filament channels above photospheric polarity inversion lines (PILs). Filament channels are the magnetic structures that support cool condensations of plasma against the downward force of solar gravity. Filament channels that contain such cool condensations of plasma are known as prominences if seen on the limb, or filaments if seen on the disk. Prominences and filaments frequently erupt, releasing their plasma into the heliosphere in the form of a coronal mass ejection (CME). These CMEs send billions of charged particles streaming towards the Earth, where they pose a tremendous hazard for our infrastructure. These eruptions arise due to the strong non-potentiality of the filament channels underlying the prominences and filaments, making understanding the source of the non-potentiality, and by extension explaining the formation mechanism of filament channels, a critically important problem for solar physicists. \par
High-resolution observations of filament channels reveal that their non-potentiality comes in the form of shear \citep{Lin05, Vourlidas10}, i.e., a displacement of the field line footpoints in the direction of the PIL. This shear represents not only free energy, but also magnetic helicity \citep{Berger99}. The free energy in filament channels is injected at the photospheric level and is studied in detail in Chapter \ref{Heating}. Equally important, however, is the problem of how filament channels obtain their magnetic helicity. Models of filament channel formation need to explain not only why helicity is observed above PILs, but also why it is apparently not observed anywhere else in the corona. High-resolution EUV and X-ray images of the closed-field corona, such as those taken by the \emph{TRACE} mission show that the corona is filled with smooth, laminar, untwisted loops everywhere except in filament channels \citep{Schrijver99}. This fact is particularly baffling, since measurements of the photosphere, including granules \citep{Bonet08, Bonet10, VD11, VD15} and supergranules \citep{Brandt88, Duvall00, Gizon03, Komm07, Attie09, Seligman14}  indicate that magnetic helicity is being injected relatively uniformly throughout the photosphere. Magnetic helicity in a high Lundquist-number system like the corona is conserved, even under magnetic reconnection \citep{Taylor74, Berger84b}, meaning that all of the helicity that is injected at the photospheric level is expected to accumulate and be observed. Instead, observations indicate that althought magnetic helicity is being injected throughout the photosphere, it is only observed in the vicinity of the PIL.\par
The `missing helicity' problem was first pointed out by \citet{Antiochos13}, who argued that magnetic helicity was being transported throughtout the corona by magnetic reconnection, before finally accumulating at PILs. In this process, known as magnetic helicity condensation, adjacent flux tubes that are being twisted up in the same sense undergo component reconnection, causing the resulting twist field to expand in scale, enclosing the combined axial fluxes of the two original flux tubes. This `inverse cascade' of magnetic helicity then continued with the resulting flux tube undergoing component reconnection with another nearby flux tube. This process continued until all of the flux in the system had been reconnected with. Since the PIL enclosed all of the flux in the system, it formed a natural boundary beyond which the twist field, and therefore the helicity, could not go. The end result of this process is that magnetic helicity accumulates at the PIL, in the form of a global shear that is localized in a small region around the PIL, leaving the rest of the corona, the `loop' portion, essentially smooth and untwisted. Crucially, the localized shear at the PIL formed precisely at the location where filament channels are known to form, making this a very promising model for explaining both the formation of filament channels above photospheric PILs and the formation of the smooth loop corona.\par
The inverse cascade of magnetic helicity was first tested by \citet{Zhao15}, who injected helicity into a plane-parallel \citet{Parker72} corona. They found that adjacent flux tubes with the same sign of magnetic helicity did undergo component reconnection, allowing the twist field to expand in scale. Furthermore, they found that randomizing the helicity injection process did not affect this result. Expanding on this, \citet{Knizhnik15} injected magnetic helicity into a flux system that was a factor of $10$ larger than the one studied by \citet{Zhao15}. They compared the results quantitatively with the predictions of the helicity condensation model and observations of filament channel formation. They found that the rate of filament channel formation was proportional to the rate of helicity injection into the corona, and derived that for a rate of helicity injection in the range observed on the Sun of about $70\%$ \citep{Martin94, Pevtsov03}, filament channels would form in about a day, more than fast enough to explain the formation and maintenance of filament channels \citep{Tandberg-Hanssen95, Anderson05, Mackay10}. This prediction was later studied in Ch. \ref{Structure}, where the effect of various helicity injection preferences on coronal structure was explored. Those simulations verified precisely the prediction that the rate of filament channel formation was proportional to the helicity injection rate, by testing several different injection rates and quantifying the formation timescales. A consequence of these simulations, reported in Ch. \ref{Heating}, was the discovery that the helicity condensation process accounts in a very natural way for the uniform heating of the closed field corona, and that, in contrast to the amount of structure, the amount of heating of the corona did not depend on the helicity injection preference.\par 
These studies have been extremely promising for understanding the formation of filament channels on the Sun. They suggest that the helicity being injected at the photospheric level does not go `missing', as postulated above, but is actually just concentrated in certain specific locations. A significant problem with the aforementioned studies, however, was the lack of a true PIL caused by the use of a plane-parallel corona. In the simulations described above, the PIL was taken, somewhat arbitrarily, to be the boundary between twisted and untwisted field. Since the field itself was everywhere uniform, there was no real `inversion' of the field, and thus no realistic PIL. The formation of filament channels, therefore, was simply inferred from the accumulation of twist at this boundary, and field lines were taken to be `sheared', when in reality they did not even start and end on the same plate. Since the magnetic field used in these simulations could not be said to resemble the coronal field in any meaningful way, the generality of the results could therefore be called into question. In this Chapter, we describe the results of numerical simulations that fix these deficiencies by introducing a more realistic coronal field that contains both a PIL and a coronal hole (CH). In this way, we will show that the results of the helicity condensation model are quite general, and we will demonstrate the formation of filament channels at PILs. We will demonstrate the formation of highly sheared, untwisted field lines overlying the PIL, and at the same time, the presence of an otherwise smooth, quasi-potential corona. Furthermore, we reproduce, for the first time, the observed $\mathrm{20^\circ}$ shear of filament channels with respect to the PIL. The presence of the CH, meanwhile, serves an important purpose. Any helicity that is passed into the CH is expected to be expelled from the system, leaving the CH completely smooth and untwisted, much like the closed loop portion of the corona. However, the dynamics of the CH boundary during this process are far from clear. In this Chapter, we explore the dynamics of the CH boundary in response to applied boundary motions, and discuss the implications of these dynamics for observations.\par

\section{Numerical Model}\label{sec:model}
For our simulations, we employ the Adaptively Refined Magnetohydrodynamics Solver \citep[ARMS; e.g.,][]{DeVore08} to solve the equations of MHD in three Cartesian dimensions. The equations have the form:

\beg{cont}
\pd{\rho}{t}+\divv{\left( \rho\vecv \right)}=0,
\done
\beg{momentum}
\pd{\rho\vecv}{t} + \divv{\left( \rho\vecv\vecv \right)} = - \nabla P + \frac{1}{4\pi} \left( \curl{\vecB} \right) \times \vecB,
\done
\beg{energy}
\pd{T}{t} + \divv{\left( T\vecv \right)} = \left( 2 - \gamma \right) T \divv{\vecv},
\done
\beg{induction}
\pd{\vecB}{t} = \curl{ \left( \vecv \times \vecB \right)}.
\done
Here $\rho$ is mass density, $T$ is temperature, $P$ is thermal pressure, $\gamma$ is the ratio of specific heats, $\vecv$ is velocity, $\vecB$ is magnetic field, and $t$ is time. We close the equations via the ideal gas equation
\beg{idealgas}
P=\rho RT,
\done
where $R$ is the gas constant. ARMS employs Flux-Corrected Transport algorithms \citep{DeVore91} and finite-volume representation of the variables to obtain its solutions. Its minimal, but finite, numerical dissipation allows reconnection to occur at electric current sheets associated with discontinuities in the direction of the magnetic field. \par
We initialize the simulation in a box of extent $[0, L_x]\times [-L_y,L_y]\times [-L_z,L_z]$, with $x$ the vertical direction between the top and bottom plates, and we choose $L_x = 2$, $L_y = L_z = 3.5$. The initial field profile is determined by specifying the value of the normal magnetic field on the bottom plate as
\beg{Bbot}
B_n(r) = B_{nu}(r) + B_0
\done
where $B_0$ is a uniform background field and
\beg{Bnu}
B_{nu}(r) = \frac{B_+}{2}\big\{1-\tanh(\frac{r^2-r_+^2}{\lambda_+^2})\big\} -  \frac{B_-}{2}\big\{1-\tanh(\frac{r^2-r_-^2}{\lambda_-^2})\big\} 
\done
is a nonuniform field distribution on the bottom plate. To obtain the potential field in the rest of the domain, we solve the Laplace equation
\beg{Laplace}
\nabla^2\phi = 0
\done
for the scalar potential $\phi$ subject to Neumann boundary conditions at the bottom plate 
\beg{neumann}
\pd{\phi}{n}\at_{x=0} = B_n,
\done
and Dirichlet boundary conditions at infinity:
\beg{dirichlet}
\phi(r\to\infty) = 0.
\done
The scalar potential is then related to the magnetic field at each point in the volume, $\vecB(x,y,z)$, via 
\beg{grad}
\vecB(x,y,z) = -\nabla\phi.
\done
$B_{nu}(r)$ has the property that the total flux through the the bottom plate, $\Phi$ vanishes in the limit of an infinite plane, i.e.
\beg{totalfluxeq0}
\Phi = \int_0^{2\pi}{d\theta}\int_0^\infty{dr\; r B_n(r)} = 0,
\done
if $r_\pm^2\gg\lambda_\pm^2$ and $B_+\lambda_+^2 = B_-\lambda_-^2$. We choose $B_0 = -4$, $B_+ = 60$, $r_+ = 2$, $\lambda_+ = 0.5$, $B_- = 15$, $r_- = 4$, and $\lambda_- = 1$. \par
\begin{figure*}[!h]
\centering\includegraphics[scale=0.60, trim=0.0cm 0.0cm 0.0cm 0.0cm]{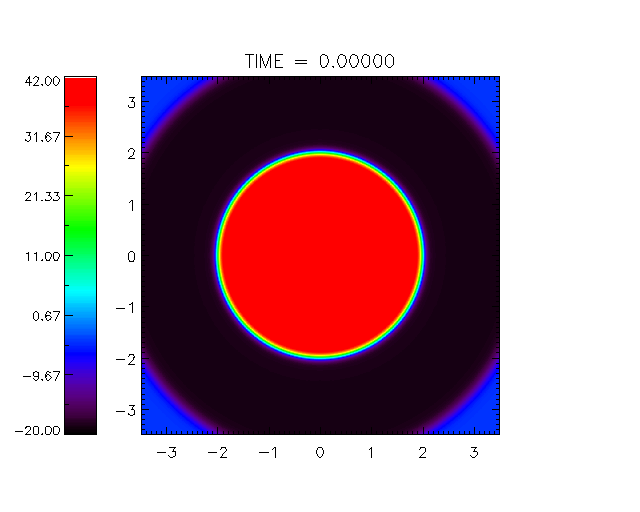}
\centering\includegraphics[scale=0.60, trim=1.0cm 0.0cm 0.0cm 0.0cm]{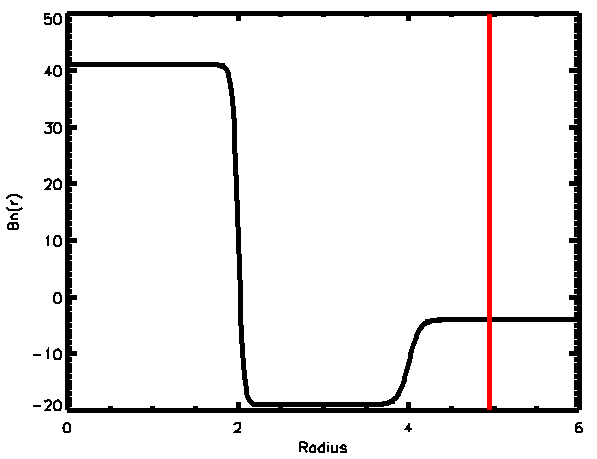}
\caption[Initial field profile]{Top: Contour plot of $B_n$ on the bottom plate. Bottom:  1-D cut of $B_n$ on the bottom plate. Solid red line marks the maximum radius $r_{max}=\sqrt{L_y^2+ L_z^2} = 3.5\sqrt{2}$ of our simulation. }
\label{fig:initial_profile}
\end{figure*}

2D and 1D plots of $B_n(r)$ on the bottom plane are shown in Fig. \ref{fig:initial_profile}.
The profile of this initial field resembles a step function. For $r<r_+$, the field has a uniform value of 
\beg{Brsmall}
B_n(r<r_+) = B_+ - B_- + B_0= 41.
\done
At $r=r_+$, the field drops sharply, over a distance $r=\lambda_+$, to a value of approximately
\beg{Brmid}
B_n(r> r_+) = B_- + B_0 = -19,
\done
At $r = r_-$, the field undergoes a steep rise, over a distance $r = \lambda_-$ after which, for $r\gg r_-$, 
\beg{Brlarge}
B_n(r > r_-) = B_0 = -4.
\done
Our simulation domain extends only to just outside the steep increase at $r=r_-$. As a result, our domain has a single PIL, located at $r=2$. This is a true PIL, in the sense that an observer looking down on the bottom plate will measure two different polarities in the flux distribution. The steep increase of the field at $r=r_-$, on the other hand, is not a true PIL, since there is no change in the sign of the field, and it only affects the far corners of the simulation domain, so the contribution to the total flux from this region is very small. Our simulation was designed so that the longest closed field lines in the system are situated at the very inner edge of the driving region. Here `closed' means that field lines both start and end on the bottom plate. \par
We employ zero-gradient conditions at all times at all six boundaries,
\beg{boundaries}
\begin{split}
\pd{\rho}{n}=0,\\
\pd{T}{n}=0,\\
\pd{\vecv}{n}=0,\\
\pd{\vecB}{n}=0,
\end{split} 
\done
where $n=x,y,z$ is the normal coordinate. With the exception of the bottom boundary, all of the boundaries are open and free slip, so that both plasma and magnetic field lines can move on and across each boundary. The bottom boundary is closed and the field is line tied, so that the field lines move only in response to the imposed boundary flows, emulating the slow driving at the dense solar photosphere, rather than in response to coronal magnetic forces.\par
The initial plasma parameters used in our dimensionless simulation are $\rho_0=1$, $T_0=1$, $P_0=1$, so that the gas constant is $R=1.0$, and the sound speed is $c_s=\sqrt{\gamma P_0/\rho_0} =1.3$ with $\gamma = 5/3$ . Based on the value of $\vecB$ determined from Eqs. (\ref{Laplace})-(\ref{grad}), the Alfv\'en speed $c_{A0}=B_0 / \sqrt{4\pi\rho_0}$ is found to range from 1.4 to 21.2, and the plasma beta, $\beta_0=8\pi P_0/B_0^2$ is found to range from $4\times10^{-3}$ to $1.0$. The $\beta\ll1$ regime corresponds to a magnetically dominated plasma, which is the environment most closely resembling the corona. In our simulation, the regions where $\beta\approx 1$ are mostly in the upper corners of the box, so that the amount of high beta plasma is very minimal, and is not found to affect the results.\par
\begin{figure*}[!h]
\centering\includegraphics[scale=0.30, trim=0.0cm 0.0cm 0.0cm 0.0cm]{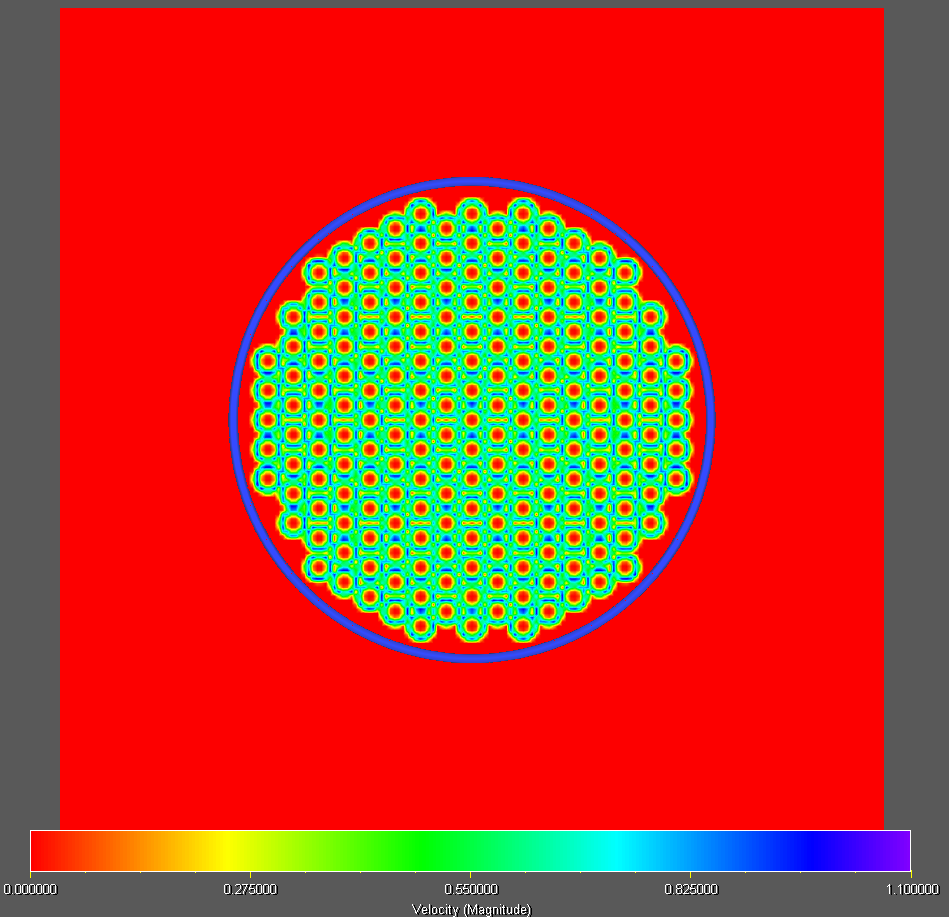}
\centering\includegraphics[scale=0.30, trim=0.0cm 0.0cm 0.0cm 0.0cm]{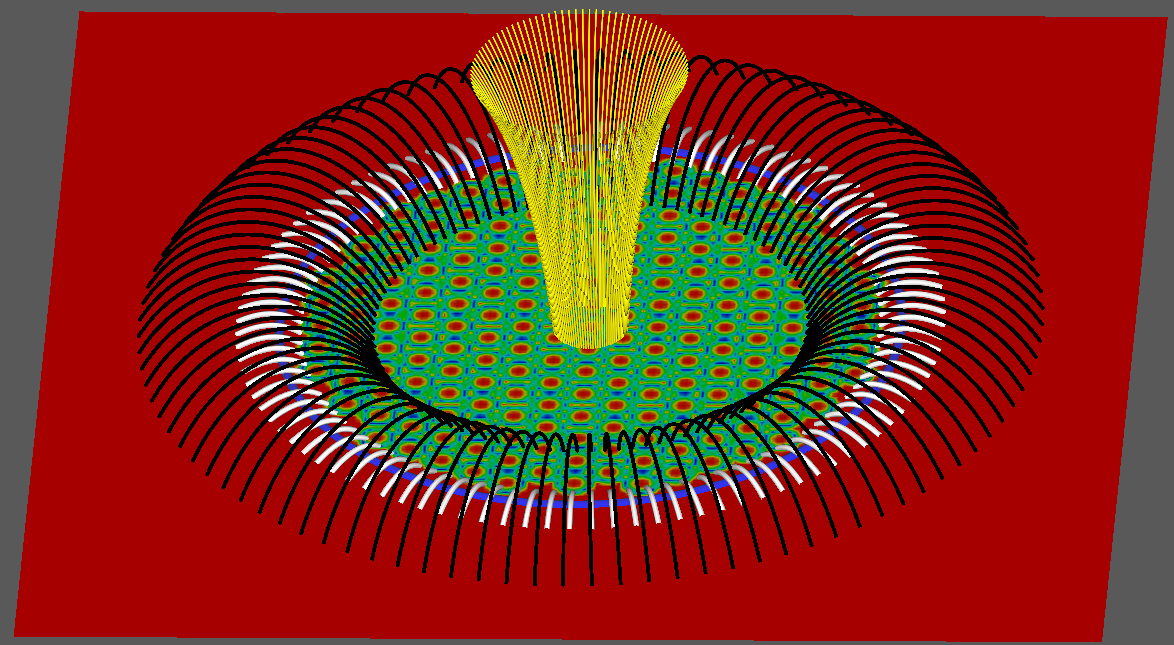}
\caption[Initial simulation setup]{Top: Velocity magnitude on the bottom plate. Red shading indicates locations where the flow speed is zero and green indicates where it is nonzero. Blue circle is the PIL. Bottom:  Initial simulation setup, showing velocity magnitude on the bottom plate, PIL (blue circle), and three sets of field lines. Coronal hole field lines (yellow) start from the bottom boundary and end on the top boundary, where they are free to move. Coronal loop field lines (black) start deep inside the region that is being twisted and terminate outside the PIL. Field lines very close to the PIL (white) are the shortest in the system. }
\label{fig:velocity}
\end{figure*}

To model the random photospheric convection, we impose convective flows on the bottom boundary, arranged in the manner shown in Fig. (\ref{fig:velocity}; top).
Each individual convective flow has an angular rotation rate given by:
\beg{omega}
\Omega(r,t) = \left\{
        \begin{array}{ll}
            -\Omega_0 f(t) g(r) & \quad r \leq a_0 \\
            0 & \quad r > a_0
        \end{array}
    \right.
\done
where
\beg{foft}
f(t) = \frac{1}{2}\Big[1-\cos(2\pi\frac{t}{\tau})\Big]
\done
and
\beg{gofr}
g(r) = \Big(\frac{r}{a_0}\Big)^4 - \Big(\frac{r}{a_0}\Big)^8.
\done
We choose $a_0 = 0.125$, $\tau = 0.572$ and $\Omega_0 = 45.75$. These choices set the peak linear velocity $\vecv_{max} = 1.2$, which is about $5\%$ of the peak Alfv\'en velocity. As argued in \citep{Knizhnik15}, this flow profile will conserve $B_n$ on the bottom boundary.\par 
The initial simulation setup is shown in Fig. (\ref{fig:velocity}; bottom). N=199 convective cells are packed inside a radius $r=r_+$, and are twisting up the flux system, which is bounded by the PIL. Yellow field lines start on the bottom plate (photosphere) and end on the top plate, where they are free to move around, emulating the open field lines inside a coronal hole (CH). Black field lines represent the `loop' portion of the corona, since they are situated deep inside the flux system, far away from the PIL, and they begin and end on the photosphere. White field lines are some of the shortest field lines in the system. starting and ending on the photosphere less than a single flow radius away from the PIL.\par
We impose $\mathrm{30}$ twist cycles on this coronal magnetic field configuration, followed by $\mathrm{5}$ cycles of relaxation. For simplicity, our model only includes convective cells inside the PIL, with no twisting occuring outside. Although on the real Sun convective motions occur throughout the photosphere, the key physics are reproduced in our simulation, since magnetic helicity is being injected into the entire flux system. Additional convective cells located outside the PIL would indeed inject more magnetic helicity into the flux system, but this scenario is physically equivalent to increasing the rate of helicity injection inside the PIL and injecting no magnetic helicity outside. This pattern of the convective cells captures the essential physics of helicity injection, since it imparts magnetic helicity to the entire flux system. Including additional convective cells outside the PIL, therefore, is not expected to change the results of the simulation.\par
\begin{figure}[!h]
\centering\includegraphics[scale=0.75, trim=0.0cm 0.0cm 0.0cm 0.0cm,clip=true]{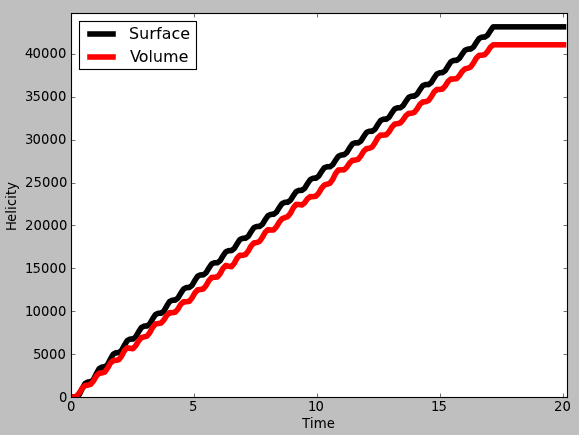}
\caption[Sunspot helicity]{Injected (black) and volume-integrated (red) magnetic helicity during the simulation.}
\label{fig:helicity}
\end{figure}

The basis for the helicity condensation model is the transport of magnetic helicity throughout the solar corona. In the corona, the gauge-invariant form of the relative magnetic helicity \citep{Finn85} and its time-derivative are defined as:
\beg{relativehelicity}
H=\int_V{(\vecA+\vecA_P)\cdot(\vecB-\vecB_P) \; dV},
\done
\beg{dHdt}
\frac{dH}{dt}=2\oint_S{\left[ \left( \vecA_P \cdot \vecv \right) \vecB - \left( \vecA_P \cdot \vecB \right) \vecv \right] \cdot \vecdS.}
\done
In these equations, $\vecB_P = \nabla\times\vecA_P$ is the potential magnetic field satisfying $\nabla\times\vecB_P = 0$, which is generated by a vector potential $\vecA_P$, $\vecB$ is the instantaneous magnetic field with vector potential $\vecA$ and $\vecv$ is the velocity on the boundary $S$ bounding the volume $V$. The first term in Equation (\ref{dHdt}) represents the effects of twisting or shearing motions on the boundary, while the second term quantifies the helicity injected or removed by flux emergence or submergence across the boundary. Since magnetic helicity is conserved under reconnection \citep{Woltjer58,Taylor74,Taylor86,Berger84b}, in the absence of numerical diffusion we should find that
\beg{Hconserved}
H(t) = \int_0^t{dt' \frac{dH(t')}{dt'}}
\done
where the LHS is calculated via Equation (\ref{relativehelicity}) and the RHS is calculated by integrating Equation (\ref{dHdt}) in time. This numerical test of helicity conservation using ARMS was performed for a plane-parallel Parker corona by \citet{Knizhnik15} and in Chapter \ref{Structure}. They confirmed Equation (\ref{Hconserved}) to better than $\mathrm{1\%}$ accuracy. In Fig. \ref{fig:helicity}, we plot the injected and volume-integrated magnetic helicity, calculated by numerically integrating Equations (\ref{relativehelicity})-(\ref{dHdt}). 
In contrast to the magnetic field at the start of the simulation, which was solved using Dirchlet conditions at infinity, the potential magnetic field at subsequent time steps depends on the normal component of the field at all of the boundaries. Thus, the scalar potential was calculated at each time step using the MUDPACK suite \citep{Adams89}, which solved the elliptic partial differential Laplace Equation (\ref{Laplace}) inside our simulation domain with the specified Neumann boundary conditions, Equation (\ref{neumann}), on all six sides using a multigrid iteration technique. The potential field was then calculated from the gradient of the scalar potential, Equation (\ref{grad}). The helicity is misconserved by about $\mathrm{4\%}$ at the end of the simulation, which is not nearly as good as \citet{Knizhnik15} and Chapter \ref{Structure}, but is remarkably good considering the amount of reconnection and complexity of the simulation. Helicity is conserved somewhat better than the helicity conservation obtained by \citet{MacNeice04}($\sim\mathrm{20\%}$) using a 2.5D code very similar to ARMS. There are two major differences between this simulation and the helicity conservation simulations which are likely to reduce the helicity measured in the volume from the value that was injected at the photospheric level. The first difference is that in the plane-parallel Parker corona, the background magnetic field was uniform and aligned with the grid, so that two components of the potential field vanish, causing any numerical errors to be confined to one component of the magnetic field. The second, probably more important difference, is the absence of a PIL in the plane-parallel Parker corona, meaning that interpolations of a (already uniform and smooth) magnetic field were not affected by any large gradients. A poorly resolved PIL, such as the one in the present simulation, will, upon interpolation, smooth out the magnetic field, likely resulting in a loss of helicity. As will be demonstrated in the next section, however, the predictions of the helicity consensation model are still confirmed despite this loss of magnetic helicity.

\section{Results}\label{sec:results}
\subsection{Helicity Condensation}
\begin{figure}[!h]
\centering\includegraphics[scale=0.35, trim=0.0cm 0.0cm 0.0cm 0.0cm]{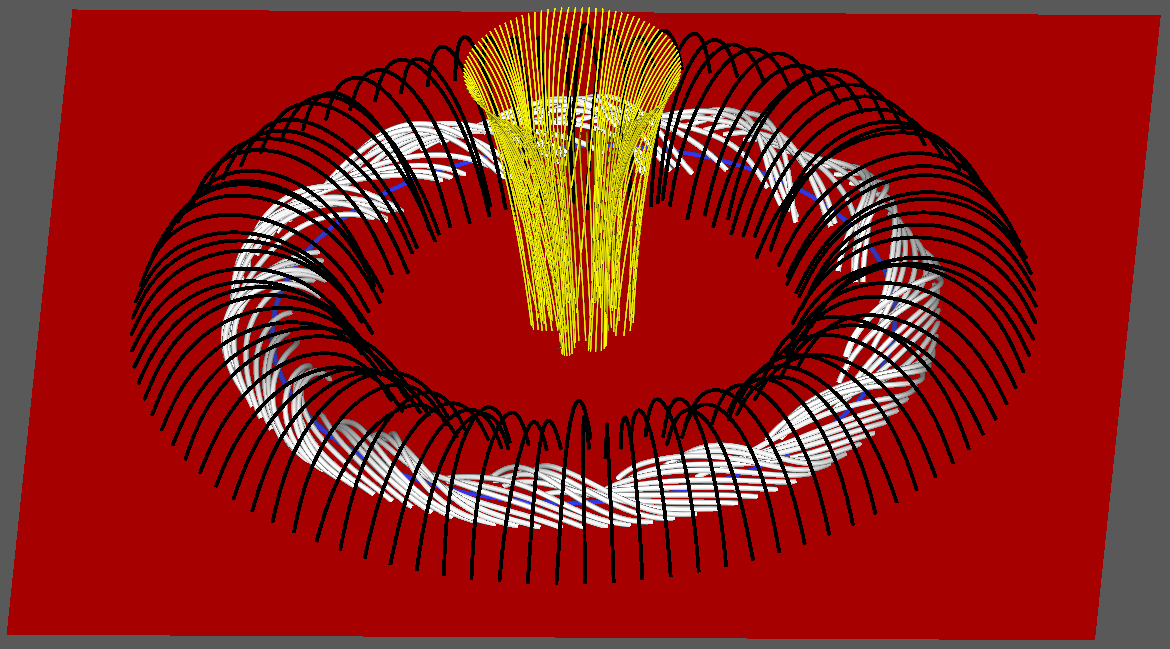}
\caption[Field lines at the end of the simulation]{Same field lines as bottom panel in Fig. \ref{fig:velocity} but at the end of the simulation.}
\label{fig:sunspot_end}
\end{figure}

\begin{figure*}[!h]
\centering\includegraphics[scale=0.40, trim=0.0cm 0.5cm 4.0cm 0.0cm,clip=true]{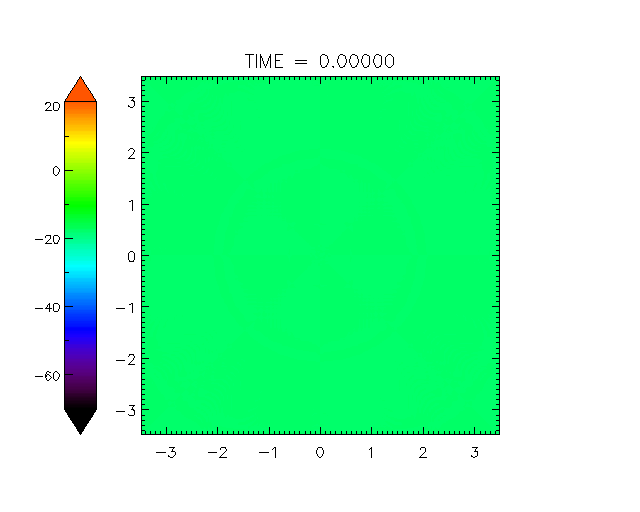}
\centering\includegraphics[scale=0.40, trim=4.0cm 0.5cm 0.0cm 0.0cm,clip=true]{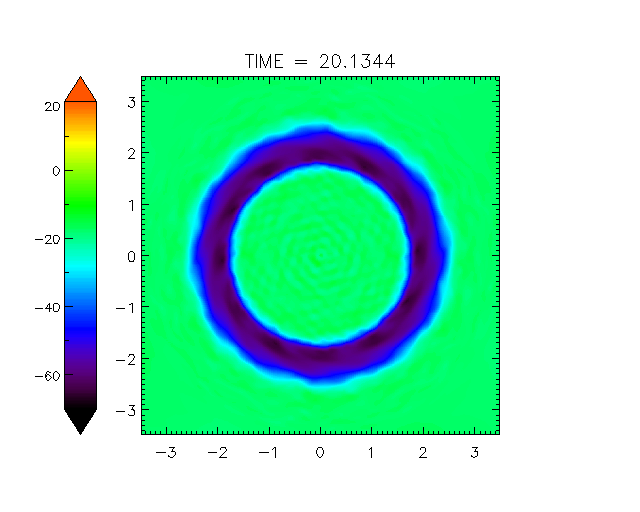}
\caption[Twist field for the sunspot simulation]{$B_\theta$ on the photosphere (bottom plane) at the beginning (left) and end (right) of the simulation.}
\label{fig:sunspot_twist}
\end{figure*}

We begin the presentation of the results of our simulation by showing, in Fig. \ref{fig:sunspot_end}, the same field lines as are shown in the bottom frame of Fig. \ref{fig:velocity} after $\mathrm{30}$ twist cycles and $\mathrm{5}$ relaxation cycles. In contrast to the initial field configuration, the field configuration at the end of the simulation has a region of localized nonpotentiality, namely the highly sheared field lines (white) near the PIL (blue circle). The overarching (black) field lines, representative of the `loop' portion of the corona, are very nearly in their original potential state. The (yellow) CH field lines are, for the most part, the same as they were at the beginning of the simulation. 
In Fig. \ref{fig:sunspot_twist}, where we plot the twist component of the field $B_\theta$ 
\beg{twist}
B_\theta = B_y \cos\theta - B_z \sin\theta,
\done
on the photosphere (bottom plane) at the beginning and end of the simulation. At the beginning of the simulation, there is, of course, no twist field anywhere in the domain. Although twist was injected everywhere inside the PIL, there is only one significant region of twist at the end of the simulation, namely, the blue/purple annular region located about the PIL. The interior of the system, including the CH, is almost entirely untwisted, since, in the open field region the twist simply leaves the system, while in the closed field region the twist is transported by reconnection to the PIL.

\begin{figure*}[!h]
\centering\includegraphics[scale=0.70, trim=0.0cm 5.2cm 0.0cm 5.5cm,clip=true]{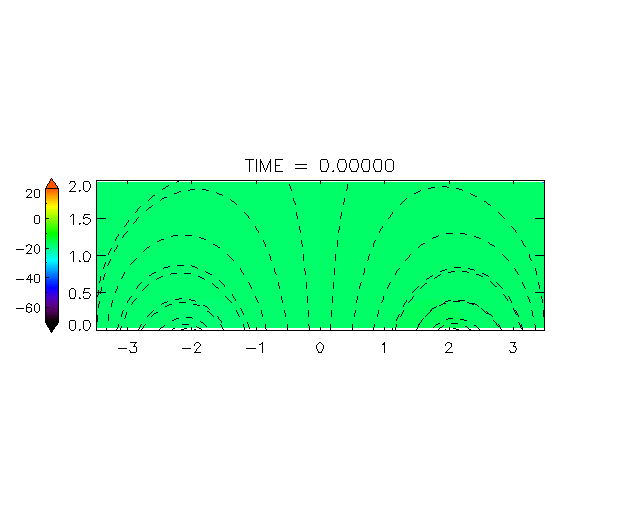}
\centering\includegraphics[scale=0.70, trim=0.0cm 5.5cm 0.0cm 5.5cm,clip=true]{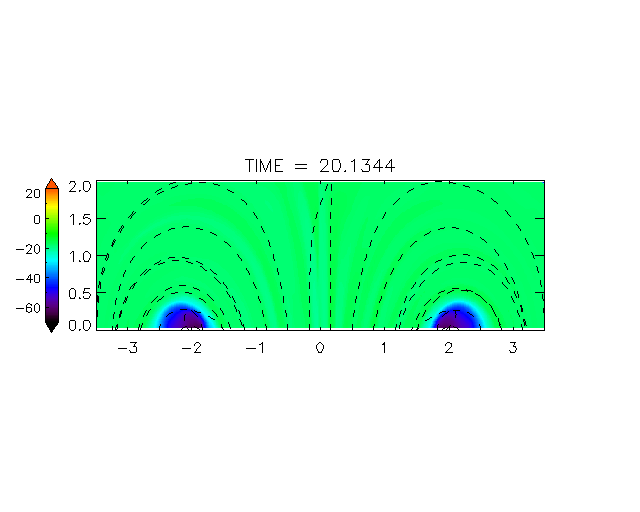}
\caption[$B_\theta$ at the beginning and end of the sunspot simulation through a vertical plane]{$B_\theta$ through the $y=0$ plane at the beginning (top) and end (bottom) of the simulation. The color scale has been saturated to bring out some of the features.}
\label{fig:sunspot_twist_vert}
\end{figure*}

In Fig. \ref{fig:sunspot_twist_vert} we plot $B_\theta$ through the $y=0$ vertical plane, along with two-dimensional projections of a set of field lines.
At the beginning of the simulation, there is no $\theta$-component of the magnetic field component through this plane, and the field lines are quasi-potential. At the end of the simulation, however, there is a strong component normal to this plane, especially located around $y=\pm2$, the location of the PIL. Since the field lines are just projections onto the 2D plane, the sheared field lines at the PIL show up as very short, apparently quasi-potential, loops, but the long field lines near the CH very clearly close back down to the bottom plate. These results indicate that, as in Chapters \ref{SimsHelicity} and \ref{Structure}, the twist field accumulates at the PIL boundary, in agreement with the prediction of \citet{Antiochos13}. \par
\subsection{Filament Channel Shear}
\begin{figure*}[!h]
\centering\includegraphics[scale=0.40, trim=0.0cm 0.5cm 4.0cm 0.0cm,clip=true]{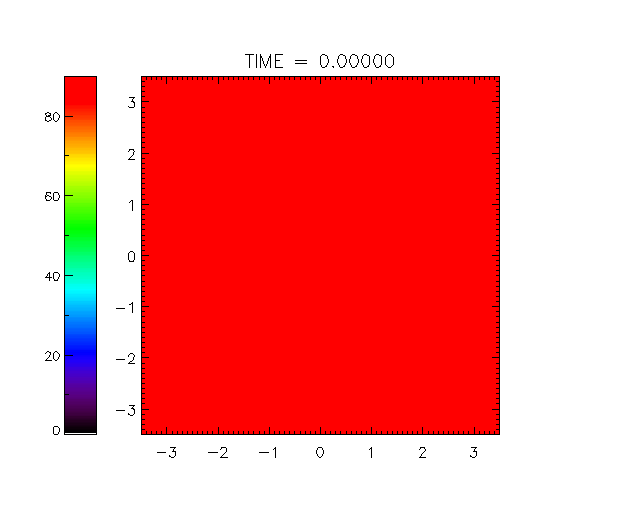}
\centering\includegraphics[scale=0.40, trim=4.0cm 0.5cm 0.0cm 0.0cm,clip=true]{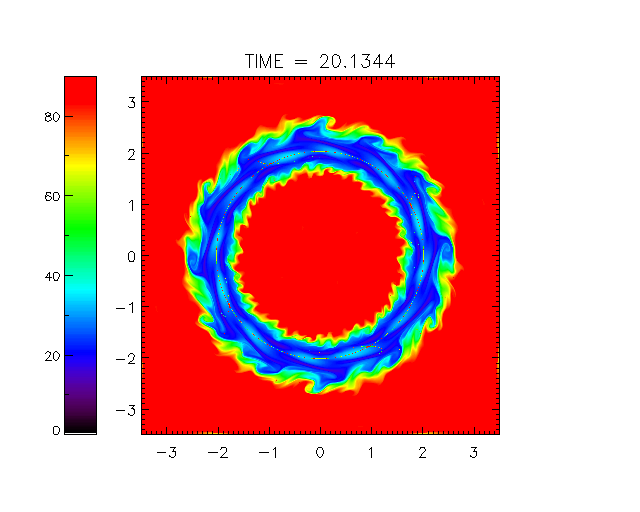}
\caption[Field line shear for the sunspot simulation]{Field line shear on the bottom plane at the beginning (left) and end (right) of the simulation.}
\label{fig:sunspot_shear}
\end{figure*}

The results described above indicate that the helicity has condensed at the PIL and has mostly been ejected out of the system at the CH. However, filament channels are known to be long, thin structures, and their helicity manifests itself primarily in the form of shear along the PIL. To this end, we plot the field line shear at the beginning and end of the simulation in Fig. \ref{fig:sunspot_shear}.
We define shear to be the angle between the straight line connecting the footpoints of a given field line and the line tangent to the PIL at the point where it is intersected by this straight line. We assign both footpoints to have the same shear angle even though they are on opposite sides of the PIL. Thus, at the start of the simulation, when the field is completely potential, the line connecting the footpoints of each field line is at a right angle to the tangent line at the intersection point with the PIL, implying a shear angle of $\theta_S=90^\circ$ everywhere. At the end of the simulation, however, there is only one region of significant shear, the band localized at the PIL. There appear to be two `bands' on either side of the PIL, but this is due to field lines crossing the PIL being assigned the same shear angle at both ends, so that the outer `band' corresponds to the outer footpoint of the same field line in the inner `band'. A striking result of this simulation is that the shear angle of field lines near the PIL appears to be approximately $\theta_S = 25^\circ$, which matches the the observed value \citep{Tandberg-Hanssen95}. 
\begin{figure}[!h]
\centering\includegraphics[scale=0.70, trim=0.0cm 0.0cm 0.0cm 0.0cm,clip=true]{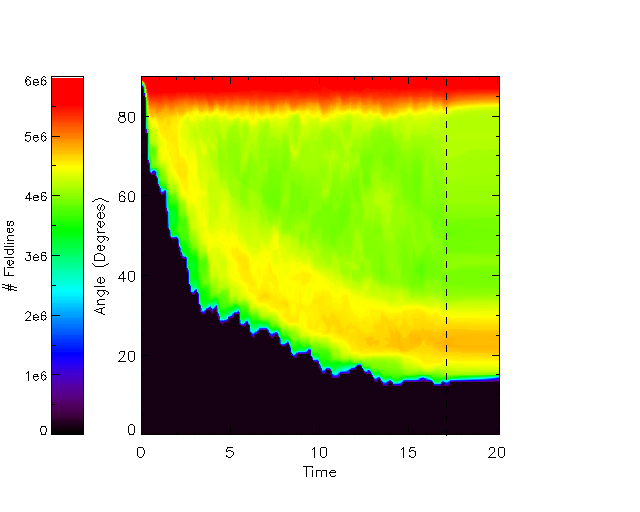}
\caption[Angle-time histogram of field line shear angle]{Angle-time histogram created by stacking in time histograms of the shear angle maps seen in Fig. \ref{fig:sunspot_shear}. The dashed line shows the time at which the twisting was stopped.}
\label{fig:histogram}
\end{figure}
Furthermore, this value appears to be the asymptotic value of for our system, as can be seen in Fig. \ref{fig:histogram}, where we calculate a time-stacked histogram of field line shear. At the beginning of the simulation, all of the field lines have a shear angle of $\mathrm{\theta_S=90^\circ}$, resulting in the vertical cut at $\mathrm{t=0}$ being black (meaning there are no field lines with this shear angle) everywhere except at this value. In fact, there are field lines with a shear angle of $\mathrm{\theta_S=90^\circ}$ throughout the simulation. As the simulation proceeds, field lines become increasing sheared (i.e. $\mathrm{\theta_S<90^\circ}$), and, in particular, at later times there is a large fraction of highly sheared field lines at a value of $\mathrm{\theta_S\approx25^\circ}$. Crucially, the shear appears to be asymptotically approaching this value. \par
\begin{figure}[!h]
\centering\includegraphics[scale=0.60, trim=0.0cm 0.5cm 4.0cm 0.0cm,clip=true]{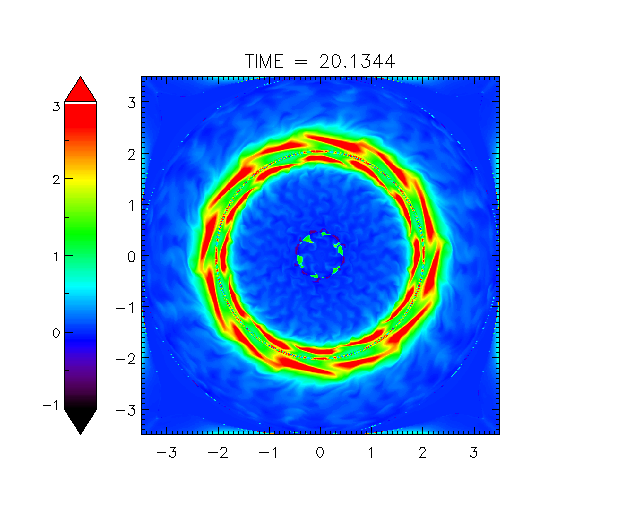}
\caption[Percent field line length difference for the sunspot simulation]{\% difference ($\times100$) in field line length between the beginning and end of the simulation.}
\label{fig:length_diff}
\end{figure}

Since the field lines closest to the PIL become highly sheared, their lengths are expected to change appreciately. Since they start out as the shortest field lines in the system, however, it is useful to plot the percent change in field line length, to account for the fact that field lines just outside the CH boundary, for example, are the longest field lines in the system. In Fig. \ref{fig:length_diff} we plot the percent difference in the length of the field lines between the beginning and end of the simulation. 
Once again, it is evident that the largest increase in length is obtained by the field lines nearest to the PIL, where a long, highly sheared filament channel is formed. There is a much larger circle where the field line length changes substantially, but this is just due to several CH field lines reconnecting with underlying field lines and closing back down, thus increasing their length.

\subsection{Coronal Hole Evolution}
\begin{figure*}[!h]
\centering\includegraphics[scale=0.22, trim=2.2cm 0.0cm 5.0cm 0.0cm,clip=true]{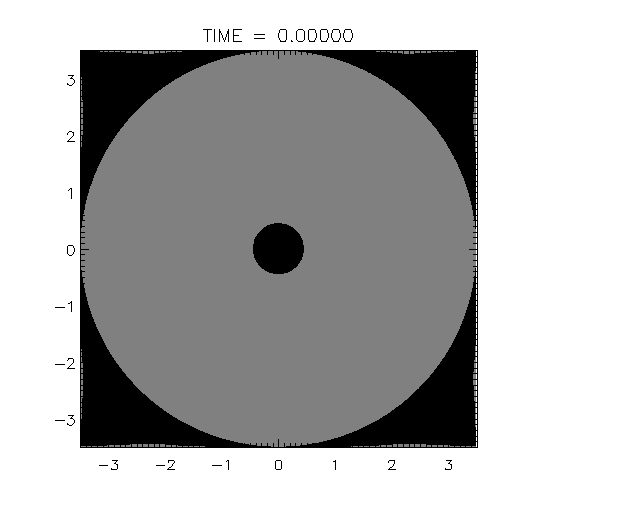}
\centering\includegraphics[scale=0.22, trim=2.2cm 0.0cm 5.0cm 0.0cm,clip=true]{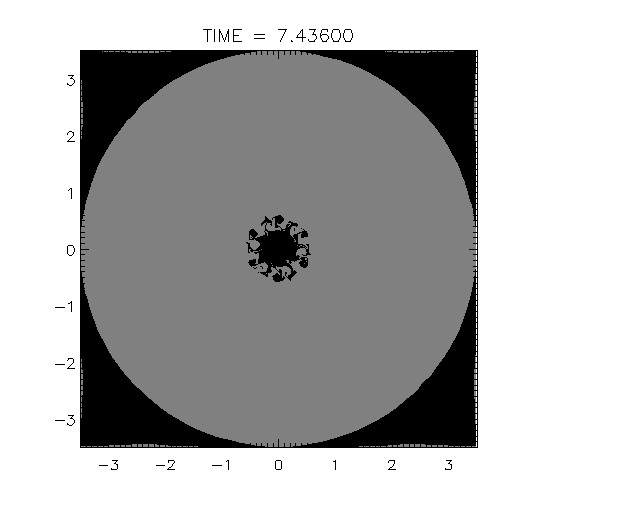}
\centering\includegraphics[scale=0.22, trim=2.2cm 0.0cm 5.0cm 0.0cm,clip=true]{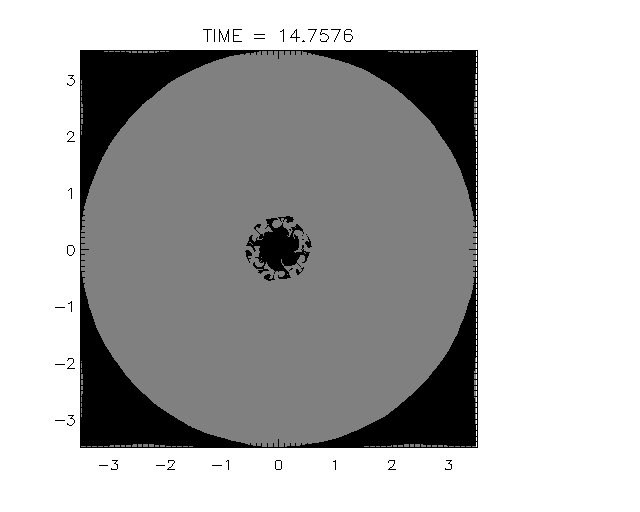}
\centering\includegraphics[scale=0.22, trim=2.2cm 0.0cm 5.0cm 0.0cm,clip=true]{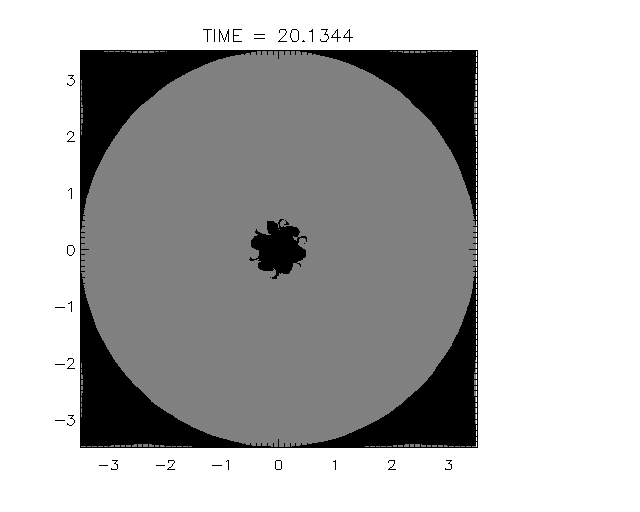}
\centering\includegraphics[scale=0.22, trim=2.2cm 0.0cm 5.0cm 0.0cm,clip=true]{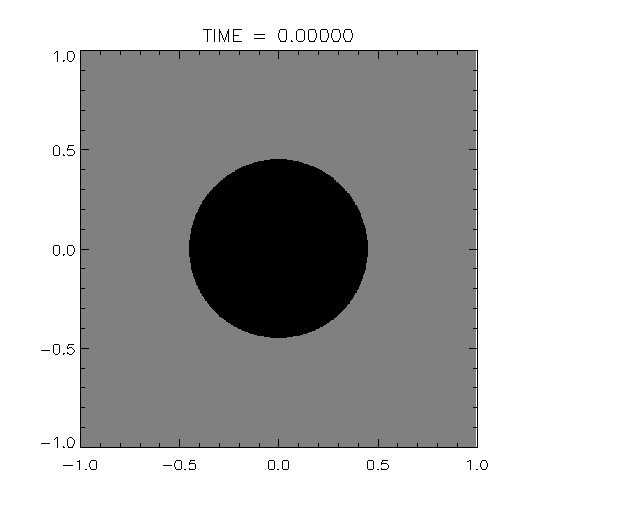}
\centering\includegraphics[scale=0.22, trim=2.2cm 0.0cm 5.0cm 0.0cm,clip=true]{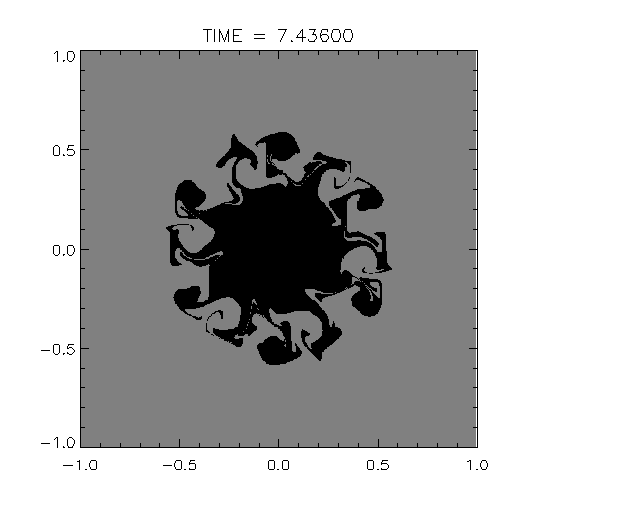}
\centering\includegraphics[scale=0.22, trim=2.2cm 0.0cm 5.0cm 0.0cm,clip=true]{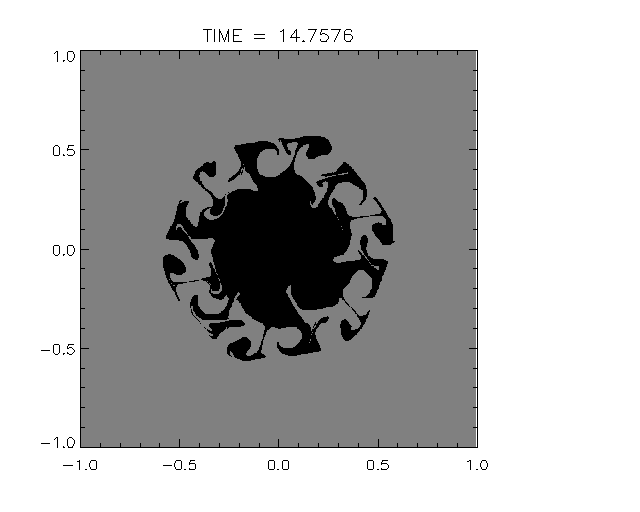}
\centering\includegraphics[scale=0.22, trim=2.2cm 0.0cm 5.0cm 0.0cm,clip=true]{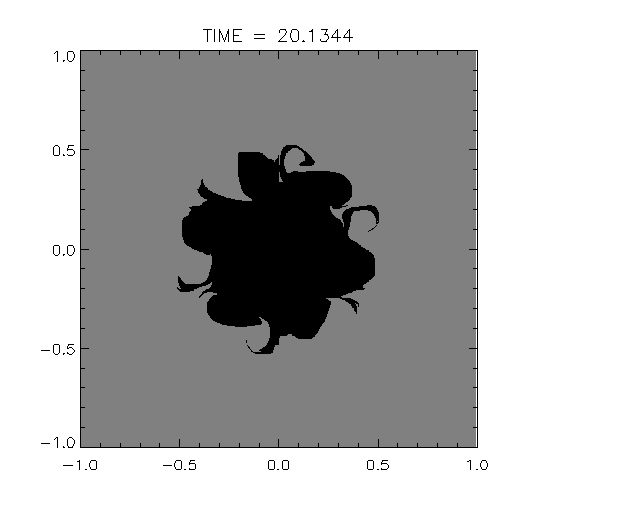}
\caption[Sunspot coronal hole]{Full view (top) and close-up (bottom) map of field line connectivity at various times throughout the simulation. Field lines are traced from the bottom to the top plate, and CH field lines (black), defined as those that terminate on a boundary other than the bottom plate, are surrounded by closed field lines (grey), defined as those which both start and end on the bottom plate.}
\label{fig:CH}
\end{figure*}

\begin{figure*}[!h]
\centering\includegraphics[scale=0.3, trim=4.0cm 0.0cm 3.0cm 0.0cm]{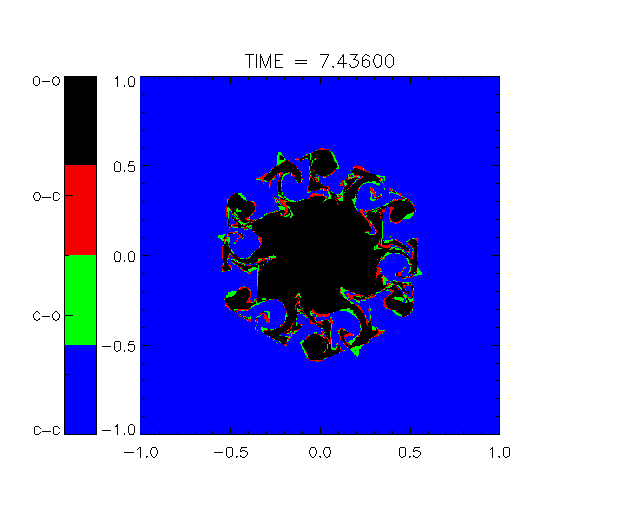}
\centering\includegraphics[scale=0.3, trim=4.0cm 0.0cm 3.0cm 0.0cm,clip=true]{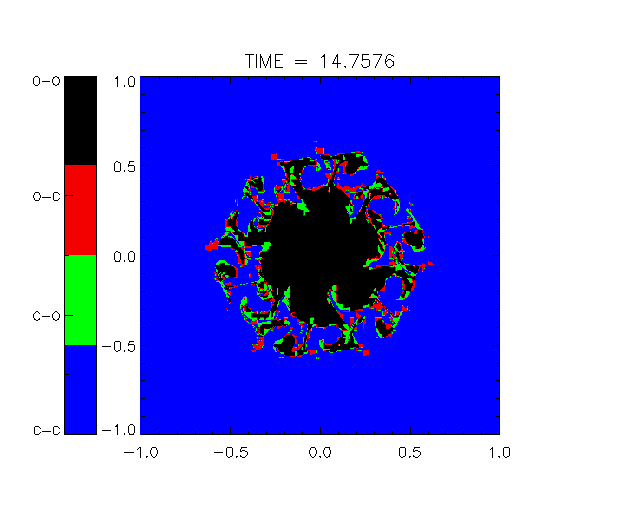}
\centering\includegraphics[scale=0.3, trim=4.0cm 0.0cm 3.0cm 0.0cm,clip=true]{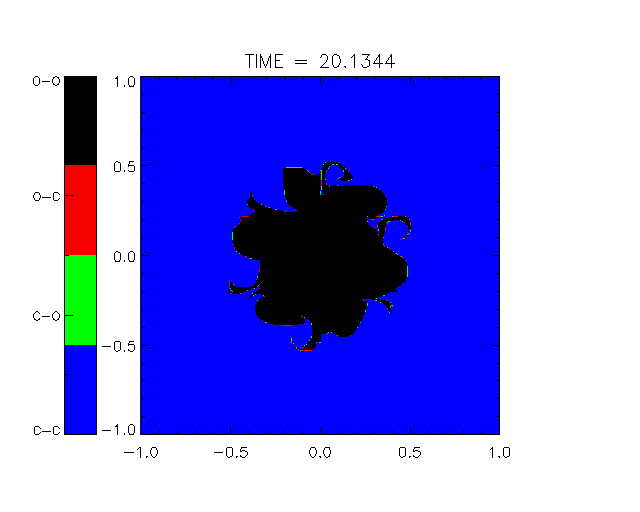}
\caption[Change in coronal hole connectivity]{Change in field line connectivity from the previous timestep for the last three snapshots of Fig. \ref{fig:CH}. Field lines that remain closed (blue) and field lines that remain open (black) and separated by regions of interchanging field lines. Red/green indicate field lines that were open (closed) that have closed (opened). }
\label{fig:change_con}
\end{figure*}

The CH in our simulation is defined as the set of field lines traced from the bottom boundary that do not connect back down to this boundary. Since field lines are able to slip at all four side walls and the top plate, any helicity injected into those field lines will simply propagate out of the simulation box. However, the location of these `open' field lines is anticipated to change with time, as the underlying closed field lines rise, due to the stresses imposed at the boundary, and either hit the top boundary or reconnect with open field lines. This process will either produce two open field lines from a single closed field line, or interchange a closed field line with an open one. In Fig. \ref{fig:CH} we plot a set of field line connectivity maps at four times during the simulation. Field lines which start and end at the bottom boundary are marked as grey. Field lines that start from the bottom boundary but terminate on a different boundary (all of which are free-slip) are marked as black and are considered to part of the CH. 
The initial open field distribution (first column) consists of a central portion of diameter $a_{CH,c}\approx0.8$, and an outer portion at a radius of approximately $a_{CH,o}\approx L_y\approx L_z$. Here $c$ and $o$ stand for the central and outer open field regions, respectively. Immediately after turning on the convective cells, the shape of the CH becomes distorted, as the boundary motions drag open and closed flux around. The convective cells that contain both open and closed flux move open flux into a previous closed region and vice versa. This creates the pinwheel shape seen in the second and third columns. At the end of the simulation (fourth column), the open field region has retracted into a more circular shape. 
Interestingly, the evolution of the interior CH boundary appears to occur without the outer open field region changing at all (top panels). This indicates that rather than evolving through a series of openings and closings which would manifest themself as a change in both the outer and inner open field boundaries, open (closed) field lines which are stressed by the flows undergo interchange reconnection with closed (open) field lines, a process which will not change the connectivity of the outer footpoint. This can be most clearly seen by plotting the change in field line connectivity. In Fig. \ref{fig:change_con}, we show, for the last three times shown in Fig. \ref{fig:CH}, zoom-in maps of field lines that were closed in the previous time step that are still closed (blue), field lines that were open in the previous time step that are still open (black), and field lines that have changed their connectivity.
During the twisting period, there is clearly a lot of interchange reconnection occurring at the CH boundary, with the convective cells bringing the open field into contact with the closed field, after which open and closed field lines exchange connectivity. This process transfers helicity into and out of the CH. During the relaxation phase of the simulation, on the other hand, interchange reconnection is very limited. The evolution of the CH appears to be mostly ideal. Exploring the dynamics of the CH in more detail will be the subject of a future work.\par

\section{Discussion}\label{sec:discussion}
In this work, we have dramatically improved and extended the results from Chapters \ref{SimsHelicity} and \ref{Structure}. We simulated the injection of magnetic helicity into a realistic coronal magnetic field, with a true PIL and CH, and allowed magnetic reconnection to transport magnetic helicity throughout the model solar corona. We found that the helicity built up at the PIL, where it manifested itself in the form of highly sheared magnetic field lines, leaving a deficit of helicity in the rest of the corona. The resulting coronal structure was one that closely matches the observations: highly sheared filament channels localized at the PIL and very little, if any, structure anywhere else. Furthermore, we showed that the field lines in the CH were almost completely untwisted, with all of the helicity that entered the CH being ejected out of the simulation domain. Nevertheless, we showed that the CH boundary was affected by the injection of magnetic helicity at the photospheric level, with the boundary responding by field line opening and interchange reconnection, transforming the CH boundary from a circular shape to more of a pinwheel one. \par
These results confirm what we found in Chapters \ref{SimsHelicity} and \ref{Structure}, but the results are much more far-reaching and much more realistic. Whereas in those simulations the model corona consisted of a uniform magnetic field that lacked a true PIL and CH, these simulations mimic the real corona because the photosphere magnetic field changes sign and the boundaries of the domain are open, allowing helicity to leave the system. As a result, at the PIL the concentrations of helicity are not manifested as simple concentration of twist field, with field lines reaching from one end of the box to the other as was the case previously, but instead are in fact highly sheared, elongated field lines that correspond to those found in filament channels. At the CH, these simulations see no accumulation of twist or shear, which is in stark contrast the results from previous Chapters, where helicity did not leave the system.\par
Filament channel formation via magnetic helicity condensation is, in some sense, a global process. Rather than requiring a region of localized shearing and converging flows (as is the case in many flux cancellation models), or a region of localized flux emergence (as is the case when filament channels form as a result of an emerging flux rope), the helicity condensation model simply requires a PIL. Since the PIL contains all of the flux in the system, it forms a natural boundary which magnetic helicity cannot cross, and so shear will quite naturally accumulate there, regardless of how the photospheric motion is distributed. This makes helicity condensation an appealling and likely candidate for explaining the filament channel formation process. \par
The results of this work as extremely interesting and immediately suggest several different avenues of further research. The physics of helicity flux leaving the CH, in particular, is intriguing because the interior of open field regions are thought to be the source of fast solar wind, while the boundary between open and closed field regions is thought to be the source of the slow solar wind. The simulations presented here suggest two important predictions for the fast and slow solar winds. First, as argued by \citep{Antiochos11} and confirmed here, the sign of the helicity injection into the CH (positive, in our simulation) is opposite to that of the helicity that is transported to its boundary (negative, in our simulation). This implies that the total helicity leaving the CH will not simply be equal to the helicity injected into the CH, nor will it be equal to the helicity that is transported to its boundary. Instead, it will be the sum of these two quantities. In the context of observations, this implies that the helicity flux measured in situ in the fast wind will be less than the helicity flux injected into the CH measured at the photospheric level. Accurate in situ measurements are needed to test this prediction.\par
A second important prediction of this model is the presence of extremely narrow corridors of open magnetic field connecting two seemingly disconnected CHs. This is most readily apparent in the second and third rows of Fig. \ref{fig:CH}, where the central CH appears to be disconnected from several of the surrounding satellite CHs. However, examining the results with high resolution reveals that the CHs are, in fact, connected by narrow lanes of open flux. Such corridors have been predicted to be crucial for explaining the presence of the high-latitude slow wind \citep{Antiochos11}. Quantifying and understanding the implications of our simulations for these two topics will be the subject of future work.\par


\chapter{Coronal Heating Via Magnetic Helicity Condensation on the Sun}\label{Heating}

\section{Introduction}\label{sec:intro}
Understanding the nature of coronal heating has been a long standing problem in solar physics. Dating back to the work of \citet{Grotrian39}, it has been known that coronal temperatures are in excess of $10^6 \; \mathrm{K}$, approximately two orders of magnitude hotter than the underlying photosphere. Calculations by \citet{Withbroe77} showed that the energy flux necessary to account for the observed temperatures was of the order of $10^7 \; \mathrm{ergs\; cm^{-2} \; s^{-1}}$ in active regions and $3\times10^5 \; \mathrm{ergs\;cm^{-2}\;s^{-1}}$ in the quiet Sun. While it is generally agreed that the energy input must come from the photosphere, there are at least two main mechanisms that are frequently used to explain the observed heating. \citet{Parker72} studied the coronal heating problem and argued that quasi-static photospheric motions tangle and braid the magnetic field, imparting it with magnetic energy, which is then converted into heat by magnetic reconnection. Such reconnection events, frequently called nanoflares \citep[e.g.][]{Klimchuk15}, are thought to occur between elemental strands less than $15 \; \mathrm{km}$ in diameter \citep{Peter13}.  An alternative explanation for the heating is that waves, such as the Alfv\'en \citep{Osterbrock61} or magnetosonic \citep{Pekunlu01} modes, generated by fast photospheric processes deposit their energy into the corona by resonant absorption or phase mixing.  \par
While these models are frequently able to quantitatively reproduce the observed heating \citep[e.g.,][see, however, \cite{Gonzalez15}]{Kumar06, Viall11, Bradshaw12, Hahn14}, there has been, to date, little consideration of the topological implications of these models. Jostling of the magnetic field footpoints on the photospheric level, whether this motion is slow (heating by nanoflare reconnection) or fast (wave heating), will inject energy in the form of twist which, in turn, is a form of magnetic helicity. In the high Lundquist number regime of the solar corona, magnetic helicity is conserved under reconnection \citep{Woltjer58,Taylor74,Taylor86,Berger84b}, so it is expected to appear in the corona as regions of localized non-potentiality. However, $\it{TRACE}$ images reveal an incredibly smooth, quasi-potential corona everywhere except in filament channels \citep{Schrijver99}. This naturally raises the question: if twist is injected at the photospheric level in order to heat the corona, where does it go? \par 
Recent theoretical work \citep{Antiochos13}, supplemented by numerical modelling \citep{Zhao15,Knizhnik15}, has resolved this question by arguing that after twist is injected into the coronal field, it is transported to polarity inversion lines (PILs), where it manifests itself as sheared filament channels. \citet{Knizhnik15} (hereafter KAD15) showed that this process, known as magnetic helicity condensation, leaves behind a generally smooth corona, with coronal loops at or near a quasi-potential state, and Chapter \ref{Structure} (hereafter KAD16) demonstrated that varying the helicity injection preference affects the timescale of filament channel formation and the amount of structure of the smooth loop corona. They found that for helicity preferences in the range observed on the Sun there is little to no twist in the smooth loop corona. The helicity condensation process, therefore, naturally explains why twist is not observed in the `loop' portion of the coronal field. \par
Although these results have shown that magnetic reconnection transports magnetic helicity to PILs by removing it from the coronal loops, it is not clear how much energy is actually converted to heat in this process. Since filament channels are highly sheared structures, they contain a tremendous amount of free energy, which must have come from the energy injected at the photospheric level. Since so much magnetic energy remains in the coronal field, in the form of filament channels, it is not obvious that the conversion of magnetic energy into heat is sufficiently efficient to account for the observed heating. Quantifying precisely how much magnetic energy is converted into heat by magnetic reconnection transporting magnetic helicity out of coronal loops, and calculating the resulting energy flux is, therefore, crucial to verifying the helicity condensation model and understanding the energy source of the multimillion degree corona. \par 
In this paper, we argue that the magnetic reconnection that is responsible for transporting magnetic helicity to the PIL converts the majority of the injected magnetic energy into plasma heating, with an energy flux sufficient to sustain the observed radiative losses. Furthermore, we argue that the coronal heating resulting from this process is completely independent of the helicity injection preference, and for large helicity preferences a significant amount of magnetic energy will be left over to be stored in the form of the large scale shear observed in filament channels. This work, combined with the results of KAD15 and KAD16, demonstrates that the helicity condensation model provides a global picture of coronal structure and heating, wherein both magnetic energy and helicity are injected at the photospheric level, but while the helicity accumulates and is observed above PILs in filament channels, the magnetic energy is everywhere converted into heat by magnetic reconnection. Thus, understanding the transport of magnetic helicity throughout the solar corona is crucial to determining the both the structure and heating of the Sun's corona.\par

The Chapter is organized as follows. In \S \ref{sec:model} we review the setup of our numerical simulations. In \S \ref{sec:Results} we discuss the results of our simulations, describing the partitioning of injected energy into magnetic, kinetic and thermal energy and relate the results to solar values. Finally, in \S \ref{sec:discussion} we discuss our conclusions and the implications for future research. 
\section{The Numerical Model}\label{sec:model}
The numerical simulations analyzed in this study are almost identical to those described in KAD16. The simulations are initialized with a plane-parallel Parker corona, with $84$ flux tubes that are being twisted on the line-tied top and bottom photospheres. The surface motions inject varying amounts of helicity, the $k=1$, $k=0.75$ and $k=0.5$ cases described in KAD16, which correspond, respectively, to a $100\%$, $50\%$ and $0\%$ helicity preference. We perform each simulation with a photospheric pattern that is randomly shifted by an arbitrary angle after each twist cycle. Each twist cycle consists of a slow ramp-up phase, followed by a slow decline phase with the angular velocity of each individual cell having the form:
\beg{Omega}
\Omega(r,t) = -\Omega_0g(r)f(t) 
\done
where
\beg{gofr}
g(r) = \Big(\frac{r}{a_0}\Big)^4 - \Big(\frac{r}{a_0}\Big)^8
\done
and
\beg{foft}
f(t) = \frac{1}{2}\Big[1-\cos\Big(2\pi\frac{t}{\tau}\Big)\Big]
\done
We set $\Omega = 7.5$ to be the flow amplitude, $a_0=0.125$ to be the flow radius, and $\tau=3.35$ to be the period during which the maximum angle of the clockwise rotation within each cell is $\phi_{max} = \pi$. Since the twisting is occurring on both the top and bottom plates, the maximum rotation angle of each flux tube is $2\pi$. Each simulation is run out for $21$ twist cycles and $5$ relaxation cycles, where no twisting motions are applied. \par 
The key difference between the simulation setup presented here and that presented in KAD16 is that the MHD equations that we solve here include the energy equation in conservative form: 
\beg{energy}
\pd{U}{t}+\divv{\Big\{\left(U+P+\frac{B^2}{8\pi}\right)\vecv-\frac{\vecB(\vecv\cdot\vecB)}{4\pi}\Big\}}=0.
\done
where
\beg{totenergydensity}
 U=\epsilon+K+W
\done
is the total energy density, the sum of the internal energy density
\beg{internal}
\epsilon=\frac{P}{\gamma-1},
\done
kinetic energy density
\beg{kinetic}
K=\frac{\rho v^2}{2},
\done
and magnetic energy density
\beg{magneticenergydensity}
W=\frac{B^2}{8\pi}.
\done
In the above equations, $\rho$ is the mass density, $P$ is the thermal pressure, $\gamma$ is the ratio of specific heats, $\vecv$ is the velocity, $\vecB$ is the magnetic field, and $t$ is time. Eq. (\ref{energy}) is in contrast to KAD15 and KAD16, in which the adiabatic temperature equation was solved, and any changes in the internal energy density were due to compression or expansion. By solving the conservative form of the energy equation, we guarantee that magnetic reconnection can be converted into both plasma heating and bulk motions. We relate the temperature to the pressure via the ideal gas law
\beg{ideal}
T = \frac{P}{nk_B}.
\done
The initial temperature, pressure, and density all have a uniform value of unity but in the simulations presented here, in contrast to KAD15 and KAD16, their values will be able to change substantially in response to stresses in the magnetic field. There are no conduction and radiation sources or sinks for the heating, so the temperature is expected to rise throughout the simulations.\par
\section{Results}\label{sec:Results}
We present below various diagnostics that quantify the heating of the corona for each of the simulations with varying helicity preferences. It should be emphasized that the results presented below are comparing simulations with different helicity budgets (i.e, amounts of helicity in the volume; see Figs. 4 \& 6 of KAD16) and preferences. In this way, both helicity preference and helicity budget parameter space are sampled most effectively.
\subsection{Energy Injection}
We begin the presentation of our results by calculating, for each simulation, the energy per unit area per unit time, known as the Poynting flux, that is injected into the volume by the surface motions:
\beg{PFeq}
S = \frac{\vecB\times(\vecv\times\vecB)}{4\pi},
\done
Since the only motions are imposed at the top and bottom $x$ boundaries, the $x$-component of this quantity,
\beg{Poynting}
\begin{split}
S_x &=  \frac{1}{4\pi}v_x(B_y^2+B_z^2) - B_x(v_zB_z + v_yB_y) \\
       &= -\frac{1}{4\pi} B_x(v_zB_z + v_yB_y),
\end{split}
\done
will dominate over $S_y$ and $S_z$. In the above, the second equality follows from the form of our imposed convective cells, which have $v_x = 0$. \par
\begin{figure*}[!h]
\centering\includegraphics[scale=0.35,trim=0cm 0cm 4.5cm 0cm, clip=true]{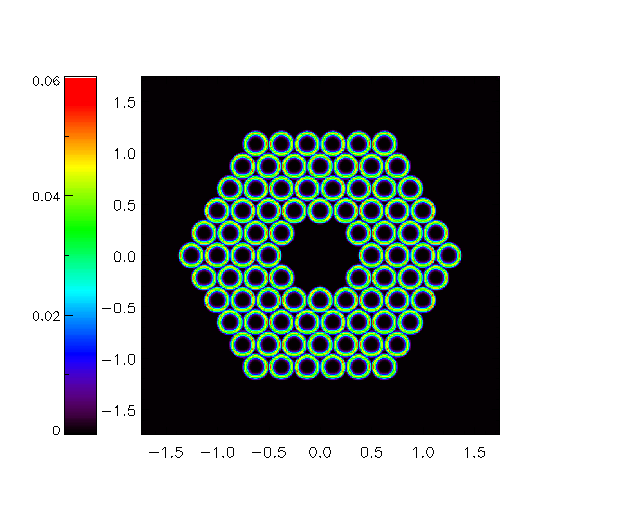}
\centering\includegraphics[scale=0.35,trim=0cm 0cm 4.5cm 0cm, clip=true]{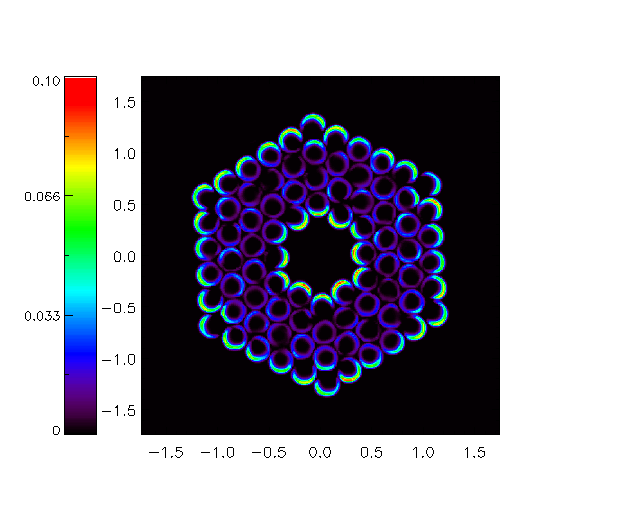}
\centering\includegraphics[scale=0.35,trim=0cm 0cm 4.5cm 0cm, clip=true]{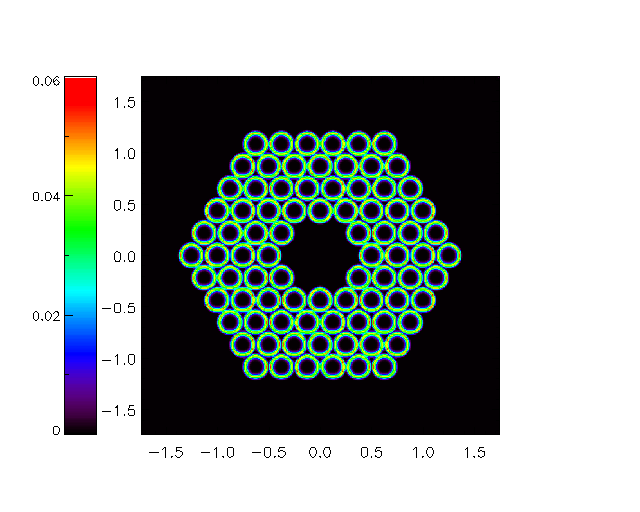}
\centering\includegraphics[scale=0.35,trim=0cm 0cm 4.5cm 0cm, clip=true]{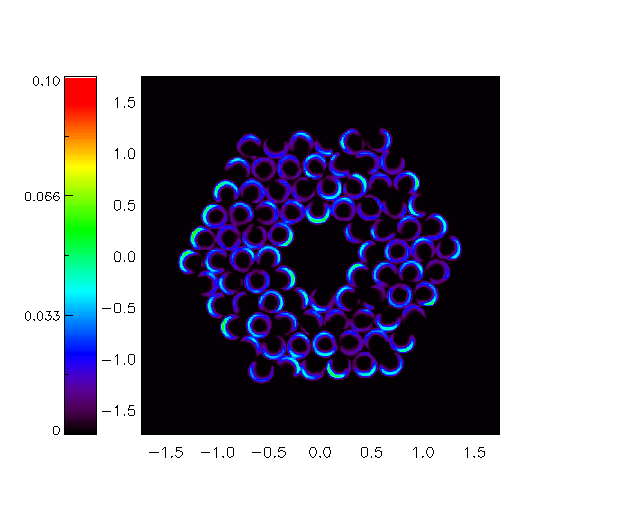}
\centering\includegraphics[scale=0.35,trim=0cm 0cm 4.5cm 0cm, clip=true]{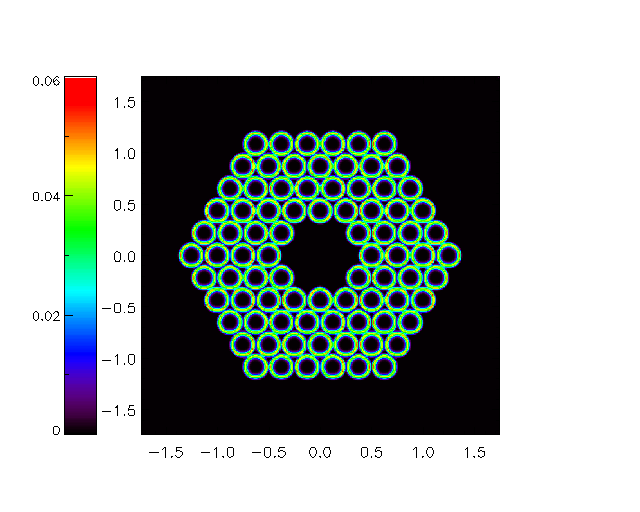}
\centering\includegraphics[scale=0.35,trim=0cm 0cm 4.5cm 0cm, clip=true]{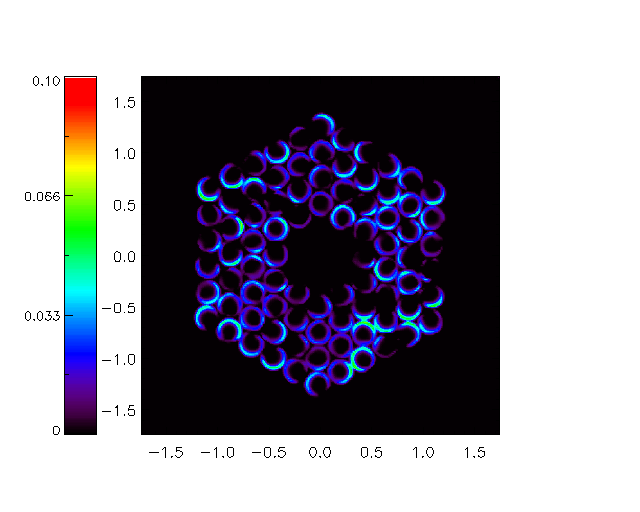}
\caption[Poynting flux map for various helicity preferences]{Poynting flux on the bottom plate in the middle of the first (left) and last (right) cycles for the $k=1$ (top), $k=0.75$ (middle), and $k=0.5$ cases.}
\label{fig:poynting}
\end{figure*}

 
We plot $S_x$ on the bottom plate during the middle of the first and last cycles for each of the three cases in Fig. \ref{fig:poynting}. The Poynting flux depends on the stress of the field lines at the surface. The contribution from this stress comes from the slow twisting of field lines due to the convective cells, as well as the rapid rearrangement of field lines caused by reconnection. The distribution of energy injection for the three cases is identical during the first cycle, since the stress on each field line is the same, regardless of its sense of twist, since reconnection has not occurred yet. The Poynting flux is distributed uniformly over the hexagonal region, with each convective cell injecting an equal amount of energy into the volume. \par 
The Poynting flux distribution at late times, by contrast, looks very different in the $k=1$ case than in the other two cases. The field lines comprising the bands of twist at the outer and inner boundaries are highly stressed, so that the amount of work the convective cells need to do to maintain their driving velocity is much larger at the boundaries than in the interior of the hexagonal region, where the field lines are twisted only weakly. The distribution of Poynting flux, therefore, is highly concentrated at the inner and outer boundaries of the hexagonal region, with much less energy being injected in its interior. The $k=0.75$ and $k=0.5$ cases, on the other hand, have not yet had time to form significant concentrations of twist. There are individual, small locations of strong twist, but no coherent structures. The distribution of Poynting flux, therefore, looks much more random and sparsely distributed. For the $k=0.75$ case, running the simulation out twice as long will result in the formation of bands of twist very similar to the $k=1$ case, as was demonstrated in KAD16. At that stage, the Poynting flux distribution will closely resemble that of the $k=1$ case. Thus, a prediction of this model is that for helicity preferences resulting in the formation of filament channels, such as is the case on the Sun, the majority of the energy injection will come from regions near the PIL, where the field is highly sheared. \par
\begin{figure}[!h]
\centering\includegraphics[scale=0.75, trim=0.0cm 6.5cm 0.0cm 5.0cm]{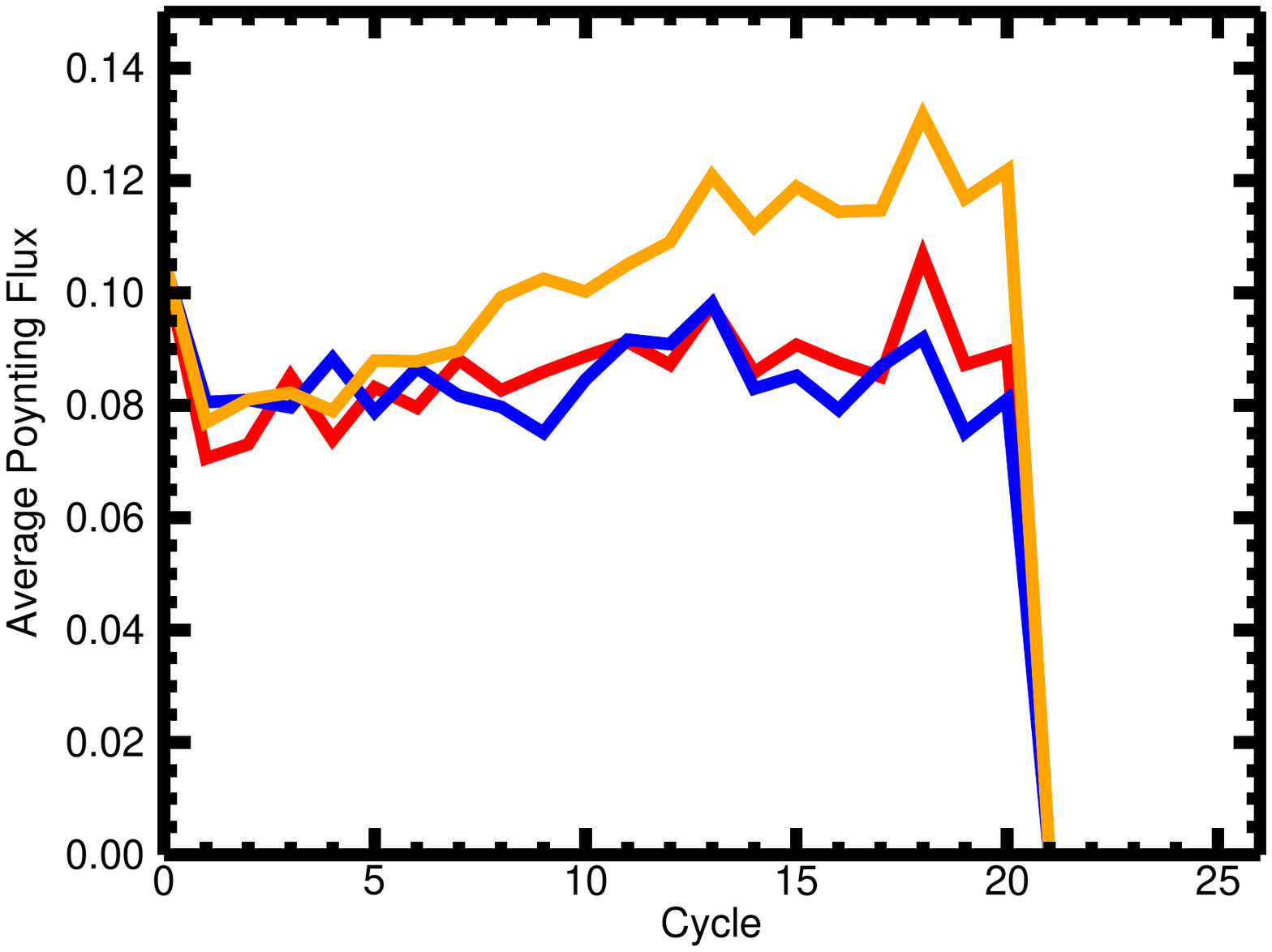}
\caption[Cycle averaged Poynting flux for various helicity preferences]{Average Poynting flux during each cycle for the $k=1$ (orange), $k=0.75$ (red), and $k=0.5$ (blue) cases.}
\label{fig:PF}
\end{figure}

Although the temporal behavior of the Poynting flux is sinusoidal, dominated by the oscillatory nature of the convective cells described in \S \ref{sec:model}, we can nevertheless observe the system reach a steady state equilibrium for each of the simulations by plotting the cycle-averaged value of $S_x$ in Fig. \ref{fig:PF}. 
Starting around cycle $15$, all three cases are in a statistically steady state, where the energy injected is approximately the same from cycle to cycle, and hovers around
\beg{PFperAtnumber}
\langle S_s \rangle \sim \left\{
        \begin{array}{ll}
            0.12 & \quad k = 1 \\
            0.08 & \quad k=0.75,0.5
        \end{array}
    \right.
\done
with the subscript $s$ standing for the steady state value.  Interestingly, the $k=1$ curve achieves a steady state at a larger value of the Poynting flux than the $k=0.75$ and $k=0.5$ cases. To understand this, it is necessary to realize that a statistically steady state is one in which energy is injected into field lines, which relieve the energy by reconnecting and relaxing. Energy is then injected again into these partially relaxed field lines, creating a constant back-and-forth between the energy injection and energy release. In Fig. \ref{fig:poynting}, it is apparent that while this process contributes to the steady state in the $k=0.75$ and $k=0.5$ cases, the energy injection in the $k=1$ case in dominated not by the convective cells in the interior of the hexagon, where field lines are most likely to reconnect and relax, but by the convective cells nearest to the outer boundary. In this region, the stress on the field is so high that even with reconnecting and relaxing field lines, the energy injection will be higher, on average, than on the weakly twisted field lines in the interior of the hexagonal region. Thus, we see that all three systems reach a statistically steady state in which the Poynting flux does not vary much with time. We will explore the consequences of this result in \S \ref{sec:Implications}.\par

\subsection{Total Energy}\label{sec:TE}
\begin{figure}[!h]
\centering\includegraphics[scale=0.75, trim=0.0cm 6.5cm 0.0cm 6.0cm]{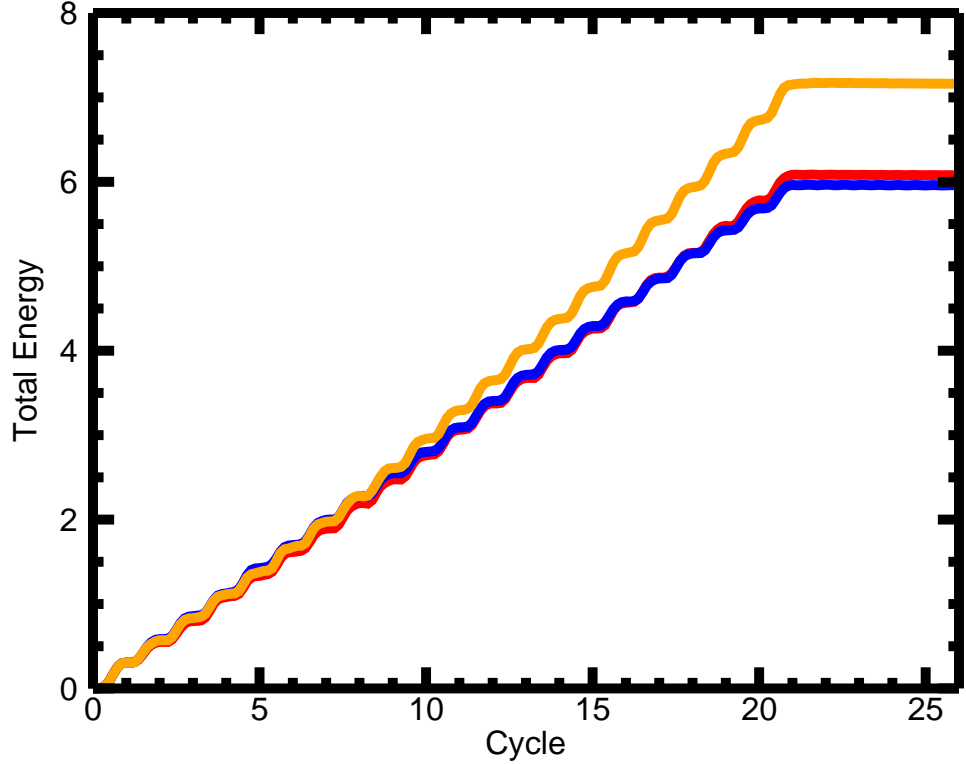}
\caption[Total energy as a function of time for various helicity preferences]{Total energy in the volume as a function of time for the different values of $k$. The orange, red, and blue curves represent the $k=1$, $k=0.75$, and $k=0.5$ cases, respectively.}
\label{fig:TE}
\end{figure}

In Fig. \ref{fig:TE} we plot the total energy in the volume as a function of time. This can be shown to be precisely equal to the time and area integrated Poynting flux injected by the convective cells. The three curves seem to track each other for approximately the first $\mathrm{8}$ cycles. At this point, the $k=1$ curve begins to significantly deviate from the other two cases, for reasons that will be discussed below. After $\mathrm{21+5}$ cycles, the most energy is present in the $k=1$ simulation, with approximately equal energies being present in the $k=0.75$ and $k=0.5$ cases. 
\subsection{Temperature}\label{sec:Temp}
\begin{figure*}[!h]
\centering\includegraphics[scale=0.35,trim=0cm 3.0cm 0.0cm 0.0cm, clip=true]{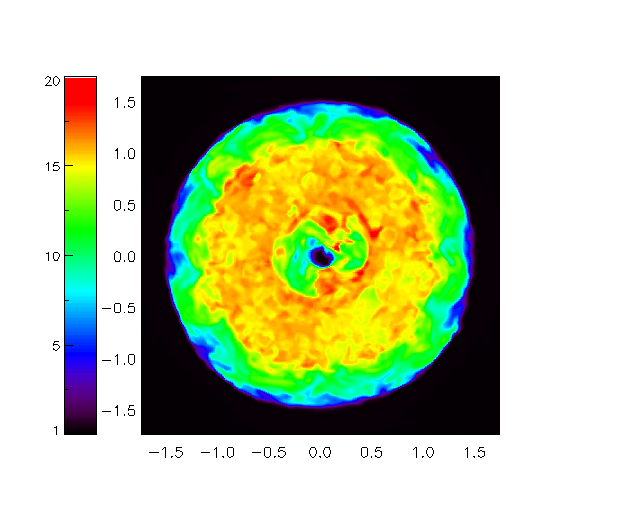}
\centering\includegraphics[scale=0.35,trim=0cm 3.0cm 0.0cm 0.0cm, clip=true]{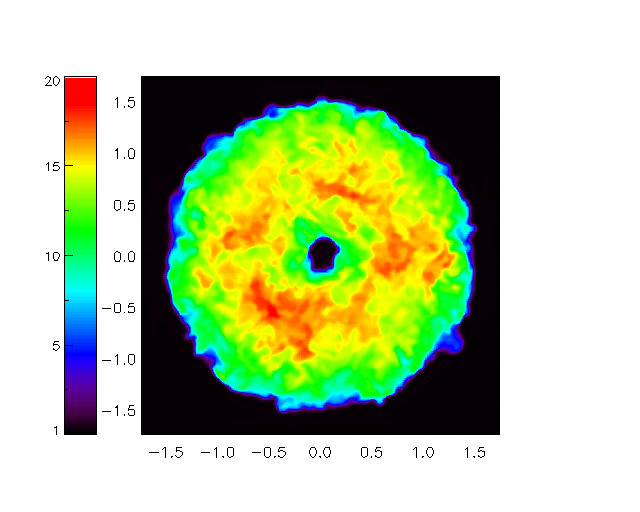}
\centering\includegraphics[scale=0.35,trim=0cm 0.0cm 0.0cm 0.0cm, clip=true]{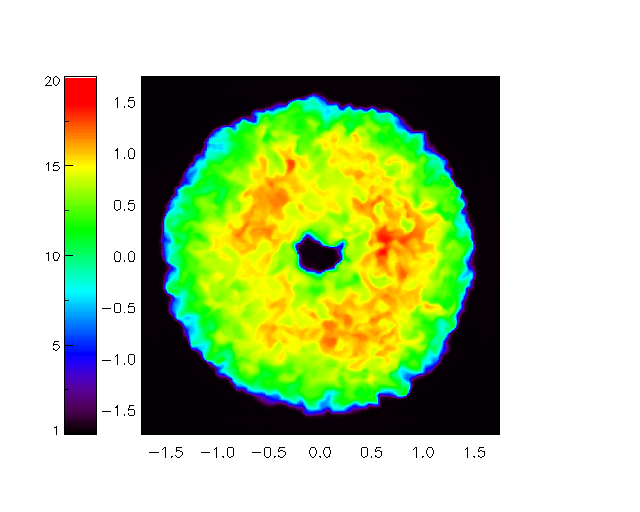}
\caption[Coronal temperature for various helicity preferences]{Temperature in the midplane at the end of the simulation for the $k=1$ (top), $k=0.75$ (middle), and $k=0.5$ (bottom) cases.}
\label{fig:Temperature}
\end{figure*}

The most obvious diagnostic to investigate coronal heating is the coronal temperature. While at the beginning of each simulation the temperature is uniform throughout the domain, the temperature increases significantly over the course of the simulation. In Fig. \ref{fig:Temperature} we show, for each simulation, the temperature in the midplane which, for the Parker model, corresponds to the corona. Evidently, the temperature is much higher than its initial, uniform value of unity, in all three cases after significant twisting has occurred. This increase in temperature occurs throughout the flux system, defined as the region that is being twisted at the photosphere. The coronal temperature profile is hot in the interior, where most of the reconnection occurs, and gets cooler at the boundaries of the flux system. At these locations, filament channels have been demonstrated to form (KAD15, KAD16), and these correspond to regions of suppressed reconnection, indicating less plasma heating.\par
A second important result that is evident from the Figure is that the coronal temperature is approximately the same for each of the helicity preferences. While there are some spatial variations, the temperature clearly reaches the same approximate value in each case.  This indicates that the coronal temperature is independent of both the helicity budget and preference.\par
\subsection{Magnetic Energy}\label{sec:ME}
\begin{figure}[!h]
\centering\includegraphics[scale=0.75, trim=0.0cm 6.5cm 0.0cm 6.0cm]{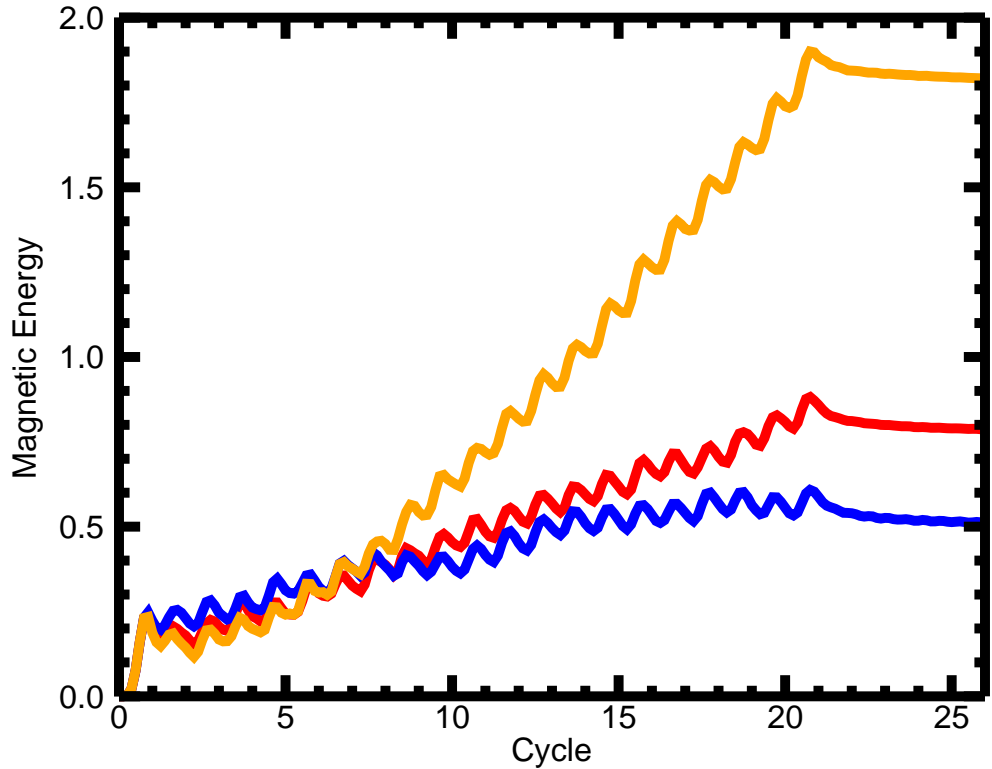}
\caption[Magnetic energy for various helicity preferences]{Magnetic energy in the volume, $W$, as a function of time for the different values of $k$. The orange, red, and blue curves represent the $k=1$, $k=0.75$, and $k=0.5$ cases, respectively.}
\label{fig:ME}
\end{figure}

We calculate the magnetic free energy in the volume in each of our simulations via  
\beg{free_energy}
W(t)=\frac{1}{8\pi}\int_V{\Big\{\vecB^2(t)-\vecB^2(0)\Big\} dV}.
\done
Strictly speaking, this is the magnetic free energy, but, since the potential field is always just the uniform background field, for brevity we will refer to the quantity in Eq. (\ref{free_energy}) as simply the magnetic energy. Fig. \ref{fig:ME} shows the magnetic energy for each of the simulations. The wave-like pattern present in each curve is due to the sinusoidal temporal driving profile. In contrast to the temperature, the magnetic energy displays a clear dependence on the helicity budget and preference. The magnetic energy in the $k=1$ case is more than double that of the $k=0.75$ case, and more than triple that of the $k=0.5$ case. \par
\begin{figure}[!h]
\centering\includegraphics[scale=0.5, trim=0.0cm 0.0cm 0.0cm 0.0cm]{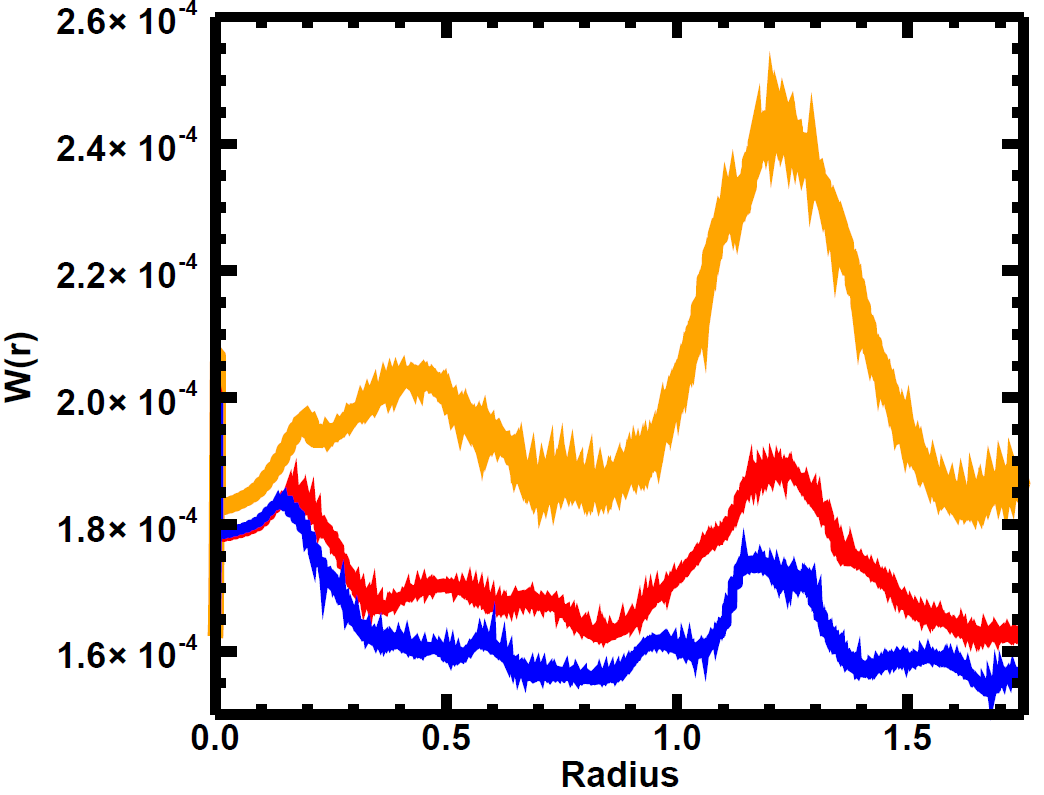}
\caption[W(r) for various helicity preferences]{W(r) integrated in the vertical direction and then binned by shells of constant radius at the end of the simulation for the $k=1$ (orange), $k=0.75$ (red), and $k=0.5$ (blue) cases. }
\label{fig:Wr}
\end{figure}

The $k=1$ case accumulates the most magnetic energy, since the field lines that make up the bands of twist at the boundaries (KAD15, KAD16) are very highly sheared, and contain a lot of stored energy. In Fig. \ref{fig:Wr} we plot the quantity
\beg{Wr}
W(r) =\int_0^{2\pi}{d\theta}\int_0^{L_x}{\Big\{\frac{\vecB^2(r,\tau)-\vecB^2(r,0)}{8\pi}\Big\} dx},
\done
for each of the three cases, with $\tau$ representing the end time of the simulations. This quantity represents the energy integrated in cylindrical shells of increasing radius, thus giving a sense of where the energy is stored. For each of the cases, most of the magnetic energy is contained at the two boundaries of the hexagonal region. For the $k=1$ case, the majority of the energy is stored in the outer boundary, corresponding to the location of the filament channels. Thus, the magnetic energy in the corona is stored overwhelmingly in the filament channels. For the other two cases, an approximately equal amount of energy is stored at the outer and inner boundaries, though in both cases it is larger at these boundaries then in the interior of the hexagonal region. In these two cases, bands of twist have not yet formed, and thus the major contributions to the stored energy are likely due to the increased magnetic pressure at the inner boundary and the uncancelled twisted field at the outer boundary.\par
After the first cycle, the magnetic energy of each of the cases is the same.  The magnetic energy of each individual flux tube is 
\beg{freenergy1tube}
W_1 \sim \frac{B_{\phi}^2}{8\pi} V_1 
\done
where $V_1=\pi a_0^2 L_x$, the subscript on $W$ and $V$ representing the fact that this is for a single tube. To determine $B_\phi$, note that after being twisted on both the top and bottom by an angle $\Theta$, a flux tube has twist flux
\beg{twist1}
\Phi_{tw}=\frac{\Theta}{\pi}\Phi_{ax},
\done
so that $\Theta=\pi$, and $\Phi_{tw}=\Phi_{ax}$.  Using $\Phi_{ax}=B_x\pi a_0^2$, Eqn. \ref{twistfield} yields:
\beg{equatetwist}
B_\phi a_0 L_x = B_x\pi a_0^2, 
\done
so that
\beg{twistfield2}
B_\phi=\frac{\pi a_0}{L_x}B_x
\done
and plugging this into Eq. \ref{freenergy1tube}, we obtain:
\beg{W1}
W_1=\frac{\pi^2 a_0^4 B^2_x}{8L_x}
\done
For $N=84$ flux tubes, the magnetic energy after $1$ cycle is
\beg{WN}
W_N^1 = \frac{N\pi^2 a_0^4 B^2_x}{8L_x} = 0.32
\done
Here the subscript $N$ and superscript $1$ represents that this is the energy of $N$ tubes after $1$ cycle.\par
For the $k=1$ case, the magnetic energy in the corona at the end of the simulation can be approximated via the magnetic helicity condensation model in the following way.  After the twist flux has accumulated at the boundaries, the final field configuration is a uniform vertical field everywhere except in two bands of width $w$ encircling the outer (o) and and inner (i) boundaries. The magnetic energy contained in the volume is dominated by the twist field at o and i, which are the only locations where the field lines are highly nonpotential. The magnetic energy of the final configuration is therefore approximately
\beg{approxfreeenergy}
W\sim\frac{B_{\phi,o}^2}{8\pi}V_o+\frac{B_{\phi,i}^2}{8\pi}V_i,
\done
with $V_o$ ($V_i$) representing the volume occupied by the outer (inner) shell of twist flux centered at a distance of $a_0$ ($a_i$) from the central axis. The twist field is given by
\beg{twistfield}
B_{\phi}=\frac{\Phi_{tw}}{wL_x}.
\done
Here $\Phi_{tw}$ is the twist flux through a vertical half plane, $w$ is the width of the band, and $L_x$ is the height of the box. In KAD15 we related the twist flux to the helicity injected into the corona, so that, at each instant, $\Phi_{tw}$ is known and is equal and opposite at o and i.  The volume of the bands is
\beg{VPIL}
\begin{split}
V_o&\sim 2\pi a_o w L_x, \\
V_i&\sim 2\pi a_i w L_x.
\end{split}
\done
Plugging these in above yields
\beg{approxfreeenergy2}
W=\frac{\Phi_{tw}^2}{4wL_x}(a_o+a_i)
\done
Using $w=0.3$, $L_x=1$, $a_o=1.2$, and $a_i=0.3$, as in our previous work, we find
\beg{Wpredict}
W=1.25\Phi_{tw}^2
\done
After $21$ cycles, for the $k=1$ case, KAD15 predicted (and found) a twist flux of $\Phi_{tw}=1.22$. Plugging this value in yields
\beg{Wfreepredictval}
W_{N,k=1}^{21}=1.86.
\done
This value agrees very well with the orange $k=1$ curve. \par
The magnetic energy in the $k=0.75$ and $k=0.5$ cases is highly dependent on the amount of twist in each individual flux tube, since bands of twist at $a_0$ and $a_i$ have not formed yet ($k=0.75$ case) or will not form at all ($k=0.5$ case). 
\subsection{Energy Released by Magnetic Reconnection}\label{sec:IE}
\begin{figure}[!h]
\centering\includegraphics[scale=0.75, trim=0.0cm 6.5cm 0.0cm 5.0cm,clip=true]{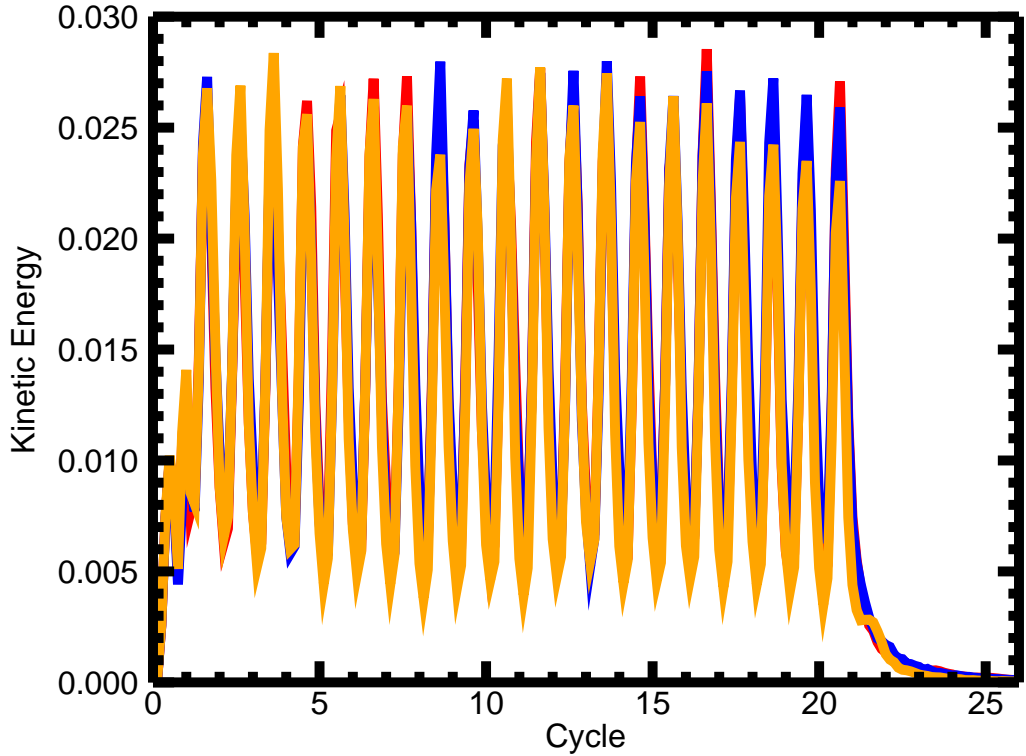}
\caption[Kinetic energy for various helicity preferences]{Kinetic Energy during simulation as a function of time for the different values of $k$. The orange, red, and blue curves represent the $k=1$, $k=0.75$, and $k=0.5$ cases, respectively.}
\label{fig:KE}
\end{figure}

\begin{figure}[!h]
\centering\includegraphics[scale=0.75, trim=0.0cm 6.5cm 0.0cm 5.0cm,clip=true]{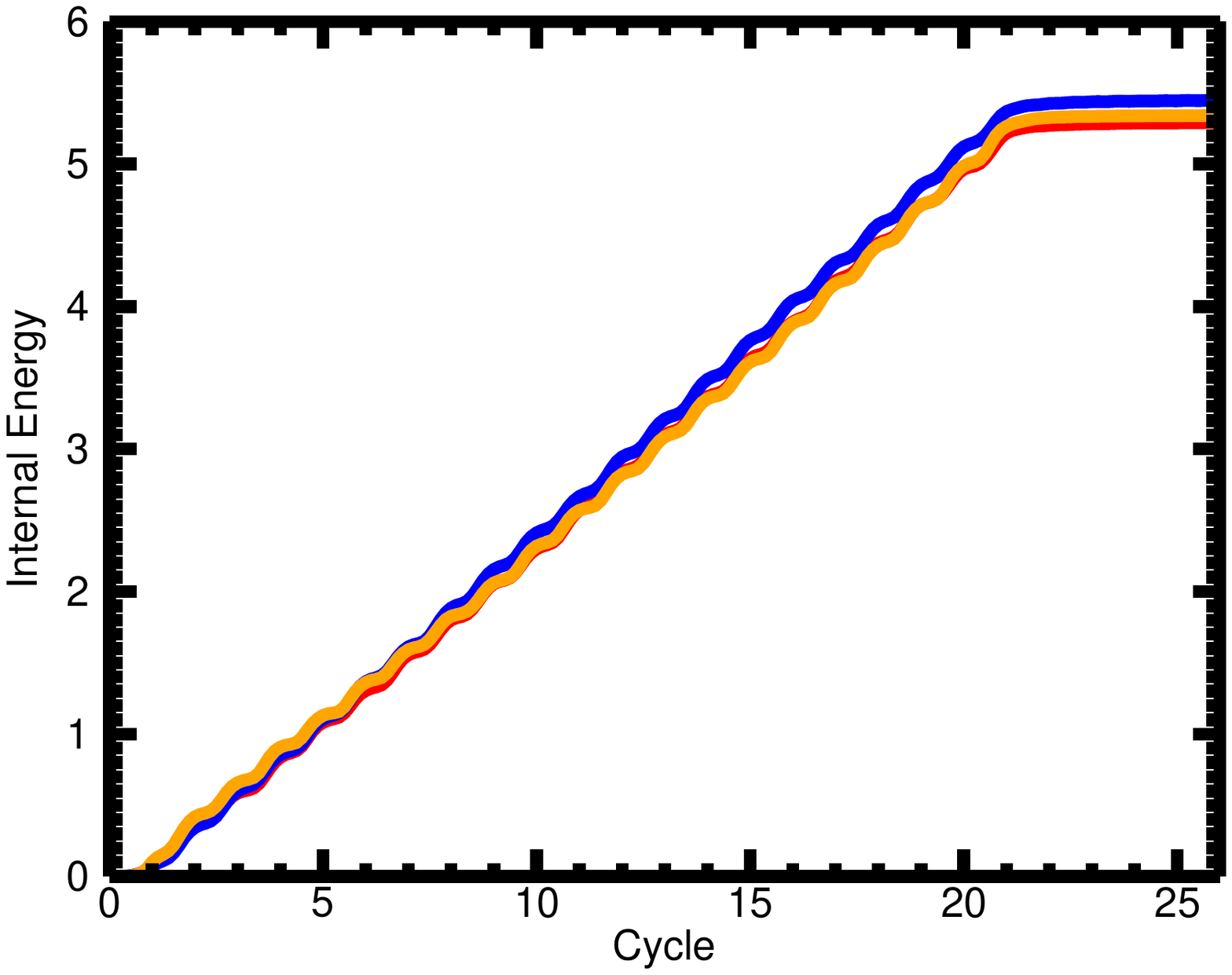}
\caption[Internal energy for various helicity preferences]{Internal Energy during simulation as a function of time for the different values of $k$. The orange, red, and blue curves represent the $k=1$, $k=0.75$, and $k=0.5$ cases, respectively.}
\label{fig:IE}
\end{figure}

Magnetic reconnection is well known to convert magnetic energy into plasma thermal and kinetic energy. Figs. \ref{fig:KE} and \ref{fig:IE} show the kinetic energy $K$ and internal energy $\epsilon$ of the plasma as a function of time for each simulation. Each curve displays the characteristic oscillation pattern, consistent with the ramping up and down of the convective cells, followed by a plateau after $21$ cycles, when the driving was stopped and the system allowed to relax. Evidently, only a negligible amount of magnetic energy is converted into the kinetic energy of the plasma, with the majority being converted into thermal energy. It is clear from Fig. \ref{fig:IE} that the internal energy is very nearly the same for each case. In fact, it is remarkable how close the curves are, given the large range of helicity preferences and budgets covered. Apparently, helicity preference plays no role in coronal heating. \par
There are two mechanism for heating the plasma in our simulations, which lack thermal conduction. The first mechanism is that numerical diffusion of the magnetic field converts magnetic energy directly into plasma heating. However, as demonstrated in both KAD15 and KAD16, magnetic helicity is conserved to better than $\mathrm{1\;\%}$ accuracy, implying that magnetic field diffusion is completely negligible in these simulations. The second mechanism is the viscous dissipation of Alfv\'enic jets caused by magnetic reconnection. In this scenario, magnetic reconnection generates strong exhaust jets and acoustic shock waves which are then damped out by numerical viscosity. The kinetic energy from these jets and shocks then goes into plasma heating. This is likely to be the mechanism by which the plasma is being heated in these simulations. \par
The above arguments indicate that there is, approximately, the same amount of magnetic reconnection occurring in each simulation. Consider a single twisted flux tube closely surrounded by $6$ twisted flux tubes. The expected number of neighbors twisted in the same sense as the interior flux is different in all three simulations. However, as long as even a single one of the exterior twisted flux tubes is twisted in the same sense as the interior flux tube, reconnection will occur, and magnetic energy will be converted into heating through viscous dissipation. Note that even if more than one of the exterior flux tubes are twisted in the same sense as the interior flux tube, the number of reconnection events with the interior flux tube does not exceed one, although which particular exterior flux tube reconnects may vary. The key point is that a given flux tube does not care whether it is surrounded by $6$ co-twisted tubes, or just one, it will reconnect with at most one of the exterior tubes. If, on the other hand, all $6$ exterior flux tubes are twisted in the opposite sense from the interior flux tube, it will not reconnect with any of them. The likelihood that the interior flux tube is surrounded by at least a single flux tube twisted in the same sense is
\beg{p1}
P_{same}^{k=1} = 1,
\done
for the $k=1$ case, while for the $k=0.5$ case it is
\beg{p0.5}
P_{same}^{k=0.5} = 1-0.5^6 = 0.98
\done
The worst case scenario in the $k=0.75$ case is
\beg{p0.75}
P_{same}^{k=0.75} = 1- 0.75^6 = 0.82.
\done
While the probability in the $k=0.75$ may seem low, it appears that there is, nevertheless, the same amount of reconnection in this case. This is likely due to the fact that even if a flux tube is surrounded on all sides by unfavorably twisted flux tubes, it is likely that it will build up enough twist to be susceptible to the kink mode, causing it to be brought into contact, and reconnect with, an adjacent flux tube anyway. Thus, all three cases are likely to experience the same amount of reconnection.

\subsection{Implications for Coronal Heating}\label{sec:Implications}
The results described above can be compared quantitatively to several aspects of coronal heating observations. In a statistically steady state, \citet{Klimchuk15} relates the average Poynting flux to the critical angle, known as the Parker angle \citep{Parker83}, at which nanoflare reconnection can sustain the observed heating via
\beg{Klim15}
\langle S_s \rangle = \frac{1}{4\pi}v_h B_v^2 \tan\theta,
\done
with $v_h$ the horizontal driving velocity, $B_v$ the vertical field, and $\theta$ the Parker angle. Therefore,
\beg{Parkangle}
\theta_P = \arctan \big(\frac{4\pi \langle S_s \rangle}{v_h B_v^2}\big).
\done
Using the simulation parameters quoted in KAD15 and KAD16, $v_h = 0.2$, $B_v = \sqrt{4\pi}$, we obtain
\beg{thetaP}
\theta_P  = \left\{
        \begin{array}{ll}
            31^\circ & \quad k = 1 \\
            22^\circ & \quad k=0.75,0.5
        \end{array}
    \right.
\done
well in line with the values estimated in \citet{Parker83}. 
\par
To convert the energy flux in Eq. \ref{PFperAtnumber} to solar values, we note that
\beg{relateStovB2}
\langle S \rangle \sim vB^2
\done
and therefore
\beg{relateSsimtoSsun}
\langle S_{\odot} \rangle = \langle S_{s} \rangle \frac{v_{\odot}}{v_{h}}\frac{B_{\odot}^2}{B_{v}^2}
\done
where the subscript $\odot$ stands for the solar value. Usingthe values of $v_h$ and $B_v$ quoted above, a photospheric velocity of $v_{\odot}=10^5\;\mathrm{cm\;s^{-1}}$, and the average quiet Sun field $B_{\odot}=10\;\mathrm{G}$, we obtain
\beg{Ssun}
\langle S_{\odot} \rangle \sim 5 \times 10^5 \; \mathrm{ergs\; cm^{-2}\; s^{-1}}
\done
well in line with canonical quiet Sun energy loss rates \citep{Withbroe77}. Active regions typically show average fluxes of an order of magnitude larger than quiet region fields \citep[e.g][]{MP97}, meaning that the injected energy per unit area per unit time will be $2$ orders of magnitude larger than that quoted above, around $10^7 \; \mathrm{ergs\; cm^2\; s^{-1}}$, in excellent agreement with the observed energy loss rate in active regions \citep{Withbroe77}.


\section{Discussion}\label{sec:discussion}
In this work, we have explored the role of helicity injection in the heating of the corona using helicity and energy conserving numerical simulations. Helicity was injected into the corona at the photospheric level via numerous convective cells, whose sense of rotation and position were randomly varied. Whereas in previous papers we showed that this helicity was transported throughout the simulation domain by magnetic reconnection, forming filament channels (KAD15) and leaving behind a smooth corona (KAD16), in this paper, we explored how magnetic reconnection distributed the injected energy between kinetic, magnetic, and thermal energy. We find that the majority of energy is converted into heating, and the rest remaining in the magnetic field, with only a negligible amount going into kinetic energy. Furthermore, we found that the amount of heating was completely independent of helicity preference, but the amount of energy remaining in the magnetic field was dependent on this preference, with the most magnetic energy remaining in the field for larger helicity preferences. We interpret this to mean that magnetic reconnection is least efficient for maximal helicity preferences, and most efficient at negligible helicity preferences. The major result of this work, shown in Fig. \ref{fig:IE}, is that the amount of heating is entirely independent of the helicity preference. Crucially, we also showed quantitatively that the heat flux is sufficient to account for the observed heating. \par
Taken together with KAD15 and KAD16, these results paint a comprehensive picture of helicity injection into the corona. In this picture, surface motions inject energy and helicity into the solar corona. The helicity is transported to the largest available scales by magnetic reconnection, while the energy is converted to heat. This process, magnetic helicity condensation, is therefore simultaneously responsible for the formation of filament channels, the smoothness of coronal loops, and the heating of the million degree solar corona. \par
These results open up many fascinating new directions of research. In a follow up study, we will explore the effect of helicity injection on a more realistic, dipolar field configration. The gradients inherent in such a system will suppress reconnection between adjacent flux tubes, inhibiting the transport of helicity by magnetic reconnection. We will investigate whether injecting helicity into such a field will result in the formation of filament channels and smooth coronal loops.\par
Our study also suggests that the reconnection between two adjacent flux tubes should be explored in more detail. Does such reconnection proceed as a series of nanoflares \citep{Klimchuk15} or is it a single, spontaneous event? The question of how much reconnection occurs between pairs of twisted flux tubes, rather than between a single twisted flux tube and the surrounding untwisted field should also be addressed. It is clear, however, that the transport of magnetic helicity plays a major role in determining the structure and heating of the solar corona.


\chapter{Conclusions and Future Work}\label{Conclusions}

\section{Conclusions}
The work presented in this thesis has focused on explaining two apparently contradictory observations. The first observation was the finding that photospheric convection injects both energy and helicity into the magnetic field. The second observation is the lack of helicity in the hot coronal loops and presence of helicity in sheared filament channels. These observations appeared contradictory since the injection of magnetic helicity, a conserved quantity, into the coronal magnetic field should lead to twisted or sheared structures throughout the corona, not just in filament channels. On the other hand, the injection of energy results in heating throughout the coronal volume. Thus, while the injection of both helicity and energy into the corona occurs uniformly throughout the photosphere, only the energy is observed uniformly throughout the corona; the helicity being observed only in local concentrations of shear in filament channels. \par
In Chapter \ref{SimsHelicity}, we demonstrated that it is possible to inject magnetic helicity throughout the photosphere, yet have it accumulate only at specific locations. We showed that the presence of helicity in filament channels follows directly from the inverse cascade of magnetic helicity due to its transport by magnetic reconnection, known as helicity condensation. In this process, magnetic helicity is injected on small scales and then transported by magnetic reconnection to the largest scales in the system, which occur at PILs, precisely where filament channels are known to form. We demonstrated that the timescale for this process depends inversely on the helicity preference, with a larger preference leading to shorter filament channel formation times. For helicity preferences in the range observed on the Sun, we showed that filament channels would form over the time scale of a couple days, in excellent agreement with observations. \par
In Chapter \ref{Structure}, we argued that the helicity condensation process naturally resulted in a smooth corona. When reconnection transfers magnetic helicity to ever-larger scales, it leaves the interior of the corona generally smooth and untwisted, explaining the observations described in Chapter \ref{Intro}. Furthermore, our results indicated that the amount of structure in the corona depended only weakly on the helicity injection preference, potentially opening new avenues into placing new bounds on the amount of structure that is expected to be observed in the corona when higher resolution data becomes available.\par
In Chapter \ref{Sunspot}, we significantly improved our results, using a much more realistic, topologically accurate, magnetic field configuration. By starting with a magnetic field containing a true PIL, we were able to show that helicity does not simply condense at an arbitrary boundary defined by the extent of the driven region (as could be inferred from the first three chapters), but that it actually accumulates at the PIL. Furthermore, since individual field lines started and ended on the photosphere, we were able to demonstrate, for the first time, the formation of sheared filament channels at PILs. The presence of a real coronal hole (CH), allowed us to observe helicity being ejected out of the system as it would be on the real Sun. The process by which magnetic helicity leaves the solar corona through the open field region of the Sun has important observational consequences that will hopefully be addressed in the future (see Future Work, below). The end result of this simulation was that we were able to qualitatively and quantitatively reproduce, for the first time, the observed properties of filament channels. Finally, we were able to demonstrate that our conclusions about the smooth loop portion of the corona were true even in a more complicated, yet realistic, magnetic field topology.\par
In Chapter \ref{Heating}, we demonstrated another major consequence of the helicity condensation model: that the magnetic reconnection that transports helicity to the PIL converts magnetic energy into heat in the process. Since this process occurs throughout the corona, it explains in a natural way the observation that the corona is uniformly heated throughout its volume. We were able to show that this process agrees both qualitatively and quantitatively with observations, with reconnection providing sufficient energy flux to account for the observed heating. In contrast to the amount of coronal structure, however, the amount of heating was shown to be entirely independent of the helicity injection preference. Although the results presented here do not definitively prove or disprove either DC or AC coronal heating models (see Chapter \ref{Intro}), they do show that sufficient heating can occur within the framework of DC heating via the slow shuffling of coronal magnetic field footpoints.\par
This work provides, for the first time, a global picture of the role that magnetic helicity plays in determining the structure and heating of the Sun's corona. In this picture, helicity is injected into the coronal magnetic field at the photospheric level by convection on the surface of the Sun. On individual flux tubes, this helicity is in the form of twist, which is quickly transported to ever-larger scales by magnetic reconnection between oppositely oriented twist components. The PIL serves as a natural boundary beyond which the helicity cannot propagate. At this scale, however, the helicity manifests itself as shear, rather than twist, resulting in the formation of long, sheared field lines known as filament channels. The shear in these filament channels, in turn, represents tremendous amounts of free energy, which is frequently expelled by the Sun in coronal mass ejections. Meanwhile, the rest of the corona remains hot and quasi-potential, with most of its free energy having been converted into heat by magnetic reconnection. \par
To summarize the results of this thesis, it is worthwhile to return to the questions that were asked in section \S \ref{sec:researchquestions} and briefly state the findings of this thesis for each of them:
\begin{enumerate}
  \item How do filament channels obtain their helicity? \newline
The helicity seen in filament channels ultimately comes from the helicity injected at the photospheric level. This helicity, injected in the form of twist, is transported by magnetic reconnection between twisted flux tubes to the largest possible scale in the system - the PIL, in a process known as helicity condensation \citep{Antiochos13}. This is a natural consequence of the inverse cascade of magnetic helicity, a process where helicity, in the form of twist, is carried to larger and larger scales. This twist manifests itself as a global shear that is localized at the PIL, appearing as a set of long, highly sheared field lines. These, in turn, are observed as filament channels. Thus, the helicity condensation process provides a natural mechanism for the formation of sheared filament channels.
  \item Why do coronal loops lack helicity, appearing unsheared and quasipotential? \newline
Coronal loops lack helicity because the inverse cascade of magnetic helicity removes it from the majority of the corona, causing it to accumulate in filament channels. The resulting `loop' portion of the corona is left smooth and untwisted, with little to no structure (for helicity preferences in the range observed on the Sun, c.f. Chapter \ref{Structure}) and very close to its potential state. Thus, both filament channel shear and coronal loop quasi-potentiality are a result of the same process - magnetic helicity condensation.
      \item What process converts energy into heat throughout the corona? \newline
The energy that is observed as heat in the solar corona ultimately comes from the energy injected at the photospheric level. This energy, injected in the form of twist, is converted by magnetic reconnection between twisted flux tubes into heating throughout the solar corona, because photospheric motions occur over the whole photosphere and twist up the entire coronal magnetic field. Thus, our results show that slow driving of magnetic field footpoints tangles and braids the coronal magnetic field, building up significant current sheets which heat the corona through Ohmic dissipation. While the work presented here did not test AC heating models (see \S \ref{sec:loops}), the results do imply that the DC model presented in Chapter \ref{Heating} is in good qualitative and quantitative agreement with observations.
\end{enumerate}

\section{Future Work}
While the results presented in this thesis have shed some light on the role of magnetic helicity in determining coronal structure and heating, several problems are still unanswered. In particular, the nature of the CH boundary in Chapter \ref{Sunspot} needs to be explored in greater detail to understand how individual field lines become open. Do closed field lines simply rise due to magnetic pressure, or do they interchange with open field lines? How does this affect the shape of the CH boundary? This has very important implications for understanding the properties of the slow solar wind, which is thought to come from the boundary between open and closed field lines.\par
An important next step that has not been addressed in this work is the role of magnetic helicity in affecting turbulence in the solar wind. We have shown in Chapters \ref{SimsHelicity} and \ref{Structure} that helicity accumulates at CH boundaries, in addition to PIL boundaries. This implies, as argued in \citet{Antiochos13}, that if the CH boundary is open to helicity leaving from the system, then the helicity flux leaving the system would be significantly reduced compared to the helicity flux injected into the CH. This means that helicity flux in the solar wind would be much smaller than the helicity flux into the CH. To test this hypothesis, a simulation could be performed in which the solar wind is included in the model, allowing the helicity flux into and out of the CH to be calculated directly. While such a simulation is outside the scope of this work, it is an important and possible next step for the development of the helicity condensation model. Confirmation of this prediction of the helicity condensation model would be an important verification of this model for the formation of filament channels and the smooth loop corona. \par
An intriguing consequence this work, not addressed in this thesis, is the idea that the coronal magnetic field is in a self-organized critical state \citep{Bak87}, and that coronal heating occurs via a nanoflare storm or avalanche. In this model, a small reconnection event in one region of the corona can cause a sequence of other nanoflares that, ultimately, affect regions of the corona very far away from the original event. In this way, the coronal field constantly keeps itself at or near a state such that a heating event in one location heats the corona in a large and extended volume. In a self-organized critical state, the distribution of sizes of such storms or avalanches follows a very specific powerlaw spectrum, one that has been observed in, for instance, solar flares \citep{Dennis85}. The presence of a powerlaw spectrum at large flare energies is very suggestive that such a process occurs also at smaller flares energies, namely, energies below our present resolving capabilities such as occur in nanoflares. High resolution, high temporal cadence simulations, such as those presented in Chapters \ref{SimsHelicity}, \ref{Structure} and \ref{Heating}, can be used to study the spectrum of energetic events. These studies are currently underway.\par
Perhaps the most important consequence of this work for the coronal heating problem is that the heating is independent of helicity preference. However, the details of how twisted flux tube reconnection converts magnetic energy into heating are still unclear. In particular, is the interaction between the twisted flux tubes and the surrounding untwisted field as important as the interaction between the twisted flux tubes themselves? While twisted flux tubes contain more magnetic energy, so reconnection between them should efficiently heat the corona, significant current sheets do build up between twisted flux tubes and the surrounding untwisted field. Our preliminary results indicate that twisted flux tubes want to reconnect with this untwisted field almost as much as, if not more than, they do with other twisted flux tubes. This seems to be a way to efficiently decrease their energy and spread twist out over a large domain. From the standpoint of coronal heating, this is an interesting question, because it implies that twisted flux tubes do not need to reconnect with each other to heat the corona. Furthermore, it suggests that they do not even need to be in close proximity to each other to reconnect, since the any untwisted field separating them could be used to transfer twist flux between them allowing their twist components to reconnect anyway and heat the corona. Extremely high resolution, detailed simulations of twisted flux tubes surrounded by an untwisted background magnetic field are therefore very important to solving the coronal heating problem. \par
The models presented in this thesis have accounted for only one way of introducing magnetic helicity into the solar corona, namely, motion of the photospheric boundary. A second, possible equally important method for injecting magnetic helicity into the corona is by emerging or submerging magnetic flux. Emerging a highly twisted flux rope into the corona is efficient way to introduce significant magnetic helicity. On the other hand, the submergence of highly sheared field lines as a result of flux cancellation, say, is a way of removing magnetic helicity from the corona. From the standpoint of helicity condensation, flux emergence or submergence is not expected to significantly affect the results. After all, the coronal field does not have any information about how it obtained its magnetic helicity (i.e. if it became twisted through shearing motions, or by flux emergence), and so any reconnection between twisted flux tubes will proceed exactly as described in this thesis. On the other hand, submergence of the sheared field lines near the filament channels might remove a significant amount of helicity from the corona, thus preventing filament channels from forming at the PIL. Furthermore, the emergence of a highly twisted flux rope into an otherwise smooth corona may affect the transport of magnetic helicity throughout the coronal volume if the sign of the emerged flux rope is not favorable for helicity transport (like some of the cases studied in Chapter \ref{Structure}). A final step to understanding helicity injection into the solar corona, therefore, is including the physics of flux emergence and submergence into these simulations.\par


\begin{appendix}\label{appendix}

\chapter{Properties of Relative Magnetic Helicity}

\section{Gauge Invariance}
Relative magnetic helicity is defined in \citet{Finn85} as
\beg{heldef}
H=\int_V{\big(\vecA+\vecA_P\Big)\cdot\Big(\vecB-\vecB_P\Big) \; dV}
\done
subject to the boundary condition that
\beg{bc}
\vecB\cdot d\textbf{S}|_S = \vecB_P\cdot d\textbf{S}|_S,
\done
i.e., that the potential field has the same normal component on the boundary enclosing the volume as does the actual field.\par
For $H$ to be a physically meaningful quantity, it is necessary to show that it is gauge invariant, i.e., that the substitution $\vecA\to\vecA+\nabla\psi$, which conserves the magnetic field $\vecB=\nabla\times\vecA$ also conserves $H$. To this end, substitute $\vecA\to\vecA+\nabla\psi$ and $\vecA_P\to\vecA_P+\nabla\chi$. Then, defining $\xi = \psi+\chi$, one obtains:
\beg{Hgauge}
\begin{split}
H &\to H + \int_V{ \nabla\xi\cdot\Big(\vecB-\vecB_P\Big)\;dV} \\
   &= H + \int_V{ \nabla\cdot\Big[\xi\Big(\vecB-\vecB_P\Big)\Big]\;dV} \\
   &= H + \int_S{ \xi\Big(\vecB-\vecB_P\Big)\cdot d\textbf{S}} \\
   &= H
\end{split}
\done
where in the second line the property $\nabla\cdot\vecB = 0$ has been used and in the fourth line the boundary condition Eq. \ref{bc} has been used. Thus, changing one or both of $\vecA$ or $\vecA_P$ by the gradient of a scalar potential leaves $H$ unchanged, meaning that it is a gauge invariant quantity.

\section{Rate of Change of Relative Magnetic Helicity}
One of the most important assumptions of this work has been the idea that the relative magnetic helicity is conserved under reconnection, and, indeed, that only surface terms cause any change in $H$. Since the gauge invariance of $H$ has already been proven above, we pick the Coloumb gauge $\nabla\cdot\vecA=\nabla\cdot\vecA_P =0$, which implies that
\beg{bcA}
\vecA\cdot d\textbf{S}|_S = \vecA_P\cdot d\textbf{S}|_S = 0.
\done
Taking the time derivative of the helicity in Eq. \ref{heldef} yields:
\beg{AdHdt}
\frac{dH}{dt} = T_1 + T_2,
\done
where
\beg{term1}
T_1 = \int_V{ \frac{\partial (\vecA+\vecA_P)}{\partial t}\cdot\Big(\vecB-\vecB_P\Big)\;dV}, 
\done
and
\beg{term2}
T_2 = \int_V{\Big(\vecA+\vecA_P\Big)\cdot\frac{\partial(\vecB-\vecB_P)}{\partial t} \; dV}.
\done
Using $(\vecB-\vecB_P) = \nabla\times(\vecA-\vecA_P)$, the first term can be rewritten using a vector identity:
\beg{term1.1}
\begin{split}
T_1 &= \int_V{\Big\{\Big(\vecA-\vecA_P\Big)\cdot\nabla\times\frac{\partial(\vecA+\vecA_P)}{\partial t}\Big\} - \nabla\cdot \Big\{\frac{\partial (\vecA+\vecA_P)}{\partial t}\times\Big(\vecA-\vecA_P\Big)\Big\}\; dV} \\
        &= \int_V{\Big\{\Big(\vecA-\vecA_P\Big)\cdot\frac{\partial(\vecB+\vecB_P)}{\partial t}\Big\} - \nabla\cdot \Big\{\frac{\partial (\vecA+\vecA_P)}{\partial t}\times\Big(\vecA-\vecA_P\Big)\Big\}\; dV} \\
        &= \int_V{\Big\{\Big(\vecA-\vecA_P\Big)\cdot\frac{\partial(\vecB+\vecB_P)}{\partial t}\Big\}\; dV} - \int_S{\Big\{\frac{\partial (\vecA+\vecA_P)}{\partial t}\times\Big(\vecA-\vecA_P\Big)\Big\}\cdot d\textbf{S}} \\
        &= \int_V{\Big\{\Big(\vecA-\vecA_P\Big)\cdot\frac{\partial(\vecB+\vecB_P)}{\partial t}\Big\}\; dV}
\end{split}
\done
where the last line follows because of the boundary condition in Eq. \ref{bcA}. Combining the expressions for $T_1$ and $T_2$ yields
\beg{AdHdt2}
\frac{dH}{dt} = 2\int_V{\Big\{\vecA\cdot\frac{\partial\vecB}{\partial t} - \vecA_P\cdot\frac{\partial\vecB_P}{\partial t}\Big\}\; dV}
\done
The second term vanishes, since $\vecB_P=-\nabla\phi$ and therefore,
\beg{term2is0}
\begin{split}
\int_V{\vecA_P\cdot\frac{\partial\vecB_P}{\partial t}\; dV} &= \int_V{\Big\{\phi\nabla\cdot\vecA_P-\nabla\cdot\big(\phi\vecA_P\big)  \Big\}\;dV}\\
& = \int_V{\phi\nabla\cdot\vecA_P\;dV} - \int_S{\phi \vecA_P\cdot d\textbf{S}} 
\end{split}
\done
which vanishes by our choice of gauge. The first term, meanwhile, can be rewritten using the induction equation of ideal MHD,
\beg{Ainduction}
\frac{\partial \vecB}{\partial t} = \nabla\times\big(\vecv\times\vecB\big) 
\done 
to become
\beg{AdHdt3}
\begin{split}
\frac{dH}{dt} &= 2\int_V{\Big\{\vecA\cdot\nabla\times\big(\vecv\times\vecB\big)\Big\}\;dV} \\
&= 2\int_V{\Big\{\big(\vecv\times\vecB\big)\cdot\big(\nabla\times\vecA\big) - \nabla\cdot\Big(\vecA\times\big(\vecv\times\vecB\big)\Big)\Big\}\;dV}
\end{split}
\done
The first term vanishes since $(\vecv\times\vecB)\cdot\vecB \equiv 0$, and applying a final vector identity to the last term yields the equation for the rate of change of the relative magnetic helicity:
\beg{dHdt_end}
\frac{dH}{dt} = 2\int_S{\Big\{\big(\vecA\cdot\vecv\big)\vecB - \big(\vecA\cdot\vecB\big)\vecv\Big\}\cdot d\textbf{S}}
\done

\end{appendix}


\begin{vita}        
Kalman Knizhnik received his B.S with high honors in physics and astronomy, \emph{magna cum laude}, from the University of Maryland, College Park in 2011, before enrolling in the Physics Ph.D program at the Johns Hopkins University. He received his M.A. in Physics and Astronomy from Johns Hopkins in 2013. 
\end{vita}
\end{document}